\documentclass[aps,amssymb,showkeys]{revtex4}

\usepackage{xcolor}
\usepackage[pdftex, bookmarks, colorlinks=true, plainpages = false, citecolor = red, urlcolor = blue, filecolor = blue]{hyperref}
\usepackage{array}
\usepackage{scalerel}
\usepackage{multirow}
\usepackage{amsmath}
\usepackage{amsthm}
\usepackage{graphicx} 
\usepackage{amssymb}
\usepackage{mathrsfs}
\usepackage{amscd}
\usepackage{afterpage}
\usepackage{enumitem}
\usepackage{soul}
\usepackage{rotating}
\newcommand{\super}[1]{^{\scaleobj{0.85}{(#1)}}}
\usepackage{tikz-cd}

\definecolor{green}{rgb}{0.1, 0.80, 0.0}

\newcommand{\red}{\textcolor{red}}
\definecolor{amber}{rgb}{1.0, 0.75, 0.0}

\newcommand{\be}{\begin{equation}}
\newcommand{\ee}{\end{equation}}
\newcommand{\bea}{\begin{eqnarray}}
\newcommand{\eea}{\end{eqnarray}}
\newcommand{\bean}{\begin{eqnarray*}}
\newcommand{\eean}{\end{eqnarray*}}
\newcommand{\SpecialZ}{\mathfrak{Z}}
\newcommand{\RCB}{\rm{RCB}}
\newcommand{\RCX}{\rm{RCX}}
\newcommand{\PMB}{\rm{PMB}}
\newcommand{\PMX}{\rm{PMX}}
\newcommand{\LGB}{\rm{LGB}}
\newcommand{\LGX}{\rm{LGX}}

\theoremstyle{plain}

\theoremstyle{definition}

\newcommand{\stZ}{%
  \ooalign{\hidewidth $Z$\hidewidth\cr\rule[.8ex]{1.5ex}{.4pt}}}

\def\clap#1{\hbox to 0pt{\hss#1\hss}}

\allowdisplaybreaks

\begin{document}
\title{A formula for crossing probabilities of critical systems inside polygons}

\date{\today}

\author{Steven M. Flores}
\email{steven.flores@helsinki.fi} 
\affiliation{Department of Mathematics \& Statistics, \\ 
Gustaf H\"allstr\"omin katu 2b, FI-00014 University of Helsinki, Finland}

\author{Jacob J.\ H.\ Simmons}
\email{jacob.simmons@mma.edu}
\affiliation{Maine Maritime Academy, Pleasant Street, Castine, ME, 04420, USA}

\author{Peter Kleban}
\email{kleban@maine.edu} 
\affiliation{LASST and Department of Physics \& Astronomy, University of Maine, Orono, Maine, 04469-5708, USA}

\author{Robert M.\ Ziff}
\email{rziff@umich.edu} 
\affiliation{Center for the Study of Complex Systems and Department of Chemical Engineering, University of Michigan, Ann Arbor, Michigan, 48109--2136, USA}

\begin{abstract}

In this article,  we use our results from \cite{florkleb,florkleb2,florkleb3,florkleb4} to generalize  known formulas for crossing  probabilities.  Prior crossing results date back to J.\ Cardy's  prediction of a formula for the probability that a percolation cluster in two dimensions connects the left and right sides of a rectangle at the percolation critical point  in the continuum limit \cite{c3}. Here, we predict a new formula for crossing probabilities of a continuum limit loop-gas model on a planar lattice inside a $2N$-sided polygon.  In this model, boundary loops exit and then re-enter the polygon through its vertices, with exactly one loop passing once through each vertex, and these loops join the vertices pairwise in some specified connectivity through the polygon's exterior.  The boundary loops also connect the vertices through the interior, which we regard as a crossing event.  For particular values of the loop fugacity, this formula specializes to FK cluster (resp.\ spin cluster) crossing probabilities of a critical $Q$-state random cluster (resp.\ Potts) model on a lattice inside the polygon in the continuum limit. 
This includes critical percolation as the $Q=1$ random cluster model. These latter crossing probabilities are conditioned on a particular side-alternating free/fixed (resp.\ fluctuating/fixed) boundary condition on the polygon's perimeter, related to how the boundary loops join the polygon's vertices pairwise through the polygon's exterior in the associated loop-gas model.  For $Q\in\{2,3,4\}$, we compare our predictions of these random cluster (resp.\ Potts) model crossing probabilities in a rectangle ($N=2$) and in a hexagon ($N=3$) with high-precision computer simulation measurements.  We find that the measurements agree  with our predictions very well for $Q\in\{2,3\}$ and reasonably well if $Q=4$.

\end{abstract}

\maketitle

{\hypersetup{linkcolor=black}
\tableofcontents
}

\section{Introduction}\label{intro}

For a statistical lattice or random-walk/loop model \cite{grim,stah,kes,brham,bax,mcwu,wu,fk,fk2,grim2,stan,jlj,blni,dmns,zcs,law1,schrsheff,weintru,madraslade} in a planar domain, a \emph{crossing event} is an event in which  a ``crossing" path along  (for example) nearest-neighbor sites in a common state connects nonadjacent segments of the domain's boundary.  The probability of such an event, called a \emph{crossing probability}, may characterize critical behavior of the system.  Indeed, in the \emph{continuum limit}, obtained by sending the lattice spacing to zero and the number of lattice sites to infinity such that the lattice fills the domain, a crossing probability often equals zero below some particular point (e.g., temperature) and equals one above it (or vice-versa), indicating a phase transition \cite{kes,lps,lpps}.  In particular, at a critical point, the crossing probability is typically neither zero nor one, and to either predict or rigorously derive its precise value is usually non-trivial \cite{c3,watts,smirnov,dub,js,bbk,argaub,kozd,argaub2}.  This class of observables has attracted considerable interest from researchers in recent decades \cite{lps,lpps,kes,c3,watts,smirnov,dub,js,bbk,dubedat,ziff,argaub,kozd,argaub2}.  In this article, we use conformal field theory (CFT) \cite{bpz,fms,henkel} and results from our previous research \cite{florkleb,florkleb2,florkleb3,florkleb4,fsk} to predict new formulas for crossing probabilities of various statistical models on a lattice inside a $2N$-sided polygon.  Our results apply in the continuum limit and at the critical point.

%%%%%%%%%%%%%%%%%%%%%%%%%%%%%%%%%%%%%%%%%%%%%%%%%%%%%%%%%%%%
\begin{figure}[t]
\centering
\includegraphics[scale=0.6]{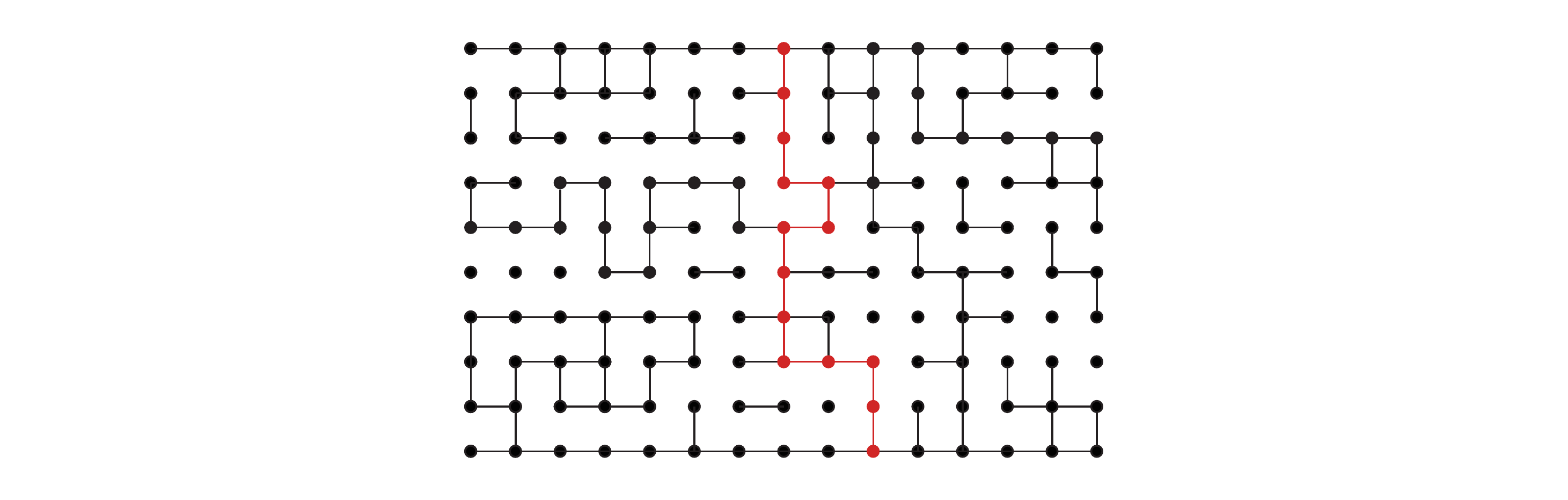}
\caption{A bond percolation sample on a square lattice inside a rectangle.  This sample illustrates a crossing event:  the collection of all activated bonds (black) contains a connected path (red) that joins the bottom side of the rectangle to the top side.}
\label{PercolationXing}
\end{figure}
%%%%%%%%%%%%%%%%%%%%%%%%%%%%%%%%%%%%%%%%%%%%%%%%%%%%%%%%%%%%

%%%%%%%%%%%%%%%%%%%%%%%%%%%%%%%%%%%%%%%%%%%%%%%%%%%%%%%%%%%%
\begin{figure}[b]
\centering
\includegraphics[scale=0.5]{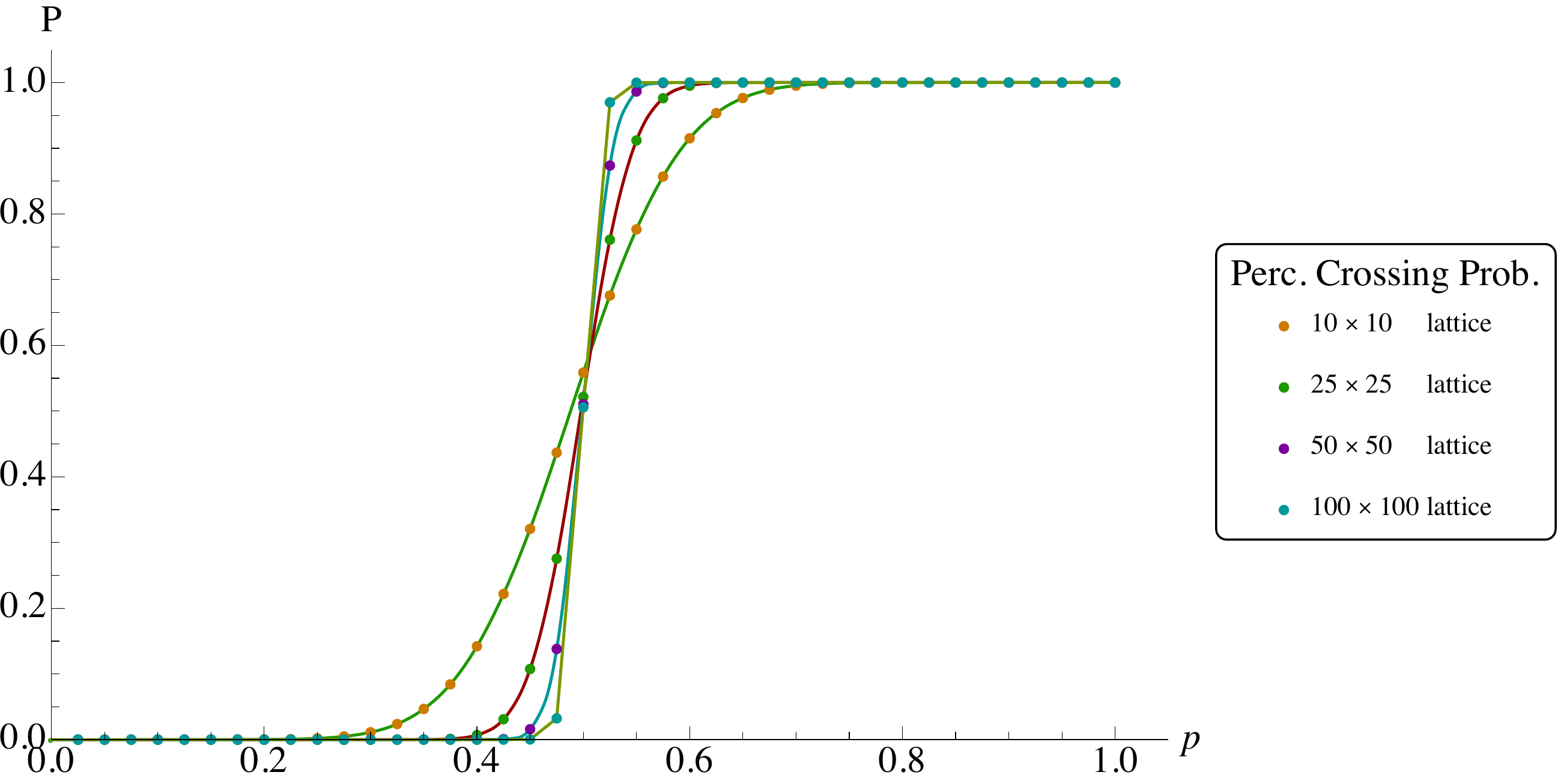}
\caption{Crossing probability $\mathsf{P}$ as a function of bond activation probability $p$ for percolation on an $M\times M$ square lattice in a square ($R=1$).  As $M\rightarrow\infty$, $\mathsf{P}$ approaches a step function (\ref{PercPhase}) that jumps at the critical probability $p_c^\text{sqr.}=0.5$.}
\label{PercProb}
\end{figure}
%%%%%%%%%%%%%%%%%%%%%%%%%%%%%%%%%%%%%%%%%%%%%%%%%%%%%%%%%%%%

 The continuum limit of Bernoulli  percolation (henceforth called ``percolation") on a  lattice  inside a rectangle gives one of the simplest examples  of a crossing probability at a critical point.  We define \emph{site} (resp.\ \emph{bond}) \emph{percolation} to be the assignment of either one (the ``activated" or ``occupied" state) or zero (the ``deactivated" or ``vacant" state) to every vertex (resp.\ edge) of a given graph.  All state assignments are i.i.d.\  Bernoulli random variables, with common probability $p\in[0,1]$ for the activated state.  Throughout this article, the vertices of the graph are the sites of a  two-dimensional lattice, and the edges of the graph are the bonds between pairs of nearest-neighbor sites.   For bond percolation, we define a (bond) \emph{cluster} to be a connected collection of activated bonds, and in site percolation, we define a (site) cluster to be a collection of activated sites such that the bonds between them, when activated, form a bond cluster.   
 
For a statistical lattice model such as percolation on a lattice inside a rectangle, we define a \emph{rectangle crossing event} to be the event that a cluster touches the top and bottom sides of a rectangle (figure \ref{PercolationXing}).  This event signals critical behavior in the continuum limit of percolation,  as follows.  First,  it is well known that, in the continuum limit, if $p<p_c$ for some (lattice-dependent) critical value $p_c$, then the probability $\mathsf{P}(p)$ of the  rectangle crossing event equals zero, and if $p>p_c$, then this probability equals one (figure \ref{PercProb}) \cite{lps,kes2,smirsch}:
\be\label{PercPhase}\mathsf{P}(p)=\begin{cases} 0,&0\leq p<p_c, \\ 1, & p_c<p\leq1,\end{cases}\quad \text{where $p_c\in(0,1)$.}\ee
The jump (\ref{PercPhase}) of $\mathsf{P}(p)$ signals a phase transition in percolation.  At the \emph{critical point} or \emph{critical value} $p=p_c$, the crossing probability $\mathsf{P}(p_c)$ is strictly between zero and one.  Using CFT, J.\ Cardy predicted a formula for this probability as a function of the rectangle's aspect ratio $R$ (bottom to left side).  (A non-trivial dependence on the geometry of the domain is typical for observables measured at critical points.)  This formula, called \emph{Cardy's formula}, is 
\be\label{RectCross}\mathsf{P}(p_c,R)=\frac{3\Gamma(2/3)}{\Gamma(1/3)^2}m^{1/3}\,_2F_1\bigg(\frac{1}{3},\frac{2}{3};\frac{4}{3}\,\bigg|\,m\bigg),\quad \text{where $R=K(m)/K(1-m)$},\ee
with $K(m)$ the elliptic function of the first kind \cite{morsefesh} (so $R\in(0,\infty)$ corresponds one-to-one with $m\in(0,1)$).  Numerical simulations of $\mathsf{P}(p_c,R)$ \cite{lpps,ziff}   verified his prediction (\ref{RectCross}) with very high accuracy for both site and bond percolation on a variety  of lattices.  Less than a decade after Cardy's prediction (\ref{RectCross}), S.\ Smirnov discovered a rigorous proof of Cardy's formula for site percolation on the triangular lattice.  His proof elucidated a key property of critical percolation in the continuum limit: the conformal invariance of certain observables, that is, invariance under conformal transformation of the system domain (while preserving the underlying lattice) \cite{dusm}.

%%%%%%%%%%%%%%%%%%%%%%%%%%%%%%%%%%%%%%%%%%%%%%%%%%%%%%%%%%%%
 \begin{figure}[t]
\resizebox{11.5cm}{2cm}{\includegraphics{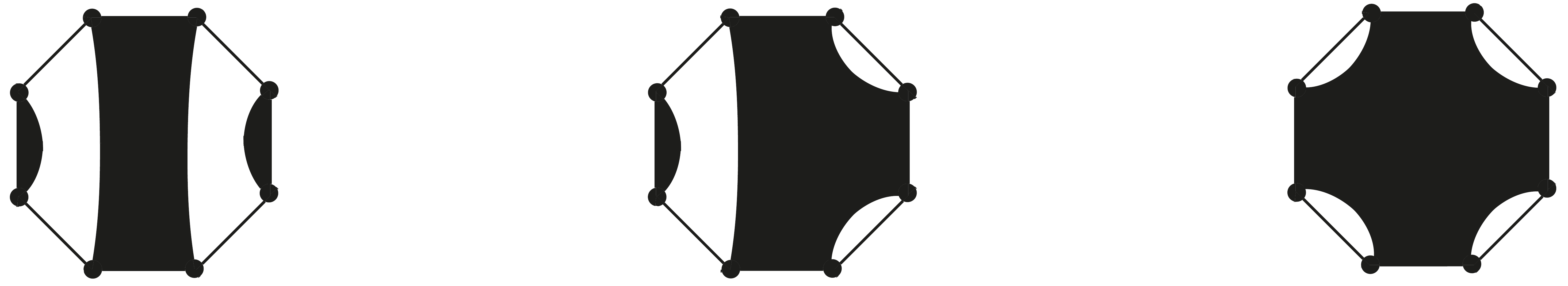}}
\caption{The three topologically distinct crossing patterns in an octagon, with all boundary clusters filled black.  We generate the $C_4-3=11$ other crossing events by rotating one of the three octagons and/or switching black and white regions.}
\label{OctXingConfigs}
\end{figure}
%%%%%%%%%%%%%%%%%%%%%%%%%%%%%%%%%%%%%%%%%%%%%%%%%%%%%%%%%%%%
%%%%%%%%%%%%%%%%%%%%%%%%%%%%%%%%%%%%%%%%%%%%%%%%%%%%%%%%%%%%

%%%%%%%%%%%%%%%%%%%%%%%%%%%%%%%%%%%%%%%%%%%%%%%%%%%%%%%%%%%%
\begin{figure}[b]
\centering
\includegraphics[scale=0.16]{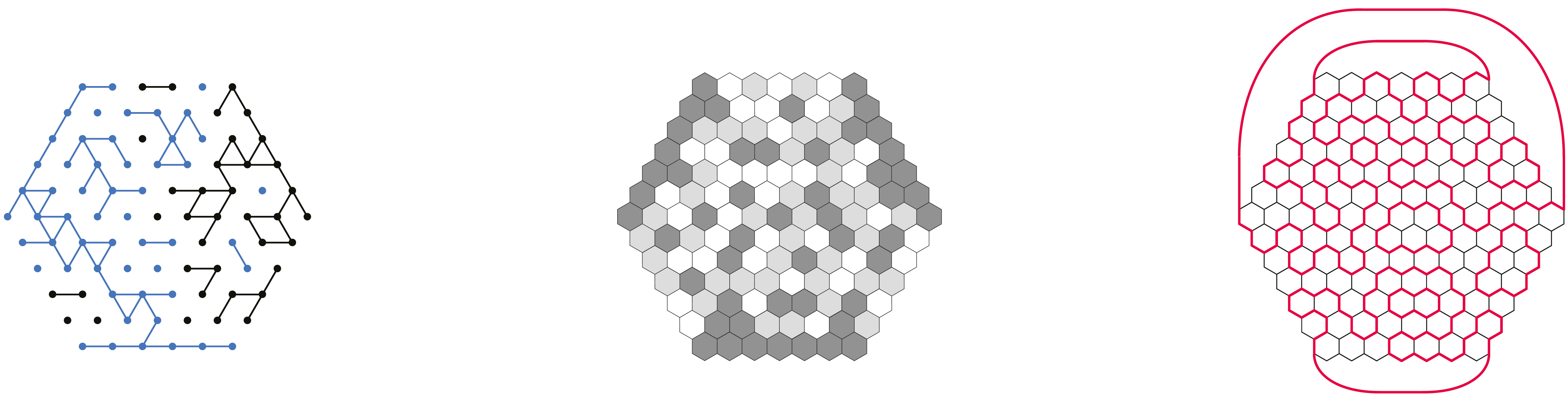}
\caption{Crossing events and side-alternating BC events of items \ref{BCitem1}--\ref{BCitem3} for the $Q=2$ random cluster model (left), the $Q=3$ Potts model (middle), and the loop-gas model (right) on a triangular (left, middle)  and hexagonal (right) lattice inside a  hexagon.} 
\label{Xings3}
\end{figure}
%%%%%%%%%%%%%%%%%%%%%%%%%%%%%%%%%%%%%%%%%%%%%%%%%%%%%%%%%%%%

 In this article, we study two generalizations of these results.  First, we consider a range of other lattice models with conformally invariant critical points, and second, we consider polygons $\mathcal{P}$ with an arbitrary even number of sides $2N$.   After enumerating the sides of $\mathcal{P}$ in counterclockwise order and ascending from one to $2N$, we allow clusters to join the odd sides of $\mathcal{P}$ in any one of the $C_N$ available topologically distinct connectivities \cite{dubedat}, called \emph{crossing patterns} (figure \ref{OctXingConfigs}).  Here, $C_N$ is the $N$th Catalan number:
\be\label{catalan}C_N=\frac{(2N)!}{N!(N+1)!}.\ee
For example, a cluster may join the two horizontal sides of a rectangle ($N=2$) or not, for $C_2=2$ crossing patterns.   We regard a ``crossing event" as the event that a particular crossing pattern occurs in $\mathcal{P}$, and we regard a ``crossing probability" as the probability of a crossing event. 

Using CFT \cite{bpz,fms,henkel} and results from our previous work \cite{florkleb,florkleb2,florkleb3,florkleb4,fsk}, we predict a formula ((\ref{xing}) of section \ref{xingsummary}) for crossing probabilities in the continuum limit of the above mentioned models inside even-sided polygons.  We always distinguish two adjacent sides of a polygon $\mathcal{P}$ by imposing different boundary conditions (BC) on those sides.  Specifically, the models, crossing events, and BCs that we consider are as follows (figure \ref{Xings3}):
\begin{enumerate}[wide, labelwidth=!, labelindent=0pt]
\item\label{BCitem1} \textbf{The $Q$-color random cluster model \cite{fk,fk2,grim2}:}  $Q\in\{1,2,3,4\}$.  This model is the same as bond percolation, except that we independently and uniformly distribute any one of $Q$ colors to all bond clusters, called  \emph{Fortuin-Kasteleyn (FK) clusters}.  ($Q=1$ gives bond percolation.)  We always condition on a certain subset of random cluster model \emph{free/fixed side-alternating boundary condition} ($\RCB$) events \cite{florkleb} (section \ref{mutsect}).  In an $\RCB$ event, all bonds adjacent to the odd sides of the polygon $\mathcal{P}$ are activated (and colored) (so each side exhibits the \emph{fixed}, or \emph{wired} BC) and the states of the bonds adjacent to the even sides are unconditioned (the \emph{free} BC).  We define a random cluster model crossing event ($\RCX$) to be the event that FK clusters join the odd sides of $\mathcal{P}$ in a particular crossing pattern.

\item\label{BCitem2} \textbf{The $Q$-state Potts model \cite{wu,bax}:} $Q\in\{2,3,4\}$.   Here, each lattice site is assigned  one of $Q$ values, called \emph{spins}, and spin configurations are Boltzmann-weighted by the total energy of nearest-neighbor spin coupling.  We  always condition on the Potts model \emph{fluctuating/fixed side-alternating boundary condition} ($\PMB$) event (section \ref{FLBCsect}).  In a $\PMB$ event, all sites adjacent to the odd sides of the polygon $\mathcal{P}$ are assigned the first spin state (so each side exhibits the ``fixed," or ``wired" BC, but fixed to the first spin state) and the sites adjacent to the even sides uniformly sample any but the first state (so each side exhibits the \emph{fluctuating} BC) \cite{gaca}.  We define a Potts model spin cluster to be a site cluster whose sites share a common spin value, and we define a Potts model crossing event ($\PMX$) to be the event that spin clusters join the odd sides of $\mathcal{P}$ in a particular crossing pattern.

\item\label{BCitem3} \textbf{The loop-gas model (or O$(n)$ model) \cite{stan,jlj,blni,dmns}:}  In this model, each configuration consists of a set of nonintersecting loops that follow the bonds of the lattice.  Every such loop configuration is weighted by a temperature parameter $x>0$ raised to the power of the total length of the loops in the configuration, multiplied by a fugacity parameter $n\in(0,2]$ raised to the power of the number of loops in the configuration.  This model includes both items \ref{BCitem1} and \ref{BCitem2} in the sense that for certain $Q$-dependent values of $n$, its continuum limit is expected to match the continuum limit of the $Q$-state random cluster model or Potts model (section \ref{Onmodel}).  We condition on a loop-gas \emph{exterior arc boundary condition} ($\LGB$) event, where ``boundary loops" (defined in section \ref{DefLoopGasSect} below) join the vertices pairwise in some connectivity via the polygon's exterior.  We define a loop-gas crossing event ($\LGX$) to be the event that boundary loops join the vertices pairwise in some particular connectivity through the polygon's interior.  This definition is natural because these connectivities bijectively correspond to the $C_N$ crossing patterns (figure \ref{OctXingConfigsBdyArcs}).
\end{enumerate}

%%%%%%%%%%%%%%%%%%%%%%%%%%%%%%%%%%%%%%%%%%%%%%%%%%%%%%%%%%%%
\begin{figure}[b]
\centering
\includegraphics[scale=0.35]{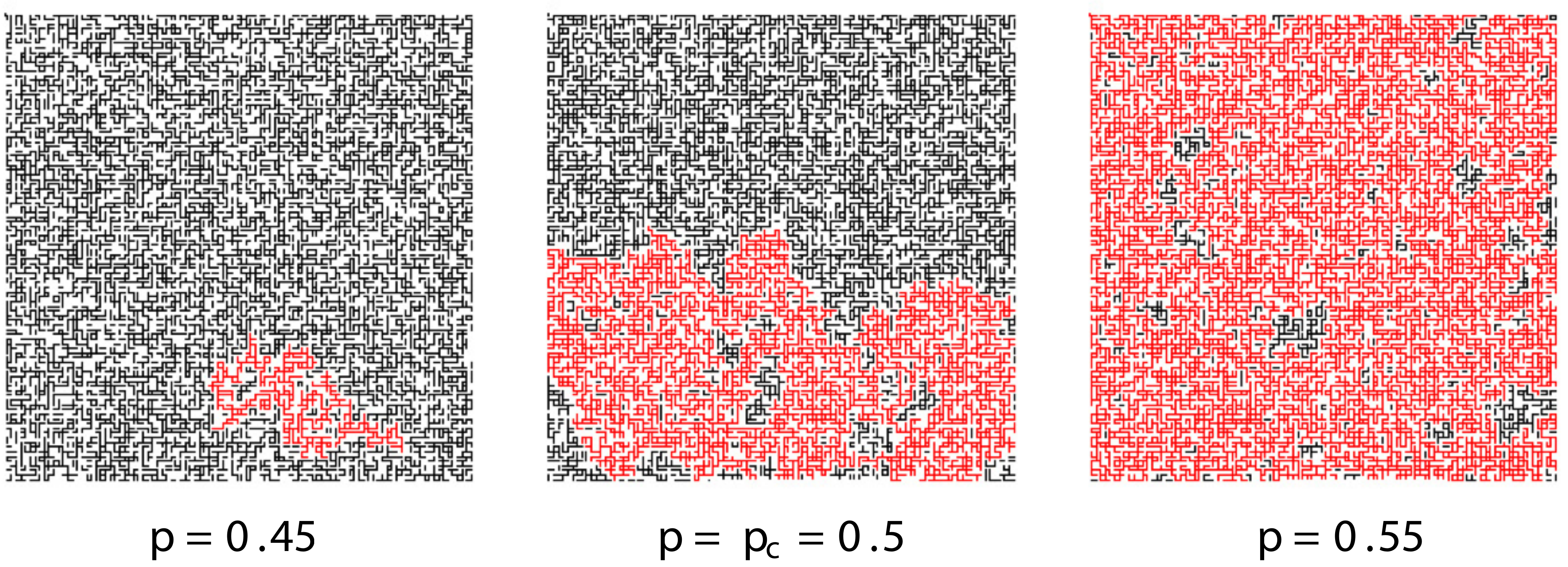}
\caption{Typical bond percolation configurations on a large square lattice with the bond activation probability $p$ slightly below, at, and slightly above the critical probability $p_c^\text{sqr.}=0.5$.  Bonds belonging to the largest cluster are colored red.}
\label{Percolating}
\end{figure}  
%%%%%%%%%%%%%%%%%%%%%%%%%%%%%%%%%%%%%%%%%%%%%%%%%%%%%%%%%%%%

 We discuss the above mentioned BCs.  In order for the side-alternating BCs of items \ref{BCitem1}--\ref{BCitem3} to be well-defined, the polygon must have an even number of sides.  Also, in items \ref{BCitem1} and \ref{BCitem2}, only one cluster, called a \emph{boundary cluster}, touches a given fixed side of the polygon, as that entire side is part of the cluster.  In section \ref{Onmodel}, we observe that items \ref{BCitem1} and \ref{BCitem2} are special cases of item \ref{BCitem3} in the continuum limit, with $Q$ and $n$ related by
\be\label{QnRelation} \text{$n=\sqrt{Q}$ in item \ref{BCitem1}},\qquad \text{$n=-2\cos\left(\dfrac{\pi^2}{\arccos(-\sqrt{Q}/2)}\right)$ in item \ref{BCitem2}}.\ee
This observation is an important step in our argument because in section \ref{Onmodel}, we connect the random cluster model and the Potts model to CFT via the loop-gas model.  (In principle, $Q$ could be any positive number in the definition of the random cluster model.  In this article, we restrict $Q$ to the range $(0,4]$, corresponding to $n\in(0,2]$ for the related loop-gas model (\ref{QnRelation}), where the random cluster model is expected to have a conformally invariant critical point \cite{dusm}, and we restrict $Q$ to positive-integer values in order to interpret it as a number of available colors. Thus, $Q\in\{1,2,3,4\}$.)

Finally, items \ref{BCitem1}--\ref{BCitem3} are the only BC scenarios that we consider in this article, but they are not the only  possibilities.  Others include:
\begin{itemize}
\item Switching the BCs (but not the models) between items \ref{BCitem1} and \ref{BCitem2}.  We are not aware of any formula for the probability of a crossing scenario with these BCs.
\item Assigning the free BC (i.e., assign no BC) to all sides of the polygon. Here, some useful results are known, and we summarize those of which we are aware at the end of section \ref{introxing} below.
\end{itemize}
We do not consider either of these options because they are beyond the scope of the CFT techniques that we use in this article. 

Next, we survey recent research on crossing probabilities.

\subsection{Crossing probabilities: a survey of known results}\label{introxing}

 In this section, we survey known results about crossing events in models with conformally invariant critical behavior.  As mentioned, percolation gives a simple example of conformally invariant critical phenomena: on an infinite square lattice, if $p>p_c$ with $p_c$ as in (\ref{PercPhase}), then a sample almost surely has a unique infinite cluster.  If $p<p_c$, then clusters are almost surely finite and typically with diameter less than some $\xi(p)<\infty$.  Finally, if $p=p_c$, then $\xi(p_c)=\infty$, meaning that there is no upper bound on the cluster size (figure \ref{Percolating}) and also no infinite cluster \cite{brham,kes,grim}.  (See \cite{aizbar} for an equivalent characterization of this phenomenon.)  Thus, as we approach the continuum limit, the maximum cluster size does not decrease if $p=p_c$, one requirement for conformal invariance at criticality.  (This pertains to Bernoulli percolation, but see \cite{wilkwill,hamwel,hin} for other percolation models with critical behavior.)

%%%%%%%%%%%%%%%%%%%%%%%%%%%%%%%%%%%%%%%%%%%%%%%%%%%%%%%%%%%%
\begin{table}[t]
\centering
\begin{tabular}{p{2.5cm}p{4.5cm}p{4.5cm}l}
lattice & site percolation threshold & bond percolation threshold \\
\hline
square & $0.592746\ldots$ \cite{jjac} & 1/2 \cite{kes2}\\
triangular & 1/2 & $2\sin(\pi/18)$ \cite{syes} \\
hexagonal & $0.697040\ldots$ \cite{jjac} & $1-2\sin(\pi/18)$ \cite{syes} \\
kagome & $1 - 2\sin(\pi/18)$ \cite{syes} & $0.524405\ldots$ \cite{zsud}
\end{tabular}
\caption{Critical points (thresholds) $p_c$ for bond and site percolation on various lattices.  Some are known exactly, and others are found via computer simulation.  For site percolation on the triangular lattice, $p_c$ follows from an easy symmetry argument.}
\label{PercThresholds}
\end{table}
%%%%%%%%%%%%%%%%%%%%%%%%%%%%%%%%%%%%%%%%%%%%%%%%%%%%%%%%%%%%

 Percolation models inside bounded planar domains can also exhibit critical behavior in the continuum limit.  Researchers have predicted that two special properties emerge in this limit at criticality.  The first property, \emph{universality} \cite{stah,lpps}, is that all observables, such as crossing probabilities, do not depend on microscopic details such as the type of percolation (e.g., site versus bond) or the lattice (although the value of $p_c$ does depend on these details \cite{kes2,syes,jjac,zsud}, see table \ref{PercThresholds}).  The second property, \emph{conformal invariance} \cite{lps,smir,dusm}, is that certain observables are invariant under conformal transformation of the system domain (leaving the underlying lattice unchanged) \cite{dusm}.  Extensive numerical simulations strongly support these predictions \cite{lps,lpps,ziff}.

Crossing probabilities are one tool for probing critical behavior, universality, and conformal invariance in percolation.  For example, the phase change (\ref{PercPhase}) of the rectangle crossing probability $\mathsf{P}(p)$ was rigorously established some time ago on certain lattices \cite{kes2}, (see also \cite{kes} and references therein), but the value of $\mathsf{P}(p_c)$ was not known in any percolation model until some years later.   Later work measured $\mathsf{P}(p_c)$ for site and bond percolation on several lattices and in many different conformal rectangles \cite{lpps,lps}.  These measurements strongly suggest that $\mathsf{P}(p_c)\in(0,1)$, with the same value for all regular  lattices (universality) in any two conformally equivalent conformal rectangles (conformal invariance).

%%%%%%%%%%%%%%%%%%%%%%%%%%%%%%%%%%%%%%%%%%%%%%%%%%%%%%%%%%%%
 \begin{figure}[b]
\includegraphics[scale=0.35]{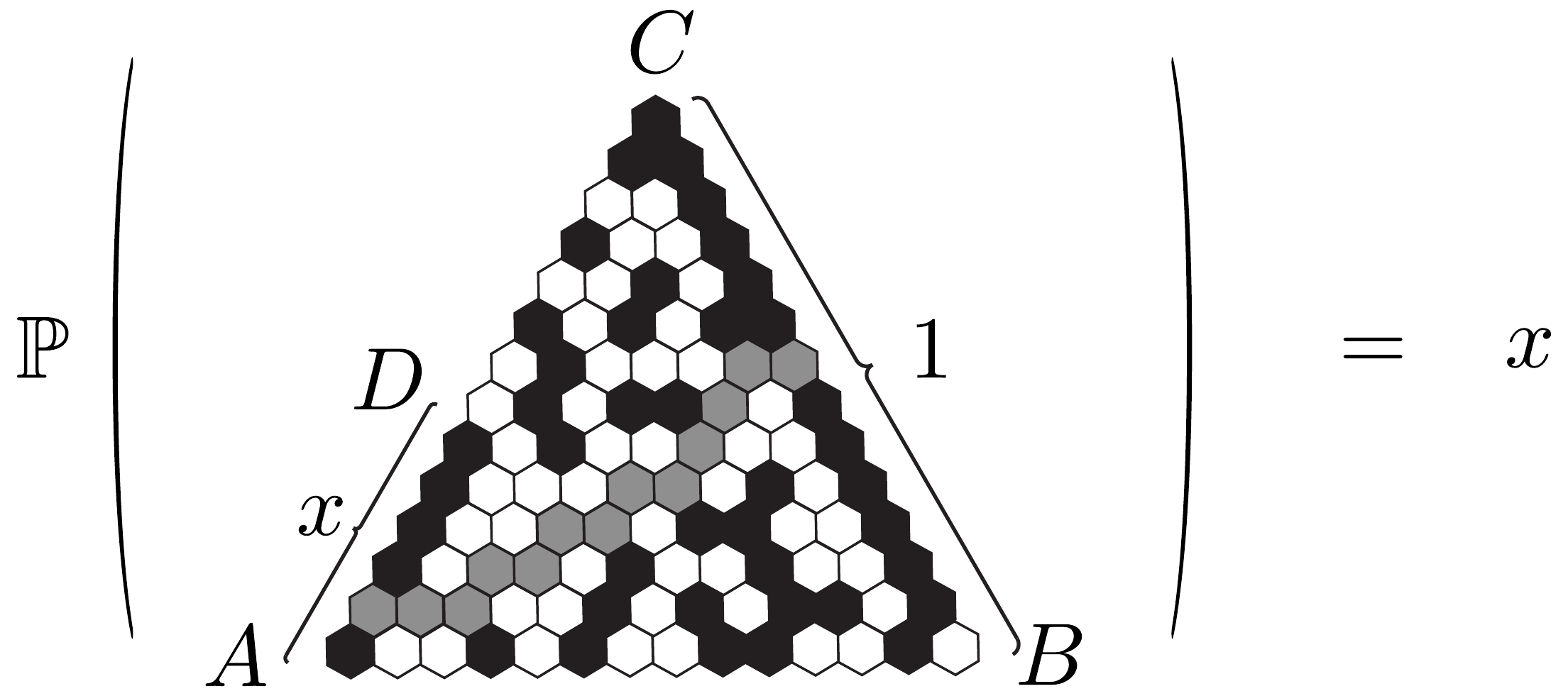}
\caption{If one conformally maps a rectangle with vertices $A$, $B$, $C$, $D$ onto the above equilateral triangle, then Cardy's formula (\ref{RectCross}) simplifies to the above.  (We illustrate with site percolation on the triangular lattice.  The crossing path is gray.) }
\label{PercolationXing2}
\end{figure}
%%%%%%%%%%%%%%%%%%%%%%%%%%%%%%%%%%%%%%%%%%%%%%%%%%%%%%%%%%%%

 As this deeper understanding of critical percolation emerged, simultaneous advancements in the application of CFT \cite{bpz,fms,henkel} to the study of related critical lattice models \cite{grim,kes,brham,bax,wu,fk,stan} were underway.  Similar to percolation, it was conjectured that many of these models have a conformally invariant phase transition, which could be exploited  to predict exact formulas for their correlation functions \cite{dots,fqsh,c1} via CFT.  One of these models, the $Q$-state random cluster model \cite{fk}, admits percolation as the special case $Q=1$.  This observation was used by J.\ Cardy to predict the formula (\ref{RectCross}) for the critical crossing probability $\mathsf{P}(p_c)$ using CFT methods \cite{c3}.  Cardy's result evinced CFT as a reliable, yet perhaps mysterious, tool for predicting this and other critical percolation observables, including Schramm's observable \cite{schr,gamcar}, cluster densities \cite{skz2,skdz,skz3,skfz}, pinch-point densities \cite{pinchpoint}, and other examples of rectangle crossing events \cite{watts,skz1}.

 Less than a decade after Cardy's prediction (\ref{RectCross}), mathematicians discovered rigorous proofs for it, as well as many others in critical percolation.  Much of this progress stemmed from O.\ Schramm's discovery of a one-parameter ($\kappa>0$) family of conformally invariant planar stochastic processes, now called \emph{Schramm-Loewner Evolution} (SLE$_\kappa$) \cite{rohshr,knk,lawler}, together with the conjecture that the law of boundary percolation cluster interfaces converges to SLE$_6$ in the continuum limit \cite{schra}.  Not long afterward, S.\ Smirnov proved this conjecture for critical site percolation on the triangular lattice, which implied conformal invariance of the law of the percolation interfaces \cite{smirnov}.  In the same article, S.\ Smirnov also proved Cardy's formula for site percolation on the triangular lattice.  (Exploiting conformal invariance, L.\ Carleson had previously noted that Cardy's formula (\ref{RectCross}) simplifies considerably to a linear function if we conformally map it to the equilateral triangle (figure \ref{PercolationXing2}).  Smirnov proved Cardy's formula in this setting.)

%%%%%%%%%%%%%%%%%%%%%%%%%%%%%%%%%%%%%%%%%%%%%%%%%%%%%%%%%%%%
 \begin{figure}[t]
\includegraphics[scale=0.16]{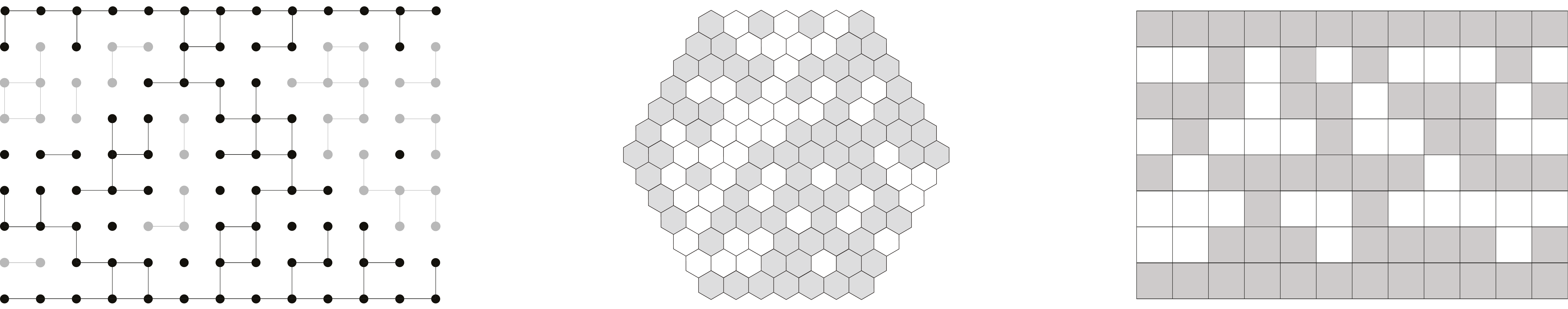}
\caption{$Q=2$ random cluster model (i.e., FK-Ising model) crossing event for (\ref{chelk}), $Q=1$ random cluster model crossing event for (\ref{DubXing}), and $Q=2$ Potts model (i.e., Ising model) crossing event for (\ref{AubinFormula}).}
\label{Xings2}
\end{figure}
%%%%%%%%%%%%%%%%%%%%%%%%%%%%%%%%%%%%%%%%%%%%%%%%%%%%%%%%%%%%

Subsequent to these discoveries about percolation crossing probabilities, some natural variations were examined.  We consider two such generalizations in this article.  First, we consider other models with conformally invariant critical points.  Here, there has been much progress.  For example, using the Ising model fermionic observable \cite{ssmi}, D.\ Chelkak and S.\ Smirnov rigorously derived this $Q=2$ FK-cluster crossing probability formula \cite{chelsmir},  
\begin{align} P(R)&=\mathbb{P}\left\{\parbox{9.8cm}{one FK cluster touches both the top and bottom of the rectangle, conditioned on both of these sides being fixed to the same color}\right\}\nonumber\\
\label{chelk}&=\frac{\sqrt{2}(1-p(m))}{p(m)+\sqrt{2}(1-p(m))},\quad \text{where $p(m):=\dfrac{\sqrt{1-\sqrt{m}}}{\sqrt{1-\sqrt{m}}+\sqrt{1-\sqrt{1-m}}}$ and $R=K(m)/K(1-m)$},\end{align}
(figure \ref{Xings2}), and another formula (\ref{chelkak}) for a similar observable.  (Recently, bounds for similar observables in conformal rectangles were derived in \cite{dhp,cdh}.)  Another example, using CFT, L.\ Arguin and Y.\ Saint-Aubin \cite{argaub} predicted this spin-cluster crossing probability formula for the $Q$-state Potts model \cite{wu} with $Q=2$ (i.e., Ising model):
\begin{align} P(R)&=\mathbb{P}\left\{\parbox{13.3cm}{one spin cluster touches both the top and bottom of the rectangle, conditioned on these two sides being fixed to the first spin value and the other two sides fixed to the second }\right\}\nonumber\\
\label{AubinFormula}&=\frac{2\Gamma(4/3)}{\Gamma(5/3)\Gamma(8/3)}\bigg(\frac{m^{5/3}}{1-m+m^2}\bigg)\,_2F_1\bigg(-\frac{1}{3},\frac{4}{3};\frac{8}{3}\,\bigg|\,m\bigg),\qquad R=K(m)/K(1-m)\end{align}
(figure \ref{Xings2}).  (This formula (\ref{AubinFormula}) is simplified here.  See beneath (\ref{ourform}).)  Subsequently, M.\ Kozdron rigorously proved (\ref{AubinFormula}) \cite{kozd}.  Interestingly, this formula (\ref{AubinFormula}) was pursued by several research groups almost simultaneously some years ago, and it appears as an original result in many articles \cite{argaub,kozd,bbk,dub,argaub2}.  Ref.\ \cite{argaub2} gives a chronological account of this work.

 As a second generalization, we consider polygons with an even number of sides greater than four.  As mentioned, a $2N$-sided polygon admits $C_N$ different crossing patterns \cite{dubedat} (figure \ref{OctXingConfigs}), where $C_N$ is the $N$th Catalan number (\ref{catalan}).  Recent results of this kind include J.\ Dub\'edat's discovery \cite{dub} that (figure \ref{Xings2})
\be\label{DubXing}\mathbb{P}\left\{\parbox{5.9cm}{a percolation cluster connects two nonadjacent odd sides of a regular hexagon}\right\}=\frac{2}{3}\left(\frac{\sqrt{3}}{2^{2/3}\pi}\right)^5\Gamma(2/3)^9\,_3F_2\Bigg(1,\frac{5}{6},\frac{5}{6};\frac{3}{2},\frac{3}{2}\,\Bigg|\,1\Bigg)\ee
and J.\ Simmons's determination of all $C_3=5$ hexagon percolation crossing probabilities with free/fixed side-alternating BCs (and free BCs) \cite{js}.  Also, K.\ Izyurov has rigorously derived some formulas for $Q=2$ FK-cluster crossing probabilities in (conformal) polygons with more general BCs \cite{izyu}.

 In all of the examples of crossing probabilities (\ref{RectCross}, \ref{chelk}--\ref{DubXing}) discussed so far, the crossing event pairs with its companion BC as in item \ref{BCitem1} or \ref{BCitem2} above.  As such, each of these is a special case of our predicted formula (\ref{xing}) for crossing probability.  However, one may consider other scenarios.  For example, it is natural to assign the free BC (i.e., no BC) to all sides of the polygon, and for this, there are many interesting results: G.\ Watts and J.\ Simmons have used CFT to predict formulas for percolation crossing probability in a rectangle and in a hexagon respectively, with the free BC \cite{watts,js}.  Later, J.\ Dub\'edat and O.\ Schramm independently discovered proofs of Watts's formula \cite{jdub,shewil}.  R.\ Langlands et al.\ found numerical evidence for the universality of the $Q=2$ Potts (i.e., Ising) model rectangle crossing probability (\ref{AubinFormula}) \cite{lls}, and S.\ Benoist et al.\ rigorously confirmed this fact \cite{bdch}.  Also, E.\ Lapalme and Y.\ Saint-Aubin used CFT and numerical methods to predict a formula for this probability \cite{ly}.  To determine such crossing probabilities with  free BCs is beyond the scope of this article.  Other studied variations that are also beyond the scope of this article include the average number of crossing clusters \cite{jcar} and crossing cluster densities \cite{skz1} for critical percolation.

\subsection{Objectives and organization}

The main purpose of this article is to predict the new formula (\ref{xing}),  introduced in section \ref{xingsummary}, for the probabilities of all $C_N$ distinct crossing events (figure \ref{OctXingConfigs}) in the critical models of items \ref{BCitem1}--\ref{BCitem3} inside a $2N$-sided polygon $\mathcal{P}$ with the specified BCs in the continuum limit.  Our prediction is exact.

 Formula (\ref{xing}), our main result, is given in terms of four quantities:
 \begin{itemize}
 \item the Coulomb gas (contour integral) function $\mathcal{F}_\vartheta$ (item \ref{step3} of section \ref{xingsummary}), whose subscript $\vartheta\in\{1,2,\ldots,C_N\}$ indexes the BC,
\item  the ``crossing weight" (i.e., ``connectivity weight" \cite{florkleb4,fsk} or ``pure SLE$_\kappa$ partition function" \cite{bbk,kype}) $\Pi_\varsigma$ (item \ref{step5} of section \ref{xingsummary}), whose subscript $\varsigma\in\{1,2,\ldots,C_N\}$ indexes the crossing pattern, 
\item the number $l_{\varsigma,\vartheta}$ of closed loops in the $(\varsigma,\vartheta)$th ``product diagram" (item \ref{step4} of section \ref{xingsummary} and figures \ref{N2loops} and \ref{innerproduct}), 
\item and the loop fugacity $n=n(\kappa)$ of the loop-gas model (item \ref{step1} of section \ref{xingsummary}).
\end{itemize}
For the random cluster model and Potts model, the loop fugacity $n$ depends on $Q$ via (\ref{QnRelation}), and throughout this article, we parameterize it (\ref{nfug}) in terms of the SLE$_\kappa$ \cite{rohshr,knk,lawler} parameter $\kappa\in[3/8,8)$.   In the random cluster model, the Potts model, and the loop-gas model respectively, the BC (resp.\ crossing) event is an $\RCB$ event, a $\PMB$ event, and an $\LGB$ event (resp.\ an $\RCX$ event, a $\PMX$ event, and an $\LGX$ event), as defined in items \ref{BCitem1}--\ref{BCitem3} above.  We first encounter the functions $\Pi_\varsigma$ and $\mathcal{F}_\vartheta$ (resp.\ the quantity $l_{\varsigma,\vartheta}$) in section \ref{CFTsect} (resp.\ section \ref{partfuncpoly}). 

Our strategy is to first relate the lattice models of items \ref{BCitem1} and \ref{BCitem2} above to the loop-gas model of item \ref{BCitem3} above.  We then consider the continuum limit of the latter, to which CFT applies.  Making use of our previous research, we then derive the crossing probability formula (\ref{xingSLE}), which is equivalent to (\ref{xing}).  Because CFT should apply directly to the original lattice models of items \ref{BCitem1} and  \ref{BCitem2}, one might wonder why we take this somewhat circuitous route.  However, the application of CFT to all of the cases we consider (various models, BCs, and number of polygon sides) is not completely straightforward.  Our route, although a bit indirect, encompasses all these cases in a single scheme.

 This article is organized as follows.  To begin, section \ref{xingsummary}, which is self-contained, gives a detailed explanation of the crossing probability formula (\ref{xing}) (but does not derive it), and section \ref{rectxingsummary} applies it to rectangles, reproducing the rectangle crossing probability for critical percolation (\ref{RectCross}), \cite{c3,smirnov}, the critical $Q=2$ random cluster (i.e., FK-Ising) model (\ref{chelk}) \cite{chelsmir}, and the critical $Q=2$ Potts (i.e., Ising) model (\ref{AubinFormula}) \cite{bbk,argaub,kozd}.

The purpose of sections \ref{partfuncpoly}--\ref{CFTsect} is to derive the predicted crossing probability formula (\ref{xing}).  To begin, in section \ref{partfuncpoly}, we give formal expressions for various partition functions of a random cluster model or a Potts model on a lattice inside $\mathcal{P}$, and we give formal expressions (\ref{chixing}, \ref{chixing2}) for crossing probabilities in terms of these partition functions.  All of these partition functions either sum exclusively over the BC events described in items \ref{BCitem1} and \ref{BCitem2} above, or they sum exclusively over these events intersected with the crossing events, also described in items \ref{BCitem1} and \ref{BCitem2} above.  For the random cluster model, in section \ref{RCxingSect}, we let $\smash{X_{\varsigma,\vartheta}^\mathcal{P}}$ (\ref{randFFBCadapt1}) denote the latter partition function, where $\varsigma,\vartheta\in\{1,2,\ldots,C_N\}$ index the $\RCX$ (crossing) event and $\RCB$ (BC) event respectively, and we let $\smash{X_\vartheta^\mathcal{P}}=\smash{X_{1,\vartheta}^\mathcal{P}}+\smash{X_{2,\vartheta}^\mathcal{P}}+\dotsm+\smash{X_{C_N,\vartheta}^\mathcal{P}}$ (\ref{randFFBCadapt2}), a sum over all crossing events, denote the former.  For the Potts model, in section \ref{FLBCsect}, we let $\smash{Y_{\varsigma,1}^\mathcal{P}}$ denote the latter partition function, where $\varsigma\in\{1,2,\ldots,C_N\}$ indexes the $\PMX$ (crossing) event  (here, the $\PMB$ event only allows the same single state on all fixed sides, which results in $\vartheta = 1$), and we let $\smash{Y_1^\mathcal{P}}=\smash{Y_{1,1}^\mathcal{P}}+\smash{Y_{2,1}^\mathcal{P}}+\dotsm+\smash{Y_{C_N,1}^\mathcal{P}}$, a sum over all crossing events, denote the former.  For the random cluster model, we only use certain $\RCB$ events called ``mutual wiring events," defined at the beginning of section \ref{mutsect}.  Furthermore, we do not use all mutual wiring events but consider only a subset of them called ``basic events," defined in section \ref{mutsect}.  (In appendix \ref{Pottsappendix}, we show that if $Q\in\{1,2,3\}$ or if $Q=4$ and $N\in\{1,2,3\}$, then in the continuum limit, any partition function summing exclusively over a mutual wiring event equals a linear combination of partition functions that all sum exclusively over a basic event.  Hence, restricting our attention to basic events does not result in a loss of generality for these $(Q,N)\in\{1,2,3,4\}\times\mathbb{Z}^+$.)  There are $C_N$ basic events, and $\vartheta\in\{1,2,\ldots,C_N\}$ indexes them.  The partition function $\smash{X_{\varsigma,\vartheta}^\mathcal{P}}$ sums over the $\vartheta$th basic event, and we readily obtain a formal expression (\ref{ZQoutFactored}) for this partition function up to a certain combinatorial quantity $\mathcal{C}_{\varsigma,\vartheta}$ that counts the number of ways to distribute color to boundary clusters in this event.  In section \ref{RCxingSect}, we use basic graph theory to find a formula (\ref{pformula}) for $\mathcal{C}_{\varsigma,\vartheta}$ in terms of the number of loops $l_{\varsigma,\vartheta}$ (figure \ref{innerproduct}).

Next, in section \ref{Onmodel}, we give formal expressions for various partition functions of a loop-gas model on a lattice inside $\mathcal{P}$, and we give a formal expression (\ref{masterxing}) for the crossing probability in terms of these functions.  All of these partition functions either sum exclusively over the BC events described in item \ref{BCitem3} above, or they sum exclusively over these events intersected with the crossing events, also described in item \ref{BCitem3} above.  For the loop-gas model, in section \ref{DefLoopGasSect}, we let $\stZ_{\varsigma,\vartheta}^\mathcal{P}$ (\ref{loopgaspart}) denote the latter partition function, where $\varsigma,\vartheta\in\{1,2,\ldots,C_N\}$ index the $\LGX$ (crossing) event and $\LGB$ (BC) event respectively, and we let $\smash{\stZ_\vartheta^\mathcal{P}}=\smash{\stZ_{1,\vartheta}^\mathcal{P}}+\smash{\stZ_{2,\vartheta}^\mathcal{P}}+\dotsm+\smash{\stZ_{C_N,\vartheta}^\mathcal{P}}$, a sum over all crossing events, denote the former. Then in sections \ref{RCLoopGasSect} and \ref{PMLoopGasSect} respectively, we argue that, at the critical point, the random cluster model partition functions and Potts model partition functions, $\smash{X_{\varsigma,\vartheta}^\mathcal{P}}$ and $\smash{Y_{\varsigma,1}^\mathcal{P}}$ respectively, are asymptotic to $\smash{\stZ_{\varsigma,\vartheta}^\mathcal{P}}$ and $\smash{\stZ_{\varsigma,1}^\mathcal{P}}$ respectively in the continuum limit (\ref{preasympXY}).  As such, to determine crossing probabilities for the critical random cluster model and the critical Potts model, it suffices to determine them for the corresponding loop-gas model.  In section \ref{XingLGSect}, we give formal expressions for the latter.
 
Finally, in section \ref{CFTsect}, we use physical and mathematical arguments to determine explicit formulas for the critical random cluster model, the critical Potts model, and the loop-gas model crossing probabilities (\ref{chixing}, \ref{chixing2}, \ref{masterxing}) in the continuum limit.  In order to do this, in section \ref{SmearedSect}, we first relate the loop-gas model partition functions $\smash{\stZ_\vartheta^\mathcal{P}}$, in the continuum limit, to boundary CFT correlation functions $\Upsilon_{\vartheta}^\mathcal{P}$ known to satisfy a particular system of $2N+3$  partial differential equations (PDEs) (\ref{nullstate}, \ref{wardid}).  The series of articles \cite{florkleb,florkleb2,florkleb3,florkleb4} determines the dimension of, as well as explicit formulas for all elements of, the system's solution space $\mathcal{S}_N$.   (As part of the definition for $\mathcal{S}_N$, we also  require that the solutions are dominated by a product of certain power functions.  See (\ref{powerlaw}) of section \ref{CFTsect} for the precise bound.)  The solutions corresponding to different BCs are distinguished by the choice of contour integrals that define the coulomb gas function $\mathcal{F}_\vartheta$ (item \ref{step3} in section \ref{xingsummary}), or alternatively, their asymptotic behavior as neighboring corners of the polygon approach each other.  (See the proof of Lemma \red{6} in \cite{florkleb3}.)   Then in section \ref{XingWeightsSect}, we argue that $\smash{\stZ_{\varsigma,\vartheta}^\mathcal{P}}$ is asymptotic to a special element $\Pi_\varsigma$ of $\mathcal{S}_N$ called a \emph{crossing weight}  (or equivalently, ``connectivity weight" \cite{florkleb4,fsk} or ``pure SLE$_\kappa$ partition function" \cite{bbk,kype}, see the remarks just below (\ref{BN}) and above (\ref{xingSLE2})).  The crossing weights are also solutions of the PDEs just mentioned, i.e., elements of $\mathcal{S}_N$, linearly related to the $\Upsilon_{\vartheta}^\mathcal{P}$ by the meander matrix $M_N$ \cite{fgg,fgut,difranc,franc}, defined in item \ref{step5} of section \ref{xingsummary}.  Finally, a simple argument gives $\Upsilon_\vartheta = \mathcal{F}_\vartheta$, which leads to our main result, the explicit formula (\ref{xing}) (or (\ref{xingSLE})) for the probability $\smash{P_{\varsigma|\vartheta}^\mathcal{P}}$ (\ref{masterxing}) of the $\varsigma$th $\LGX$ (crossing) event conditioned on the $\vartheta$th $\LGB$ (BC) event for the loop-gas model in the continuum limit (item \ref{BCitem3} above).  As special cases, this includes the probability $\smash{P_{\varsigma|\vartheta}^\mathcal{P}}$ (\ref{chixing}) of the $\varsigma$th $\RCX$ (crossing) event conditioned on the $\vartheta$th basic $\RCB$ (BC) event for the critical random cluster model (item \ref{BCitem1} above), and the probability $\smash{P_{\varsigma|1}^\mathcal{P}}$ (\ref{chixing2}) of the $\varsigma$th $\PMX$ (crossing) event conditioned on the $\PMB$ (BC) event for the critical Potts model (item \ref{BCitem2} above), both in the continuum limit.  The discussion around (\ref{xingSLE2}) also relates (\ref{xingSLE}) to a similar formula  for the  probability of eventual curve connectivity in multiple SLE$_\kappa$ \cite{dub2,graham,kl,sakai}.

To support our predictions, we present high-resolution computer results in section \ref{simxing}.  The simulations measure crossing probabilities of the $Q\in\{2,3,4\}$ random cluster (resp.\ Potts) model on a large square lattice inside a rectangle (section \ref{RecSimSect}) and on a large triangular lattice inside a hexagon (section \ref{HexSimSect}), with the side-alternating BCs of item \ref{BCitem1} (resp.\ item \ref{BCitem2}).  We compare these measurements with our prediction (\ref{xing}), finding very good agreement for $Q\in\{2,3\}$ and reasonable agreement for $Q=4$.  (For the latter case, much larger lattices and many more samples are required for better agreement \cite{car1,cziff}.)  (Similar results for the $Q=1$ critical random cluster model, i.e., critical percolation, already appear  in \cite{fzs,c3}.)

%%%%%%%%%%%%%%%%%%%%%%%%%%%%%%%%%%%%%%%%%%%%%%%%%%%%%%%%%%%%%%%%%%%%%%%%%%%%%%%%%%%%%%%%%%%%%%%%%%%%%%%%%%%%%%%%%%%%%%%%%%%%%%%%%%%%%%%%%%%%%%%%%%%%%%%%%%
\begin{figure}[b]
\centering
\includegraphics[scale=0.24]{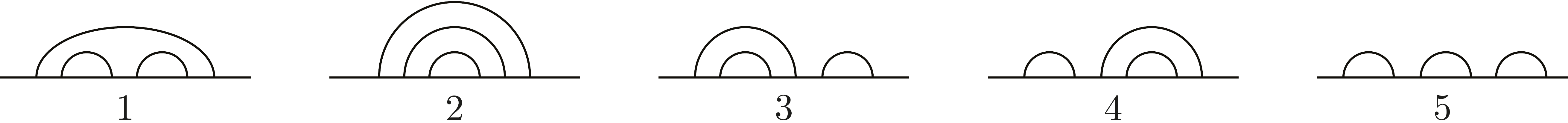}
\caption{An enumeration of the $C_3=5$ arc connectivities involving six points.}
\label{Connectivities}
\end{figure}
%%%%%%%%%%%%%%%%%%%%%%%%%%%%%%%%%%%%%%%%%%%%%%%%%%%%%%%%%%%%%%%%%%%%%%%%%%%%%%%%%%%%%%%%%%%%%%%%%%%%%%%%%%%%%%%%%%%%%%%%%%%%%%%%%%%%%%%%%%%%%%%%%%%%%%%%%%

Three appendices accompany this article.  In appendix \ref{transformxing}, we use CFT ``corner operators" to regularize the conformal transformation of the boundary CFT correlation functions in section \ref{CFTsect} from the upper half-plane onto the polygon.  The need for regularization arises because this transformation is  not conformal  at the polygon's vertices.  In appendix \ref{Pottsappendix},  we study random cluster model partition functions $W_c^\mathcal{P}$ that sum exclusively over ``color schemes" $c$.  These are mutual wiring events in which the color of each fixed side of $\mathcal{P}$ is specified.  Any partition function summing over an $\RCB$ event is a sum of color scheme partition functions $W_c^\mathcal{P}$.  We show that, in the continuum limit, the span of the partition functions $W_c^\mathcal{P}$ (as functions of the shape of $\mathcal{P}$) is the same as that of the $X_\vartheta^\mathcal{P}$ for $Q\in\{1,2,3\}$ and any $N\in\mathbb{Z}^+$ or $Q=4$ and $N\in\{1,2,3\}$.  Thus, for these $(Q,N)$, working with $X_\vartheta^\mathcal{P}$ over $W_c^\mathcal{P}$, as we do in this article, does not results in loss of generality.  In appendix \ref{equivappendix}, we give some simpler formulas for the Coulomb gas function $\mathcal{F}_\vartheta$ (\ref{Fexplicit}) appearing in the denominator of the crossing probability formula (\ref{xing}).  They include an alternative formula  (\ref{Fother}) used in \cite{florkleb3,florkleb4} for all $\kappa>0$ and some explicit algebraic formulas (\ref{Fkappa6}, \ref{Fkappa3}, \ref{Fkappa=16/3}, \ref{Fkappa4}) for $\mathcal{F}_\vartheta(\kappa)$ that hold at certain rational values of $\kappa$.  Some of these algebraic formulas already appear in the literature ($Q=2$ \cite{guim,kype}), although not in direct connection with $\mathcal{F}_\vartheta(\kappa)$, and others are, to our knowledge, completely new ($Q\in\{1,4\}$).  For $Q\in\{1,2\}$, we obtain our algebraic formulas not by explicit evaluation of the multiple contour integrals that appear in $\mathcal{F}_\vartheta(\kappa)$, but by using a new if indirect method making use of our previous results in \cite{florkleb,florkleb2,florkleb3,florkleb4} and other results already in the literature \cite{guim,kype}.  Thus we rigorously derive algebraic formulas for some rather complicated multiple contour integrals   based on a full analysis of the solutions of the system of PDEs (\ref{nullstate}, \ref{wardid}) that  $\mathcal{F}_\vartheta(\kappa)$ satisfies.  

\section{Formula for crossing probability}\label{xingsummary} 

This section contains a step-by-step but unmotivated presentation and explanation  of our central result, the formula (\ref{xing}) for the particular  crossing probabilities that we consider in various critical statistical mechanics models inside polygons.  It is completely self-contained (although reading the introduction \ref{intro} of this article first may be useful) and is illustrated for the rectangle case in section \ref{rectxingsummary}.   We emphasize that this section simply states and explains (\ref{xing});  its derivation only appears much later at the end of section \ref{CFTsect}.

In steps \ref{step1}--\ref{step6} below, we identify the interior of a $2N$-sided polygon with the upper half-plane, and we identify the polygon's vertices with $2N$ marked points $x_1<x_2<\ldots<x_{2N}$ on the real axis.  In step \ref{step7}, we conformally map the upper half-plane onto a conformal  $2N$-sided polygon $\mathcal{D}$ in the complex plane.

%%%%%%%%%%%%%%%%%%%%%%%%%%%%%%%%%%%%%%%%%%%%%%%%%%%%%%%%%%%%%%%%%%%%%%%%%%%%%%%%%%%%%%%%%%%%%%%%%%%%%%%%%%%%%%%%%%%%%%%%%%%%%%%%%%%%%%%%%%%%%%%%%%%%%%%%%%
\begin{figure}[b]
\centering
\includegraphics[scale=0.28]{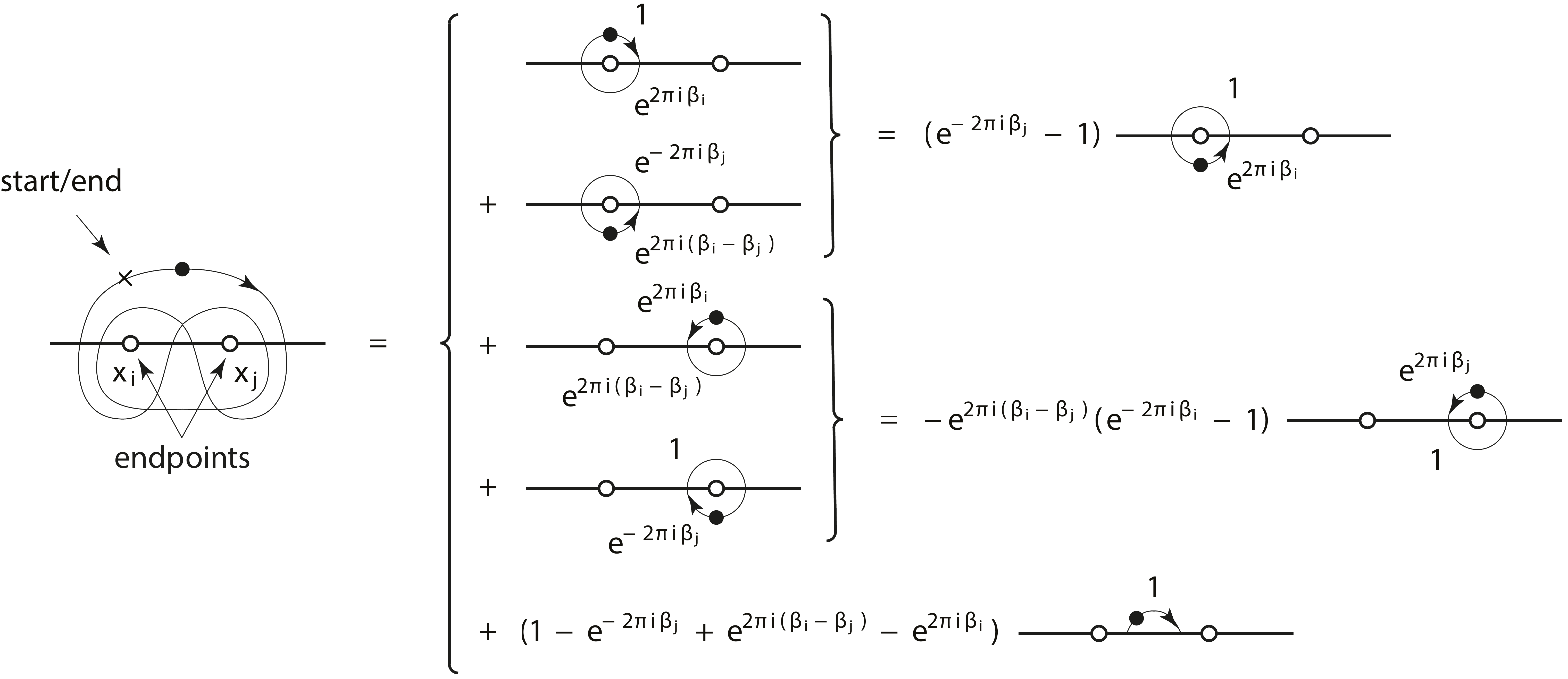}
\caption{The Pochhammer contour $\Gamma:=\mathscr{P}(x_i,x_j)$, and its decomposition in $\oint_{\Gamma}(u-x_i)^{\beta_i}(x_j-u)^{\beta_j}f(u)\,{\rm d}u$ for $f(u)$ analytic in a domain containing $\Gamma$.  The phase factor of the integrand at the start point and end point (arrow) of each contour is shown.  We call the points $x_i$ and $x_j$ ``endpoints" of $\mathscr{P}(x_i,x_j)$.}
\label{BreakDown}
\end{figure}
%%%%%%%%%%%%%%%%%%%%%%%%%%%%%%%%%%%%%%%%%%%%%%%%%%%%%%%%%%%%%%%%%%%%%%%

\begin{enumerate}[wide, labelwidth=!, labelindent=0pt]
\item\label{step1} We let $x_1<x_2<\ldots<x_{2N}$ be $2N$ marked points on the real axis, let $\kappa\in(0,8)$ (the \emph{SLE$_\kappa$ parameter}) \cite{rohshr,knk,lawler}, and let $n(\kappa)=-2\cos(4\pi/\kappa)$ (the \emph{loop-gas model fugacity formula}) \cite{nein,smir}.
\item\label{step2} We enumerate 1 through $C_N$ (with $C_N:=(2N)!/N!(N+1)!$ the $N$th Catalan number) the distinct arc  ``connectivities" or ``crossing patterns" (figure \ref{Connectivities}).  Each is a collection of $N$ curves (\emph{arcs}) in the upper half-plane
\begin{enumerate}[leftmargin=*]
\item that do not intersect themselves or each other,
\item with all of their endpoints among the $x_j$,
\item with two distinct endpoints per arc, and
\item with no two arcs sharing any endpoints.
\end{enumerate}
In particular, we let the first (resp.\ $C_N$th) connectivity be such that the $j$th arc has its endpoints at $x_{2j}$ and $x_{2j+1}$ for $1\leq j<N$ and the $N$th arc has its endpoints at $x_1$ and $x_{2N}$ (resp.\ $j$th arc has its endpoints at $x_{2j-1}$ and $x_{2j}$).

\item\label{step3} For each $\vartheta\in\{1,2,\ldots,C_N\}$, we let $\mathcal{F}_\vartheta$ be the $\vartheta$th \emph{Coulomb gas function}, given by the following formula (with explanation of the integration contours and the symbol $\mathcal{N}[\,\,\ldots\,\,]$ beneath it):
\begin{multline}\label{Fexplicit}\mathcal{F}_\vartheta(\kappa\,|\,x_1,x_2,\ldots,x_{2N}):=\left[\frac{n(\kappa)\Gamma(2-8/\kappa)}{4\sin^2(4\pi/\kappa)\Gamma(1-4/\kappa)^2}\right]^N\Bigg(\prod_{j<k}^{2N}(x_k-x_j)^{2/\kappa}\Bigg)\\ 
\times\,\,\oint_{\Gamma_N}{\rm d}u_N\,\,\dotsm\oint_{\Gamma_2}{\rm d}u_2\,\,\oint_{\Gamma_1}{\rm d}u_1\,\,\mathcal{N}\Bigg[\Bigg(\prod_{l=1}^{2N}\prod_{m=1}^N(x_l-u_m)^{-4/\kappa}\Bigg)\Bigg(\prod_{p<q}^N(u_p-u_q)^{8/\kappa}\Bigg)\Bigg].\end{multline}
\begin{enumerate}[leftmargin=*]
\item\label{step3a} $\Gamma_1$, $\Gamma_2,\ldots,\Gamma_N$ are nonintersecting Pochhammer contours with endpoints among $x_1$, $x_2,\ldots,x_{2N}$ (figure \ref{BreakDown}) and bent to lie completely in the upper half-plane (except for where they wrap around their endpoints).
\item\label{step3b} The two endpoints of $\Gamma_m$ (i.e., the two points surrounded by a cycle of $\Gamma_m$) are the endpoints of the $m$th arc in the $\vartheta$th connectivity (as we have enumerated the arc connectivities in item \ref{step2} above).
\item\label{step3c} We choose the branch of the logarithm for the power functions in (\ref{Fexplicit}) so $-\pi<\arg z\leq\pi$ for all $z$ in the complex plane.  Thus, the branch cut of $f(z)=z^\beta$ for $\beta\not\in\mathbb{Z}$ is $(-\infty,0]$.
\item\label{step3e} If $\kappa>4$, then we simplify (\ref{Fexplicit}) by replacing each contour by a simple curve (that bends into the upper half-plane) with the same endpoints, and we drop all factors of $4\sin^2(4\pi/\kappa)$ from (\ref{Fexplicit}) (figure \ref{PochhammerContour}).
\item\label{step3d}We order the differences in the integrand of (\ref{Fexplicit}) as figure \ref{Orderings} shows, and we indicate this ordering by enclosing the integrand for (\ref{Fexplicit}) between the square brackets of $\mathcal{N}[\,\,\ldots\,\,]$.  Thus, $\mathcal{F}_\vartheta$ is real-valued \cite{florkleb4}.
\end{enumerate} 
Alternatively, we may use the formula (\ref{Fother}) for $\mathcal{F}_\vartheta$ given in appendix \ref{equivappendix}.  Although the latter appears more complicated, it is really shorter and has one less integration contour.  In appendix \ref{equivappendix}, we show that these formulas are equivalent.

%%%%%%%%%%%%%%%%%%%%%%%%%%%%%%%%%%%%%%%%%%%%%%%%%%%%%%%%%%%%%%%%%%%%%%%%%%%%%%%%%%%
\begin{figure}[t]
\centering
\includegraphics[scale=0.28]{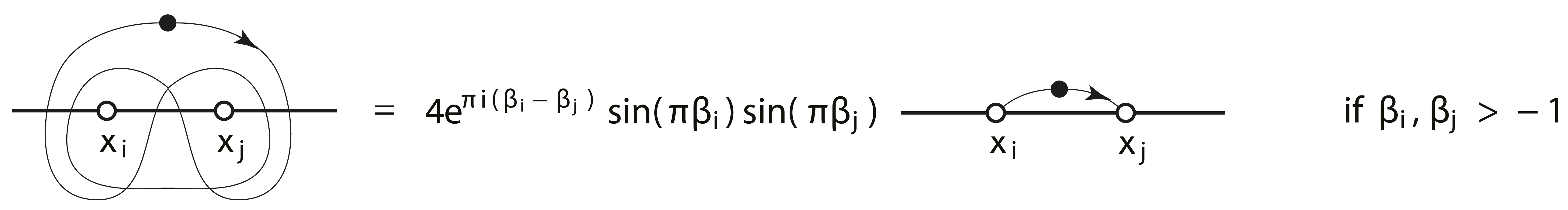}
\caption{If $e^{2\pi i\beta_i}$ and $e^{2\pi i\beta_j}$ are the monodromy factors associated with $x_i$ and $x_j$ respectively, and $\beta_i,\beta_j> -1 $, then $\mathscr{P}(x_i,x_j)$ may be replaced with the simple contour shown on the right.}
\label{PochhammerContour}
\end{figure}
%%%%%%%%%%%%%%%%%%%%%%%%%%%%%%%%%%%%%%%%%%%%%%%%%%%%%%%%%%%%%%%%%%%%%%%

%%%%%%%%%%%%%%%%%%%%%%%%%%%%%%%%%%%%%%%%%%%%%%%%%%%%%%%%%%%%%%%%%%%%%%%%%%%%%%%%%%%
\begin{figure}[b]
\centering
\includegraphics[scale=0.28]{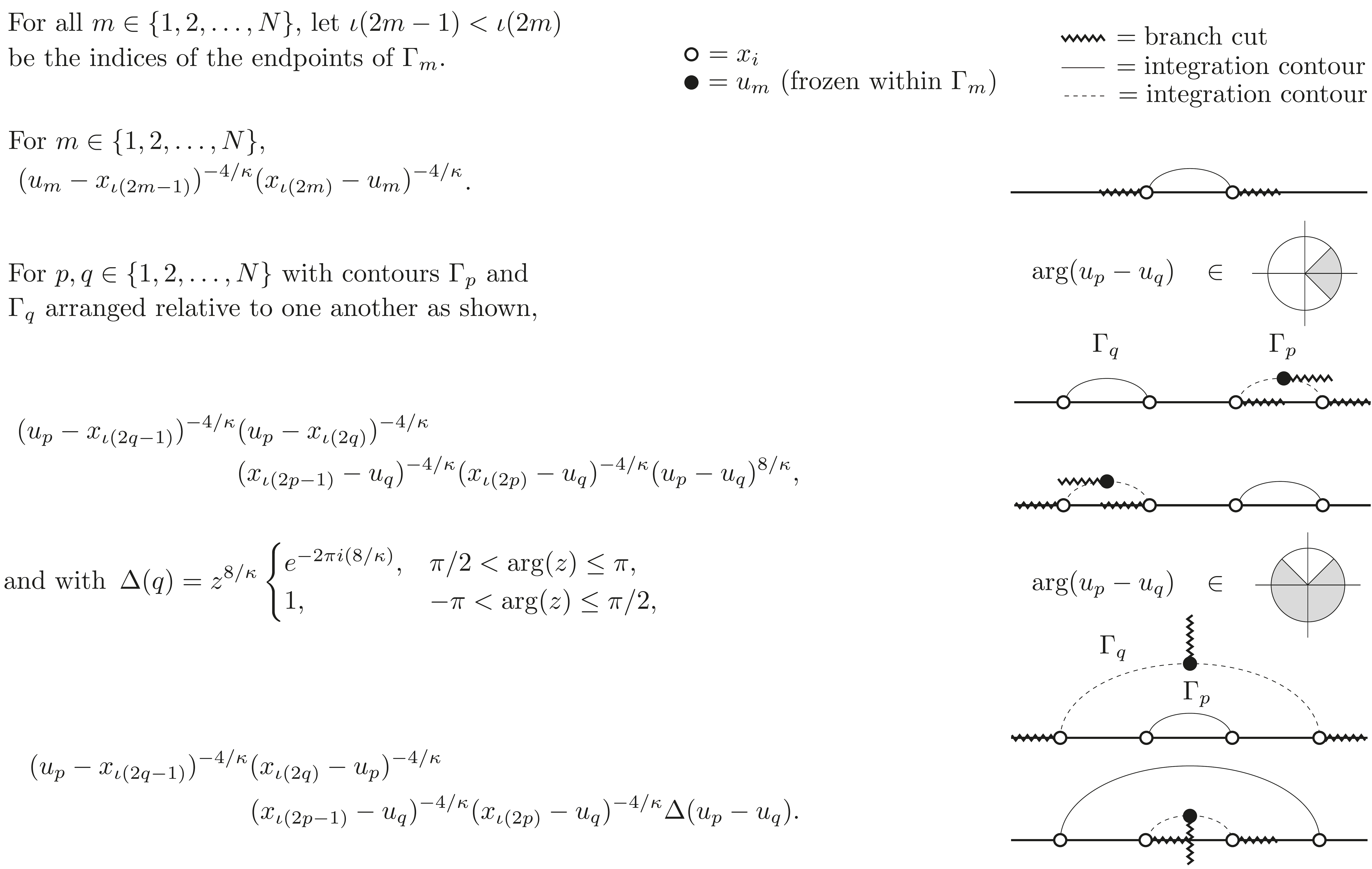}
\caption{Factors of the integrand for the contour integral appearing in the formula (\ref{Fexplicit}) for $\mathcal{F}_\vartheta$.  To indicate that we are using the conventions shown in this figure, we enclose the integrand of (\ref{Fexplicit}) between the brackets of $\mathcal{N}[\,\,\ldots\,\,]$.}
\label{Orderings}
\end{figure}
%%%%%%%%%%%%%%%%%%%%%%%%%%%%%%%%%%%%%%%%%%%%%%%%%%%%%%%%%%%%%%%%%%%%%%%%%%%%%%%%%%%%%%%%%%%%%%%%%%%%%%%%%%%%%%%%%%%%%%%%%%%%%%%%%%%%%%%%%%%%%%%%%%%%%%%%%%

\item\label{step4} For each $\varsigma,\vartheta\in\{1,2,\ldots,C_N\}$, we draw the $\varsigma$th connectivity in the upper half-plane, draw the $\vartheta$th connectivity in the lower half-plane, delete the real axis, and let $l_{\varsigma,\vartheta}$ be the number of loops left (figure \ref{N2loops} shows the case $N=2$).
\item\label{step5} We let $M_N(n)$ be the $C_N\times C_N$ \emph{meander matrix} \cite{fgg,fgut,difranc,franc} whose $(\varsigma,\vartheta)$th element is $n^{l_{\varsigma,\vartheta}}$.  If $\kappa\neq 4q/q'$ for any coprime pair of positive integers $q$ and $q'$ such that $q'>1$ and $q\leq N+1$, then $\det(M_N\circ n)(\kappa)\neq0$, and we solve
\be\label{F=MPi}\left(\begin{array}{l}\mathcal{F}_1\\ \mathcal{F}_2\\\,\,\vdots\\ \mathcal{F}_{C_N}\end{array}\right)=(M_N\circ n)\left(\begin{array}{l}\Pi_1\\ \Pi_2 \\ \,\,\vdots\\ \Pi_{C_N}\end{array}\right),\qquad [M_N\circ n]_{\varsigma,\vartheta}(\kappa)=n(\kappa)^{l_{\varsigma,\vartheta}},\ee
for the $C_N$ unknown functions $\Pi_\varsigma$, called ``crossing weights."  (If $\kappa=4q/q'$, then the system is not invertible.  Instead, we replace $\kappa$ with $\varkappa\approx\kappa$, solve the system, and send $\varkappa\rightarrow\kappa$ to find the crossing weights for this $\kappa$, or we use \cite{fsk}.)

%%%%%%%%%%%%%%%%%%%%%%%%%%%%%%%%%%%%%%%%%%%%%%%%%%%%%%%%%%%%%%%%%%%%%%%%%%%%%%%%%%%%%%%%%%%%%%%%%%%%%%%%%%%%%%%%%%%%%%%%%%%%%%%%%%%%%%%%%%%%%%%%%%%%%%%%%%
\begin{figure}[t]
\centering
\includegraphics[scale=0.35]{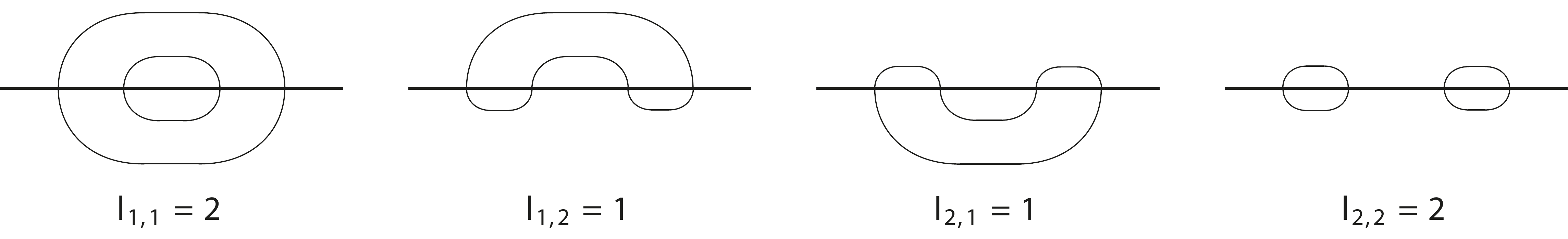}
\caption{In the left (resp.\ right) two figures, upper half-plane arcs join their endpoints in the first (resp.\ second) connectivity.  The number of loops formed by joining $\varsigma$th-connectivity upper half-plane arcs to $\vartheta$th-connectivity lower half-plane arcs is $l_{\varsigma,\vartheta}$.}
\label{N2loops}
\end{figure}
%%%%%%%%%%%%%%%%%%%%%%%%%%%%%%%%%%%%%%%%%%%%%%%%%%%%%%%%%%%%%%%%%%%%%%%%%%%%%%%%%%%%%%%%%%%%%%%%%%%%%%%%%%%%%%%%%%%%%%%%%%%%%%%%%%%%%%%%%%%%%%%%%%%%%%%%%%
\item\label{step6}\textbf{Formula for the crossing probability:}  As we will derive in section \ref{CFTsect}, the following function gives the probability of the $\varsigma$th crossing event conditioned on the $\vartheta$th BC event for models described below:  
\be\label{xing}\boxed{P_{\varsigma|\vartheta}(\kappa\,|\,x_1,x_2,\ldots,x_{2N})=n(\kappa)^{l_{\varsigma,\vartheta}}\frac{\Pi_\varsigma(\kappa\,|\,x_1,x_2,\ldots,x_{2N})}{\mathcal{F}_\vartheta(\kappa\,|\,x_1,x_2,\ldots,x_{2N})}}\ee
\begin{enumerate}[leftmargin=*]
\item\label{step6a}\emph{$Q$-state random cluster model:} With $Q\in\{1,2,3,4\}$, we suppose that the continuum limit of a critical random cluster model on a lattice fills the upper half-plane and that the boundary segments $(x_j,x_{j+1})$ with $j$ odd exhibit the fixed BC (but with the state of each fixed-BC segment unspecified).  With
\be\label{Qdense}Q=4\cos^2(4\pi/\kappa),\quad\kappa\in[4,8),\quad Q\in\{1,2,3,4\}\quad\Longrightarrow\quad \begin{cases} Q=1\quad\Longrightarrow& \kappa=6, \\
Q=2\quad\Longrightarrow& \kappa=16/3, \\ 
Q=3\quad\Longrightarrow& \kappa=24/5, \\  
Q=4\quad\Longrightarrow&\kappa=4,\end{cases}\ee
the formula (\ref{xing}) gives the probability that the FK clusters connect the fixed-BC segments together in such a way that the cluster boundaries join the points $x_1$, $x_2,\ldots,x_{2N}$ pairwise in the $\varsigma$th connectivity, conditioned on the event that any two fixed-BC segments joined together by an arc in the $\vartheta$th connectivity exhibit the same state.
\item\label{step6b} \emph{$Q$-state Potts model:} With $Q\in\{2,3,4\}$, we suppose that the continuum limit of a critical Potts model on a lattice fills the upper half-plane and that the boundary segments $(x_j,x_{j+1})$ with $j$ odd exhibit the first spin state while sites in all other boundary segments sample all but the first spin state uniformly.  With 
\be\label{Qdilute}Q=4\cos^2(\pi\kappa/4),\quad\kappa\in(0,4],\quad Q\in\{2,3,4\}\quad\Longrightarrow\quad\begin{cases} 
Q=2\quad\Longrightarrow& \kappa=3,\\ 
Q=3\quad\Longrightarrow& \kappa=10/3,\\ 
Q=4\quad\Longrightarrow& \kappa=4,\end{cases}\ee
the formula (\ref{xing}) with $\vartheta=1$ gives the probability that the spin clusters connect the fixed-BC segments together in such a way that the cluster boundaries join the points $x_1$, $x_2,\ldots,x_{2N}$ pairwise in the $\varsigma$th connectivity, conditioned on the BC event described.  (Having all fixed boundary segments exhibit the same state corresponds to $\vartheta=1$.)
\item\label{step6c} \emph{Loop-gas model:} With $n\in(0,2]$, we suppose that the continuum limit of a dense-phase or dilute-phase loop-gas model on a lattice fills the upper half-plane.  In addition, ``boundary loops" exit and then re-enter the upper half-plane through $x_1$, $x_2,\ldots,x_{2N}$, with exactly one loop passing once through each point.  With
\be\label{nfug} n(\kappa)=-2\cos(4\pi/\kappa),\qquad\begin{cases}\kappa\in(8/3,4],& \text{dilute phase}, \\ 
\kappa\in(4,8),& \text{dense phase}, 
\end{cases}\ee
formula (\ref{xing}) gives the probability that the parts of the boundary loops in the upper half-plane (``interior arcs") connect the points $x_1$, $x_2,\ldots,x_{2N}$ pairwise in the $\varsigma$th connectivity, conditioned on the event that the parts of the boundary loops inside the lower-half plane (``exterior arcs") join these points pairwise in the $\vartheta$th connectivity.
\end{enumerate}
The formula (\ref{xing}) for crossing probability is M\"obius invariant in the sense that if $f$ is a conformal bijection of the half-plane onto itself such that $x_i'<x_j'$ if $i<j$, where $x_j':=f(x_j)$, then
\be\label{halfplaneconformalinvar} P_{\varsigma|\vartheta}(\kappa\,|\,x_1',x_2',\ldots,x_{2N}')=P_{\varsigma|\vartheta}(\kappa\,|\,x_1,x_2,\ldots,x_{2N}).\ee
(If $f$ is not conformal, then this invariance is generally violated.)  Hence, (\ref{xing}) depends only on the $2N-3$ independent cross-ratios that we may form from $x_1$, $x_2,\ldots,x_{2N}$.  For example, we may choose these cross-ratios to be
\be\label{lambda}\lambda_j:=\frac{(x_{j+1}-x_1)(x_{2N}-x_{2N-1})}{(x_{2N-1}-x_1)(x_{2N}-x_{j+1})},\quad j\in\{1,2,\ldots,2N-3\}.\ee
\item\label{step7} If $f$ is a conformal map from the upper half-plane onto a Jordan domain $\mathcal{D}$ (such as a polygon) with $w_j:=f(x_j)$, and if the same statistical mechanics model lives on the same lattice in $\mathcal{D}$ as in item \ref{step6} above, then similar to (\ref{halfplaneconformalinvar}),
\be\label{confPext2}P_{\varsigma|\vartheta}^{\mathcal{D}}(\kappa\,|\,w_1,w_2,\ldots,w_{2N})=P_{\varsigma|\vartheta}(\kappa\,|\,x_1,x_2,\ldots,x_{2N})\ee
gives the crossing probability for the above model in $\mathcal{D}$, with the $\vartheta$th BC or boundary-loop condition involving $x_j$ sent to the same condition involving $w_j$.  In this sense, the crossing probability is conformally invariant.
\end{enumerate}

\section{Random cluster model and Potts model with boundary conditions}\label{partfuncpoly}

\begin{figure}[b]
\centering
\includegraphics[scale=0.35]{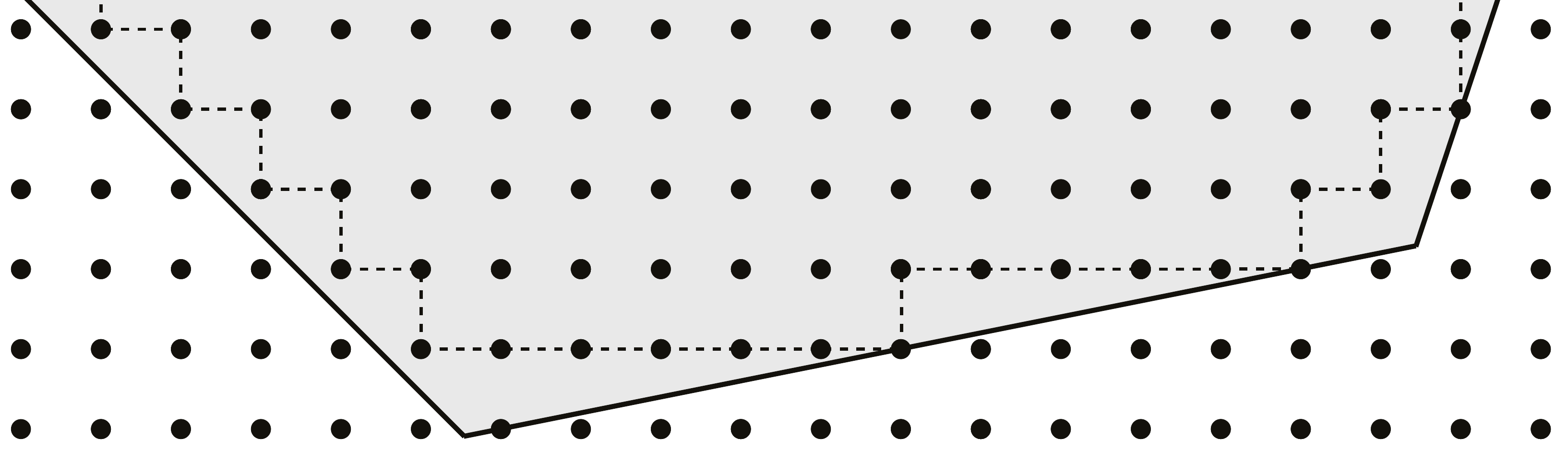}
\caption{The collection of bonds (dashed) and lattice sites (connected by dashed bonds) adjacent to the boundary of a polygon (partly shown, gray) superimposed on a square lattice. The polygon remains fixed as we take the continuum limit.}
\label{AdjacentSites}
\end{figure}

Now we begin to study specific models.  This, combined with the continuum limit, will eventually lead to our main result (\ref{xing}).  In this section, we give formal expressions for various partition functions of the random cluster (resp.\ Potts) model on a lattice inside a $2N$-sided polygon $\mathcal{P}$, summing exclusively over the BC events and/or crossing events in item \ref{BCitem1} (resp.\ item \ref{BCitem2}) of the introduction \ref{intro}.  We assign BCs exclusively to the sites and/or bonds adjacent to the sides of $\mathcal{P}$, as illustrated in figure \ref{AdjacentSites}. (We note that lattice sites and bonds adjacent to a side of $\mathcal{P}$ approach that side in the continuum limit.)  Anticipating the continuum limit, we regard the lattice spacing $a$ as very small compared to the lengths of these sides.  As we discuss in section \ref{introxing}, universality obviates the need to specify the lattice in the continuum limit, but for clarity, our figures use either the square lattice or the triangular lattice.

%%%%%%%%%%%%%%%%%%%%%%%%%%%%%%%%%%%%%%%%%%%%%%%%%%%%%%%%%%%%%%%%%%%%%%%%%%%%%%%%%%%%%%%%%%%%%%%%%%%%%%%%%%%%%%%%%%%%%%%%%
%%%%%%%%%%%%%%%%%%%%%%%%%%%%%%%%%%%%%%%%%%%%%%%%%%%%%%%%%%%%%%%%%%%%%%%%%%%%%%%%%%%%%%%%%%%%%%%%%%%%%%%%%%%%%%%%%%%%%%%%%

%%%%%%%%%%%%%%%%%%%%%%%%%%%%%%%%%%%%%%%%%%%%%%%%%%%%%%%%%%%%%%%%%%%%%%%%%%%%%%%%%%%%%%%%%%%%%%%%%%%%%%%%%%%%%%%%%%%%%%%%%
\begin{figure}[t]
\centering
\includegraphics[scale=0.16]{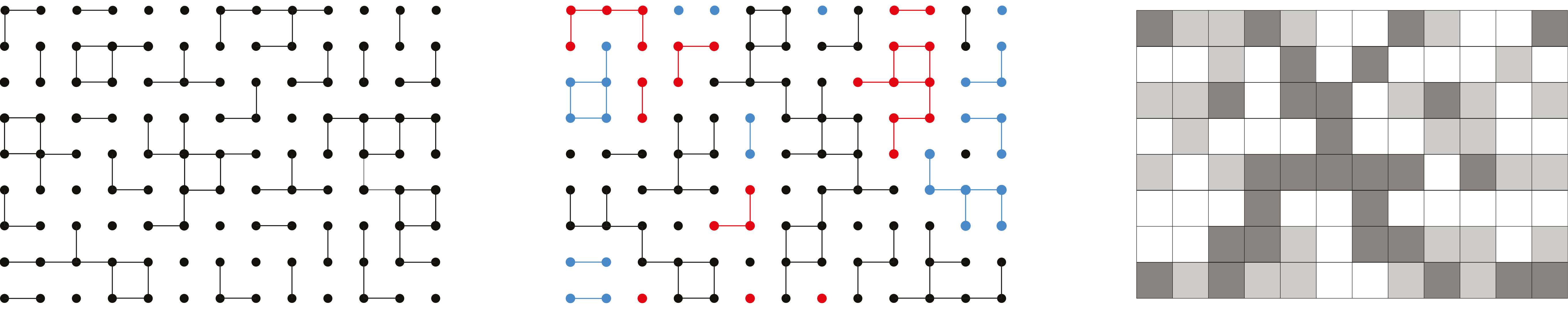}
\caption{ Sample configurations of bond percolation (left), the $Q=3$ random cluster model (middle), and the $Q=3$ Potts model (right) on the square lattice.  (In the right illustration, the centers of the tiles are the sites of the square lattice.)}
\label{PercRandPotts}
\end{figure}
%%%%%%%%%%%%%%%%%%%%%%%%%%%%%%%%%%%%%%%%%%%%%%%%%%%%%%%%%%%%%%%%%%%%%%%%%%%%%%%%%%%%%%%%%%%%%%%%%%%%%%%%%%%%%%%%%%%%%%%%%

\subsection{Random cluster model and Potts model with free boundary conditions}

Among the random cluster models, (Bernoulli) bond percolation on a lattice is the simplest.  Here, the bonds between nearest-neighbor sites (i.e., two sites separated by exactly one lattice space) are activated, each with probability $p$ and independently of the other bonds (figure \ref{PercRandPotts}).  The bond percolation partition function is 
\be\label{percpart}
X^\mathcal{P}(p,Q=1)=\sum_{\text{B}}p^{N_\beta}(1-p)^{N_\text{b}-N_\beta}=1,\ee
where B is the entire sample space of the model (i.e., the collection of all possible nearest-neighbor bond configurations on the lattice inside $\mathcal{P}$ (figure \ref{PercRandPotts})), and
\be\label{Nb}N_\text{b}=\left\{\parbox{6cm}{total number of bonds between nearest-neighbor sites in the entire system}\right\},\qquad N_\beta=\left\{\parbox{3.8cm}{total number of activated bonds in sample $\{\beta\}\in{\rm B}$}\right\}.\ee
(We include the $Q=1$ in (\ref{percpart}) to conform with notation for the $Q$-state random cluster model partition function.)

The \emph{$Q$-state random cluster model} generalizes bond percolation.  In this model, all bond clusters in every percolation sample are randomly assigned, uniformly and independently, one of $Q$ possible colors (figure \ref{PercRandPotts}).  As mentioned in the introduction \ref{intro}, we call these colored bond clusters ``FK clusters."  The partition function for this model is
\be\label{ZQ}X^\mathcal{P}(p,Q)=\sum_{\text{B}}p^{N_\beta}(1-p)^{N_\text{b}-N_\beta}Q^{N_\text{c}},\quad \text{where}\,\,N_\text{c}=\left\{\parbox{8.1cm}{number of activated bond clusters in sample $\{\beta\}\in{\rm B}$ (including size-zero clusters, i.e., isolated lattice sites)}\right\}.\ee
The new factor of $Q^{N_\text{c}}$ that is absent from (\ref{percpart}) uniformly distributes the $Q$ colors to the bond clusters in every sample of B, including to size-zero clusters (or isolated lattice sites).  We recover bond percolation in the special case $Q=1$.
 
The random cluster model is closely related to the Potts model \cite{wu} on the same lattice.  In the $Q$-state Potts model, we randomly assign any one of $Q$ available values, called ``spins," to each lattice site (figure \ref{PercRandPotts}).  In a given spin configuration (i.e., sample) $\{\sigma\}$, each pair of lattice sites $i$ and $j$, with respective spins $\sigma_i$ and $\sigma_j$, contribute energy 
\be\label{energy}E_{ij}=\begin{cases}-J\delta_{\sigma_i,\sigma_j},& \text{sites $i$ and $j$ are nearest-neighbors}, \\ 0, & \text{sites $i$ and $j$ are not nearest-neighbors},\end{cases}\quad \text{where $\sigma_j,\sigma_j$}\,\in\{e^{2\pi i\theta/Q}\,|\,\theta\in\{1,2,\ldots,Q\}\},\ee
to the configuration, where $J\in\mathbb{R}$ is the spin-coupling parameter.

The \emph{$Q$-state Potts model} weights each spin configuration by the Boltzmann distribution, with the total energy for its Hamiltonian.  Then the Potts model partition function is
\begin{alignat}{2}\label{ZPotts} 
Y^\mathcal{P}(K,Q)&=\sum_\text{S} e^{-\beta\sum_{\langle ij\rangle} E_{ij}},&&\quad \text{where $\beta>0$,}\\
\label{ZPotts2}&=\sum_\text{S}\prod_{\langle ij\rangle}\Big(1+(e^K-1)\delta_{\sigma_i,\sigma_j}\Big),&&\quad  \text{where $K:=\beta J$, and}\end{alignat}
where S is the collection of all spin configurations, $\beta>0$ is the (inverse) temperature, and the  sum $\smash{\sum_{\langle ij\rangle}}$ and product $\smash{\prod_{\langle ij\rangle}}$ are over all pairs $\langle ij\rangle$ of nearest-neighbor sites $i$ and $j$.  After expanding the product in (\ref{ZPotts2}), we find \cite{wu}
\be\label{relation}
 Y^\mathcal{P}(K,Q)=(1-p)^{-N_\text{b}} 
 X^\mathcal{P}(p,Q),\quad \text{where $e^K=1/(1-p)$}.\ee
 This identifies the ferromagnetic ($J>0$, so $p\in(0,1)$) $Q$-state Potts model with the $Q$-state random cluster model.
 
 %%%%%%%%%%%%%%%%%%%%%%%%%%%%%%%%%%%%%%%%%%%%%%%%%%%%%%%%%%%%%%%%%%%%%%%%%%%%%%%%%%%%%%%%%%%%%%%%%%%%%%%%%%%%%%%%%%%%%%%%%%%%%%%%%%%%%%%
\begin{figure}[b]
\centering
\includegraphics[scale=0.27]{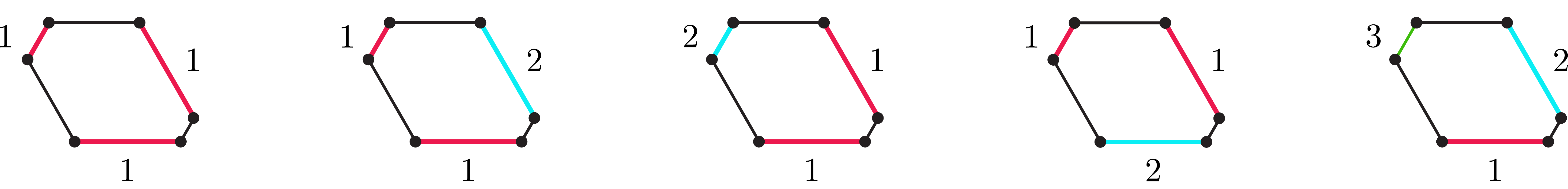}
\caption{Color schemes for the $Q=3$ random cluster model inside the hexagon.  We label the color $c_i\in\{1,2,3\}$ next to the $i$th fixed side of the hexagon.}
\label{ColorSchemeFig}
\end{figure}
%%%%%%%%%%%%%%%%%%%%%%%%%%%%%%%%%%%%%%%%%%%%%%%%%%%%%%%%%%%%%%%%%%%%%%%%%%%%%%%%%%%%%%%%%%%%%%%%%%%%%%%%%%%%%%%%%%%%%%%%%%%%%%%%%%%%%%%

 For every $Q\in\{1,2,3,4\}$, the ferromagnetic $Q$-state Potts model undergoes a second-order phase transition at a critical temperature $K_c$.  This critical behavior manifests itself in many ways.  For example, on an infinite lattice, the typical spin cluster size (i.e., collection of sites in a common spin state and such that the bonds between them form a connected cluster) is finite for $K<K_c$ but becomes infinite at $K=K_c$ \cite{grim2}.  Alternatively, for the $Q=2$ Potts (i.e., Ising) model inside a Jordan domain $\mathcal{D}$ with all boundary spins fixed to the $\sigma=+1$ state, the magnetization (i.e., mean spin) vanishes for $K\leq K_c$ but is positive for $K>K_c$ \cite{mcwu,per,ons}.  Explicit formulas for the critical temperature as a function of $Q$ are known for the square lattice \cite{befdu} and the triangular lattice \cite{kim}: 
\be\label{pc}K_c^{\text{sqr.}}=\log(1+\sqrt{Q}),\qquad K_c^{\text{tri.}}=\log\left\{1+\left[\frac{2}{\sqrt{Q}}\cos\left(\frac{1}{3}\arctan\sqrt{\frac{4}{Q}-1}\right)\right]^{-1}\right\}.\ee
From (\ref{relation}), we infer that the FK-cluster size in the $Q$-state random cluster model also undergoes a phase transition at critical probability $p_c$ related to $K_c$ via (\ref{relation}).  For the square lattice \cite{befdu} and the triangular lattice \cite{kim}, these are 
\begin{alignat}{3}\label{criticalpt}&e^{K_c^{\text{sqr.}}}=1/(1-p_c^{\text{sqr.}})&&\qquad\Longrightarrow\qquad && p_c^\text{sqr.}=\frac{\sqrt{Q}}{1+\sqrt{Q}},\\
\label{pc2}&e^{K_c^{\text{tri.}}}=1/(1-p_c^{\text{tri.}})&& \qquad\Longrightarrow\qquad &&p_c^{\text{tri.}}=\left[1+\frac{2}{\sqrt{Q}}\cos\left(\frac{1}{3}\arctan\sqrt{\frac{4}{Q}-1}\right)\right]^{-1}.
\end{alignat}
In the partition functions (\ref{percpart}, \ref{ZQ}, \ref{ZPotts}) encountered so far, the sums are not restricted to events with a BC imposed on lattice sites adjacent to a boundary.  Because we call this lack of a BC constraint the ``free BC" in item \ref{BCitem1} of the introduction \ref{intro}, we naturally call the partition functions of this section \emph{free BC partition functions.}

\subsection{Random cluster model with free/fixed boundary conditions}\label{mutsect}

%%%%%%%%%%%%%%%%%%%%%%%%%%%%%%%%%%%%%%%%%%%%%%%%%%%%%%%%%%%%%%%%%%%%%%%%%%%%%%%%%%%%%%%%%%%%%%%%%%%%%%%%%%%%%%%%%%%%%%%%%
\begin{figure}[t]
\centering
\includegraphics[scale=0.35]{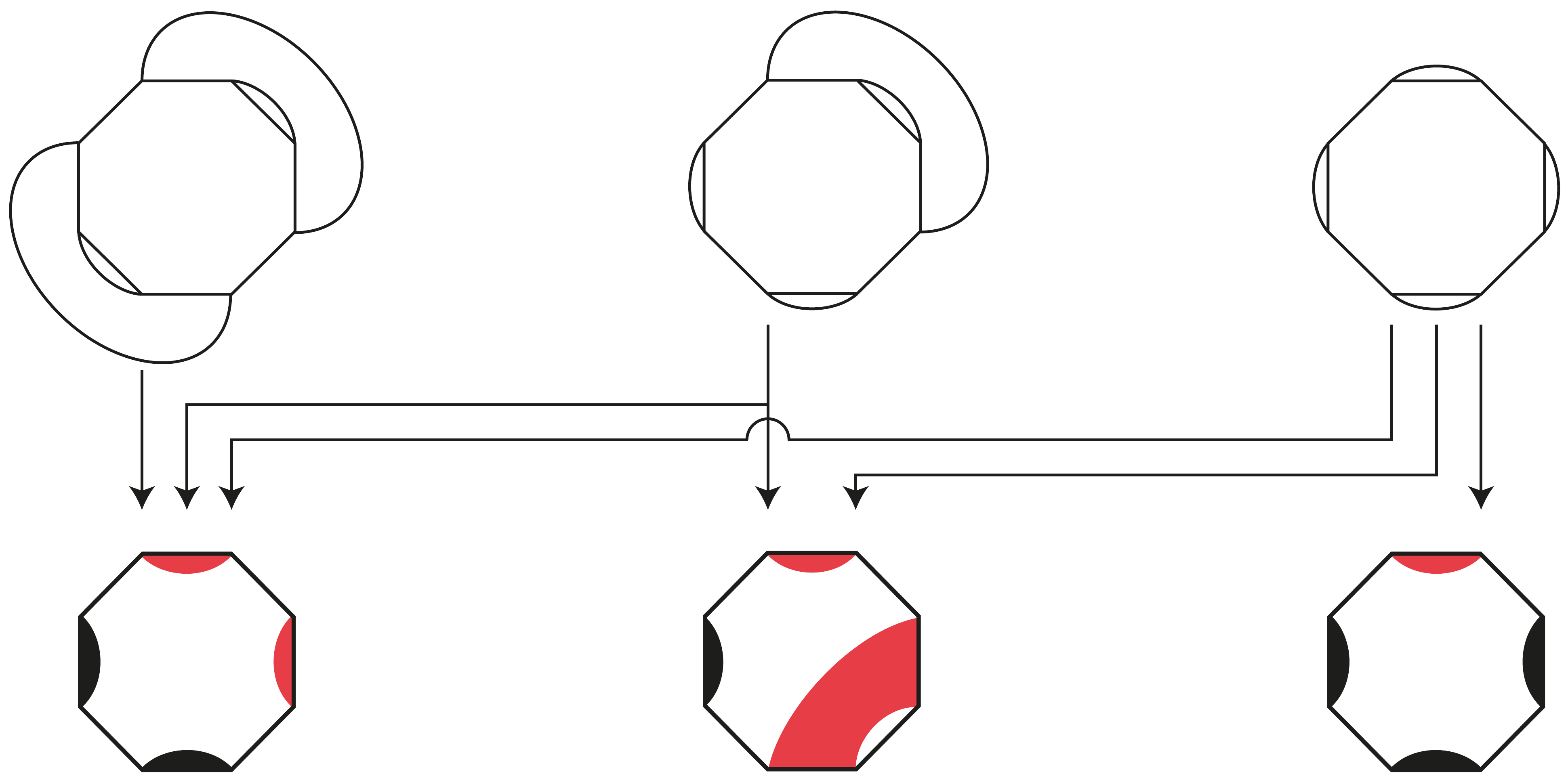}
\caption{We represent the $\vartheta$th $\RCB$ event by the $\vartheta$th exterior arc connectivity in the top row.  Arrows indicate which colorings of boundary clusters in the bottom row are consistent with these $\RCB$ events: exterior arcs can only connect like colors.}
\label{IndMut}
\end{figure}
%%%%%%%%%%%%%%%%%%%%%%%%%%%%%%%%%%%%%%%%%%%%%%%%%%%%%%%%%%%%%%%%%%%%%%%%%%%%%%%%%%%%%%%%%%%%%%%%%%%%%%%%%%%%%%%%%%%%%%%%%

In this section, we give formal expressions for \emph{$\RCB$ partition functions}, or random cluster model partition functions that sum exclusively over an $\RCB$ event (as defined in item \ref{BCitem1} in the introduction \ref{intro}) inside a $2N$-sided polygon $\mathcal{P}$.  A natural $\RCB$ event is a \emph{color scheme} $c=(c_1,c_2,\ldots,c_N)$.  This event comprises all samples where the $i$th fixed side has color $c_i\in\{1,2,\ldots,Q\}$ for each $i\in\{1,2,\ldots,N\}$ (figure \ref{ColorSchemeFig}).  There are $Q^N$ different color schemes, and color schemes are disjoint.  We define a \emph{color scheme partition function} $\smash{W_c^\mathcal{P}(p,Q)}$ to be an $\RCB$ partition function that sums exclusively over a given color scheme $c$.  Any random cluster model $\RCB$ event is a union of color schemes, so any $\RCB$ partition function is a sum of color scheme partition functions.  Thus, it is ostensibly fruitful to study the latter.  However, in this article, we instead work with a certain subset of $\RCB$ partition functions that we call ``basic" partition functions.  In appendix \ref{Pottsappendix}, we show that for $Q\in\{1,2,3\}$, any color scheme partition function equals a linear combination of basic partition functions.  Furthermore, in section \ref{RCLoopGasSect}, we show that basic partition functions are more tractable to use in the continuum limit.  For this reason, we work with basic partition functions.

The basic partition functions are examples of \emph{mutual wiring partition functions}, or $\RCB$ partition functions that sum exclusively over a mutual wiring event.  A \emph{mutual wiring event} is an $\RCB$ event where every fixed side of $\mathcal{P}$ belongs to a collection of fixed sides that are \emph{mutually wired}, i.e., constrained to have the same color (figure \ref{IndMut}).  In such an event, the actual color of a collection of mutually wired sides is not specified; it can by anything.  (If one such collection has only one fixed side of $\mathcal{P}$, then that side is not constrained to exhibit the color of any other fixed side, and we say that it is \emph{independently wired}.)  A mutual wiring event $\mathcal{E}$ samples (i.e., is a union of) all color schemes $c$ consistent with it.  By ``consistent," we mean $c\leftrightsquigarrow\mathcal{E}$, where by definition,
\be\label{squiqequiv} c=(c_1,c_2,\ldots,c_N)\leftrightsquigarrow\mathcal{E}  \qquad\Longrightarrow\qquad\parbox{2.48in}{if the $i$th and $j$th fixed sides of $\mathcal{P}$ are mutually wired together in $\mathcal{E}$, then $c_i=c_j$.}\ee
Thus, the mutual wiring partition function $\smash{X_\mathcal{E}^\mathcal{P}}$ summing exclusively over $\mathcal{E}$ equals a sum of all color scheme partition functions consistent with $\mathcal{E}$:
\be\label{XEsum}X_\mathcal{E}^\mathcal{P}(p,Q)=\sum_{c\leftrightsquigarrow\mathcal{E}} W_c^\mathcal{P}(p, Q).\ee
Writing $N_\text{c}(\mathcal{E})$ for the total number of independent clusters in a given sample, that is the number of clusters after identifying all boundary clusters that touch a common collection of mutually wired sides as one, (\ref{ZQ}) becomes
\be X_\mathcal{E}^\mathcal{P}(p, Q)=\sum_{\text{B}}p^{N_\beta}(1-p)^{N_\text{b}-N_\beta}Q^{N_\text{c}(\mathcal{E})}.\ee

\begin{figure}[t]
\centering
\includegraphics[scale=0.35]{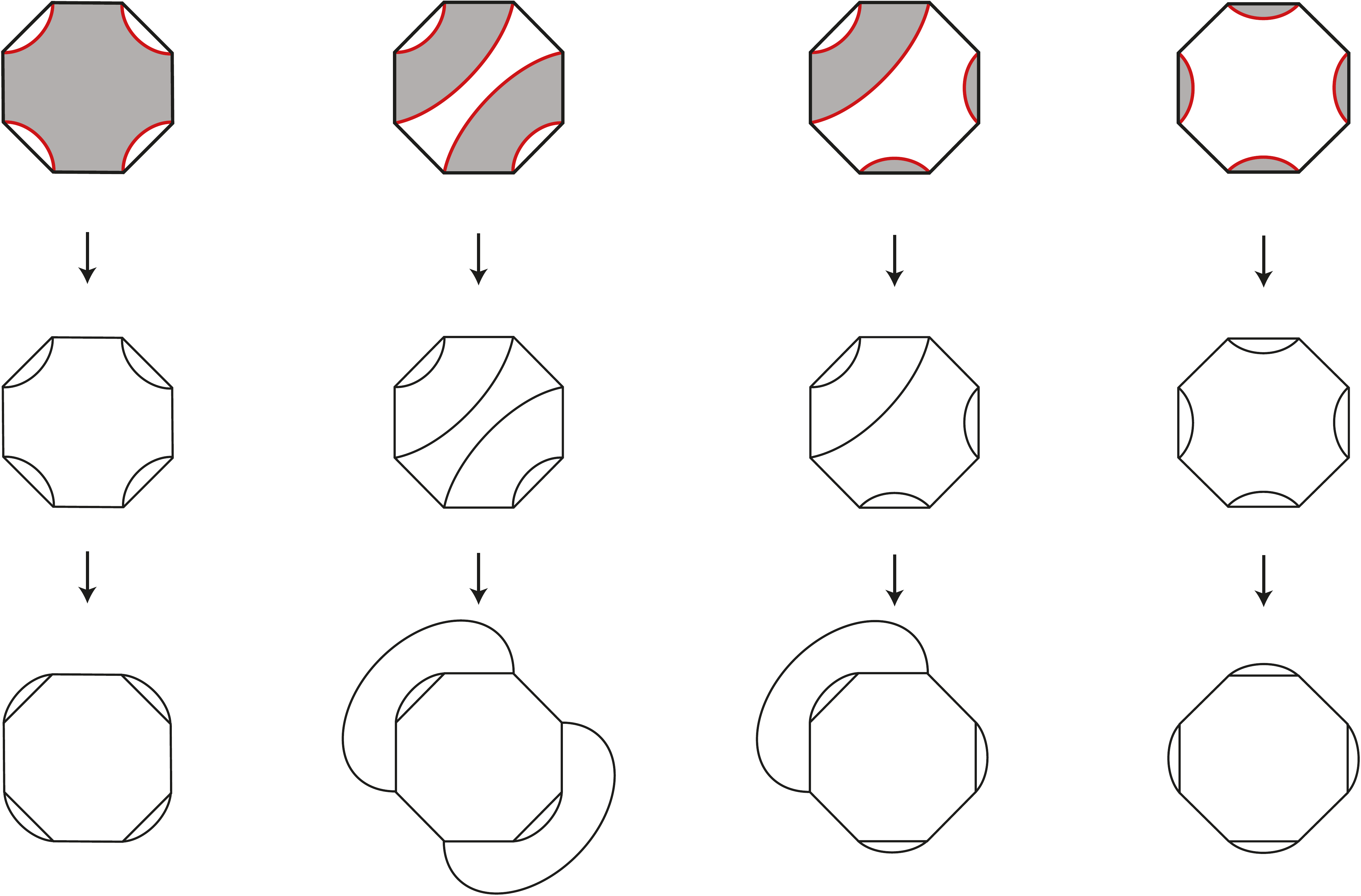}
\caption{Crossing patterns (top) bijectively correspond to interior arc connectivities (middle), which bijectively correspond to exterior arc connectivities (bottom).  The bijection respects our enumeration, one through $C_N$, of these three entities.}
\label{OctXingConfigsBdyArcs}
\end{figure}

In this article, we consider only a special subset of mutual wiring events $\RCB_1$, $\RCB_2,\ldots,\RCB_{\text{$C_N$}}$, called \emph{basic events}.  To define them, we consider all of the distinct ways (up to homotopy) to join the vertices of $\mathcal{P}$ pairwise with $N$ nonintersecting planar curves (\emph{exterior arcs}), drawn outside $\mathcal{P}$.  We call each possibility an \emph{exterior arc connectivity}, and figure \ref{OctXingConfigsBdyArcs} shows that they bijectively correspond to the $C_N$ (\ref{catalan}) crossing patterns in $\mathcal{P}$.  (We interpret exterior arc connectivities as crossing events in the exterior of $\mathcal{P}$.  This analogy is not obvious, but the work that we do below justifies it.)  After enumerating these connectivities one through $C_N$, we let the basic event $\RCB_\vartheta$ be the mutual wiring event where any two fixed sides joined by an exterior arc in the $\vartheta$th connectivity are mutually wired (figure \ref{IndMut}).  Then the \emph{basic partition functions} are
\be\label{BasicX} X_\vartheta^\mathcal{P}:=X_{\RCB_\vartheta}^\mathcal{P},\quad\text{for all $\vartheta\in\{1,2,\ldots,C_N\}$}.\ee
We note that $\{\RCB_1,\RCB_2,\ldots,\RCB_{\text{$C_N$}}\}$ comprises all mutual wiring events only if $N<3$.  Indeed, for $N=4$, no exterior arc connectivity generates the mutual wiring event in which the octagon's top and bottom sides are mutually wired together while the octagon's left and right sides are also mutually wired together, independent of the former.  Therefore, we do not include this mutual wiring event.  For larger $N$, even more possibilities are not included.

%%%%%%%%%%%%%%%%%%%%%%%%%%%%%%%%%%%%%%%%%%%%%%%%%%%%%%%%%%%%%%%%%%%%%%%%%%%%%%%%%%%%%%%%%%%%%%%%%%%%%%%%%%%%%%%%%%%%%%%%%
%%%%%%%%%%%%%%%%%%%%%%%%%%%%%%%%%%%%%%%%%%%%%%%%%%%%%%%%%%%%%%%%%%%%%%%%%%%%%%%%%%%%%%%%%%%%%%%%%%%%%%%%%%%%%%%%%%%%%%%%%

\subsection{Random cluster model crossing events with free/fixed boundary conditions}\label{RCxingSect}

In this section, we give formal expressions for the probabilities of random cluster model crossing events conditioned on the mutual wiring events $\RCB_1$, $\RCB_2,\ldots,\RCB_{\text{$C_N$}}$ defined in the previous section \ref{mutsect}.  In order to do this, we restrict the partition function sums of the previous section to sums over a ``crossing event," or the event that boundary FK clusters join the sides of $\mathcal{P}$ in a particular crossing pattern.  For this, we introduce some convenient notation:

%%%%%%%%%%%%%%%%%%%%%%%%%%%%%%%%%%%%%%%%%%%%%%%%%%%%%%%%%%%%%%%%%%%%%%%%%%%%%%%%%%%%%%%%%%%%%%%%%%%%%%%%%%%%%%%%%%%%%%%%%
\begin{figure}[b]
\centering
\includegraphics[scale=0.25]{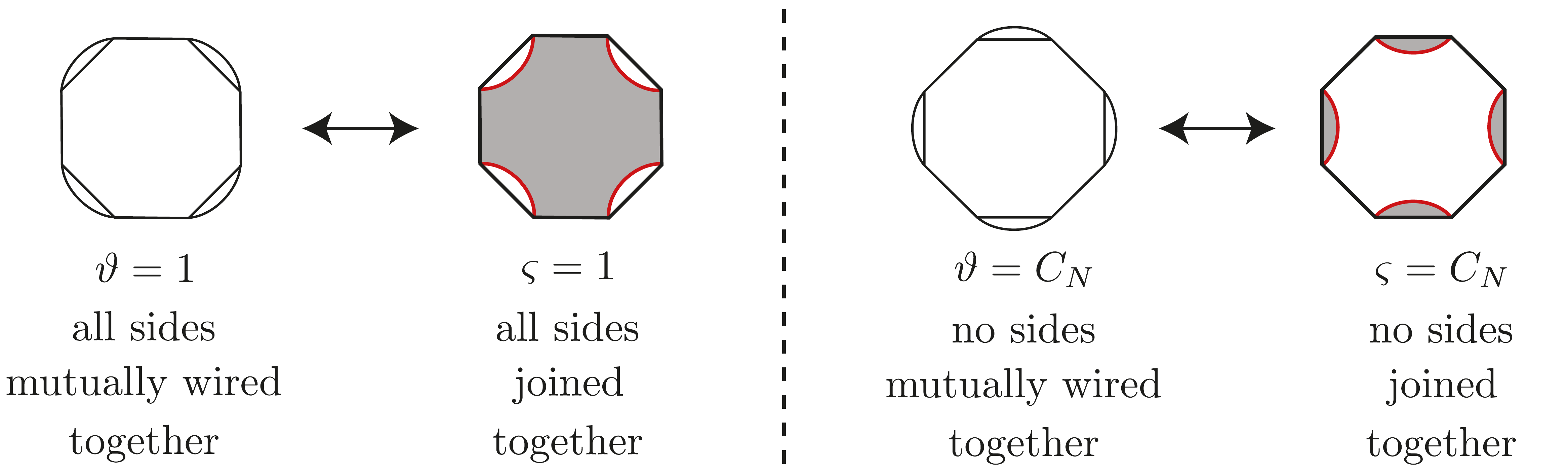}
\caption{ The first exterior (resp.\ interior) arc connectivity corresponds to the event $\RCB_1$ (resp.\ $\RCX_1$) in which all fixed sides are mutually wired together (resp.\ a single cluster connects all fixed sides).  The $C_N$th exterior (resp.\ interior) arc connectivity corresponds to the event $\RCB_{\text{$C_N$}}$ (resp.\ $\RCX_{\text{$C_N$}}$) in which all fixed sides are independently wired (resp.\ no cluster connects any two fixed sides).}
\label{ExtremeCases}
\end{figure}
%%%%%%%%%%%%%%%%%%%%%%%%%%%%%%%%%%%%%%%%%%%%%%%%%%%%%%%%%%%%%%%%%%%%%%%%%%%%%%%%%%%%%%%%%%%%%%%%%%%%%%%%%%%%%%%%%%%%%%%%%

\begin{figure}[t]
\centering
\includegraphics[scale=0.3]{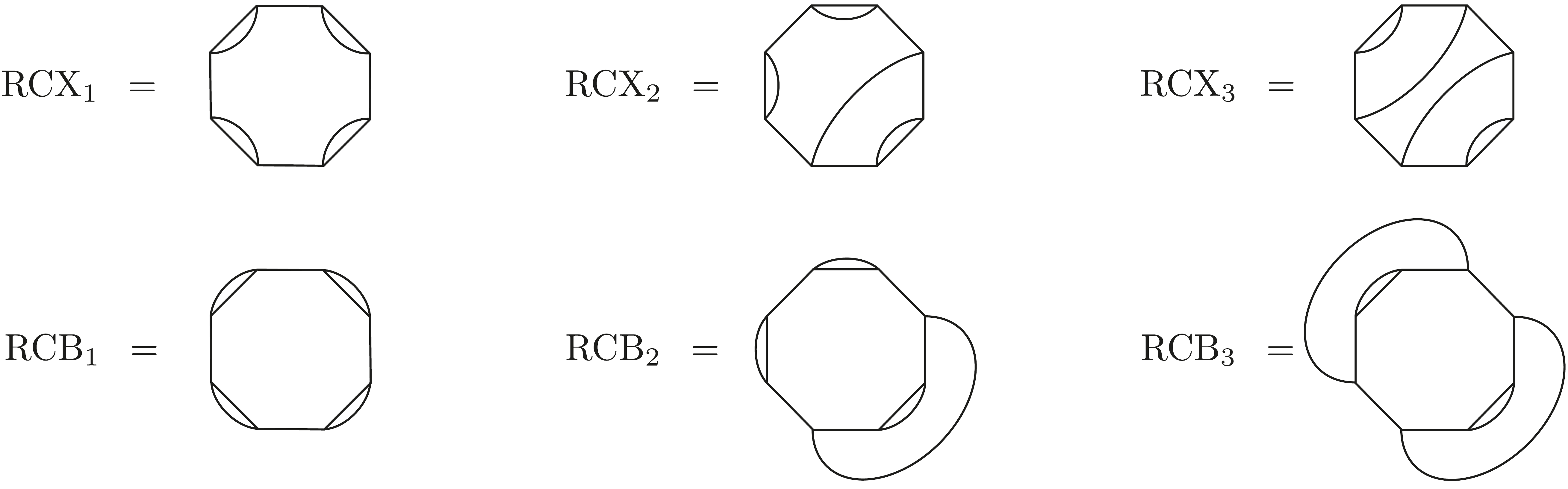}
\caption{Diagrams for $\RCX$ events  (item \ref{it4}) (top) and $\RCB$ events  (item \ref{it5}) (bottom).  Taking the top, bottom, left, and right sides to be  fixed, two fixed sides are connected by an exterior arc if and only if they are mutually wired together.}
\label{BCfigs}
\end{figure}

\begin{enumerate}
\item\label{it1}  We enumerate the $C_N$ crossing events (and corresponding arc connectivities (figure \ref{OctXingConfigsBdyArcs})) such that a single (resp.\ no) boundary cluster connects all (resp.\ any two) fixed sides of $\mathcal{P}$ in the first (resp.\ $C_N$th) event (figure \ref{ExtremeCases}).
\item\label{it4}  We let $\RCX_\varsigma$ denote the $\varsigma$th  FK-cluster crossing event, and we represent it by the $\varsigma$th \emph{interior arc connectivity} (i.e., the $\varsigma$th exterior arc connectivity, but with \emph{interior arcs} drawn inside $\mathcal{P}$) (figures \ref{OctXingConfigsBdyArcs}, \ref{BCfigs}).
\item\label{it5}  We let $\RCB_\vartheta$ denote the $\vartheta$th random cluster model mutual wiring event (fixed sides joined by an arc in the $\vartheta$th connectivity are mutually wired, figure \ref{IndMut}) and represent it by the $\vartheta$th exterior arc connectivity (figures \ref{OctXingConfigsBdyArcs}, \ref{BCfigs}).
\item\label{it7} We let $X_{\varsigma,\vartheta}^\mathcal{P}(\text{$p,Q$})$ denote the partition function for  the $Q$-state random cluster model on  a lattice in $\mathcal{P}$, summing exclusively over the intersection $\RCX_\varsigma\cap\RCB_\vartheta$ of the $\varsigma$th crossing event with the $\vartheta$th $\RCB$ event.
\item\label{it8} We let $X_\vartheta^\mathcal{P}(\text{$p$}, Q)$  denote the partition function for the $Q$-state random cluster model on a lattice in $\mathcal{P}$, summing exclusively over the $\vartheta$th $\RCB$ event $\RCB_\vartheta$.  Hence,  $X_\vartheta^\mathcal{P}=X_{1,\vartheta}^\mathcal{P}+X_{2,\vartheta}^\mathcal{P}+\dotsm+X_{C_N,\vartheta}^\mathcal{P}.$
\end{enumerate}
 Although section \ref{mutsect} already has item \ref{it5}, we reiterate it here for easy reference.  We infer from item \ref{it1} with figure \ref{ExtremeCases} that $\RCB_1$ (resp.\ $\RCB_{\text{$C_N$}}$) is the event with all fixed sides of $\mathcal{P}$ mutually wired together (resp.\ independently wired).  

Turning to crossing probabilities of the critical random cluster model, we write a formal expression for the probability of the $\varsigma$th FK-cluster crossing event $\RCX_\varsigma$ conditioned on the $\vartheta$th $\RCB$ event $\RCB_\vartheta$.  First defining
\be\label{Bsigma}\text{B}_\varsigma=\{\text{$\varsigma$th percolation (i.e., $Q=1$ FK cluster) crossing event} \}\subset\text{B},\ee
and then replacing $\text{B}\mapsto\text{B}_\varsigma$ and $\mathcal{E}\mapsto\RCB_\vartheta$ in (\ref{ZQ}), we obtain these formal expressions for $\smash{X_{\varsigma,\vartheta}^\mathcal{P}}$ and $\smash{X_\vartheta^\mathcal{P}}$:
\begin{align}\label{randFFBCadapt1} X_{\varsigma,\vartheta}^\mathcal{P}(\text{$p$}, Q)&=\sum_{\text{B}_\varsigma}p^{N_\beta}(1-p)^{N_\text{b}-N_\beta}Q^{N_{\text{c}}(\RCB_\vartheta)},\\
\label{randFFBCadapt2} X_\vartheta^\mathcal{P}(\text{$p$}, Q)&=X_{1,\vartheta}^\mathcal{P}(p, Q )+X_{2,\vartheta}^\mathcal{P}(p, Q )+\dotsm+X_{C_N,\vartheta}^\mathcal{P}(p, Q ).\end{align}
Then the probability of the $\varsigma$th  FK-cluster crossing event $\RCX_\varsigma$ conditioned on the $\vartheta$th $\RCB$ event $\RCB_\vartheta$ is given  by the ratio of these partition functions (abbreviating  ``random cluster" with ``RC"):
\be\label{chixing} P_{\varsigma|\vartheta}^\mathcal{P}(p, Q ):=\mathbb{P}_{\text{RC}}(\RCX_\varsigma\,|\,\RCB_\vartheta)=\frac{X_{\varsigma,\vartheta}^\mathcal{P}(\text{$p,Q$})}{X_\vartheta^\mathcal{P}(\text{$p,Q$} )}.\ee
Our primary goal in this article is to find a formula for the continuum limit of (\ref{chixing})  at the critical point $p=p_c$.  This poses two tasks: the first is to  determine the numerator and denominator of (\ref{chixing}), which by (\ref{randFFBCadapt1}, \ref{randFFBCadapt2}), amounts to determining $ N_{\text{c}}(\RCB_\vartheta) $.  The second task, taken up in section \ref{CFTsect}, is to determine the limit of (\ref{chixing}) as the lattice spacing $a$ vanishes. 

 %%%%%%%%%%%%%%%%%%%%%%%%%%%%%%%%%%%%%%%%%%%%%%%%%%%%%%%%%%%%%%%%%%%%%%%%%%%%%%%%%%%%%%%%%%%%%%%%%%%%%%%%%%%%%%%%%%%%%%%%%
%%%%%%%%%%%%%%%%%%%%%%%%%%%%%%%%%%%%%%%%%%%%%%%%%%%%%%%%%%%%%%%%%%%%%%%%%%%%%%%%%%%%%%%%%%%%%%%%%%%%%%%%%%%%%%%%%%%%%%%%%

Taking up the first task, we let $N_\text{c}^\text{free}$ denote the number of \emph{free clusters}, or clusters of activated bonds in  configuration $\{\beta\}\in\text{B}_\varsigma$ that are not boundary clusters.  As such, no bond in a free cluster touches a fixed side of $\mathcal{P}$.  (These include \emph{bulk clusters}, which do not touch any side of $\mathcal{P}$.)  Furthermore, we let $\mathcal{C}_{\varsigma,\vartheta}$ denote the number of boundary clusters in $\{\beta\}$, with all boundary clusters attached to a common collection of mutually wired sides identified as one.  Then 
\be\label{Ndecomp} N_{\text{c}}(\RCB_\vartheta) =\mathcal{C}_{\varsigma,\vartheta}+\text{$N$}_\text{c}^\text{free}\quad\text{for each $\{\beta\}\in\text{B}_\varsigma$}.\ee
Because $\mathcal{C}_{\varsigma,\vartheta}$ is the same for all bond configurations in the $\varsigma$th crossing event, (\ref{randFFBCadapt1}) factors into
\be\label{ZQoutFactored}X_{\varsigma,\vartheta}^\mathcal{P}(\text{$p$}, Q )=Q^{\mathcal{C}_{\varsigma,\vartheta}}\sum_{\text{B}_\varsigma}p^{N_\beta}(1-p)^{N_\text{b}-N_\beta}Q^{N^\text{free}_{\text{c}}}.\ee
Although the prefactor $Q^{\mathcal{C}_{\varsigma,\vartheta}}$ is not physically relevant to this partition function (\ref{ZQoutFactored}), we need it in order to compute the sum (\ref{randFFBCadapt2}), which appears in the denominator of (\ref{chixing}).

%%%%%%%%%%%%%%%%%%%%%%%%%%%%%%%%%%%%%%%%%%%%%%%%%%%%%%%%%%%%%%%%%%%%%%%%%%%%%%%%%%%%%%%%%%%%%%%%%%%%%%%%%%%%%%%%%%%%%%%%%
\begin{figure}[b]
\centering
\includegraphics[scale=0.3]{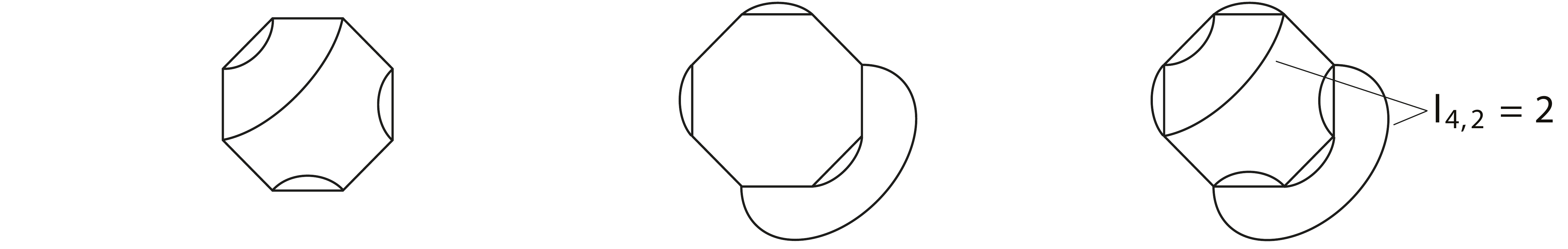}
\caption{The $(4,2)$th product diagram.  The arcs of the fourth interior arc connectivity join with the arcs of the second exterior arc connectivity to give $l_{4,2}=2$ loops.}
\label{innerproduct}
\end{figure}
%%%%%%%%%%%%%%%%%%%%%%%%%%%%%%%%%%%%%%%%%%%%%%%%%%%%%%%%%%%%%%%%%%%%%%%%%%%%%%%%%%%%%%%%%%%%%%%%%%%%%%%%%%%%%%%%%%%%%%%%% 

\begin{figure}[b]
\centering
\includegraphics[scale=0.34]{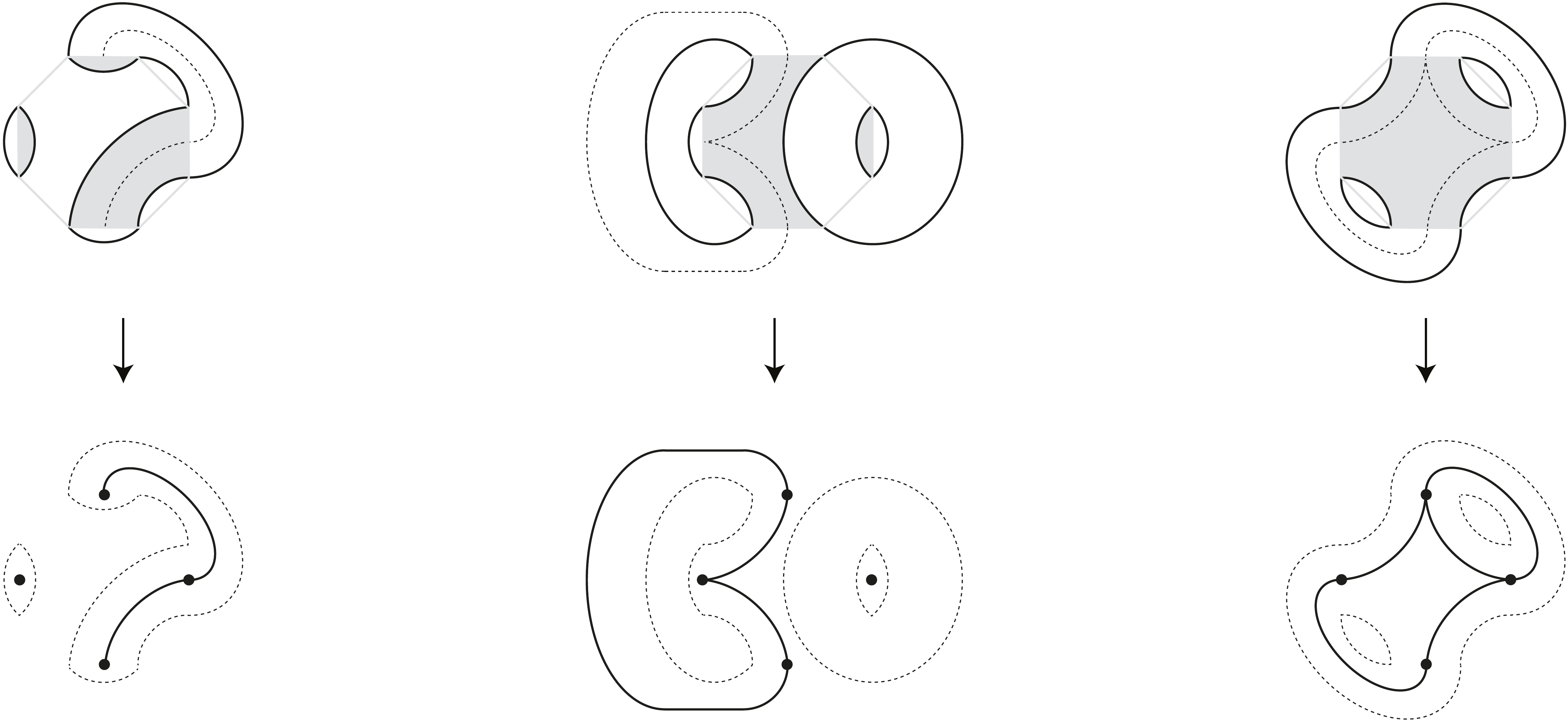}
\caption{We associate to the  ($\varsigma,\vartheta)$th product diagram (i.e., collection of $l_{\varsigma,\vartheta}$ black loops in the top row) a graph $\mathcal{G}_{\varsigma,\vartheta}$  (collection of black vertices and edges in the bottom row), constructed in steps \ref{G1}--\ref{G4} beneath (\ref{pformula}).  Gray regions indicate boundary clusters.}
\label{PolyGraph}
\end{figure}

Next, we determine a formula for $\mathcal{C}_{\varsigma,\vartheta}$ in terms of the number $l_{\varsigma,\vartheta}$ of loops in the $(\varsigma,\vartheta)$th \emph{product diagram}, created by superimposing the  $\varsigma$th interior arc connectivity and $\vartheta$th exterior arc connectivity on the same polygon, joining the interior arcs to the exterior arcs at the vertices of the polygon to form loops, and deleting the polygon, leaving only the loops  (figure \ref{innerproduct}).   The formula that we find is
\be\label{pformula}\mathcal{C}_{\varsigma,\vartheta}=(l_{\varsigma,\vartheta}-l_{1,\vartheta}+l_{\varsigma,C_N}+1)/2.\ee
To  derive (\ref{pformula}), we identify the $(\varsigma,\vartheta)$th product diagram with a graph $\mathcal{G}_{\varsigma,\vartheta}$ in which the quantity $\mathcal{C}_{\varsigma,\vartheta}$ naturally arises.  This identification goes as follows:
\begin{enumerate}
\item\label{G1} We enumerate the fixed sides of the $2N$-sided polygon $\mathcal{P}$ one through $N$ in counterclockwise order.  (By convention, we take the lowest side of $\mathcal{P}$ to be the first fixed side.)
\item We identify each fixed side of $\mathcal{P}$ with a vertex of the graph $\mathcal{G}_{\varsigma,\vartheta}$ under construction, and we enumerate these vertices so the $i$th vertex corresponds with the $i$th fixed side.
\item\label{G3} If $\{s_1,s_2,\ldots,s_k\}\subset\mathbb{Z}^+$ is the set of labels for all fixed sides joined by a common boundary cluster in $\RCX_\varsigma$ (with $1<k\leq N$), then we join the $s_i$th vertex and the $s_{i+1}$th vertex of $\mathcal{G}_{\varsigma,\vartheta}$ with an edge for each $i\in\{1,2,\ldots,k-1\}$.
\item\label{G4} If $\{s_1,s_2,\ldots,s_k\}\subset\mathbb{Z}^+$ is the set of labels for all fixed sides of a collection of mutually wired sides in $\RCB_\vartheta$ (with $1<k\leq N$), then we join the $s_i$th vertex and the $s_{i+1}$th vertex of $\mathcal{G}_{\varsigma,\vartheta}$ with an edge for each $i\in\{1,2,\ldots,k-1\}$.
\end{enumerate}
 (We note that an edge joins the first and $N$th vertices of $\mathcal{G}_{\varsigma,\vartheta}$ via item \ref{G3} (resp.\ item \ref{G4}) if and only if a boundary cluster in $\RCX_\varsigma$ joins only these two fixed sides (resp.\ these sides are mutually wired to only each other in $\RCB_\vartheta$).)  Figure \ref{PolyGraph} gives three examples of product diagrams in its top row and their associated graphs beneath.  Now we let
\begin{itemize}
\item $\mathcal{C}_{\varsigma,\vartheta}=$ number of connected components in $\mathcal{G}_{\varsigma,\vartheta}$,
\item $\mathcal{I}_{\varsigma,\vartheta}=$ number of internal faces in $\mathcal{G}_{\varsigma,\vartheta}$,
\item $\mathcal{E}_{\varsigma}^i=$ number of ``internal" edges in $\mathcal{G}_{\varsigma,\vartheta}$, i.e., the edges arising from step \ref{G3} in the construction of $\mathcal{G}_{\varsigma,\vartheta}$,
\item $\mathcal{E}_{\vartheta}^e=$ number of ``external" edges in $\mathcal{G}_{\varsigma,\vartheta}$, i.e., the edges arising from step \ref{G4} in the construction of $\mathcal{G}_{\varsigma,\vartheta}$,
\item $\mathcal{V}_{\varsigma,\vartheta}=$ number of vertices in $\mathcal{G}_{\varsigma,\vartheta}$.
\end{itemize}

%%%%%%%%%%%%%%%%%%%%%%%%%%%%%%%%%%%%%%%%%%%%%%%%%%%%%%%%%%%%%%%
\begin{figure}[t]
\centering
\includegraphics[scale=0.34]{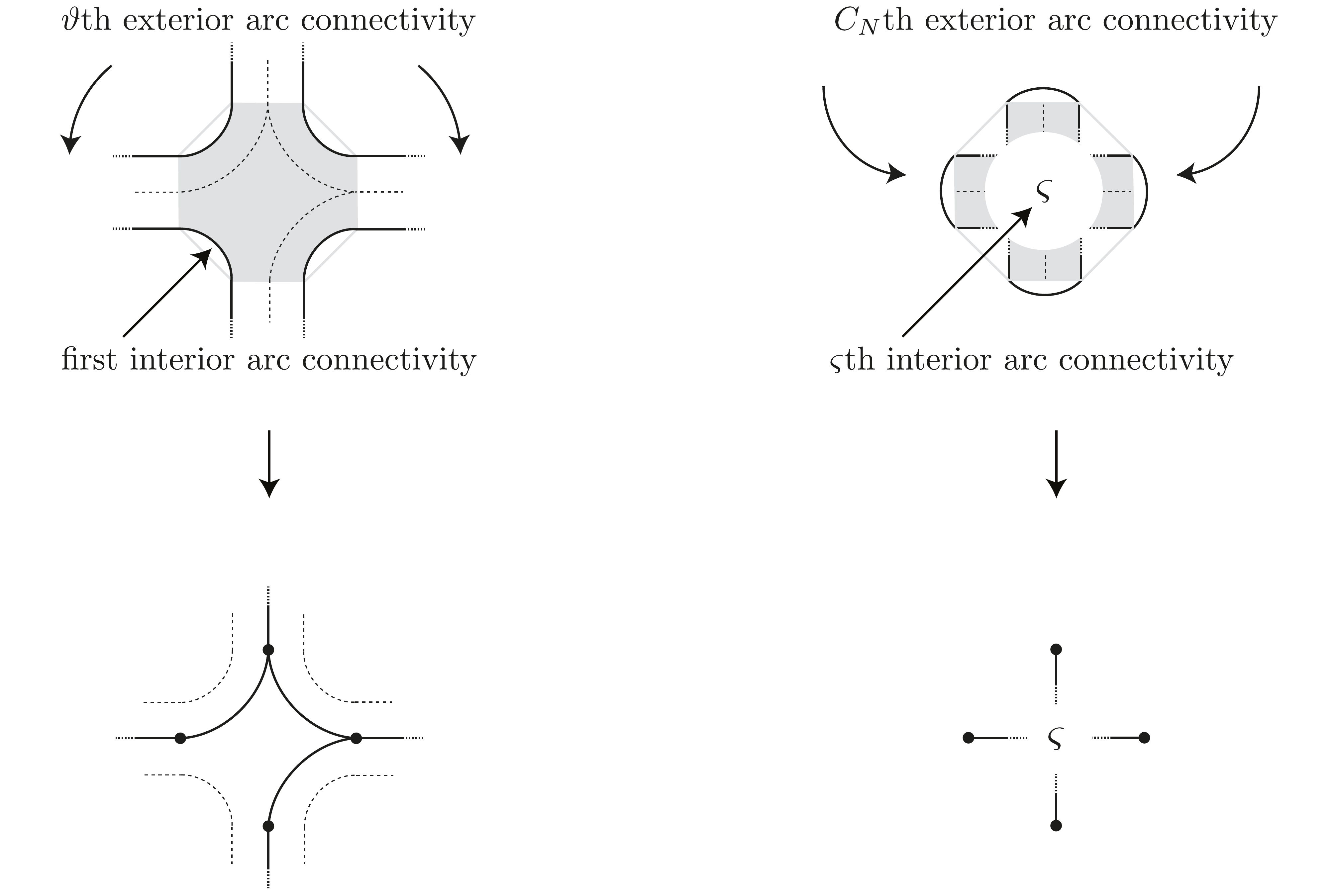}
\caption{The first (resp. $\varsigma$th)  interior arc connectivity  pairs with the $\vartheta$th (resp.\ $C_N$th)  exterior arc connectivity   in the upper-left (resp.\ upper-right).  Its graph $\mathcal{G}_{1,\vartheta}$ (resp.\ $\mathcal{G}_{\varsigma,C_N}$) appears beneath it.  Dashed ends continue into the part of the graph not shown. }
\label{PolyGraph2}
\end{figure}

%%%%%%%%%%%%%%%%%%%%%%%%%%%%%%%%%%%%%%%%%%%%%%%%%%%%%%%%%%%%%%%
It is evident from steps \ref{G3} and \ref{G4} that the number $\mathcal{C}_{\varsigma,\vartheta}$ of components in $\mathcal{G}_{\varsigma,\vartheta}$ equals the number of  independent  boundary clusters in the event $\RCX_\varsigma\cap\RCB_\vartheta$, also denoted by $\mathcal{C}_{\varsigma,\vartheta}$.  Euler's formula relates all of these quantities together as
\be\label{EulerG} \mathcal{C}_{\varsigma,\vartheta}-\mathcal{I}_{\varsigma,\vartheta}+\mathcal{E}_{\varsigma,\vartheta}-\mathcal{V}_{\varsigma,\vartheta}=0,\quad \text{where $\mathcal{E}_{\varsigma,\vartheta}:=\mathcal{E}_{\varsigma}^i+\mathcal{E}_{\vartheta}^e$}.\ee
Figure \ref{PolyGraph} shows that the loops in the $(\varsigma,\vartheta)$th product diagram trace all of the internal faces and surround all of the connected components of $\mathcal{G}_{\varsigma,\vartheta}$ (seen as embedded on the Riemann sphere), with one loop per face and one loop per component.  Also, it is obvious that the number $\mathcal{V}_{\varsigma,\vartheta}$ of vertices in $\mathcal{G}_{\varsigma,\vartheta}$ is $N$.   Thus,
\be\label{lp} l_{\varsigma,\vartheta}=\mathcal{C}_{\varsigma,\vartheta}+\mathcal{I}_{\varsigma,\vartheta},\qquad\mathcal{V}_{\varsigma,\vartheta}=N.\ee
To derive formula (\ref{pformula}) for $\mathcal{C}_{\varsigma,\vartheta}$ from (\ref{EulerG}, \ref{lp}) we consider two extreme cases.  In the first case, we take $\varsigma=1$.  According to item \ref{it1}  in section \ref{RCxingSect}, a single boundary cluster connects all fixed sides of $\mathcal{P}$ in the first  crossing  event $\RCX_1$.  The left column of figure \ref{PolyGraph2} depicts this event, with the $(1,\vartheta)$th product diagram on top of it and its associated graph $\mathcal{G}_{1,\vartheta}$ beneath.  This graph has one component with $N-1$ internal edges joining the vertices.  Hence,
\be \mathcal{C}_{1,\vartheta}=1,\qquad \mathcal{E}_1^i=N-1.\ee
In the second case, we take $\vartheta=C_N$.  According to items \ref{it1} and \ref{it5}  in section \ref{RCxingSect}, all fixed sides of $\mathcal{P}$ are independently wired in the event $\RCB_{\text{$C_N$}}$.  The right column of figure \ref{PolyGraph2} depicts this event, with the $(\varsigma,C_N)$th product diagram on top of it and its associated graph $\mathcal{G}_{\varsigma,C_N}$ beneath.  This graph has no external edges.  Furthermore, it has no internal faces because any face is formed by at least one external edge meeting one internal edge.  Hence,
\be\label{2ndExtreme}\mathcal{I}_{\varsigma,C_N}=0,\qquad \mathcal{E}_{C_N}^e=0.\ee
Combining (\ref{EulerG}--\ref{2ndExtreme})  gives the expression  (\ref{pformula}) for $\mathcal{C}_{\varsigma,\vartheta}$, relating the effective number of independent boundary clusters in the event $\RCX_\varsigma\cap\RCB_\vartheta$ with the number $l_{\varsigma,\vartheta}$ of loops in the $(\varsigma,\vartheta)$th product diagram (figure \ref{innerproduct}).  In section \ref{Onmodel}, we use this connection to interpret the partition functions (\ref{randFFBCadapt2}, \ref{ZQoutFactored}) as  instances of loop-gas model partition functions.

%%%%%%%%%%%%%%%%%%%%%%%%%%%%%%%%%%%%%%%%%%%%%%%%%%%%%%%%%%%%%%%
%%%%%%%%%%%%%%%%%%%%%%%%%%%%%%%%%%%%%%%%%%%%%%%%%%%%%%%%%%%%%%%

\subsection{Potts model crossing events with fluctuating/fixed boundary conditions}\label{FLBCsect}

 Although it seems natural to consider free/fixed side-alternating BC events in the Potts model, determining crossing probabilities conditioned on these events is beyond the scope of the methods used in this article.  Instead, our methods (section \ref{CFTsect}) are suited to $\PMB$ events for the Potts model, as  defined in item \ref{BCitem2} in the introduction \ref{intro}.  These are side-alternating BC events in which  every fixed side exhibits the first spin state and the remaining sides exhibit the ``fluctuating BC," where sites uniformly sample all but the first spin state \cite{gaca}.  (See the left illustration of figure \ref{SpinSea}.)   In parallel with items \ref{it1}--\ref{it8} at the beginning of section \ref{RCxingSect}, we define the following for the Potts model:
\begin{enumerate}
\item\label{it13}  For the Potts model, crossing events are now events in which spin clusters join the fixed sides of $\mathcal{P}$ in some particular  crossing pattern.  We enumerate these events as in item \ref{it1}  in section \ref{RCxingSect}.
\item\label{it23} We let $\PMX_\varsigma$ denote the $\varsigma$th spin-cluster crossing event, and we represent it by the $\varsigma$th interior arc connectivity (figures \ref{OctXingConfigsBdyArcs}).
\item\label{it33} We let $\PMB$ denote the Potts model fluctuating/fixed side-alternating BC event (we note that, as defined, there is only one), and we  represent it by the first exterior arc connectivity (figure \ref{ExtremeCases}).
\item\label{it43} We let $Y_{\varsigma,1}^\mathcal{P}( K,Q)$ denote the partition function for  the $Q$-state Potts model on  a lattice inside $\mathcal{P}$, summing exclusively over the intersection $\PMX_\varsigma\cap\PMB$ of the $\varsigma$th crossing event with the $\PMB$ event.
\item\label{it53} We let $Y_1^\mathcal{P}(K,Q)$ denote the partition function for  the $Q$-state Potts model on  a lattice  inside $\mathcal{P}$, summing exclusively over the $\PMB$ event.  Hence,  $Y_1^\mathcal{P}=Y_{1,1}^\mathcal{P}+Y_{2,1}^\mathcal{P}+\dotsm+Y_{C_N,1}^\mathcal{P}.$
\end{enumerate}
(The subscript 1 in $\smash{Y_{\varsigma,1}^\mathcal{P}(K,Q)}$ and $Y_1^\mathcal{P}(K,Q)$ of items \ref{it43} and \ref{it53} is natural to use, as will become evident in section \ref{Onmodel}.)  
In terms of these quantities, the probability of the $\varsigma$th spin-cluster crossing event conditioned on the $\PMB$ event is 
\be\label{chixing2} P_{\varsigma|1}^\mathcal{P}(K,Q):=\mathbb{P}_{\text{Potts}}(\PMX_\varsigma\,|\,\PMB)=\frac{Y_{\varsigma,1}^\mathcal{P}(\text{$K,Q$})}{Y_1^\mathcal{P}(\text{$K,Q$})}.\ee
In section \ref{CFTsect}, we determine an explicit formula for the continuum limit of the crossing probabilities (\ref{chixing}, \ref{chixing2}) encountered so far.  For the former, we condition the random cluster model on an $\RCB$ event, and for the latter, we condition the Potts model on a $\PMB$ event.   As mentioned, the Potts model with $\RCB$ conditioning is outside the scope of our methods.  The same is true for the random cluster model with $\PMB$ conditioning.

\section{Convergence to loop-gas models}\label{Onmodel}

In this section, we show that partition functions of the random cluster model and the Potts model that we encountered in sections \ref{mutsect}, \ref{RCxingSect}, and \ref{FLBCsect} are asymptotic to partition functions of certain loop-gas models \cite{stan,nein,nein2,nein3} in the continuum limit.  In section \ref{CFTsect}, these identifications lead us, finally, to the formula (\ref{xing}) for the crossing probabilities (\ref{chixing}, \ref{chixing2}) in the continuum limit.  (Section \ref{xingsummary} explains how to make this formula completely explicit for the random cluster model and the Potts model.  Section \ref{rectxingsummary} and section \ref{simxing} give these explicit formulas if the polygon is a rectangle or a hexagon respectively.)

\subsection{Definition of loop-gas models}\label{DefLoopGasSect}

\begin{figure}[t]
\centering
\includegraphics[scale=0.27]{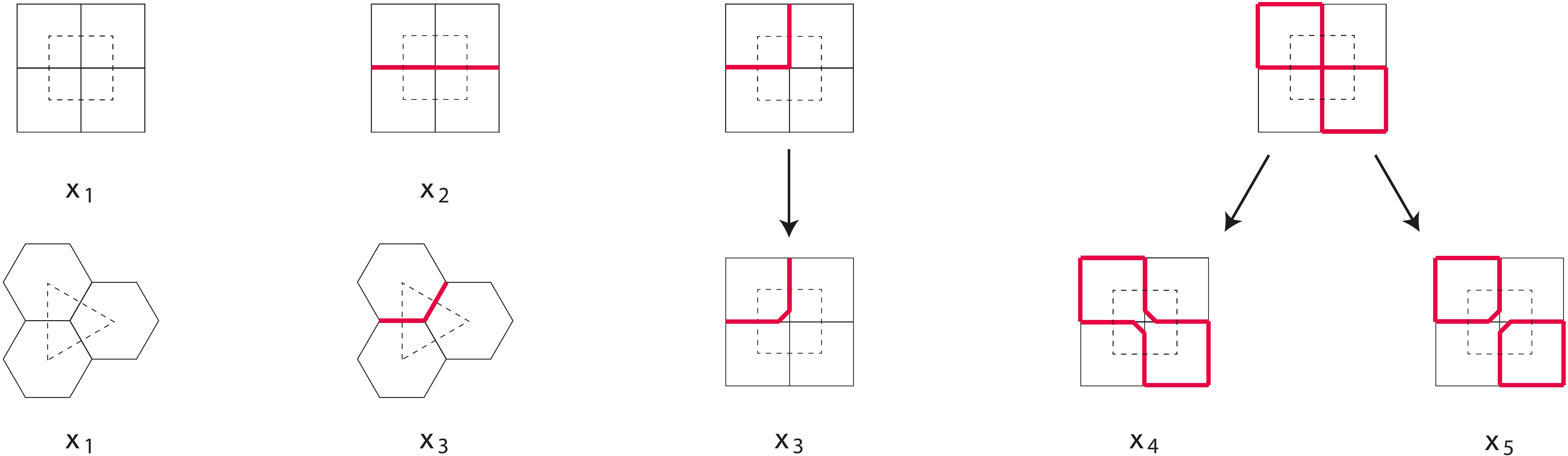}
\caption{Scenarios in which one or two loops comprising bonds on the square lattice or hexagonal lattice may interact with a single lattice site.  Dashed segments and their endpoints correspond with bonds and lattice sites on the dual lattice.}
\label{PullLoop}
\end{figure}

We begin with a definition of a loop-gas model.  The sample space of a loop-gas model on a given graph inside $\mathcal{P}$ is the collection of all sets $\{\gamma\}$ of non-crossing and non-self-crossing loops $\gamma$ that trace the graph's edges \cite{jlj,blote}.  In this section, we assume that the graph is a square or hexagonal lattice, with edges being bonds between nearest-neighbor lattice sites.
Figure \ref{PullLoop} shows the five scenarios in which loops may interact with a site of the square lattice:
\begin{enumerate}
\item\label{loop1} No loop passes through the lattice site.
\item A loop passes through the site without turning.
\item\label{loop3} A loop touches the site, turns, and never returns to the site. 
\item\label{loop4} A loop touches the site, turns, and eventually returns to the site.
\item\label{loop5} Two distinct loops touch a common site and turn.
\end{enumerate}
In a typical \emph{loop-gas model}, we endow the $k$th scenario with some weight $x_k\geq0$, and we endow a sample collection of loops $\{\gamma\}$ with weight
\be\label{weights}\omega(\{\gamma\})=n^{N_\ell}x_1^{N_1}x_2^{N_2}x_3^{N_3}x_4^{N_4}x_5^{N_5},\quad\text{where}\,\,\begin{cases}\text{$N_\ell$ is the number of loops in the sample $\{\gamma\}$}, \\ \text{$N_k$ is the number of $k$th scenario sites in the sample $\{\gamma\}$}, \\ \text{$n>0$ is the \emph{loop fugacity} of the loop-gas model.}\end{cases}\ee
(One may consider loop-gas models with $n\leq0$, but we do not use them here.)  Because scenarios \ref{loop4} and \ref{loop5} appear identical, (i.e., the information contained within the dashed box of either scenario in figure \ref{PullLoop} is the same), we typically set $x_4=x_5$.  If we also set $x_1=1$, $x_2=x_3=x$, and $x_4=x_5=x^2$ for some  \emph{temperature} $x\geq0$, then (\ref{weights}) simplifies to
\be\label{weights2}\omega(\{\gamma\})=n^{N_\ell}x^\rho,\quad \text{where $\rho:=$ total length of (i.e., number of bonds in) all loops of $\{\gamma\}$}.\ee
Figure \ref{PullLoop} also shows the two different ways in which the loops may interact with a single site on the hexagonal lattice.  Because only scenarios \ref{loop1} and \ref{loop3} occur, we set $x_2=x_4=x_5=1$ in the expression (\ref{weights}) for the weight of the loop sample $\{\gamma\}$.  Setting $x_1=1$ and $x_3=x$, this expression again reduces to (\ref{weights2}).  The partition functions for these models are
\be\label{lgpart}
 Z_f^\mathcal{P}(x,n)=\sum_\text{L}\omega(\{\gamma\}),\quad \text{where L $:=$ the sample space of the loop-gas model.}\ee
The subscript $f$  (for ``free") means that we sum over the entire sample space, without imposing any BCs such as those described below.  (We mention that this loop-gas model (\ref{weights2}, \ref{lgpart}) is closely related to the O$(n)$ model \cite{stan} of statistical physics.)  In what follows, we identify the loop-gas model partition function with the random cluster model and Potts model partition functions previously encountered in sections \ref{mutsect} and \ref{FLBCsect}.

%%%%%%%%%%%%%%%%%%%%%%%%%%%%%%%%%%%%%%%%%%%%%%%%%%%%%%%%%%%%%%%%%%%%%%%%%%%%%%%%%%%%%%%%%%%%%%%%%%%%%%%%%%%%%%%%%%%%%%%%%
%%%%%%%%%%%%%%%%%%%%%%%%%%%%%%%%%%%%%%%%%%%%%%%%%%%%%%%%%%%%%%%%%%%%%%%%%%%%%%%%%%%%%%%%%%%%%%%%%%%%%%%%%%%%%%%%%%%%%%%%%

The loop-gas model (\ref{weights2}) exhibits critical behavior in its continuum limit.  This critical behavior is well understood on the hexagonal lattice \cite{blni,nein,smir,rbgw}.  Under the renormalization group flow induced by shrinking the lattice spacing $a$, the temperature $x$ of (\ref{weights2}) flows to either zero if it is less than some critical temperatures $x_c$ and $n\in[-2,2]$,  or to another value $\tilde{x}_c>x_c$ if it is greater than $x_c$ and $n\in[0,2]$.  On the hexagonal lattice, we have \cite{smir}
\be\label{fixedpts}x_c(n)=(2+\sqrt{2-n})^{-1/2},\quad \tilde{x}_c(n)=(2-\sqrt{2-n})^{-1/2}.\ee
If $x<x_c$ and $n\in[0,2]$, then as $a\downarrow0$, the model strongly favors small-diameter loops over large-diameter loops,  and its continuum limit contains only infinitesimal loops.  However, if $x>x_c$, then the continuum limit contains loops of all sizes.  Moreover, the fraction of all lattice sites that a typical loop visits does not vanish as $a\downarrow0$.  For this reason, we call the interval $x>x_c$ the \emph{dense phase}.  In the continuum limit, the loops are conjectured to have the conformally invariant law of CLE$_\kappa$ \cite{shefwer,sheffield,doyon}, with $\kappa\in(4,8]$ and 
\be\label{fugacity} n=n(\kappa):=-2\cos(4\pi/\kappa).\ee
 Finally, if $x=x_c$ and $n\in[-2,2]$, then $x$ is invariant under the renormalization group flow.  As $a\downarrow0$, the model still strongly favors large-diameter loops over small-diameter loops, but the fraction of all lattice sites that a typical loop visits vanishes.  For this reason, we call $x=x_c$ the \emph{dilute phase}.  In the continuum limit, the loops are conjectured to  have  the conformally invariant law of CLE$_\kappa$ \cite{shefwer,sheffield,doyon}, for $\kappa\in(8/3,4]$ and $n\in(0,2]$ related by (\ref{fugacity}).
 
 %%%%%%%%%%%%%%%%%%%%%%%%%%%%%%%%%%%%%%%%%%%%%%%%%%%%%%%%%%%%%%%%%%%%%%%%%%%%%%%%%%%%%%%%%%%%%%%%%%%%%%%%%%%%%%%%%%%%%%%%%%
\begin{figure}[t]
\centering
\includegraphics[scale=0.16]{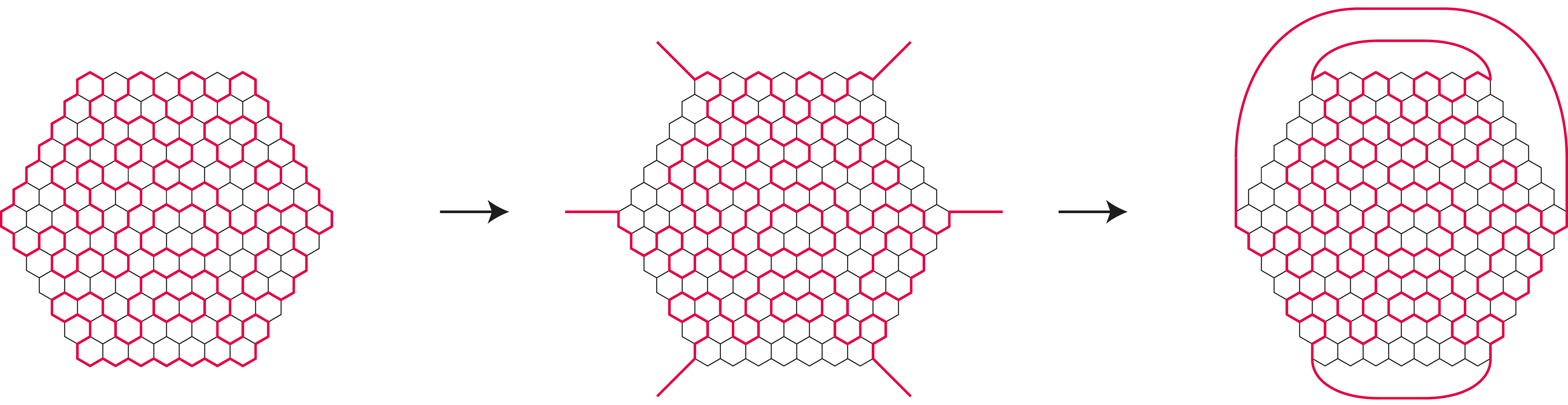}
\caption{In the left hexagon, each odd side is traced by a boundary loop.  In the middle hexagon, we cut the part of each boundary loop that traces a side, and in the right hexagon, we rejoin their ends in an exterior arc connectivity.}
\label{CutForm}
\end{figure}
%%%%%%%%%%%%%%%%%%%%%%%%%%%%%%%%%%%%%%%%%%%%%%%%%%%%%%%%%%%%%%%%%%%%%%%%%%%%%%%%%%%%%%%%%%%%%%%%%%%%%%%%%%%%%%%%%%%%%%%%%%
%
 
A loop-gas model on a lattice inside an even-sided polygon $\mathcal{P}$ admits certain events that are analogous to the random cluster model $\RCX$ and $\RCB$ events in $\mathcal{P}$.  To realize this analogy, we restrict the loop-gas sample space to all configurations such that each odd side of $\mathcal{P}$  (see above (\ref{catalan}) for the enumeration of the sides) is traced from vertex to vertex by one loop, called a \emph{boundary loop} (figure \ref{CutForm}).  These boundary loops necessarily join the vertices of $\mathcal{P}$ pairwise through its interior in one of the $C_N$ available connectivities.  If it is the $\varsigma$th connectivity, then we regard this event as the analog of the $\varsigma$th crossing event $\RCX_\varsigma$ in the random cluster model.  Furthermore, we may bisect the part of the boundary loops that follow the odd sides of $\mathcal{P}$ and rejoin the $2N$ dangling ends with nonintersecting exterior arcs outside $\mathcal{P}$ (figure \ref{CutForm}).  If the exterior arcs join the vertices of $\mathcal{P}$ in the $\vartheta$th connectivity, then we regard this event as the analog of the $\vartheta$th mutual wiring event $\RCB_\vartheta$ in the random cluster model.  Thus, we define the following:
%%%%%%%%%%%%%%%%%%%%%%%%%%%%%%%%%%%%%%%%%%%%%%%%%%%%%%%%%%%%%%%%%%%%%%%%%%%%%%%%%%%%%%%%%%%%%%%%%%%%%%%%%%%%%%%%%%%%%%%%%%
\begin{figure}[b]
\centering
\includegraphics[scale=0.3]{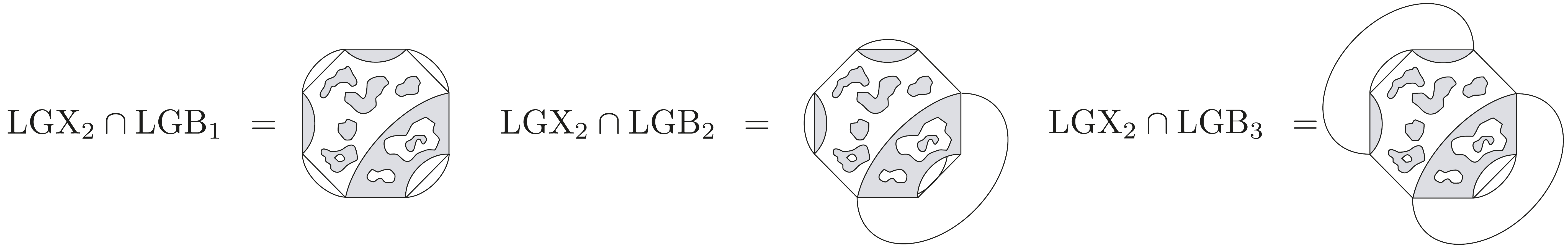}
\caption{Typical loop-gas model samples in the event $\LGX_\varsigma\cap\LGB_\vartheta$.  For loop fugacities among (\ref{Qdense}) (resp.\ (\ref{Qdilute})), gray regions correspond to FK (resp.\ spin) boundary clusters in the event $\RCX_\varsigma\cap\RCB_\vartheta$ (resp.\ $\PMX_\varsigma\cap\PMB_\vartheta$).}
\label{On}
\end{figure}
%%%%%%%%%%%%%%%%%%%%%%%%%%%%%%%%%%%%%%%%%%%%%%%%%%%%%%%%%%%%%%%%%%%%%%%%%%%%%%%%%%%%%%%%%%%%%%%%%%%%%%%%%%%%%%%%%%%%%%%%%%
%
\begin{enumerate}
\item  For the loop-gas model, the $\varsigma$th crossing ($\LGX$) event (resp.\ $\vartheta$th exterior arc BC ($\LGB$) event) is the event that the boundary loops join  the vertices of $\mathcal{P}$ via the $\varsigma$th interior (resp.\ $\vartheta$th exterior) arc connectivity.
\item\label{it22}  We let $\LGX_\varsigma$ denote the $\varsigma$th loop-gas model crossing ($\LGX$) event, and we represent it by the $\varsigma$th interior arc connectivity (figure \ref{OctXingConfigsBdyArcs}).
\item\label{it32}  We let $\LGB_\vartheta$ denote the $\vartheta$th exterior arc BC ($\LGB$) event, and we represent it by the $\vartheta$th exterior arc connectivity (figure \ref{OctXingConfigsBdyArcs}).
\item\label{it42}  We let $\smash{\stZ_{\varsigma,\vartheta}^\mathcal{P}(x,n)}$ denote the partition function (\ref{lgpart}) for the loop-gas model on a lattice inside $\mathcal{P}$, summing exclusively over the intersection $\LGX_\varsigma\cap\LGB_\vartheta$ of the $\varsigma$th $\LGX$ event with the $\vartheta$th $\LGB$ event.
\item\label{it52}  We let $\smash{\stZ_\vartheta^\mathcal{P}(x,n)}$ denote the partition function (\ref{lgpart}) for the loop-gas model on a lattice inside $\mathcal{P}$, summing exclusively over the $\vartheta$th $\LGB$ event $\LGB_\vartheta$.  Hence, $\smash{\stZ_\vartheta^\mathcal{P}=\stZ_{1,\vartheta}^\mathcal{P}+\stZ_{2,\vartheta}^\mathcal{P}+\dotsm+\stZ_{C_N,\vartheta}^\mathcal{P}}.$
\end{enumerate}
Figure \ref{On} illustrates the event $\LGX_\varsigma\cap\LGB_\vartheta$.  The number of boundary loops in this event always equals $l_{\varsigma,\vartheta}$ (figures \ref{innerproduct}, \ref{On}).  Hence, writing the total number of loops as a the sum of boundary loops and the remaining \emph{free loops} gives
\be\label{loopgaspart}\stZ_{\varsigma,\vartheta}^\mathcal{P}(x,n)=\sum_{\LGX_\varsigma\cap\LGB_\vartheta}\omega(\{\gamma\})=n^{l_{\varsigma,\vartheta}}\sum_{\LGX_\varsigma}n^{N_\ell^\text{free}}x^\rho.\ee
The rightmost expression in (\ref{loopgaspart}) only depends on our choice of $\LGB$ event (i.e., on $\vartheta$) via the factor $n^{l_{\varsigma,\vartheta}}$.   The sum that multiplies this factor depends on the crossing event $\LGX_\varsigma$, and we interpret it as a partition function with the fugacities of all boundary loops equaling one.

%%%%%%%%%%%%%%%%%%%%%%%%%%%%%%%%%%%%%%%%%%%%%%%%%%%%%%%%%%%%%%%%%%%%%%%%%%%%%%%%%%%%%%%%%%%%%%%%%%%%%%%%%%%%%%%%%%%%%%%%%
\begin{figure}[t]
\centering
\includegraphics[scale=0.34]{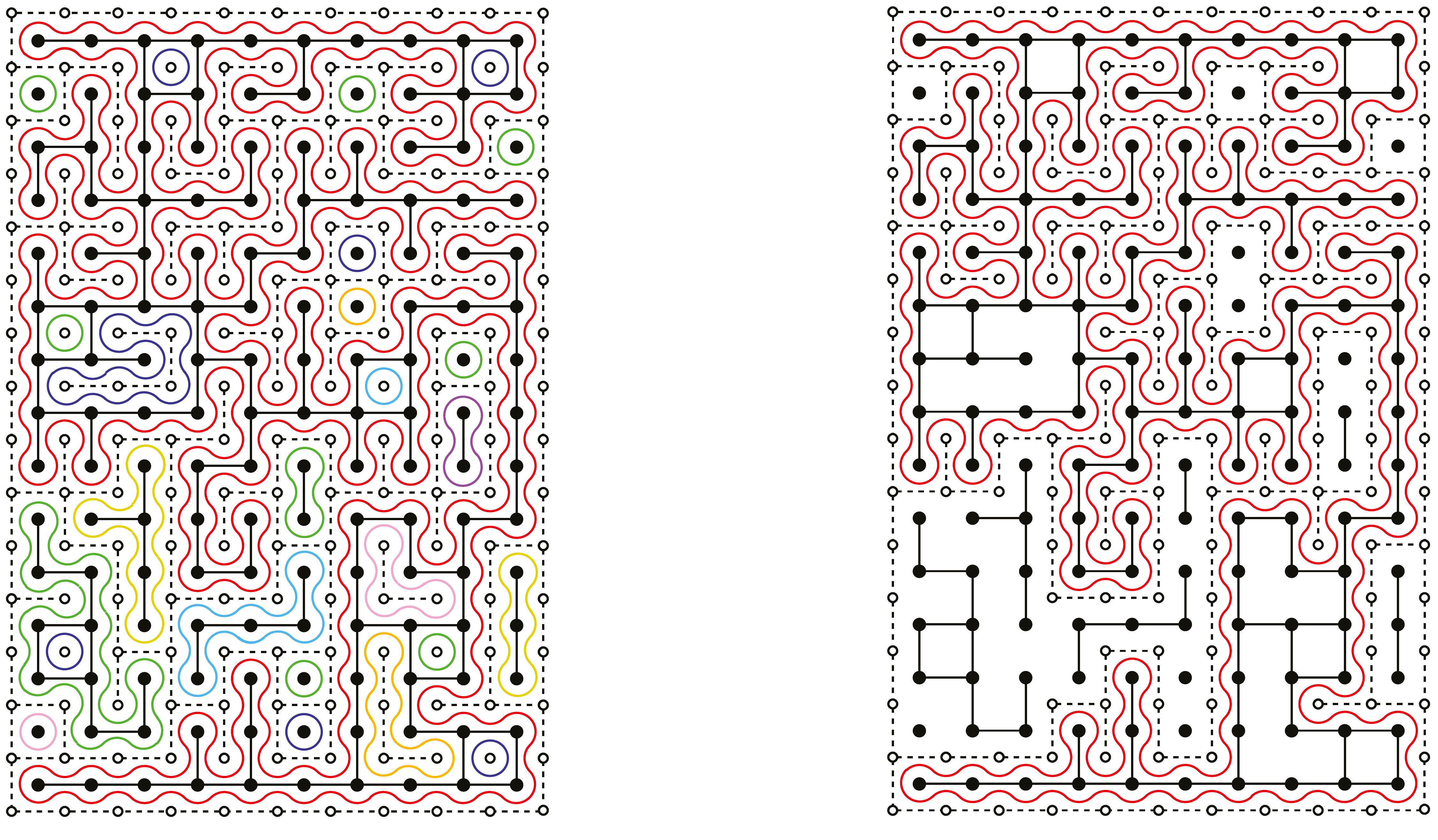}
\caption{Free loops (any color but red) surround free bond clusters and dual-bond clusters (i.e., internal faces of bond clusters), and boundary loops (red) surround boundary bond clusters.  The right illustration shows only the boundary loops.}
\label{NoExternalBonds}
\end{figure}
%%%%%%%%%%%%%%%%%%%%%%%%%%%%%%%%%%%%%%%%%%%%%%%%%%%%%%%%%%%%%%%%%%%%%%%%%%%%%%%%%%%%%%%%%%%%%%%%%%%%%%%%%%%%%%%%%%%%%%%%%

\subsection{Critical random cluster models in relation to loop-gas models}\label{RCLoopGasSect}

Turning to the critical random cluster model, our next goal is to identify each basic partition function $\smash{X_{\varsigma,\vartheta}^\mathcal{P}(\text{$p,Q$})}$ (\ref{randFFBCadapt1}) with the corresponding partition function $\stZ_{\varsigma,\vartheta}^\mathcal{P}(x,n)$ of a loop-gas model on the square lattice.  This identification is well known for the free partition function (\ref{ZQ}, \ref{lgpart}) \cite{dupl}; we adapt this argument to our purpose here.  To begin, we map each bond configuration in $\text{B}_\varsigma$  (\ref{Bsigma}) onto a unique loop configuration in which  the perimeter of each bond cluster and dual-bond cluster is traced by one loop particular to that cluster (figure \ref{NoExternalBonds}).  (This includes size-zero clusters, which are isolated lattice sites and isolated dual lattice sites.)  Ref.\ \cite{dupl} describes this map  in detail.   As in section \ref{mutsect}, Euler's formula for planar graphs  gives
\be\label{eulerformula}2N_\text{c}+N_\beta=N_\ell+N_\text{s},\quad\text{where}\quad\begin{cases}N_\text{c} = \text{number of activated bond clusters}, \\ \text{$N_\beta$ = number of activated bonds}, \\ N_\ell = \text{number of loops surrounding clusters}, \\ N_\text{s} = \text{number of lattice sites}.\end{cases}
 \ee
In every bond configuration, we distinguish boundary clusters, which touch the fixed sides of $\mathcal{P}$, from the remaining free clusters,  which do not, and we distinguish boundary loops, which surround these boundary clusters, from  free loops, which surround the remaining  free  clusters (figure \ref{NoExternalBonds}).  Denoting the number of each of these as 
 $\smash{N_\text{c}^\text{bdry}}$, $\smash{N_\text{c}^\text{free}}$,  $\smash{N_\ell^\text{bdry}}$, and $\smash{N_\ell^\text{free}}$ respectively, we have
\be\label{Nrelations} N_\text{c}=\smash{N_\text{c}^\text{bdry}}+\smash{N_\text{c}^\text{free}},\qquad N_\ell=\smash{N_\ell^\text{bdry}}+\smash{N_\ell^\text{free}},\qquad \smash{N_\text{c}^\text{bdry}}=\smash{N_\ell^\text{bdry}}=l_{\varsigma,C_N}\quad \text{for $\{\beta\}\in\text{B}_\varsigma$}.\ee
The last equality of (\ref{Nrelations}) follows from the facts that, in the $\varsigma$th connectivity, a unique boundary loop surrounds every  boundary cluster and that these loops are topologically identical to those in the $\smash{(\varsigma,C_N)}$th product diagram (figure \ref{innerproduct}), where interior  arcs close into boundary loops by wrapping outside and around the fixed (i.e., odd-numbered) sides of $\mathcal{P}$.  After inserting (\ref{criticalpt}, \ref{Ndecomp}) into the partition function formula (\ref{randFFBCadapt1}), we find
\be\label{preXgas}X_{\varsigma,\vartheta}^\mathcal{P}(\text{$p$}_c,Q )=(1+\sqrt{Q})^{N_\text{b}}Q^{\mathcal{C}_{\varsigma,\vartheta}}\sum_{\text{B}_\varsigma}Q^{N^\text{free }_{\text{c}}+N_\beta/2}.\ee
The factor of $(1+\sqrt{Q})^{N_\text{b}}$ is physically irrelevant, so we drop it.  Then after inserting (\ref{eulerformula}, \ref{Nrelations}) into (\ref{preXgas}) and using (\ref{pformula}) to simplify the result, we find
\be\label{Xgas}X_{\varsigma,\vartheta}^\mathcal{P}(\text{$p$}_c,Q )=Q^{(l_{\varsigma,\vartheta}-l_{1,\vartheta}+1)/2}\sum_{\text{B}_\varsigma}Q^{(N_\text{s}+N^\text{free }_\ell)/2}.\ee
Finally, after dropping irrelevant factors and straightening  loops  into zig-zag paths along bonds on the lattice medial \cite{jlj} to the square lattice in $\mathcal{P}$ (figure \ref{NoExternalBonds}),  (\ref{Xgas}) becomes  a partition function for a loop-gas model on the medial lattice:
\begin{align}\label{Xgas1}X_{\varsigma,\vartheta}^\mathcal{P}(\text{$p$}_c,Q )&\propto n^{l_{\varsigma,\vartheta}}\sum_{\text{B}_\varsigma}n^{N_\ell^\text{free}},\qquad n:=\sqrt{Q},\quad\text{$N_\ell^\text{free}$ defined above (\ref{Nrelations}),}&&\parbox{1.3in}{random cluster model on the square lattice,}\\
\label{Xgas2}&=n^{l_{\varsigma,\vartheta}}\sum_{\LGX_\varsigma}n^{N_\ell^\text{free}},\qquad\begin{gathered}\text{$x_1=x_2=x_3=0$ for bulk sites},\\ \text{$x_i=1$ otherwise},\end{gathered}&&\parbox{1.1in}{loop-gas model on the medial lattice.}\end{align}
The condition that $x_1=x_2=x_3=0$ for all bulk (i.e., interior) sites implies that only loop configurations with each bulk site visited twice by loops contributes to the sum (\ref{Xgas2}) (figure \ref{PullLoop}).  Because the other loop configurations do not contribute, we cannot equate (\ref{Xgas2}) with $\smash{\stZ_{\varsigma,\vartheta}^\mathcal{P}}(x,n)$ for $x=1$.  However, the temperature $x=1$ is greater than the critical temperature $x_c(n)$ of the loop-gas model on the square lattice \cite{blonien}, so it flows under renormalization to the dense-phase fixed point $\tilde{x}_c(n)$ as we take the continuum limit.  Because scenarios \ref{loop4} and \ref{loop5} statistically dominate scenarios \ref{loop1}--\ref{loop3} in the dense phase, (\ref{Xgas2}) is at least asymptotic to (a multiple of) $\smash{\stZ_{\varsigma,\vartheta}^\mathcal{P}}(1,n)$ in the continuum limit.
 
\subsection{Critical Potts model in relation to loop-gas models}\label{PMLoopGasSect}

Turning next to the critical Potts model on some given lattice in $\mathcal{P}$, our next goal is to show that the $\PMB$ partition function $\smash{ Y_{\varsigma,1 }^\mathcal{P}(K_c,Q)}$ is asymptotic to (a multiple of) the partition function $\smash{\stZ_{\varsigma,\vartheta}^\mathcal{P}(x,n)}$ for the loop-gas model on the dual lattice.  This is easy to do rigorously at any temperature if $Q=2$ (i.e., the Ising model), but if $Q\in\{3,4\}$, the arguments are not rigorous and are somewhat imprecise.  The simulation results of section \ref{simxing}, however, strongly suggest that these arguments are in fact correct, especially if $Q=3$.

%%%%%%%%%%%%%%%%%%%%%%%%%%%%%%%%%%%%%%%%%%%%%%%%%%%%%%%%%%%%%%%%%%%%%%%%%%%%%%%%%%%%%%%%%%%%%%%%%%%%%%%%%%%%%%%%%%%%%%%%%
\begin{figure}[b]
\centering
\includegraphics[scale=0.27]{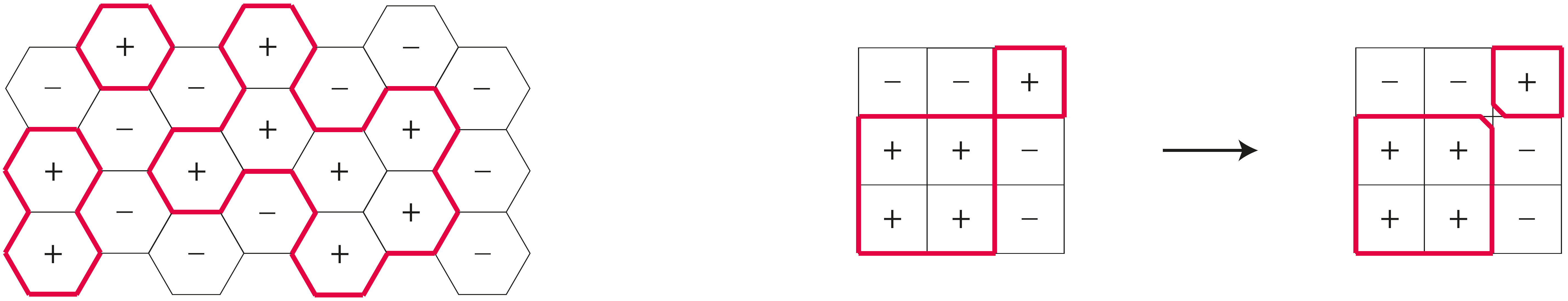}
\caption{An Ising model sample on the triangular lattice (left) and on the square lattice (right) and their respective loop-gas samples.  In the latter, we retract the loops from their intersection to clarify that a unique loop surrounds each spin $+$ cluster.}
\label{Q2Loops}
\end{figure}
%%%%%%%%%%%%%%%%%%%%%%%%%%%%%%%%%%%%%%%%%%%%%%%%%%%%%%%%%%%%%%%%%%%%%%%%%%%%%%%%%%%%%%%%%%%%%%%%%%%%%%%%%%%%%%%%%%%%%%%%%

We begin with the Ising model.  Here, the arguments are already well known \cite{fms}. If the Ising model lives on the triangular lattice, then the loops live on the (dual) hexagonal lattice, and they surround clusters of spin $+$ lattice sites.  Hence, spin configurations on the triangular lattice bijectively correspond to loop-gas configurations on the hexagonal lattice (figure \ref{Q2Loops}).  With this bijection, it is easy to see that 
\be\label{Isinghex}\begin{aligned}
 Y^\mathcal{P}(K,Q=2)&\propto\sum_{\text{L}}x^\rho,\quad x:=\exp(-K),\quad\text{$\rho$ as in (\ref{weights2}),}&&\text{Ising model on the triangular lattice,}\\
&=\sum_{\text{L}}\omega(\{\gamma\}),\quad\begin{gathered}x_1=x_2=x_4=x_5=1,\\ x_3=x,\quad n=1,\end{gathered}&&\text{loop-gas model on the hexagonal lattice.}\end{aligned}\ee
Because (\ref{Isinghex}) equates the Ising model on the triangular lattice with the loop-gas model on the hexagonal lattice, the critical points of these two models must match.  And indeed, they do: setting $Q=2$ in (\ref{pc}) and $n=1$ in (\ref{fixedpts}) gives
\be\label{Kc}K_c^{\text{tri.}}=\log\sqrt{3}\quad \Longrightarrow\quad x(K_c^{\text{tri.}})=\exp(-K_c^{\text{tri.}})=1/\sqrt{3}=x_c(n=1).\ee
 Similarly, every spin configuration on a square lattice corresponds to a unique loop-gas configuration on its dual, also the square lattice, and vice versa, if we exclude scenario \ref{loop4} (figures \ref{PullLoop} and \ref{Q2Loops}).   Thus, for the square lattice,
\be\label{Isingsq}\begin{aligned}
 Y^\mathcal{P}(K,Q=2) &\propto\sum_{\text{L}'}x^\rho,\quad x:=\exp(-K),\quad\text{$\rho$ as in (\ref{weights2}),}&&\text{Ising model on the square lattice,}\\
&=\sum_{\text{L}}\omega(\{\gamma\}),\quad \begin{gathered}x_1=1,\quad x_2=x_3=x,\\ x_4=0,\quad x_5=x^2,\quad n=1,\end{gathered} &&\text{loop-gas model on the dual square lattice,}\end{aligned}\ee
where $\text{L}'$ is the subset of L  that excludes  from L all samples with a scenario \ref{loop4}  vertex.

%%%%%%%%%%%%%%%%%%%%%%%%%%%%%%%%%%%%%%%%%%%%%%%%%%%%%%%%%%%%%%%%%%%%%%%%%%%%%%%%%%%%%%%%%%%%%%%%%%%%%%%%%%%%%%%%%%%%%%%%%
\begin{figure}[t]
\centering
\includegraphics[scale=0.27]{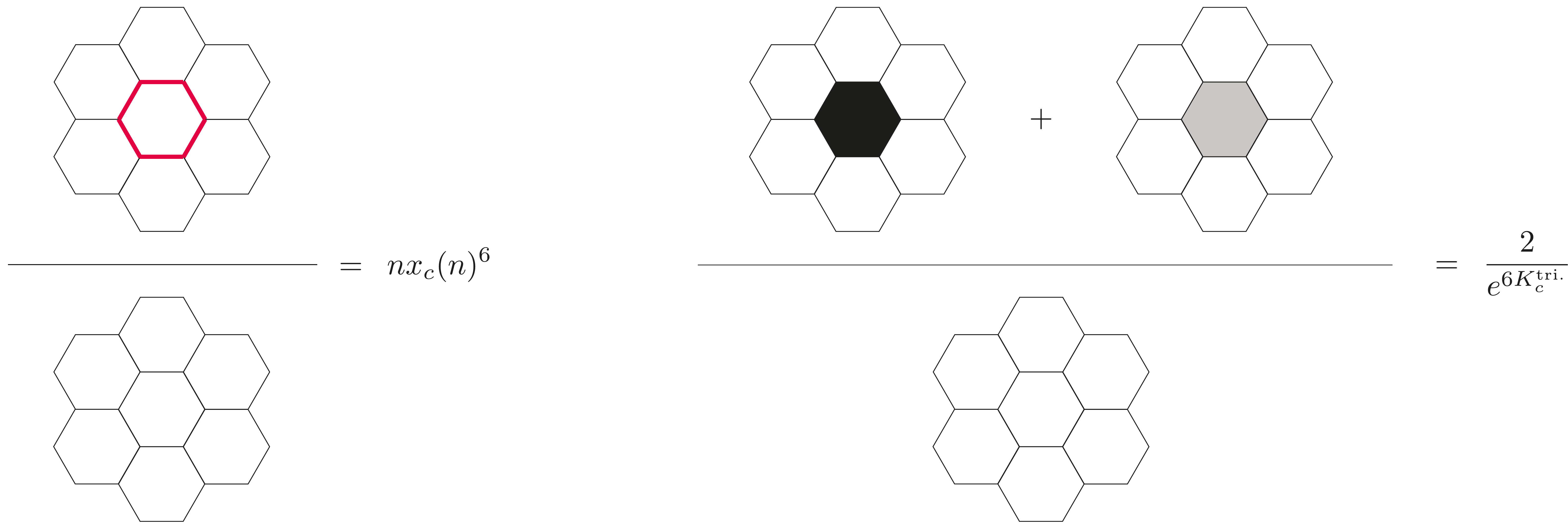}
\caption{Graphical illustrations of (\ref{firstratio}) (left) and (\ref{secondratio}) (right).  In both numerators (resp.\ denominators), the center site exhibits the second (black) or third (gray) (resp.\ first (white)) spin state, and all of the other sites exhibit the first (white) spin state.}
\label{Ratios}
\end{figure}
%%%%%%%%%%%%%%%%%%%%%%%%%%%%%%%%%%%%%%%%%%%%%%%%%%%%%%%%%%%%%%%%%%%%%%%%%%%%%%%%%%%%%%%%%%%%%%%%%%%%%%%%%%%%%%%%%%%%%%%%%

By contrast, equating the partition function (\ref{ZPotts}) for the critical $Q=3$ Potts model with that (\ref{lgpart}) for a loop-gas model in its dilute phase appears to be impossible for finite systems.  To understand why, we consider a sample $\{\sigma_1\}$ of the $Q=3$ Potts model on a triangular lattice in which some given site $P$ and its six neighbors exhibit the first spin state.  From this sample, we generate two more samples $\{\sigma_2\}$ and $\{\sigma_3\}$ by changing the spin at $P$ to the second and third state respectively.  Then the ratio of the total weight of the latter two samples to the first sample's weight is
\be\label{secondratio}\frac{(\dotsm\text{factors from interactions between all sites except $P$}\dotsm)(Q-1)}{(\dotsm\text{factors from interactions between all sites except $P$}\dotsm)\exp(K_c^{\text{tri.}})^6}=0.0453874\ldots,\quad \text{where $Q=3$}.\ee
Next, we translate this scenario into the loop-gas picture.  Numerical results \cite{gaca} indicate that, at least in the continuum limit, the loops  trace the boundaries of clusters with all sites in the first spin state.  Therefore, $\{\sigma_1\}$ corresponds to a loop configuration $\{\gamma\}$ with no loop separating $P$ from any of its neighbors, and the event that either $\{\sigma_2\}$ or $\{\sigma_3\}$ occurs corresponds to the loop configuration $\{\gamma\}\cup\{\gamma_0\}$, where $\{\gamma\}$ is the loop configuration for $\{\sigma_1\}$ and $\gamma_0$ is a loop that surrounds $P$, separating it from its neighbors.  By (\ref{weights2}, \ref{fixedpts}, \ref{pc}), the ratio of the weights of these configurations is
\be\label{firstratio}\frac{\omega(\{\gamma\}\cup\{\gamma_0\})}{\omega(\{\gamma\})}=nx_c(n)^6=\frac{4(1+\sqrt{5})}{(4 +\sqrt{6 - 2 \sqrt{5}})^3}= 0.0901699\ldots,\quad \text{where $n=(1+\sqrt{5})/2$.}\ee
(figure \ref{Ratios}).  (See (\ref{dilute}) for the value of $n$ in (\ref{firstratio}).)  If the identification of the critical $Q=3$ Potts model on the triangular lattice with the dilute-phase loop-gas model on the hexagonal lattice exists, then the ratios (\ref{secondratio}) and (\ref{firstratio}) should be equal.  Because they are not, such an identification does not exist, at least in the discrete setting.  (However, it is true that (\ref{secondratio}, \ref{firstratio}) are equal if $Q=2$ and $n=1$, so $x_c(n)=\smash{1/\sqrt{3}}$, and $\smash{K_c^\text{tri}=\log\sqrt{3}}$ (\ref{Kc}), as must be the case by (\ref{Isinghex}).)

\begin{figure}[t]
\centering
\includegraphics[scale=0.4]{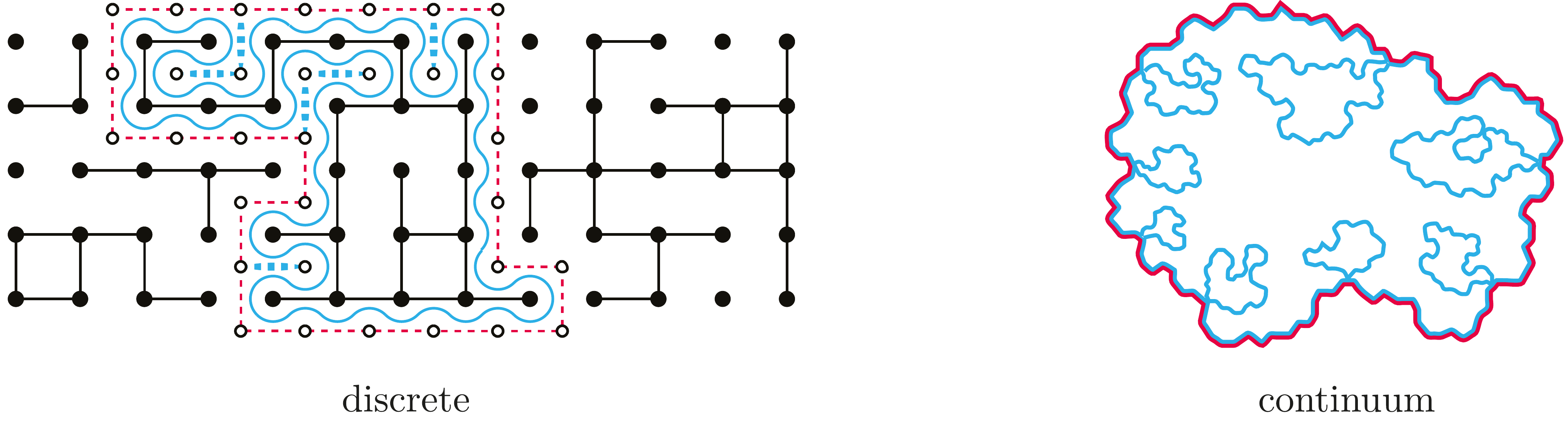}
\caption{The blue loop follows the perimeter of an FK cluster, venturing into its fjords (bloated blue bonds), and the red loop follows the perimeter of a spin cluster, bypassing the fjords.  We form the latter loop by removing fjords from the former loop.}
\label{FKBoundaryLoop}
\end{figure}

Because a dense-phase loop-gas representation of the critical random cluster model emerges in the continuum limit for $Q\in\{1,2,3,4\}$, the association (\ref{relation}) between the random cluster model and the Potts model suggests that a dilute-phase loop-gas representation of the latter,  although apparently  nonexistent in the discrete setting for $Q>2$, emerges in the continuum limit too.   As evidence for this claim, we consider a bulk FK cluster on the square lattice (figure \ref{FKBoundaryLoop}).  Typically, this cluster has fjords entering it, and a loop on the medial lattice that surrounds this cluster visits the mouth of each fjord twice.  Now, as we approach the continuum limit, the loop that surrounds the FK cluster is conjectured to  have  the law of CLE$_\kappa$ \cite{sheffield,shefwer,doyon} for some $\kappa>0$.  Also, the fjords into the FK cluster do not vanish in this limit, so this loop must retain the self-intersection points at the fjord openings.   Now, a loop in a CLE$_\kappa$ sample has self-intersection points only if $\kappa\in(4,8)$ (and has points where the loop very nearly self-intersects at $\kappa=4$).  Indeed, this property follows from the fact that the fractal geometry of a CLE$_\kappa$ loop is locally that of an SLE$_\kappa$ curve \cite{sheffield,shefwer,doyon}, for which this property holds in the SLE$_\kappa$ dense phase $\kappa\in(4,8)$ \cite{rohshr,knk,lawler}.  These facts together with (\ref{fugacity}, \ref{Xgas1}) give
\be\label{dense}\text{$\sqrt{Q}=n(\kappa)=-2\cos(4\pi/\kappa)$ with $\kappa\in[4,8)$}\quad\Longrightarrow\quad \begin{cases} Q=1\quad\Longrightarrow\quad n=1,& \kappa=6, \\
Q=2\quad\Longrightarrow\quad n=\sqrt{2},& \kappa=16/3, \\ 
Q=3\quad\Longrightarrow\quad n=\sqrt{3},& \kappa=24/5, \\  
Q=4\quad\Longrightarrow\quad n=2,&\kappa=4.\end{cases}\ee
Returning to the discrete setting, if we activate the bond at the mouth of each fjord into the FK cluster, then the loop surrounding this cluster becomes simple, and we interpret it as the boundary of the smallest Potts model spin cluster that may contain the given FK cluster.  In the continuum limit,  the simple loop thus created  goes to the previously encountered CLE$_\kappa$ loop,  but  with its fjords removed (figure \ref{FKBoundaryLoop}).  By SLE$_\kappa$ duality \cite{rohshr,knk,dubdul,zhan1,zhan2}, this  latter simple loop must have  the law of CLE$_{\hat{\kappa}}$ with parameter $\hat{\kappa}=16/\kappa\in(8/3,4]$, where $\kappa$ is the parameter associated with the CLE$_\kappa$ loop having fjords (\ref{dense})  (figure \ref{FKBoundaryLoop}).  For the above Potts models, these parameters are
\be\label{dilute}\hat{\kappa}=16/\kappa\quad\overset{(\ref{dense})}{\Longrightarrow}\quad \begin{cases} 
Q=2\quad\Longrightarrow\quad \hat{\kappa}=3, &n=n(\hat{\kappa})=1, \\ 
Q=3\quad\Longrightarrow\quad \hat{\kappa}=10/3, &n=n(\hat{\kappa})=(1+\sqrt{5})/2, \\ 
Q=4\quad\Longrightarrow\quad \hat{\kappa}=4, &n=n(\hat{\kappa})=2.\end{cases}\ee
 In general, we see from (\ref{dense}) that the relation between $Q$ and $\hat{\kappa}=16/\kappa$ in (\ref{dilute}) is $\sqrt{Q}=-2\cos(\hat{\kappa}\pi/4)$ with $\hat{\kappa}\in(8/3,4]$.  (The $\hat{\kappa}=8/3$ cutoff arises because CLE$_\kappa$ is defined  only for $\kappa\in(8/3,8]$ \cite{sheffield,doyon}.)   Combining this with (\ref{fugacity}) gives 
\be\label{ndilute}n:=-2\cos\left(\frac{\pi^2}{\arccos(-\sqrt{Q}/2)}\right),\quad \text{where $Q\in\{2,3,4\}$},\ee 
which relates the loop fugacity $n$ of the dilute-phase loop-gas model to the number of states $Q$ in the corresponding critical Potts model in the continuum limit.  Thus, we expect that a dilute-phase loop-gas representation of the critical $Q\in\{ 2,3,4\}$ Potts model exists in the continuum limit as CLE$_{\hat{\kappa}}$, with $\hat{\kappa}$ given by (\ref{dilute}).  

%%%%%%%%%%%%%%%%%%%%%%%%%%%%%%%%%%%%%%%%%%%%%%%%%%%%%%%%%%%%%%%%%%%%%%%%%%%%%%%%%%%%%%%%%%%%%%%%%%%%%%%%%%%%%%%%%%%%%%%%%
%%%%%%%%%%%%%%%%%%%%%%%%%%%%%%%%%%%%%%%%%%%%%%%%%%%%%%%%%%%%%%%%%%%%%%%%%%%%%%%%%%%%%%%%%%%%%%%%%%%%%%%%%%%%%%%%%%%%%%%%%

 Our identification of the critical Potts model with the dilute-phase loop-gas model in the continuum limit does not involve BCs, so next, we modify this identification to include the event $\PMB$ (item \ref{it33} of section \ref{FLBCsect}) if the Potts model is inside an even-sided polygon $\mathcal{P}$.   Naively, because all fixed sides of $\mathcal{P}$ exhibit the same state in $\PMB$, the $\LGB$ event corresponding to $\PMB$, should one exist, must be $\LGB_1$.  Indeed, the latter corresponds  to the random cluster model mutual  wiring event $\RCB_1$  (figure \ref{ExtremeCases}). However,  verifying this idea  is difficult because the  identification  of the Potts model with the loop-gas model  holds only in the continuum limit.  (The $Q=2$ exception does not help because, here with $n=1$ (\ref{dilute}), the loop-gas  model  partition functions (\ref{loopgaspart}) are identical for all $\LGB$ events $\LGB_\vartheta$.)

\begin{figure}[t]
\centering
\includegraphics[scale=0.15]{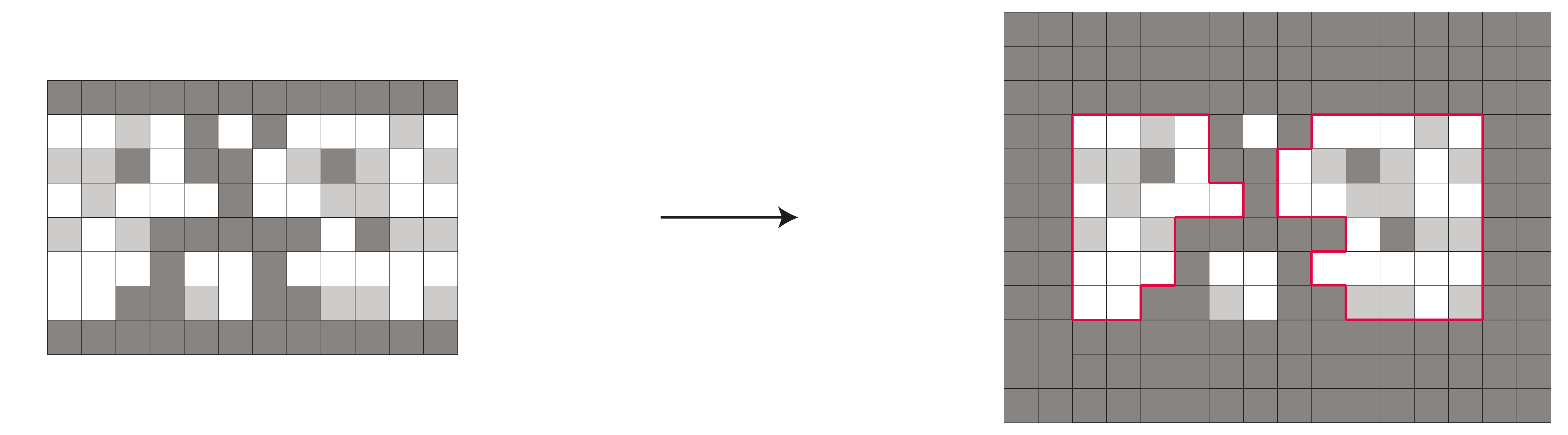}
\caption{$Q=3$ Potts model in a rectangle with the $\PMB$ (fixed BC on top/bottom, fluctuating BC on the sides).  Surrounding the sample with lattice sites whose states match the fixed BC reveals how boundary loops (red) close outside the rectangle.}
\label{SpinSea}
\end{figure}

In spite of this difficulty, there is a  simple argument that identifies $\PMB$ with $\LGB_1$.   To make it, we extend the lattice inside $\mathcal{P}$ so it fills a very large planar region $\mathcal{R}$ containing $\mathcal{P}$, with the boundary of $\mathcal{R}$ far from the boundary of $\mathcal{P}$.  Next, we restrict the state of the lattice sites inside $\mathcal{R}\setminus\mathcal{P}$ to exhibit only the state common to all fixed sides of $\mathcal{P}$.  This addition alters the Potts model partition function $Y_1^\mathcal{P}$, summing over $\PMB$, by only a  physically irrelevant  constant.  Furthermore, the boundary loops, surrounding clusters anchored to the fixed sides of $\mathcal{P}$, close through exterior arcs that pass just behind the fluctuating sides of $\mathcal{P}$ (figure \ref{SpinSea}).  Because $\LGB_1$ is the only  $\LGB$  event with its boundary loops  arranged outside $\mathcal{P}$ in this way, we identify $\PMB$ with $\LGB_1$.

%%%%%%%%%%%%%%%%%%%%%%%%%%%%%%%%%%%%%%%%%%%%%%%%%%%%%%%%%%%%%%%%%%%%%%%%%%%%%%%%%%%%%%%%%%%%%%%%%%%%%%%%%%%%%%%%%%%%%%%%%%
%%%%%%%%%%%%%%%%%%%%%%%%%%%%%%%%%%%%%%%%%%%%%%%%%%%%%%%%%%%%%%%%%%%%%%%%%%%%%%%%%%%%%%%%%%%%%%%%%%%%%%%%%%%%%%%%%%%%%%%%%%
%

\subsection{Crossing probability for loop-gas models}\label{XingLGSect}

Above, we identify $\RCB$ random cluster model partition functions (items \ref{it7} and \ref{it8}  in section \ref{RCxingSect}) and $\PMB$ Potts model partition functions (items \ref{it43} and \ref{it53}  in section \ref{FLBCsect}) with $\LGB$ loop-gas model partition functions (items \ref{it42} and \ref{it52} of this section).  To summarize, for the dense and dilute phases, we respectively have
\be\label{preasympXY}\left\{\begin{array}{lll} X_{\varsigma,\vartheta}^{\mathcal{P}}(p_c,Q)\underset{a\to0}{\sim} \stZ_{\varsigma,\vartheta}^{\mathcal{P}}(\tilde{x}_c(n),n)&\text{and}\,\,X_{\vartheta}^{\mathcal{P}}(p_c,Q)\underset{a\to0}{\sim} \stZ_{\vartheta}^{\mathcal{P}}( \tilde{x}_c(n),n), & \text{dense, $n=\sqrt{Q}$ and $Q\in\{1,2,3,4\}$}, \\ Y_{\varsigma, 1 }^{\mathcal{P}}(K_c,Q)\underset{a\to0}{\sim}\stZ_{\varsigma, 1}^{\mathcal{P}}(x_c(n),n)&\text{and}\,\,Y_{ 1 }^{\mathcal{P}}(K_c,Q)\underset{a\to0}{\sim} \stZ_1^{\mathcal{P}}(x_c(n),n), & \text{dilute, $n$ given by (\ref{ndilute}) and $Q\in\{2,3,4\}$},\end{array}\right.\ee
as the lattice spacing $a$ vanishes in the continuum limit, where we define the symbol $\sim$ as follows:
\be  f(x)\underset{x\to c}{\sim} g(x)\qquad\Longrightarrow\qquad\lim_{x\to c}\frac{f(x)}{g(x)}=  \text{$\lambda$ for some $\lambda\in\mathbb{R}\setminus\{0\}$.}\ee
In (\ref{preasympXY}), the constant $\lambda$ is not universal, i.e., it depends on microscopic details of the model such as the lattice.  The loop fugacity of the dilute phase $\{n\in[-2,2],x=x_c(n)\}$ overlaps that of the dense phase $\{n\in[0,2],x=\tilde{x}_c(n)\}$, but  conveniently, we may  simultaneously  represent both phases with a single parameter, the SLE$_\kappa$/CLE$_\kappa$ parameter $\kappa$, by replacing $n$ with (\ref{fugacity}).  After doing this, restricting to $n\in(0,2]$ as usual, and using (\ref{dense}, \ref{dilute}), we write (\ref{preasympXY}) as
\be\label{asympXY}\begin{cases} X_{\varsigma,\vartheta}^{\mathcal{P}}(\kappa)\underset{a\to0}{\sim} Z_{\varsigma,\vartheta}^{\mathcal{P}}(\kappa)\,\,\text{and}\,\, X_{\vartheta}^{\mathcal{P}}(\kappa)\underset{a\to0}{\sim} Z_{\vartheta}^{\mathcal{P}}(\kappa), & \text{dense, $Q=4\cos^2(4\pi/\kappa)$, $\kappa\in(4,8)$ and $Q\in\{1,2,3,4\}$}, \\ Y_{\varsigma,1}^{\mathcal{P}}(\kappa)\hspace{0.1cm}\underset{a\to0}{\sim} Z_{\varsigma,1}^{\mathcal{P}}(\kappa)\,\,\text{and}\,\, Y_{ 1 }^{\mathcal{P}}(\kappa)\hspace{0.03cm}\underset{a\to0}{\sim} Z_1^{\mathcal{P}}(\kappa), & \text{dilute, $Q=4\cos^2(\pi\kappa/4)$, $\kappa\in(8/3,4]$, and $Q\in\{2,3,4\}$},\end{cases}\ee
where we have combined the dense phase and dilute phase loop-gas model partition functions together into a single function, defined by 
\be\label{Znk}Z_{\varsigma,\vartheta}^{\mathcal{P}}(\kappa):=\begin{cases} \stZ_{\varsigma,\vartheta}^{\mathcal{P}}( \tilde{x}_c(n),n ), \\ \stZ_{\varsigma,\vartheta}^{\mathcal{P}}( x_c(n),n ),\end{cases}
 Z_{\vartheta}^{\mathcal{P}}(\kappa):=\begin{cases} \stZ_{\vartheta}^{\mathcal{P}}( \tilde{x}_c(n),n ),& \text{\text{dense}, $n=-2\cos(4\pi/\kappa)$ and $\kappa\in(4,8)$}, \\ \stZ_{\vartheta}^{\mathcal{P}}( x_c(n),n ),& \text{\text{dilute,} $n=-2\cos(4\pi/\kappa)$ and $\kappa\in(8/3,4]$}.\end{cases}\ee
As such, the two crossing probabilities (\ref{chixing}, \ref{chixing2}) unify under the following ratio of loop-gas model partition functions, which we may  propose as a ``crossing probability"  for a loop-gas model in $\mathcal{P}$ conditioned on an $\LGB$ event:
\be\label{masterxing}P_{\varsigma|\vartheta}^\mathcal{P}(\kappa):= \mathbb{P}_{\text{loop-gas}}(\LGX_\varsigma\,|\,\LGB_\vartheta)=  \frac{Z_{\varsigma,\vartheta}^\mathcal{P}(\kappa)}{Z_\vartheta^\mathcal{P}(\kappa)}.\ee
In the next section, we use methods of CFT and results of \cite{florkleb,florkleb2,florkleb3,florkleb4} to predict a formula for the continuum limit of this ratio (\ref{masterxing}) as a function of $\kappa\in(8/3,8)$.

\section{The continuum limit and conformal field theory}\label{CFTsect}

In this section, we use boundary CFT to predict a formula for the continuum limit of the crossing probability (\ref{masterxing}).  For this, we conformally map the interior of the $2N$-sided polygon $\mathcal{P}$ onto the upper half-plane via a Schwarz-Christoffel transformation.  This map extends to a continuous bijection that sends the sides and vertices $w_1$, $w_2,\ldots,w_{2N}$ of $\mathcal{P}$ to the real axis and the respective points $x_1<x_2<\ldots<x_{2N}$.  It is widely believed \cite{lps,smir,dusm}, and in some cases rigorously proven \cite{smirnov,ssmi,chelsmir}, that in the continuum limit, the statistics of each of the lattice models considered in this article (Potts model, random cluster model, and loop-gas model) are invariant under a conformal transformation of the model's domain (but not its lattice).  Hence, this transformation does not  affect our determination of crossing probabilities, and it has the advantage of sending all polygon interiors to a common domain, the upper half-plane.

\subsection{Smeared partition functions and conformal field theory}\label{SmearedSect}

If a Potts model or random cluster model occupies  a lattice  inside $\mathcal{P}$, then  under a conformal transformation sending the interior of $\mathcal{P}$ onto the upper half-plane,  the BC associated with each side carries over to its image segment on the real axis, and the boundary condition changes (BCCs) occurring at the vertices carry over to their image points.  In the loop-gas representation of these models, the BCCs mark points along the real axis where loops pass into or out of the upper half-plane.  We let $Z_{\varsigma,\vartheta}$ (resp.\ $Z_\vartheta$) denote the partition function (\ref{Znk}) for the loop-gas model in the upper half-plane, summing exclusively over the event $\LGX_\varsigma\cap\LGB_\vartheta$ (resp.\ $\LGB_\vartheta$)  also sent to the upper half-plane  (items \ref{it22} and \ref{it32} beneath (\ref{loopgaspart})),  and we let $Z_f$ denote the free partition function (\ref{lgpart}) for the critical loop-gas model in the upper half-plane.  As usual, we have $Z_\vartheta=Z_{1,\vartheta}+Z_{2,\vartheta}+\dotsm+Z_{C_N,\vartheta}.$

\begin{figure}[b]
\centering
\includegraphics[scale=0.24]{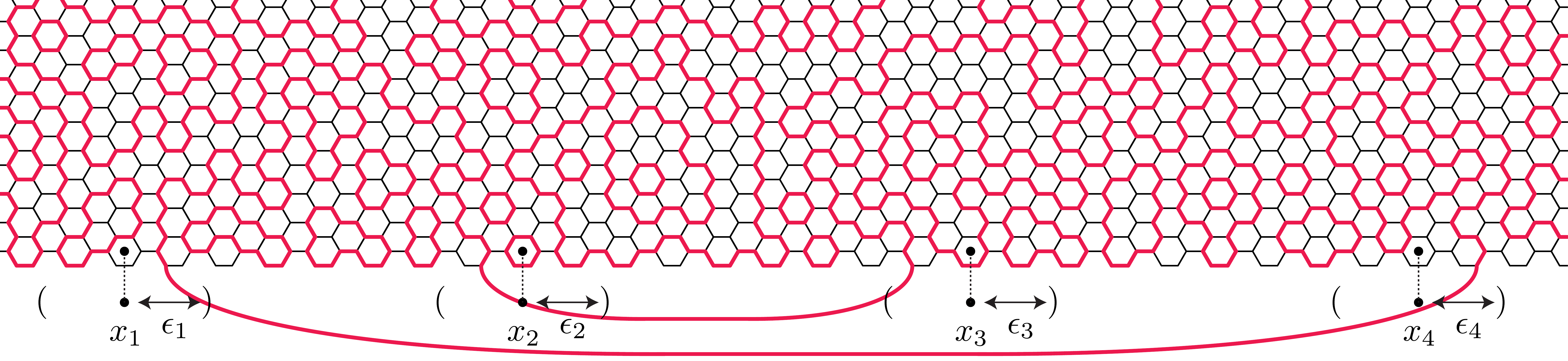}
\caption{Sample of a loop-gas model on the  hexagonal lattice in the upper half-plane that contributes to the  smeared  partition function $Z_2(x_1,x_2,x_3,x_4;\epsilon_1,\epsilon_2,\epsilon_3,\epsilon_4)$.  Only one boundary loop enters the system within distance $\epsilon_j$ from boundary point $x_j$.}
\label{HalfPlane}
\end{figure}

Before continuing, we address a subtlety regarding the BCC locations on the real axis.  In the discrete setting, the BCCs occur exactly at the lattice sites nearest to the image points $x_j$.  However, in the continuum limit where the lattice spacing vanishes, the BCCs occur exactly at the $x_j$, an event whose probability is zero.  To avoid summing over events with vanishing probabilities, we replace our partition function $Z_{\varsigma,\vartheta}$ (resp.\ $Z_\vartheta$) for the discrete system with a \emph{smeared} partition function  $Z_{\varsigma,\vartheta}(\epsilon_1,\epsilon_2,\ldots,\epsilon_{2N})$ (resp.\ $Z_\vartheta(\epsilon_1,\epsilon_2,\ldots,\epsilon_{2N})$) that sums over all $\LGB$ events of the type $\LGB_\vartheta$ that have the $j$th BCC occurring within some small distance $\epsilon_j\ll |x_j-x_{j\pm1}|$ of $x_j$ (figure \ref{HalfPlane}).  

%%%%%%%%%%%%%%%%%%%%%%%%%%%%%%%%%%%%%%%%%%%%%%%%%%%%%%%%%%%%%%%%%%%%%%%%%%%%%%%%%%%%%%%%%%%%%%%%%%%%%%%%%%%%%%%%%%%%%%%%%
%%%%%%%%%%%%%%%%%%%%%%%%%%%%%%%%%%%%%%%%%%%%%%%%%%%%%%%%%%%%%%%%%%%%%%%%%%%%%%%%%%%%%%%%%%%%%%%%%%%%%%%%%%%%%%%%%%%%%%%%%

A standard prediction of physics literature \cite{fms,cardydomb} is that, after shrinking the lattice spacing $a$ to zero to arrive with the continuum limit, the loop-gas model partition function behaves as 
\be\label{Upsilondefn}\lim_{a\rightarrow0}Z_\vartheta(\{x_j\};\{\epsilon_j\})/Z_f\underset{\epsilon_j\rightarrow0}{\sim}
\epsilon_1^{\theta_1}\epsilon_2^{\theta_1}\dotsm\epsilon_{2N}^{\theta_1}\Upsilon_\vartheta(\{x_j\}),\ee
where $\Upsilon_\vartheta$ is some universal function, and where $\theta_1$ is given in (\ref{theta1}) below.  Throughout this article, we abbreviate (\ref{Upsilondefn}) by writing
\be\label{simeqdefn} Z_\vartheta( \epsilon )\simeq\Upsilon_\vartheta\quad \text{as $\epsilon\downarrow0$,}\ee
where the precise meaning of the symbol $\simeq$ follows from rewriting (\ref{Upsilondefn}) in the form (\ref{simeqdefn}).   Another well-known prediction of the physics literature \cite{fms,cardydomb,c1,c3,c4,bauber,bauber2,bbk} is that $\Upsilon_\vartheta$ equals the $2N$-point boundary CFT correlation function 
\be\label{corrfunc}\Upsilon_\vartheta(x_1,x_2,\ldots,x_{2N})=\langle\psi_1(x_1)\psi_1(x_2)\dotsm\psi_1(x_{2N})\rangle_\vartheta.\ee
The \emph{one-leg boundary operator} $\psi_1(x_j)$, so-called because it induces the event that a boundary loop passes into or out of the upper half-plane through (or near) the point $x_j$, is a primary operator in a CFT with central charge \cite{bauber}
\be\label{central}c(\kappa)=(6-\kappa)(3-8\kappa)/2\kappa,\ee
where $\kappa$ is the SLE$_\kappa$ parameter.  The conformal weight $\theta_1$ of $\psi_1$ equals the $(1,2)$ (resp.\ $(2,1)$) Kac weight in the dense (resp.\ dilute) phase of SLE$_\kappa$ \cite{florkleb}:
\be\label{theta1}\theta_1=\frac{6-\kappa}{2\kappa}=\begin{cases} h_{1,2}(c),\quad \kappa>4, \\ h_{2,1}(c),\quad \kappa\leq 4.\end{cases}\ee
 The  $(1,2)$ (resp.\ $(2,1)$) CFT null-state condition  implies  that (\ref{corrfunc}) satisfies the  following elliptic system of  \emph{null-state PDEs}  \cite{bpz}
\be\label{nullstate}\Bigg[\frac{\kappa}{4}\partial_j^2+\sum_{k\neq j}^{2N}\left(\frac{\partial_k}{x_k-x_j}-\frac{(6-\kappa)/2\kappa}{(x_k-x_j)^2}\right)\Bigg]\Upsilon_\vartheta(x_1,x_2,\ldots,x_{2N})=0,\quad \text{for all  $j\in\{1,2,\ldots,2N\}$}.\ee
Also, $\Upsilon_\vartheta$ (\ref{corrfunc}) satisfies three \emph{conformal Ward identities}  that require  it to be invariant under uniform translation of the points $x_1$, $x_2,\ldots,x_{2N}$ and covariant, with weight $\theta_1$, under dilations  and special conformal transformations \cite{florkleb}: 
\be\label{wardid}\begin{gathered}\sum_{k=1}^{2N}\partial_k\Upsilon_\vartheta(x_1,x_2,\ldots,x_{2N})=0,\qquad \sum_{k=1}^{2N}\,[x_k\partial_k+\theta_1]\Upsilon_\vartheta(x_1,x_2,\ldots,x_{2N})=0,\\
 \sum_{k=1}^{2N}\,[x_k^2\partial_k+2\theta_1x_k]\Upsilon_\vartheta(x_1,x_ 2,\ldots,x_{2N})=0.\end{gathered}\ee
Altogether, (\ref{wardid}) implies that if $f$ is a conformal bijection of the upper half-plane onto itself (it is easy to show that this is a M\"obius transformation with real coefficients) such that $x_i'<x_j'$ if $i<j$, where $x_j':=f(x_j)$, then $\Upsilon_\vartheta$ satisfies
\be\label{confcov}\Upsilon_\vartheta(x_1',x_2',\ldots,x_{2N}')=\partial f(x_1)^{-\theta_1}\partial f(x_2)^{-\theta_1}\dotsm\partial f(x_{2N})^{-\theta_1}\Upsilon_\vartheta(x_1,x_2,\ldots,x_{2N}),\ee
where $\partial f$ is the derivative of $f$.  (We note that $\partial f(x)>0$ for all $x\in\mathbb{R}$.)  We call this property (\ref{confcov}) \emph{M\"obius covariance}.  In light of (\ref{Upsilondefn}, \ref{corrfunc}), this transformation rule (\ref{confcov}) is natural.  Indeed, under $f$, the radius of the $\epsilon_j$-ball centered on $x_j$ dilates to $\partial f(x_j)\epsilon_j$.  Combining this fact with (\ref{confcov}) gives the conformal invariance of the continuum limit of the ratio $Z_\vartheta(\{\epsilon_j\})/Z_f$ on the left side of (\ref{Upsilondefn}).  This is consistent with our expectations of conformal invariance in these models.

A typical indication of a critical point in a statistical mechanics  system  such as the Potts model or random cluster model is the existence of certain correlation functions with power-law decay or growth (exclusively) at that critical point \cite{henkel}.  Because we are working only at the critical point of these models, where conformal invariance and non-trivial crossing observables are expected (see the introduction \ref{intro}),  it is reasonable to assume that $\Upsilon_\vartheta$ satisfies  the bound
\be\label{powerlaw} |\Upsilon_\vartheta(x_1,x_2,\ldots,x_{2N})|\leq C\prod_{i<j}^{2N}|x_j-x_i|^{\mu_{ij}(p)},\quad\text{with}\,\,\mu_{ij}(p):=\begin{cases}-p, & |x_j-x_i|<1, \\ +p, & |x_j-x_i|\geq1,\end{cases}\ee
for all $x_1<x_2<\ldots<x_{2N}$ and some positive constants $p$ and $C$ that may depend on $\Upsilon_\vartheta$.  As in \cite{florkleb,florkleb2,florkleb3,florkleb4}, we let $\mathcal{S}_N$ denote the vector space (over the real numbers) of solutions to the system (\ref{nullstate}, \ref{wardid}) that satisfy the power-law bound (\ref{powerlaw}) for some constants $p$ and $C$ (that may depend on the solution).  Thus, we have $\Upsilon_\vartheta\in\mathcal{S}_N$.  

In \cite{florkleb,florkleb2,florkleb3,florkleb4}, two authors of this article completely determine the solution space $\mathcal{S}_N$ for all $\kappa\in(0,8)$.  In \cite{florkleb,florkleb2}, they rigorously prove that $\dim\mathcal{S}_N\leq C_N$, with $C_N$ the $N$th Catalan number (\ref{catalan}), and in \cite{florkleb3}, they construct a linearly independent collection of $C_N$ explicit solutions via the CFT Coulomb gas (contour integral) formalism \cite{df1,df2}, proving that $\dim\mathcal{S}_N=C_N$ and that such solutions span $\mathcal{S}_N$.  With the solution space $\mathcal{S}_N$ completely understood, we seek a property of $\Upsilon_\vartheta$ that identifies it with a particular explicit element of $\mathcal{S}_N$.

\subsection{Crossing weights and crossing probability}\label{XingWeightsSect}

An explicit formula for $\Upsilon_\vartheta$ is not sufficient to determine the continuum limit of the crossing probability (\ref{masterxing}); we also need a formula for the asymptotic behavior of $Z_{\varsigma,\vartheta}(\{\epsilon_j\})$ as $\epsilon_j\rightarrow0$.  Similar to (\ref{Upsilondefn}), we expect
\be\label{contZsigthet}Z_{\varsigma,\vartheta} (\epsilon) \simeq\Upsilon_{\varsigma,\vartheta}\quad \text{as $\epsilon\downarrow0$ }\ee
for some universal function $\Upsilon_{\varsigma,\vartheta}$ (where the meaning of the symbol $\simeq$ is given in and explained beneath (\ref{simeqdefn})).   Then  the continuum limit of the crossing probability (\ref{masterxing}), now in the upper half-plane setting, is given by
\be\label{2ndchi}P_{\varsigma|\vartheta}=\lim_{\epsilon_j\rightarrow0}\lim_{a\rightarrow0}\frac{Z_{\varsigma,\vartheta}(\{\epsilon_j\})/Z_f}{\,\,\,\,Z_\vartheta(\{\epsilon_j\})/Z_f}=\frac{\Upsilon_{\varsigma,\vartheta}}{\Upsilon_\vartheta}.\ee
Furthermore, we only need to determine the numerator $\Upsilon_{\varsigma,\vartheta}$ of (\ref{2ndchi}), as summing over the crossing index $\varsigma$ determines the denominator:
\be\label{ZUps} Z_\vartheta=Z_{1,\vartheta}+Z_{2,\vartheta}+\dotsm+Z_{C_N,\vartheta}\qquad\stackrel{(\ref{simeqdefn}, \ref{contZsigthet})}{\Longrightarrow}\qquad\Upsilon_\vartheta=\Upsilon_{1,\vartheta}+\Upsilon_{2,\vartheta}+\dotsm+\Upsilon_{C_N,\vartheta}\quad\text{for all  $\vartheta\in\{1,2,\ldots,C_N\}$}.\ee

To determine the universal functions $\Upsilon_{\varsigma,\vartheta}$, it is useful to study the quantity $\SpecialZ_\varsigma:=n^{-l_{\varsigma,\vartheta}}Z_{\varsigma,\vartheta}$, which we identify with the second sum in (\ref{loopgaspart}) in the continuum limit.  As mentioned beneath (\ref{loopgaspart}), we interpret $\SpecialZ_\varsigma$ as the usual loop-gas model partition function $Z_{\varsigma,\vartheta}$, except that the boundary loops of each sample have fugacity $n=1$.  As such, $\SpecialZ_\varsigma$ does not depend on the $\LGB$ index $\vartheta$.  If we define a  smeared  version $\SpecialZ_\varsigma(\{\epsilon_j\})$ relative to $\SpecialZ_\varsigma$ in the usual way,  then
\be\label{contscriptZsig}
\SpecialZ_\varsigma (\epsilon) \simeq\Pi_\varsigma\quad \text{as $\epsilon\downarrow0$ },\quad\text{where $\Pi_\varsigma:=n^{-l_{\varsigma,\vartheta}}\Upsilon_{\varsigma,\vartheta}$}.\ee
We call the new universal function $\Pi_\varsigma$ the \emph{$\varsigma$th crossing weight}.  Altogether, the partition functions $Z_{\varsigma,\vartheta}$ and $\SpecialZ_\varsigma$ and  the universal part of their respective asymptotic behaviors  $\Upsilon_{\varsigma,\vartheta}$ and $\Pi_\varsigma$ are related via the diagram 
\be\label{diagram}
\begin{tikzcd}[column sep=0.25cm, row sep=.7cm]
&\arrow{d}{}[swap]{(\epsilon_j>a)\rightarrow0} Z_{\varsigma,\vartheta} & = &\arrow{d}{(\epsilon_j>a)\rightarrow0}[swap]{} n^{l_{\varsigma,\vartheta}}\SpecialZ_\varsigma \\ 
&\Upsilon_{\varsigma,\vartheta}  & = & n^{l_{\varsigma,\vartheta}}\Pi_\varsigma
\end{tikzcd}.
\ee
The crossing weights $\Pi_\varsigma$ are useful because we may express all of the universal functions $\Upsilon_{\varsigma,\vartheta}$ and $\Upsilon_\vartheta$ and the crossing probability (\ref{2ndchi}) in terms of them.  Indeed, from (\ref{contscriptZsig}), we have
\begin{align}\label{UpsilonPiSum}\Upsilon_{\varsigma,\vartheta}=n^{l_{\varsigma,\vartheta}}\Pi_\varsigma\qquad&\stackrel{(\ref{ZUps})}{\Longrightarrow}\qquad\Upsilon_\vartheta=n^{l_{1,\vartheta}}\Pi_1+n^{l_{2,\vartheta}}\Pi_2+\dotsm+n^{l_{C_N,\vartheta}}\Pi_{C_N},\quad\text{for all  $\vartheta\in\{1,2,\ldots,C_N\}$},\\
\label{halfplanexing}&\stackrel{(\ref{2ndchi})}{\Longrightarrow}\qquad P_{\varsigma|\vartheta}=\frac{n^{l_{\varsigma,\vartheta}}\Pi_\varsigma}{n^{l_{1,\vartheta}}\Pi_1+n^{l_{2,\vartheta}}\Pi_2+\dotsm+n^{l_{C_N,\vartheta}}\Pi_{C_N}}.\end{align}
(This formula (\ref{halfplanexing}) immediately gives the necessary property $P_{1|\vartheta}+P_{2|\vartheta}+\dotsm+P_{C_N|\vartheta}=1$ for all $\vartheta\in\{1,2,\ldots,C_N\}$.)

Our goal now is to find explicit formulas for all crossing weights $\Pi_\varsigma$.   We first  show that these functions are elements of $\mathcal{S}_N$  by solving the system (\ref{UpsilonPiSum})  for all of the $\Pi_\varsigma$ and expressing each as a linear combination of $\Upsilon_1$, $\Upsilon_2,\ldots,\Upsilon_{C_N}\in\mathcal{S}_N$.  This approach is possible if and only if the $C_N\times C_N$ meander matrix $(M_N\circ n)(\kappa)$ (\ref{F=MPi}) \cite{fgg,fgut,difranc,franc}, whose $(\varsigma,\vartheta)$th  entry  is by definition $n(\kappa)^{l_{\varsigma,\vartheta}}$ (figure \ref{innerproduct}), is invertible.  Ref.\ \cite{fgg} calculates the determinant of this matrix, and from this result, we find \cite{florkleb2} that $(M_N\circ n)(\kappa)$ is invertible if and only if $n(\kappa)$ (\ref{fugacity}) is not among
\be\label{nqq}n_{q,q''}=-2\cos(\pi q''/q),\quad \text{with $q,q''\in\mathbb{Z}^+$, $1\leq q''< q\leq N+1,$ and $q,q''$ coprime.}\ee
Hence, if $\kappa\in(0,8)$ does not satisfy $n(\kappa)=n_{q,q''}$ for any $q\leq N+1$, then the crossing weights $\Pi_\varsigma$ are elements of $\mathcal{S}_N$.  The solutions to this equation are the \emph{exceptional speeds} \cite{florkleb3}, given by
\be\label{exceptional}\kappa_{q,q'}:=4q/q',\quad \text{with $q,q'\in\mathbb{Z}^+$, $q,q'$ coprime, and $q>1$}.\ee
These speeds inconveniently include the random cluster model with $Q=1$ ($\kappa=\kappa_{3,2}=1$), $Q=2$ ($\kappa=\kappa_{4,3}=16/3$), and $Q=3$ ($\kappa=\kappa_{6,5}=24/5$) (\ref{dense}), and other conformally invariant critical models \cite{florkleb3}.  We assume for now that $\kappa$ is not an exceptional speed.  Later, in the paragraph beneath (\ref{BN}), we extend the results obtained under this assumption to the case that $\kappa$ is an exceptional speed.

Asymptotic arguments identify each crossing weight with a unique element of $\mathcal{S}_N$.  We begin with the case $N=1$.  With only two BCCs on the real axis, at $x_1$ and $x_2$ respectively, there is $C_1=1$  arc connectivity, in which an arc joins $x_1$ with $x_2$, and the corresponding $(1,1)$th product diagram (figure \ref{innerproduct}) has  $l_{1,1}=1$ boundary loop that passes into the upper half-plane through $x_1$ and out of the upper half-plane through $x_2$.   Thus, there is only one  $N=1$  crossing weight $ \Pi_1\super{1}$, related to the loop-gas partition functions via (\ref{ZUps}, \ref{diagram}):
\be\label{ZN=1}Z_1\super{1}=Z_{1,1}\super{1}=n\SpecialZ_1\super{1}\qquad\Longleftrightarrow\qquad\Upsilon_1\super{1}=\Upsilon_{1,1}\super{1}=n \Pi_1\super{1}.\ee
 In (\ref{ZN=1}) and below, the superscript 1 emphasizes that $N=1$.
According to previous arguments, $\smash{ \Pi_1\super{1}}$ solves the system (\ref{nullstate}, \ref{wardid}) with $N=1$, the solution space $\mathcal{S}_1$ of which we easily see to be all multiples of $(x_2-x_1)^{1-6/\kappa}$ \cite{florkleb}.  So far, $\SpecialZ_1$ and $\Pi_1$ are defined  only  up to normalization, so we choose
\be\label{PiN=1} \Pi_1\super{1}(x_1,x_2)=(x_2-x_1)^{1-6/\kappa}.\ee

%%%%%%%%%%%%%%%%%%%%%%%%%%%%%%%%%%%%%%%%%%%%%%%%%%%%%%%%%%%%%%%%%%%%%%%%%%%%%%%%%%%%%%%%%%%%%%%%%%%%%%%%%%%%%%%%%%%%%%%%%%
\begin{figure}[t]
\centering
\includegraphics[scale=0.3]{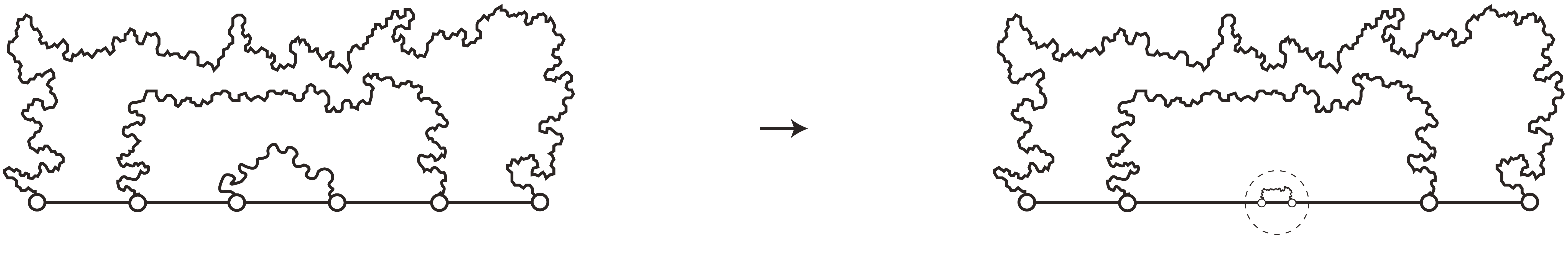}
\caption{If an interior arc $\gamma$ joins two adjacent boundary points, then bringing those points together decouples the system into two independent subsystems, one microscopic with $\gamma$ contracted (circled), and the other macroscopic with $\gamma$ deleted entirely. }
\label{Micro}
\end{figure}
%%%%%%%%%%%%%%%%%%%%%%%%%%%%%%%%%%%%%%%%%%%%%%%%%%%%%%%%%%%%%%%%%%%%%%%%%%%%%%%%%%%%%%%%%%%%%%%%%%%%%%%%%%%%%%%%%%%%%%%%%%
%

For $N>1$ there are $C_N>1$ crossing events to consider, and thus $C_N$ crossing weights to determine.  We determine all of the latter from the $N=1$ case and an asymptotic argument.  To begin, we note that for any $\varsigma\in\{1,2,\ldots,C_N\}$, the $\varsigma$th  interior arc connectivity has at least one interior arc with endpoints $x_i<x_{i+1}$ for some $i<2N$.  This interior arc corresponds to part of a boundary loop $\gamma$ joining $x_i$ to $x_{i+1}$ in the underlying loop-gas model.  As such, if $x_i$ and $x_{i+1}$ are very close, then  $\gamma$  only explores a region very close to its two endpoints with high probability.  Thus, as $x_{i+1}\rightarrow x_i$, the system fragments into two statistically independent subsystems: one microscopic with an infinitesimal-sized  curve  joining $x_i$ with $x_{i+1}$, and the other macroscopic with the  boundary loops  joining the remaining $2N-2$ points as before (figure \ref{Micro}).  Because these two subsystems are expected to be statistically independent, the partition function $\smash{\SpecialZ_\varsigma\super{N}}$ for the entire system factors into 
\begin{multline}\label{factor0}
\SpecialZ\super{N}_\varsigma(\{x_j\};\{\epsilon_j\})\underset{\substack{x_{i+1}\rightarrow x_i \\ \epsilon_1,\epsilon_2\ll x_{i+1}-x_i}}{\sim} \\\SpecialZ_1\super{1}( x_i,x_{i+1};\epsilon_i,\epsilon_{i+1})\SpecialZ_{\varsigma'}\super{N-1}(\{x_j\}_{j\neq i,i+1};\{\epsilon_j\}_{j\neq i,i+1}),\quad\text{where $\varsigma'\in\{1,2,\ldots,C_{N-1}\}$},
\end{multline}
(the superscript $N$ indicates the number of points $2N$) and where  the $\varsigma'$th  connectivity  follows from removing the interior arc joining $x_i$ with $x_{i+1}$ from the $\varsigma$th connectivity.  Similarly, if $ \smash{\Pi\super{N-1}_{\varsigma'} }\in\mathcal{S}_{N-1}$  (resp.\ $\smash{\Pi\super{N}_\varsigma}\in\mathcal{S}_N)$  gives the universal part of the asymptotic behavior of $\SpecialZ_{\varsigma'}(\{\epsilon_j\})$  (resp.\ $\SpecialZ_\varsigma(\{\epsilon_j\})$)  as $\epsilon_j\rightarrow0$, then by (\ref{contscriptZsig}, \ref{PiN=1}, \ref{factor0}), we have
\be\label{factor1}
 \Pi_\varsigma\super{N} (x_1,x_2,\ldots,x_{2N})\underset{x_{i+1}\rightarrow x_i}{\sim} \overbrace{(x_{i+1}-x_i)^{1-6/\kappa}}^{\Pi_1\super{1}( x_i,x_{i+1} )}\Pi_{\varsigma'}\super{N-1} (x_1,x_2,\ldots,x_{i-1},x_{i+2},\ldots,x_{2N}).
\ee
Because $\smash{ \Pi_{\varsigma'}\super{N-1} }$ is a crossing weight, we may apply this same asymptotic argument to it and repeat.  
 Altogether, we find that if we enumerate the arcs in the $\varsigma$th connectivity such that the $j$th arc nests the $k$th arc only if $j>k$ and let $x_{i_{2j-1}}$ and $x_{i_{2j}}$ be the endpoints of the $j$th arc, then for any such enumeration,
\be\label{allowablelim}[\mathscr{L}_\varsigma]\Pi_\varsigma\super{N}:=\lim_{x_{i_{2N}}\rightarrow x_{i_{2N-1}}}\ldots\lim_{x_{i_4}\rightarrow x_{i_3}}\lim_{x_{i_2}\rightarrow x_{i_1}}\prod_{j=1}^{N}(x_{i_{2j}}-x_{i_{2j-1}})^{6/\kappa-1}\Pi_{\varsigma}\super{N}(x_1,x_2,\ldots,x_{2N})=1,\ee 
where $[\mathscr{L}_\varsigma]$ denotes the sequence of limits (actually, an equivalence class of them \cite{florkleb3}) shown above.  In \ \cite{florkleb3}, theorem \red{8} states that the collection of all $C_N$ of these functionals is a basis for the dual space $\mathcal{S}_N^*$, and we denote this basis as
\be\label{BNstar}\mathscr{B}_N^*:=\{[\mathscr{L}_1],[\mathscr{L}_2],\ldots,[\mathscr{L}_{C_N}]\}\subset\mathcal{S}_N^*.\ee
 By lemma \red{12} of \cite{florkleb},  reordering the limits in (\ref{allowablelim}) does not change the value of the limit, assuming that the reordering does not cause points to collide (as happens if, e.g., $x_{i_{2k-1}}<x_{i_{2j-1}}<x_{i_{2j}}<x_{i_{2k}}$ and the $k$th limit precedes the $j$th).

With the quantity $[\mathscr{L}_\varsigma]\Pi_\varsigma$ known (\ref{factor1}), we  next determine $[\mathscr{L}_\vartheta]\Pi_\varsigma$ for all $\vartheta\neq\varsigma$.  Combined with (\ref{allowablelim}), this information is sufficient to determine $\Pi_\varsigma$ because $\mathscr{B}_N^*$ is a basis for $\mathcal{S}_N^*$.  To this end, we note that for all $\vartheta\in\{1,2,\ldots,C_N\}$, $P_{\varsigma|\vartheta}$ (\ref{2ndchi}) goes to the probability of the $\varsigma'$th crossing event conditioned on some $\LGB$ event for a $(2N-2)$-sided polygon as $x_{i+1}\rightarrow x_i$, where $x_i$ and $x_{i+1}$ are still endpoints of a common arc in the $\varsigma$th connectivity.   Then (\ref{UpsilonPiSum}, \ref{halfplanexing}, \ref{factor1}) imply 
\be\label{factor2}
\Upsilon_\vartheta(x_1,x_2,\ldots,x_{2N})\underset{x_{i+1}\rightarrow x_i}{\sim} (x_{i+1}-x_i)^{1-6/\kappa}\Psi_\vartheta(x_1,x_2,\ldots,x_{i-1},x_{i+1},\ldots,x_{2N})\quad\text{for all $\vartheta\in\{1,2,\ldots,C_N\}$},
\ee
for some function $\Psi_\vartheta\neq0$ because $P_{\varsigma|\vartheta}$ does not vanish as $x_{i+1}\rightarrow x_i$.  (Recalling the identification of $\Upsilon_\vartheta$ with the CFT correlation function in (\ref{corrfunc}), the power-law behavior in (\ref{factor2}), with its power $-2\theta_1+\theta_0=1-6/\kappa$, implies the Kac fusion rule $\psi_1\times\psi_1=\psi_0$, where $\psi_0$ is the identity operator with conformal weight $\theta_0=0$ \cite{fms,henkel,bpz}.)  The index $i$ depends on our choice of crossing weight $\Pi_\varsigma$, but because this choice is arbitrary, it  is evident that (\ref{factor2}) holds for any $i\in\{1,2,\ldots,2N-1\}$ (although the function $\Psi_\vartheta$, while not zero, may depend on $i$). 

Meanwhile, $P_{\varsigma|\vartheta}$ (\ref{2ndchi}) should vanish as we bring together two endpoints of two different arcs  in the $\varsigma$th connectivity, such as by sending  $x_i\rightarrow x_{i-1}$ (assuming that $i>1$).   Indeed,  if $x_{i-1}$ and $x_i$ are close, then it is very probable that a  boundary loop  joins them together through the upper half-plane in the underlying loop-gas model, making the $\varsigma$th crossing event very unlikely.  In order for $P_{\varsigma|\vartheta}$ (\ref{halfplanexing}) to vanish as $x_i\rightarrow x_{i-1}$, the asymptotic behavior (\ref{factor2}) of its denominator (\ref{UpsilonPiSum}) must dominate that of its numerator  $\Pi_\varsigma$.  According  to theorem \red{2} of \cite{florkleb4}, $\Pi_\varsigma$  then behaves  as
\be\label{factor3}\Pi_\varsigma(x_1,x_2,\ldots,x_{2N})\underset{x_i\rightarrow x_{i-1}}{\sim} (x_i-x_{i-1})^{2/\kappa}\Lambda_\varsigma(x_1,x_2,\ldots,x_{i-1},x_{i+1},\ldots,x_{2N}),\ee
for some function $\Lambda_\varsigma\neq0$.  (Here, the power-law behavior in (\ref{factor3}), with its power $-2\theta_1+\theta_2=2/\kappa$, implies the Kac fusion rule $\psi_1\times\psi_1=\psi_2$, where $\psi_2$ is the two-leg boundary operator with conformal weight $\theta_2=8/\kappa-1$ \cite{florkleb2}.) 

The asymptotic behaviors (\ref{factor1}) and (\ref{factor3}) of $\Pi_\varsigma$ let us  determine $[\mathscr{L}_\vartheta]\Pi_\varsigma$ for all $\vartheta,\varsigma\in\{1,2,\ldots,C_N\}$.  To begin,  by (\ref{factor1}, \ref{allowablelim}), $[\mathscr{L}_\vartheta]\Pi_\varsigma=1$ if $\vartheta=\varsigma$.  And if $\vartheta\neq\varsigma$, then at least one limit of $[\mathscr{L}_\vartheta]$ brings together two adjacent endpoints of different interior arcs in the $\varsigma$th arc connectivity.  If the $j$th limit of $[\mathscr{L}_\vartheta]$ is the first to do this, then (\ref{factor3}) gives
\be[\mathscr{L}_\vartheta]\Pi_\varsigma\quad=\quad\underbrace{\dotsm \quad \dotsm \quad \dotsm}_{\substack{\text{$(j+1)$th through} \\ \text{last limit of $[\mathscr{L}_\vartheta]$}}}\quad\times\quad\underbrace{\lim_{x_{i_{2j}}\rightarrow x_{i_{2j-1}}}(x_{i_{2j}}-x_{i_{2j-1}})^{6/\kappa-1}}_{\text{$j$th limit in $[\mathscr{L}_\vartheta]$}}\quad\times\quad\underbrace{O((x_{i_{2j}}-x_{i_{2j-1}})^{2/\kappa})}_{\substack{\text{crossing weight} \\ \text{with $2(N-j+1)$ points}}}\quad=\quad0\ee
for $\vartheta\neq\varsigma$ and $\kappa\in(0,8)$.  In summary, we see that the complete collection of crossing weights is dual to the collection (\ref{BNstar}) of (equivalence classes of) sequences of limits in $\mathscr{B}_N^*$  in the sense that
\be\label{dual}[\mathscr{L}_\vartheta]\Pi_\varsigma=\delta_{\vartheta,\varsigma}. \ee
Because  $\mathscr{B}_N^*$ (\ref{BNstar}) is  a basis for $\mathcal{S}_N^*$ \cite{florkleb3},  the  duality relation (\ref{dual}) is sufficient to identify each crossing weight with a unique element of $\mathcal{S}_N$ and  proves that  the collection $\mathscr{B}_N$ of all $C_N$ crossing weights  is  a basis for $\mathcal{S}_N$:
\be\label{BN}\mathscr{B}_N:=\{\Pi_1,\Pi_2,\ldots,\Pi_{C_N}\}\subset\mathcal{S}_N.\ee
Ref.\ \cite{florkleb4} calls $\Pi_\varsigma\in\mathcal{S}_N$ satisfying the duality condition (\ref{dual}) the \emph{$\varsigma$th connectivity weight}.   Thus, the terms ``crossing weight" and ``connectivity weight"  are synonymous (as are ``crossing" and ``connectivity" by figure \ref{OctXingConfigsBdyArcs}).  

So far, our results presume that the crossing weights are elements of $\mathcal{S}_N$, a fact that we proved only for $\kappa\in(0,8)$ not an exceptional speed (\ref{exceptional}) with $q\leq N+1$.  To extend our results to these speeds, we reasonably assume that each crossing weight is a continuous function of $\kappa$.  Ref.\ \cite{florkleb4} shows that each connectivity weight is continuous, so by continuous extension,  crossing weights are identified with connectivity weights  for exceptional speeds also.

To find explicit formulas for all crossing weights from (\ref{dual}), we use the collection of $C_N$ elements $\mathcal{F}_1$, $\mathcal{F}_2,\ldots,\mathcal{F}_{C_N}$ of $\mathcal{S}_N$ given by item \ref{step3} of section \ref{xingsummary}.   Their formulas  (\ref{Fexplicit}) follow from the CFT Coulomb gas (contour integral) formalism \cite{fms,henkel,df1,df2}.  (J.\ Dub\'edat constructed such solutions in \cite{dub}.)  Ref.\ \cite{florkleb3} proves that the set 
\be\label{TLBN}\mathcal{B}_N:=\{\mathcal{F}_1, \mathcal{F}_2,\ldots,\mathcal{F}_{C_N}\}\subset\mathcal{S}_N\ee
is a basis for $\mathcal{S}_N$ if and only if $\kappa\in(0,8)$ is not an exceptional speed (\ref{exceptional}) with $q\leq N+1$ and that $[\mathscr{L}_\varsigma]\mathcal{F}_\vartheta =n(\kappa)^{l_{\varsigma,\vartheta}}$ for $\kappa\in(0,8)$  (figure \ref{innerproduct}).  As such, we have (see also (\ref{F=MPi}) in item \ref{step5} of section \ref{xingsummary})
\be\label{Fsum}[\mathscr{L}_\varsigma]\mathcal{F}_\vartheta=n^{l_{\varsigma,\vartheta}}\quad\stackrel{(\ref{dual})}{\Longrightarrow}\quad \mathcal{F}_\vartheta=n^{l_{1,\vartheta}}\Pi_1+n^{l_{2,\vartheta}}\Pi_2+\dotsm+n^{l_{C_N,\vartheta}}\Pi_{C_N},\quad\text{for all  $\varsigma,\vartheta\in\{1,2,\ldots,C_N\}$.}\ee 
As stated above (\ref{nqq}),  this  system (\ref{Fsum}) is invertible if and only if $\kappa\in(0,8)$ is not an exceptional speed (\ref{exceptional}) with $q\leq N+1$ \cite{florkleb3}, so for all such $\kappa$,  we obtain  formulas for all crossing weights $\Pi_\varsigma$ and the crossing probability (\ref{halfplanexing}).

If $\kappa$ is an exceptional speed with $q\leq N+1$, then the system (\ref{Fsum}) is not invertible.  In this case, we may either isolate the crossing weights for some SLE$_\kappa$ parameter $\varkappa$ near $\kappa$ and then send $\varkappa\rightarrow\kappa$, or we may verify that a certain choice of integration contours in (\ref{Fexplicit}) gives a solution satisfying (\ref{dual}) for all $\vartheta\in\{1,2,\ldots,C_N\}$ and some $\varsigma$ in that range.
Both approaches have drawbacks.  The first is tedious, as the finiteness of the limit of $\Pi_\varsigma(\varkappa)$ as $\varkappa\rightarrow\kappa$ relies on the cancelation of many infinite quantities.  The second is used in \cite{fsk} for $N\in\{1,2,3,4\}$, but it is impractical for $N\geq5$. 

Equation (\ref{Fsum}) offers a useful physical interpretation of the Coulomb gas functions in (\ref{TLBN}).  Comparing (\ref{Fsum}) to (\ref{UpsilonPiSum}) gives an explicit formula for the universal function $\Upsilon_\vartheta$ (item \ref{step3} of section \ref{xingsummary}):
\be\label{UpsF}\Upsilon_\vartheta=\mathcal{F}_\vartheta\,\,(\ref{Fexplicit}),\quad \text{for all $\vartheta\in\{1,2,\ldots,C_N\}$.}\ee  
 After inserting (\ref{UpsF}) into (\ref{Upsilondefn}), we find a physical interpretation of the Coulomb gas function $\mathcal{F}_\vartheta$.  Moreover, using (\ref{Fsum}, \ref{UpsF}), we can sum the denominator of (\ref{halfplanexing}) to find  
\be\label{xingSLE}\boxed{P_{\varsigma|\vartheta}=\frac{n^{l_{\varsigma,\vartheta}}\Pi_\varsigma}{\mathcal{F}_\vartheta}.}\ee 
This is a very important step because it justifies the crossing probability formula (\ref{xing}) previously stated in section \ref{xingsummary}.  This summation is a significant simplification because the formulas for the crossing weights are often considerably more complicated than those of $\mathcal{F}_\vartheta$.  

Ref.\ \cite{florkleb4} posits that each crossing weight $\Pi_\varsigma(\kappa)$ is positive-valued for all $\kappa\in(0,8)$.  Assuming that this is true,  the crossing probability formula (\ref{halfplanexing}, \ref{xingSLE}) satisfies $0<P_{\varsigma|\vartheta}<1$  if $n(\kappa)>0$, or $\kappa\in(8/3,8)$.  This restricts the fugacity of the loop-gas model to $n\in(0,2]$, or $\kappa\in(8/3,4]$, cutting off the $n\in[-2,0]$ range of the dilute phase.   To our knowledge, (\ref{xingSLE}) has no application if $n(\kappa)\leq0$.  Indeed, we cannot interpret it as a probability if it is negative.

\begin{figure}[t]
\centering
\includegraphics[scale=0.3]{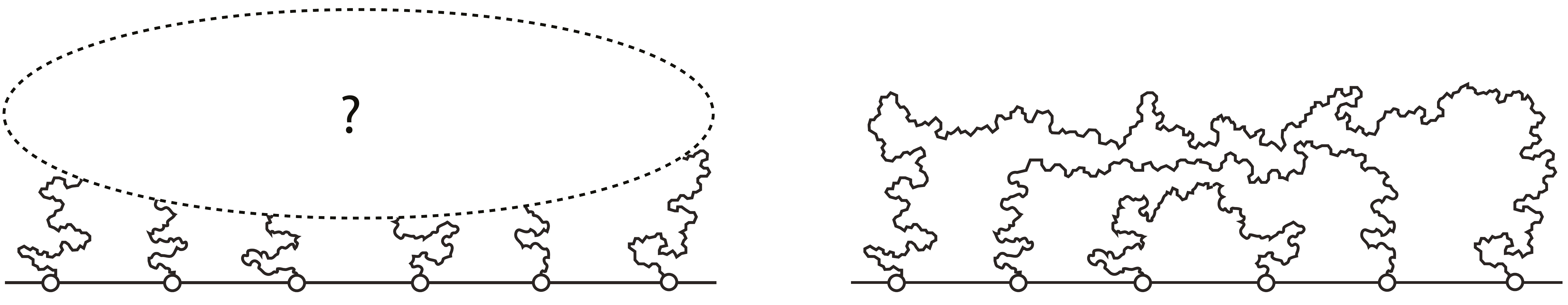}
\caption{Ref.\ \cite{bauber2} predicts a formula (\ref{xingSLE2}) for the probability that the growing curves of a multiple-SLE$_\kappa$ process (left) join pairwise in the $\varsigma$th connectivity (right).  This formula includes the crossing probability formula (\ref{xingSLE}) as a special case.}
\label{Connect}
\end{figure}

It is interesting to interpret (\ref{xingSLE}) in terms of multiple-SLE$_\kappa$ \cite{bbk}, the random process  that is expected to give the law of boundary cluster interfaces in the continuum limit of the critical random cluster model and the critical Potts model  (figure \ref{Connect}).  Ref.\ \cite{bauber2,florkleb4} predict that in a multiple-SLE$_\kappa$ process with ``SLE$_\kappa$ partition function" $F\in\mathcal{S}_N$, the $2N$ curves exploring the upper half-plane join together pairwise in the $\varsigma$th connectivity with probability
\be\label{xingSLE2}P_\varsigma=\frac{a_\varsigma\Pi_\varsigma}{a_1\Pi_1+a_2\Pi_2+\dotsm+a_{C_N}\Pi_{C_N}}=\frac{a_\varsigma\Pi_\varsigma}{F},\quad a_\varrho:=[\mathscr{L}_\varrho]F.\ee
If $F=\mathcal{F}_\vartheta$, then (\ref{xingSLE2}) matches (\ref{xingSLE}).   This implies the following: if we choose $F=\mathcal{F}_\vartheta$ for the SLE$_\kappa$ partition function with $\kappa$ among (\ref{dense}), then the generated multiple-SLE$_\kappa$ curves match the FK boundary cluster interfaces in the corresponding (continuum limit) critical random cluster model with the $\vartheta$th mutual wiring event $\RCB_\vartheta$.  Or if we choose $F=\mathcal{F}_1$ for the SLE$_\kappa$ partition function with $\kappa$ among (\ref{dilute}), then the generated multiple-SLE$_\kappa$ curves  match the spin boundary cluster interfaces in the corresponding (continuum limit) critical Potts model with the $\PMB$ event.

To finish, we note that the half-plane crossing probability formulas (\ref{halfplanexing}, \ref{xingSLE}) are manifestly M\"obius invariant.  Indeed, the crossing weights, as elements of $\mathcal{S}_N$, are M\"obius covariant, satisfying the functional equation (\ref{confcov}) for any conformal bijection $f$ of the upper half-plane onto itself that preserves coordinate order.  Therefore, for any such $f$, (\ref{halfplanexing}, \ref{xingSLE}) give
\be\label{confP}P_{\varsigma|\vartheta}(x_1',x_2',\ldots,x_{2N}')=P_{\varsigma|\vartheta}(x_1,x_2,\ldots,x_{2N}),\quad \text{where  $x_j':=f(x_j)$}.\ee
 An example  of this is the following M\"obius transformation $f$ that sends the points $x_1<x_2<x_3<\ldots<x_{2N-2}<x_{2N-1}<x_{2N}$ in that order to the points $0<\lambda_1<\lambda_2<\ldots<\lambda_{2N-3}<1<\infty$ respectively.  Here, (\ref{confP}) gives
\begin{multline}\label{crossratio}f(x_j)=\frac{(x_j-x_1)(x_{2N}-x_{2N-1})}{(x_{2N-1}-x_1)(x_{2N}-x_j)},\quad \lambda_j:=f(x_{j+1})\\
\Longrightarrow\quad P_{\varsigma|\vartheta}(x_1,x_2,x_3\ldots,x_{2N-2},x_{2N-1},x_{2N})=\lim_{w\rightarrow\infty}P_{\varsigma|\vartheta}(0,\lambda_1,\lambda_2,\ldots,\lambda_{2N-3},1,w).\end{multline}
This shows that the crossing probability $P_{\varsigma|\vartheta}$ depends only on the $2N-3$ cross-ratios $\lambda_j$ (\ref{crossratio}) that can be formed from the points $x_1$, $x_2,\ldots,x_{2N}$.  These cross-ratios are M\"obius invariants.

Numerical evidence \cite{lps,lpps} and analytic results \cite{smirnov,chelsmir}  promote  M\"obius invariance (\ref{confP}) of crossing probabilities to conformal invariance.  (Alternatively, we may posit, as we do in appendix \ref{transformxing}, that the partition functions appearing in (\ref{2ndchi}) are conformally invariant, which immediately implies conformal invariance of crossing probabilities.)  Indeed, if $f$ is a conformal map from the upper half-plane onto a Jordan domain $\mathcal{D}$, then
\be\label{confPext}P_{\varsigma|\vartheta}^{\mathcal{D}}(\kappa\,|\,w_1,w_2,\ldots,w_{2N})=P_{\varsigma|\vartheta}(\kappa\,|\,x_1,x_2,\ldots,x_{2N}),\quad \text{where  $w_j:=f(x_j)$},\ee
gives the crossing probability for the same model (on the  same lattice) in $\mathcal{D}$, with the BC or boundary-loop condition at $x_j$ sent to the same condition at $w_j\in\partial\mathcal{D}$.  Unlike (\ref{confP}), (\ref{confPext}) is not a functional equation for $P_{\varsigma|\vartheta}$ because the left side of (\ref{confPext}) is a function completely different from that on the right side.  Indeed, $\smash{P_{\varsigma|\vartheta}^\mathcal{D}}$ and $P_{\varsigma|\vartheta}$ have different domains.  Thanks to (\ref{crossratio}), we see that $\smash{P_{\varsigma|\vartheta}^\mathcal{D}}$ depends on only  $2N-3$  degrees of freedom that determine the shape, but not the position, orientation, or scale, of $\mathcal{D}$ in the complex plane.

Appendix \ref{transformxing} shows that conformal covariance of boundary CFT correlation functions is consistent with the conformal invariance law (\ref{confPext}).  This justification encounters a difficulty in our application: if $f$ is a conformal map that sends the upper half-plane onto the interior of a polygon, then its continuous extension to the real axis is not conformal at the vertex pre-images $x_j$.   In appendix \ref{transformxing}, we circumvent these problems formally with the introduction of CFT corner operators \cite{c1,simkleb}.

\section{Formula for rectangle crossing probability}\label{rectxingsummary}

Section \ref{xingsummary} gives a step-by-step construction of the formula (\ref{xing}) for crossing probability.  As an example,  we calculate  all crossing probabilities (\ref{xing}) for the rectangle ($N=2$), generalizing Cardy's formula for percolation (\ref{RectCross}).  At times, we introduce some helpful but nonessential simplifications that are typically not available for $N>2$.  The steps of this section mirror those of section \ref{xingsummary}, and except for this reliance, this section is completely self contained.  To keep it so, we avoid any notations and terminology  not introduced in  section \ref{xingsummary}.
\begin{enumerate}[wide, labelwidth=!, labelindent=0pt]
\item\label{step12} We let $x_1<x_2<x_3<x_4$ be four marked points on the real axis, let $\kappa\in(0,8)$ (the SLE$_\kappa$ parameter) \cite{rohshr,knk,lawler}, and let $n(\kappa)=-2\cos(4\pi/\kappa)$ (the ``loop-gas model fugacity formula") \cite{nein,smir}.
\item\label{step22} We let the first arc connectivity comprise an arc joining $x_2$ and $x_3$ and another arc joining $x_1$ with $x_4$.  We let the second  arc connectivity comprise an arc joining $x_1$ with $x_2$ and another arc $x_3$ with $x_4$.
\item\label{step32} We use the formulas for $\mathcal{F}_\vartheta$ given in section \red{III} of \cite{florkleb3} (or (\ref{Fother}) of appendix \ref{equivappendix}) instead of those in item \ref{step3} of section \ref{xingsummary} because the former are a little simpler.  According to  our enumeration of  connectivities in step \ref{step22}, we have
\begin{multline}\label{FN2}\mathcal{F}_\vartheta(\kappa\,|\,x_1,x_2,x_3,x_4)=\frac{n(\kappa)^2\Gamma(2-8/\kappa)}{4\sin^2(4\pi/\kappa)\Gamma(1-4/\kappa)^2}[x_{21}x_{31}x_{32}]^{2/\kappa}[x_{41}x_{42}x_{43}]^{1-6/\kappa}\\
\times\,\oint_\Gamma\,(u-x_1)^{-4/\kappa}\left\{\begin{array}{ll}(u-x_2)^{-4/\kappa}, & \vartheta=1 \\ (x_2-u)^{-4/\kappa}, & \vartheta=2\end{array}\right\}(x_3-u)^{-4/\kappa}(x_4-u)^{12/\kappa-2}\,{\rm d}u,\quad \Gamma=\begin{cases} \mathscr{P}(x_2,x_3), &\vartheta=1, \\ \mathscr{P}(x_1,x_2), & \vartheta=2,\end{cases}\end{multline}
where $x_{ji}:=x_j-x_i$ and $\mathscr{P}(x_i,x_j)$ is the Pochhammer contour of figures \ref{BreakDown} and \ref{PochhammerContour}.   It  is useful to express these functions (\ref{FN2}) in terms of hypergeometric functions \cite{morsefesh}.  To begin, the quantities
\be\label{Iinvar} I_1(\kappa\,|\,x_1,x_2,x_3,x_4):=\frac{\mathcal{F}_1(\kappa\,|\,x_1,x_2,x_3,x_4)}{[x_{31}x_{42}]^{1-6/\kappa}},\quad I_2(\kappa\,|\,x_1,x_2,x_3,x_4):=\frac{\mathcal{F}_2(\kappa\,|\,x_1,x_2,x_3,x_4)}{[x_{31}x_{42}]^{1-6/\kappa}}\ee
are M\"obius invariants because their numerators and denominators transform as in (\ref{confcov})  (using (\ref{theta1})) for a  conformal bijection $f$ of the upper half-plane onto itself preserving coordinate order.  So with $\lambda:=\lambda_1=x_{21}x_{43}/x_{31}x_{42}$ (\ref{lambda}), 
\be\begin{aligned}\label{Ivar}I_\vartheta(\kappa\,|\,x_1,x_2,x_3,x_4)&=I_\vartheta(\kappa\,|\,0,\lambda,1,\infty)=\frac{n(\kappa)^2\Gamma(2-8/\kappa)}{4\sin^2(4\pi/\kappa)\Gamma(1-4/\kappa)^2}\\
&\hspace{0.6cm}\times\,\oint_\Gamma u^{-4/\kappa}\left\{\begin{array}{ll}(u-\lambda)^{-4/\kappa}, & \vartheta=1 \\ (\lambda-u)^{-4/\kappa}, & \vartheta=2\end{array}\right\}(1-u)^{-4/\kappa}\,{\rm d}u,\quad \Gamma=\begin{cases} \mathscr{P}(\lambda,1), &\vartheta=1, \\ \mathscr{P}(0,\lambda), & \vartheta=2,\end{cases}\\
&=n(\kappa)^2\,\eta^{2/\kappa}(1-\eta)^{1-6/\kappa}\,_2F_1\bigg(\frac{4}{\kappa},1-\frac{4}{\kappa};2-\frac{8}{\kappa}\,\bigg|\,1-\eta\bigg),\quad\eta=\begin{cases} \lambda, &\vartheta=1, \\ 1-\lambda, & \vartheta=2,\end{cases}\end{aligned}\ee
which, when inserted into (\ref{Iinvar}), gives a formula for $\mathcal{F}_\vartheta$ in terms of hypergeometric functions.
\item\label{step42} We have $l_{1,1}=l_{2,2}=2$ and $l_{1,2}=l_{2,1}=1$ (figure \ref{N2loops}).
\item \label{step52} Using our results from items \ref{step32} and \ref{step42} (figure \ref{N2loops}), we may calculate the crossing weights $\Pi_\varsigma$ by inverting the $2\times2$ meander matrix $M_2(n)$, as follows:
\be\label{F=MPi2}\left(\begin{array}{l}\mathcal{F}_1\\ \mathcal{F}_2\end{array}\right)=\overbrace{\left(\begin{array}{ll}n^2 & n \\ n & n^2\end{array}\right)}^{M_2(n)}\left(\begin{array}{l}\Pi_1\\ \Pi_2 \end{array}\right)\quad\Longrightarrow\quad
\Pi_1=\dfrac{n\mathcal{F}_1-\mathcal{F}_2}{n(n^2-1)},\quad \Pi_2=\dfrac{n\mathcal{F}_2-\mathcal{F}_1}{n(n^2-1)}.\ee
By using a  standard  linear transformation formula of the hypergeometric function \cite{absteg}, we may express these crossing weights in terms of another hypergeometric function:
\begin{multline}\label{Pivarsigma}\Pi_\varsigma(\kappa\,|\,x_1,x_2,x_3,x_4):=\frac{\Gamma(12/\kappa-1)\Gamma(4/\kappa)}{\Gamma(8/\kappa)\Gamma(8/\kappa-1)}[x_{31}x_{42}]^{1-6/\kappa}\\
\times\,\eta^{2/\kappa}(1-\eta)^{1-6/\kappa}\,_2F_1\bigg(\frac{4}{\kappa},1-\frac{4}{\kappa};\frac{8}{\kappa}\,\bigg|\,\eta\bigg),\quad \eta=\begin{cases} \lambda, &\varsigma=1, \\ 1-\lambda, & \varsigma=2.\end{cases}\end{multline}
The meander matrix $(M_2\circ n)(\kappa)$ is not invertible if $n(\kappa)=1$, that is, if $\kappa=6$ or $\kappa=6/(3m\pm1)<4$ for some $m\in\mathbb{Z}^+$.  If $\kappa$ is any of these values, then the limit of $\Pi_\varsigma(\varkappa)$ as $\varkappa\rightarrow\kappa$ is subtle to calculate from the formula given in (\ref{F=MPi2}) because $n(\kappa)=1$,  but trivial  to find from the formula in (\ref{Pivarsigma}).
\item\label{step62}We assemble these results as in (\ref{xing}) to find formulas for crossing probabilities of the models described in item \ref{step6} of section \ref{xingsummary}.  Because these formulas depend only on $\lambda$, we express all probabilities as functions of only this parameter:
\begin{align}\label{N2explicit1} P_{1|1}(\kappa\,|\,\lambda)&=P_{2|2}(\kappa\,|\,1-\lambda)=\frac{\Gamma(12/\kappa-1)\Gamma(4/\kappa)}{\Gamma(8/\kappa)\Gamma(8/\kappa-1)}\frac{_2F_1(\frac{4}{\kappa},1-\frac{4}{\kappa};\frac{8}{\kappa}\,|\,\lambda)}{_2F_1(\frac{4}{\kappa},1-\frac{4}{\kappa};2-\frac{8}{\kappa}\,|\,1-\lambda)},\\
\label{N2explicit2}P_{1|2}(\kappa\,|\,\lambda)&=P_{2|1}(\kappa\,|\,1-\lambda)=-\frac{\Gamma(12/\kappa-1)\Gamma(4/\kappa)}{2\cos(4\pi/\kappa)\Gamma(8/\kappa)\Gamma(8/\kappa-1)}\frac{\lambda^{8/\kappa-1}\,_2F_1(\frac{4}{\kappa},1-\frac{4}{\kappa};\frac{8}{\kappa}\,|\,\lambda)}{(1-\lambda)^{8/\kappa-1}\,_2F_1(\frac{4}{\kappa},1-\frac{4}{\kappa};2-\frac{8}{\kappa}\,|\,\lambda)}.\end{align}
We interpret these probabilities as follows.  Our descriptions here are equivalent to those in items \ref{step6a}--\ref{step6c} of section \ref{xingsummary} above, but we state them a little differently.
%%%%%%%%%%%%%%%%%%%%%%%%%%%%%%%%%%%%%%%%%%%%%%%%%%%%%%%%%%%%%%%%%%%%%%%%%%%%%%%%%%%%%%%%%%%%%%%%%%%%%%%%%%%%%%%%%%%%%%%%%%%%%%%%%%%%%%%
\begin{figure}[b]
\centering
\includegraphics[scale=0.4]{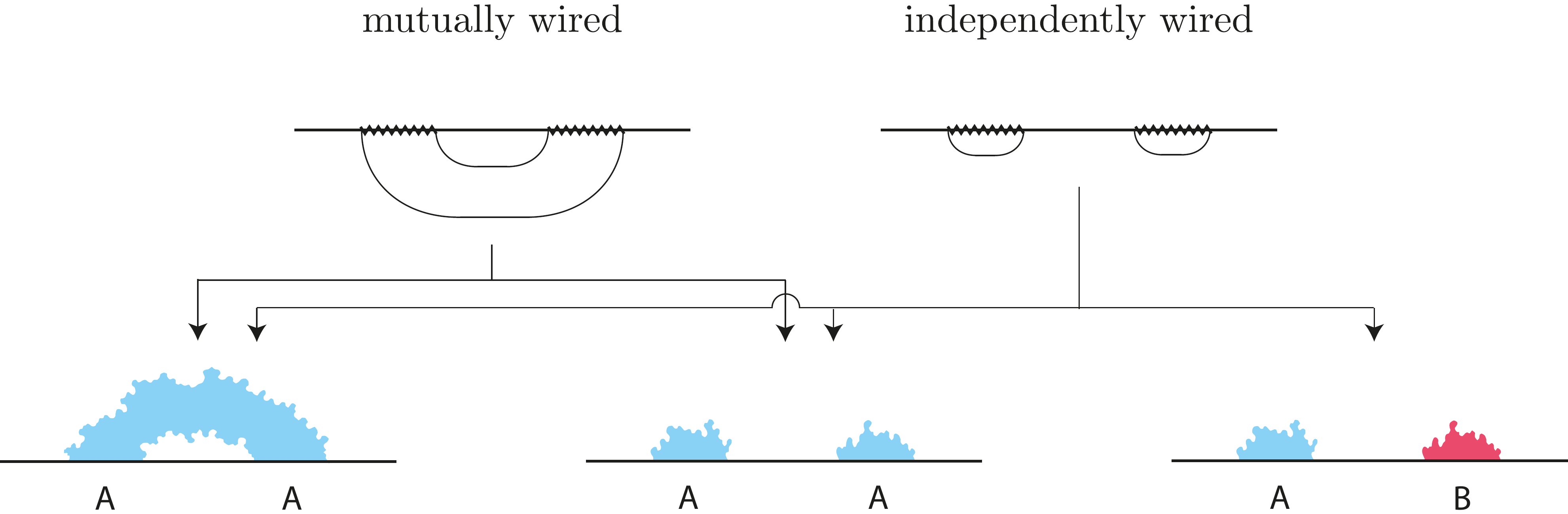}
\caption{The fixed segments $(x_1,x_2)$ and $(x_3,x_4)$ (zig zags) are mutually wired if an exterior arc joins them and are independently wired otherwise.  These $\RCB$ events contribute to the FK crossing events as shown, with $A\neq B$ denoting color. }
\label{IndMutWired}
\end{figure}
%%%%%%%%%%%%%%%%%%%%%%%%%%%%%%%%%%%%%%%%%%%%%%%%%%%%%%%%%%%%%%%%%%%%%%%%%%%%%%%%%%%%%%%%%%%%%%%%%%%%%%%%%%%%%%%%%%%%%%%%%%%%%%%%%%%%%%%
\begin{enumerate}[leftmargin=*]
\item\label{step6a2}\emph{$Q$-state random cluster model:} With $Q\in\{1,2,3,4\}$, we suppose that the continuum limit of a critical random cluster model on a lattice fills the upper half-plane and that the segments $(x_1,x_2)$  and $(x_3,x_4)$ are fixed to some state.  With (\ref{Qdense}) relating $Q$ and $\kappa\in[4,8)$, we have (figure \ref{IndMutWired})
\begin{align}\label{firstinterp1}
P_{1|1}(\kappa\,|\,\lambda)&=\left\{\parbox{11cm}{the probability that one FK cluster touches both $(x_1,x_2)$ and $(x_3,x_4)$, conditioned on both of these segments being fixed to the same state}\right\},\\
P_{1|2}(\kappa\,|\,\lambda)&=\left\{\parbox{11cm}{the probability that one FK cluster touches both $(x_1,x_2)$ and $(x_3,x_4)$, conditioned on both of these segments being independently fixed}\right\},\\
P_{2|1}(\kappa\,|\,\lambda)&=\left\{\parbox{11cm}{the probability that no FK cluster touches both $(x_1,x_2)$ and $(x_3,x_4)$, conditioned on both of these segments being fixed to the same state}\right\},\\
\label{lastinterp1}P_{2|2}(\kappa\,|\,\lambda)&=\left\{\parbox{11cm}{the probability that no FK cluster touches both $(x_1,x_2)$ and $(x_3,x_4)$, conditioned on both of these segments being independently fixed }\right\}.\end{align}
It is interesting to examine some special cases.  First, the case $Q=1$ corresponds to critical percolation.  Setting $\kappa=6$ in (\ref{N2explicit1}, \ref{N2explicit2}), we recover Cardy's formula (\ref{RectCross}):
\be P_{1|1}(6\,|\,\lambda)=P_{1|2}(6\,|\,\lambda)=P_{2|1}(6\,|\,1-\lambda)=P_{2|2}(6\,|\,1-\lambda)=\frac{3\Gamma(2/3)}{\Gamma(1/3)^2}\lambda^{1/3}\,_2F_1\bigg(\frac{1}{3},\frac{2}{3};\frac{4}{3}\,\bigg|\,\lambda\bigg).\ee
These formulas are independent of the second index $\vartheta$ because, with one color distributed to all bond clusters, there is no difference between the BC for $\vartheta=1$ and the BC for $\vartheta=2$.  Next, the case $Q=2$ corresponds to the $Q=2$ random cluster (i.e., FK-Ising) model.  Setting $\kappa=16/3$ in (\ref{N2explicit1}, \ref{N2explicit2})  gives
\be\label{ours}\begin{aligned}
P_{1|1}(16/3\,|\,\lambda)&=P_{2|2}(16/3\,|\,1-\lambda)=\frac{\sqrt{2\lambda}}{\sqrt{1+\sqrt{1-\lambda}}\sqrt{1+\sqrt{\lambda}}},\\
P_{1|2}(16/3\,|\,\lambda)&=P_{2|1}(16/3\,|\,1-\lambda)=\frac{\sqrt{1-\lambda}}{1+\sqrt{\lambda}}.
\end{aligned}\ee
In \cite{chelsmir}, D.\ Chelkak and S.\ Smirnov rigorously calculated the above Ising-FK crossing probabilities, and in our notation, their results read
\be\label{chelkak}\begin{aligned}P_{1|1}(16/3\,|\,\lambda)&=P_{2|2}(16/3\,|\,1-\lambda)=\frac{\sqrt{2}q(\lambda)}{p(\lambda)+\sqrt{2}q(\lambda)},\\
P_{1|2}(16/3\,|\,\lambda)&=P_{2|1}(16/3\,|\,1-\lambda)=\frac{q(\lambda)}{\sqrt{2}p(\lambda)+q(\lambda)},\end{aligned} \quad\begin{aligned}\quad p(\lambda)&=1-q(\lambda)\\ 
&:=\frac{\sqrt{1-\sqrt{\lambda}}}{\sqrt{1-\sqrt{\lambda}} + \sqrt{1-\sqrt{1-\lambda}}}.\end{aligned}\ee
In order to verify that these formulas (\ref{chelkak}) agree with ours (\ref{ours}), we consider the difference of these two formulas for $P_{1|1}(16/3\,|\,\lambda)$:
\be\label{difference}\overbrace{\frac{\sqrt{2}q(\lambda)}{p(\lambda)+\sqrt{2}q(\lambda)}}^{(\ref{chelkak})}-\overbrace{\frac{\sqrt{2\lambda}}{\sqrt{1+\sqrt{1-\lambda}}\sqrt{1+\sqrt{\lambda}}}}^{(\ref{ours})}=\frac{\sqrt{\lambda}\Big(\overbrace{\sqrt{2}\sqrt{1+\sqrt{\lambda}}-\sqrt{2}\sqrt{1-\sqrt{\lambda}}}^{=:g(\lambda)}-\overbrace{2\sqrt{1-\sqrt{1-\lambda}}}^{=:h(\lambda)}\Big)}{\Big(\sqrt{2}\sqrt{1-\sqrt{1-\lambda}}+\sqrt{1-\sqrt{\lambda}}\Big)\sqrt{1+\sqrt{1-\lambda}}\sqrt{1+\sqrt{\lambda}}}.\ee
To show that this difference vanishes, we may either note that for all $\lambda\in[0,1]$, $g(\lambda)^2=h(\lambda)^2$ and $g(1)=h(1)$, from which $g(\lambda)=h(\lambda)$ follows, or we may observe that $f=g+h$ satisfies
\be \lambda(1-\lambda)\partial_\lambda^2f(\lambda)+(1/2-\lambda)\partial_\lambda f(\lambda)+f(\lambda)/16=0\ee
and the initial conditions $f(1/2)=\partial_\lambda f(1/2)=0$, which leads directly to $f=0$.  Equivalence of the two formulas for $P_{2|1}(16/3\,|\,\lambda)$ then follows from the fact that the first (\ref{ours}) satisfies 
\be\label{imposible} P_{1|1}(16/3\,|\,\lambda)+P_{2|1}(16/3\,|\,\lambda)=\frac{\sqrt{2\lambda}}{\sqrt{1+\sqrt{1-\lambda}}\sqrt{1+\sqrt{\lambda}}}+\frac{\sqrt{1-\lambda}}{1+\sqrt{\lambda}}=1.\ee
We may also verify this algebraic identity (\ref{imposible}) by moving the second term to the right side, cross-multiplying the denominators, and squaring both sides of the resulting equation.  
\item\label{step6b2}\emph{$Q$-state Potts model:} With $Q\in\{2,3,4\}$, we suppose that the continuum limit of a critical Potts model on a lattice fills the upper half-plane and that the segments $(x_1,x_2)$ and $(x_3,x_4)$ are fixed to the first state while boundary sites outside these segments freely sample any but the first state.  With (\ref{Qdilute}) relating $Q$ and $\kappa\in(0,4]$, 
\begin{align}\label{PottsXing1} 
P_{1|1}(\kappa\,|\,\lambda)&=\left\{\parbox{11cm}{the probability that one spin cluster touches both $(x_1,x_2)$ and $(x_3,x_4)$, conditioned on both of these segments being fixed to the first spin state and the rest of the boundary freely sampling any but the first spin state}\right\},\\
\label{PottsXing2} 
P_{2|1}(\kappa\,|\,\lambda)&=\left\{\parbox{11cm}{the probability that no spin cluster touches both $(x_1,x_2)$ and $(x_3,x_4)$, conditioned on both of these segments being fixed to the first spin state and the rest of the boundary freely sampling any but the first spin state}\right\}.\end{align}
It is interesting to look at the special case $Q=2$ corresponding to the Ising model.  Setting $\kappa=3$ in (\ref{N2explicit1}, \ref{N2explicit2}), we find the following formula for crossing probability of Ising spin clusters:

\begin{figure}[b]
\centering
\includegraphics[scale=0.16]{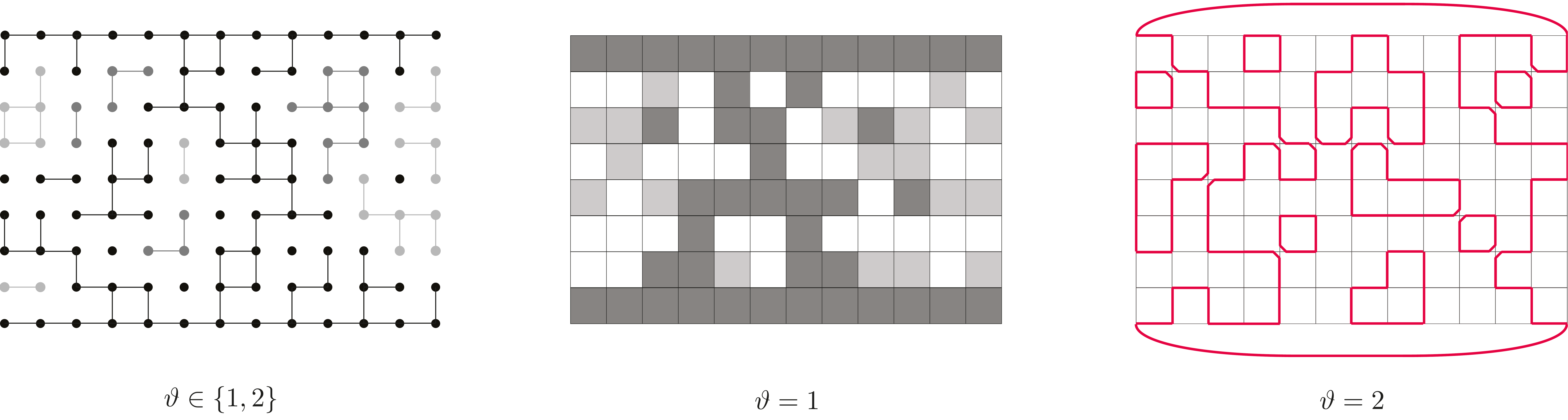}
\caption{$Q=3$ random cluster model sample of either  event $\RCB_1$ or $\RCB_2$ (left), $Q=3$ Potts model sample of the event $\PMB$ (center), and a loop-gas model sample of the event $\LGB_2$, all in the rectangle (with its bottom and top sides fixed).}
\label{Xings}
\end{figure}
\be\label{ourform}P_{1|1}(3\,|\,\lambda)=P_{2|1}(3\,|\,1-\lambda)=\frac{2\Gamma(4/3)}{\Gamma(5/3)\Gamma(8/3)}\bigg(\frac{\lambda^{5/3}}{1-\lambda+\lambda^2}\bigg)\,_2F_1\bigg(-\frac{1}{3},\frac{4}{3};\frac{8}{3}\,\bigg|\,\lambda\bigg).\ee
The literature \cite{argaub,bbk,kozd,dub,argaub2} contains at least three different formulas for this probability (\ref{ourform}), summarized in section \red{7} of \cite{kozd} and discussed in \cite{argaub2}.  In particular, the second formula in \cite{kozd} simplifies to (\ref{ourform}) via linear transformation formulas \cite{absteg} of the hypergeometric function.
\item\label{step6c2}\emph{Loop-gas model:} With $n\in(0,2]$, we suppose that the continuum limit of a dense-phase or dilute-phase loop-gas model on  a lattice fills the upper half-plane.  In addition, ``boundary" loops exit and then re-enter the upper half-plane through $x_1$, $x_2$, $x_3$, and $x_4$, with exactly one loop passing once through each point.  With
\be n(\kappa)=-2\cos(4\pi/\kappa),\quad\begin{cases}\kappa\in(8/3,4],& \text{dilute phase},\\ 
\kappa\in(4,8),&\text{dense phase}, 
\end{cases}\ee
formulas (\ref{N2explicit1}, \ref{N2explicit2}) for $P_{\varsigma|\vartheta}$ give the probability that the parts of the boundary loops in the upper half-plane  connect the points $x_1$, $x_2,$ $x_3$, and $x_4$ pairwise in the $\varsigma$th connectivity, conditioned on the event that the parts of the boundary loops inside the lower-half plane join these points pairwise in the $\vartheta$th connectivity.
\end{enumerate}
Figure \ref{Xings} illustrates some of these interpretations for the random cluster model, the Potts model, and the loop-gas model.  (We note  that our formulas apply only in the continuum limit, not shown in the figure.)
\item\label{step72} We map the upper half-plane onto the rectangle $\mathcal{R}=\{x+iy\,|\,0<x<R, 0<y<1\}$ via the conformal bijection that sends $x_1<x_2<x_3<x_4$ to the corners $w_1=0$, $w_2=R$, $w_3=R+i$, and $w_4=i$ respectively.  With $K$ and $\vartheta_l$ denoting the complete elliptic integral of the first kind and the $l$th Jacobi theta function respectively \cite{morsefesh}, we have
\be\label{RecExplicitR} R=K(\lambda)/K(1-\lambda)\quad\Longleftrightarrow\quad \lambda=\frac{\vartheta_4\left(0,e^{- \pi R} \right)^4} {\vartheta_3\left(0,e^{- \pi R} \right)^4}.\ee
The fixed segments $(x_1,x_2)$ and $(x_3,x_4)$ go to the bottom and top sides of $\mathcal{R}$ respectively, and the BC of the half-plane system goes to the equivalent BC for the image system too.  Because the crossing probability (\ref{xing}) is (predicted to be) conformally invariant, we have the rectangle crossing probability
\be\label{confxing}P_{\varsigma|\vartheta}^\mathcal{R}(\kappa\,|\,R)=P_{\varsigma|\vartheta}\bigg(\kappa\,\bigg|\,\frac{\vartheta_4\left(0,e^{- \pi R} \right)^4} {\vartheta_3\left(0,e^{- \pi R} \right)^4}\bigg).\ee
After replacing ``$(x_1,x_2)$" with ``the bottom side of $\mathcal{R}$" and ``$(x_3,x_4)$" with ``the top side of $\mathcal{R}$," (\ref{firstinterp1}--\ref{lastinterp1}) and (\ref{PottsXing1}, \ref{PottsXing2}) interpret (\ref{confxing}) for the random cluster model and the Potts model in $\mathcal{R}$ respectively.  Also, after replacing ``$x_j$" with ``$w_j$" and ``upper half-plane" with ``$\mathcal{R}$,"  item \ref{step6c2} interprets (\ref{confxing}) for the loop gas model in $\mathcal{R}$.
\end{enumerate}

\section{Simulation results for rectangle and hexagon crossing probability}\label{simxing}

In this section, we present computer simulation  results for  FK cluster (resp.\ spin cluster) crossing probabilities for the critical $Q\in\{2,3,4\}$ random cluster (resp.\ Potts) model inside a rectangle $\mathcal{R}$ and inside a hexagon $\mathcal{H}$.  Item \ref{BCitem1} (resp.\ item \ref{BCitem2}) of the introduction \ref{intro} explains the corresponding side-alternating BCs.  Here, we assign the top and bottom sides of $\mathcal{R}$ the fixed BC, and we assign the upper left/right and bottom sides of $\mathcal{H}$ the fixed BC.  (Ref.\ \cite{fzs} presents simulation  results  for critical percolation ($Q=1$) crossing probabilities for the hexagon.)

\subsection{Simulation results for rectangle crossing probability}\label{RecSimSect}

\begin{figure}[b]
\centering
\includegraphics[scale=0.27]{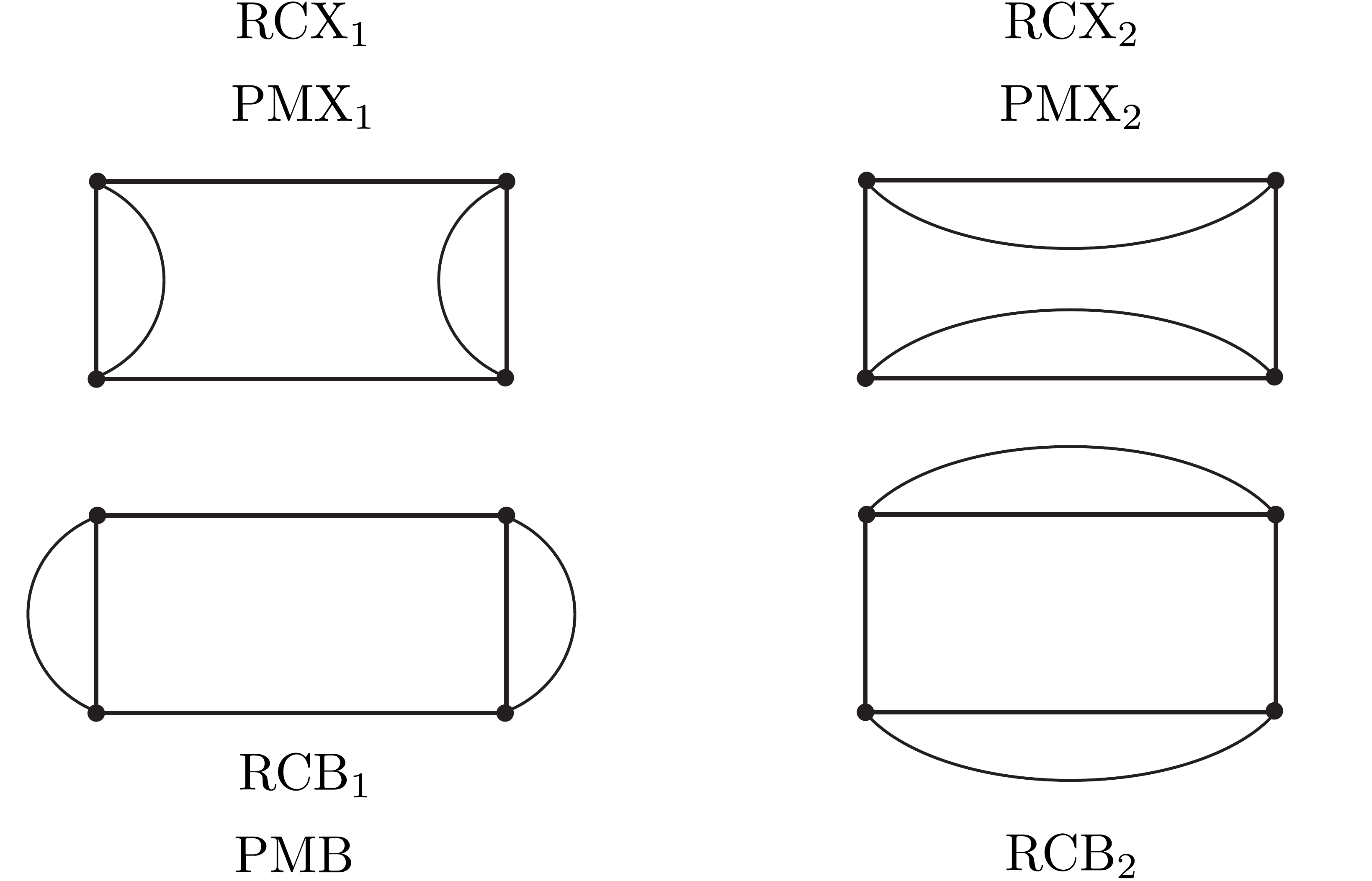}
\caption{ Rectangle crossing events (top row) and  side-alternating BC events (bottom row).  With its top and bottom sides fixed, the bottom left (resp.\ right) rectangle is mutually (resp.\ independently) wired.}
\label{RecXings}
\end{figure}

\begin{figure}[p]
\centering
\includegraphics[scale=0.5]{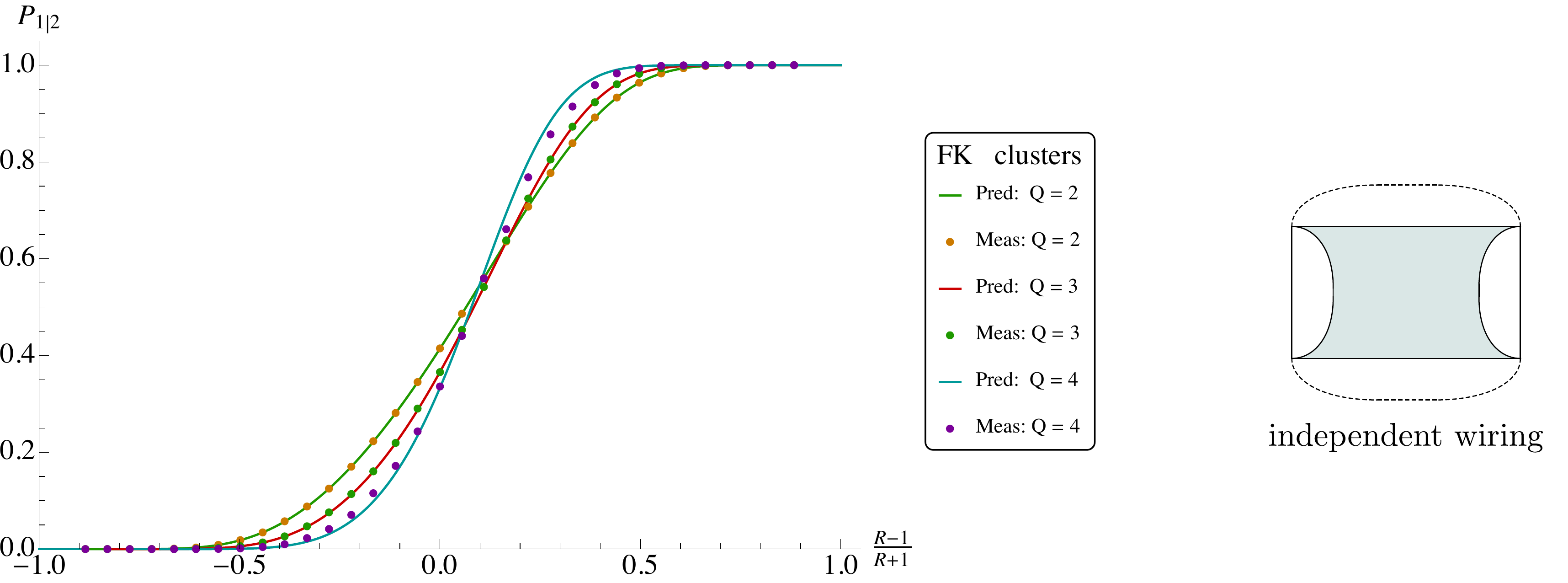}
\vspace{1cm}\\
\includegraphics[scale=0.5]{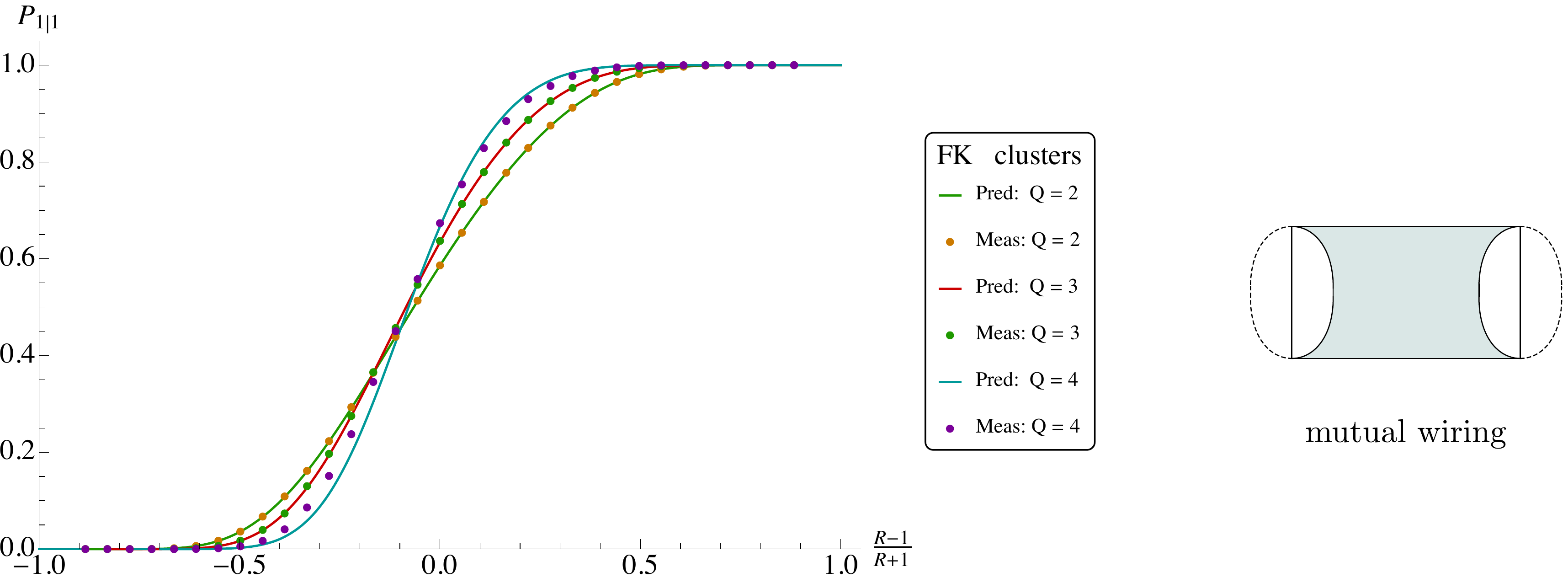}
\vspace{1cm}\\
\includegraphics[scale=0.5]{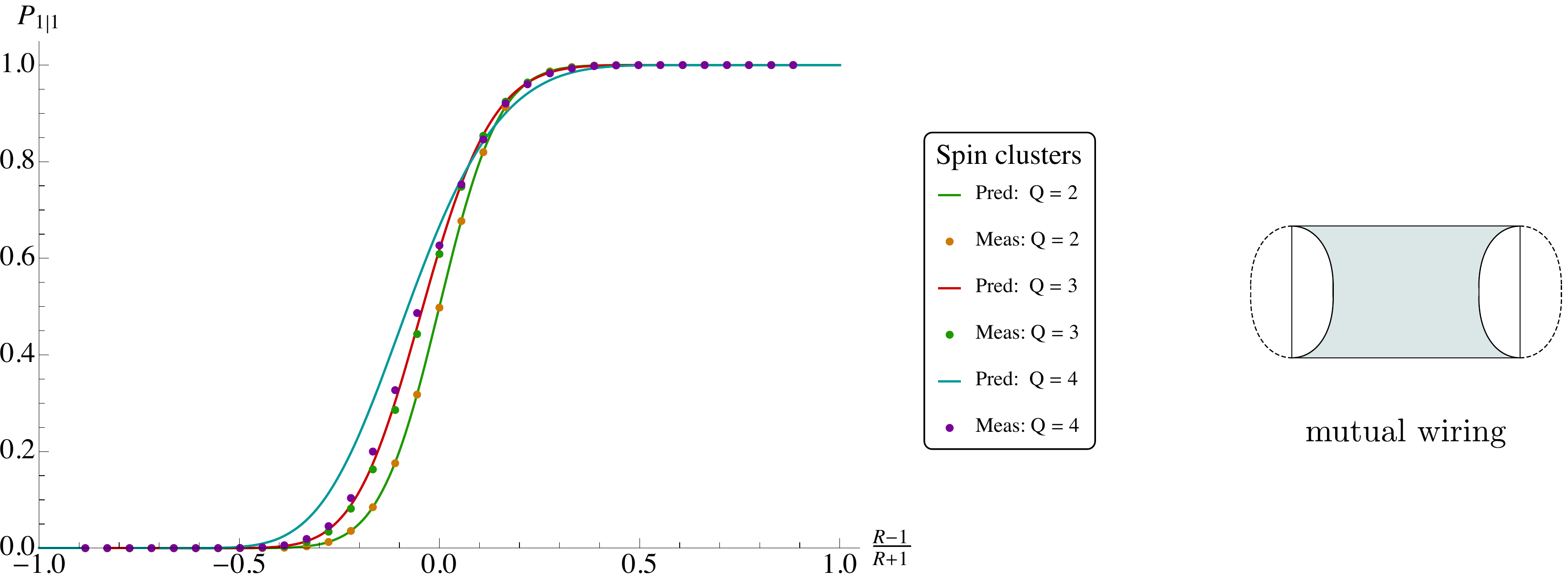}
\vspace{1cm}\\
\caption{Comparison of computer simulation  results  to our prediction (\ref{xing}, \ref{N2explicit1}, \ref{N2explicit2}) of vertical cluster crossing probability in a rectangle with aspect ratio (bottom side to left side) $R$.  Item \ref{BCitem1} (resp.\ item \ref{BCitem2}) of the introduction \ref{intro} defines the BCs and crossing events for random cluster model FK (resp.\ Potts model spin) clusters.  The rectangle to the right of each graph shows the vertical $(\varsigma=1)$ crossing event whose probability $P_{1|\vartheta}$ appears in the graph.  In it, the crossing boundary cluster is colored gray, sides touching a boundary cluster are fixed, and any two fixed sides (one fixed side) with endpoints joined by a dashed exterior  arc are mutually wired together (resp.\ wired independently of the other fixed side).}
\label{RecXingPlots}
\end{figure}

In this section, we present computer simulation  results  of FK cluster (resp.\ spin cluster) crossing probabilities for the critical $Q\in\{2,3,4\}$ random cluster (resp.\ Potts) model on a square lattice inside a rectangle $\mathcal{R}$.  (Ref.\ \cite{c3} presents  similar results for  critical percolation ($Q=1$) crossing probabilities for the rectangle.)  Our simulations sampled thirty-three rectangles,  each with two successive side lengths, $\ell$ for the length of (i.e., number of lattice sites in) the top and bottom sides, and $\ell'$ for the length of the left and right sides.  For the $i$th rectangle, we chose
\be\label{ellRec}\ell_i=\lfloor 120.471+56.4706 (i-1)\rfloor,\quad \ell'_i\approx2048-\ell_i\approx\ell_{34-i},\quad \text{for $i\in\{1,2,\ldots,33\}$}.\ee
These side-lengths (\ref{ellRec}) are almost uniformly distributed between $\ell_1$ and $\ell_{33}$, and with $R_i:=\ell_i/\ell_i'$ the aspect ratio of the $i$th rectangle, we have $R_1\approx1/16,$ $R_{17}=1$, and $R_{33}\approx16$.  

The formula (\ref{xing}) for crossing probability presumes that we condition the system inside the rectangle on a side-alternating BC event, as described in items \ref{BCitem1} or \ref{BCitem2} of the introduction \ref{intro}.  For the random cluster model, these are $\RCB$ events, and the bottom row of figure \ref{RecXings} illustrates them with exterior arc rectangle diagrams (item \ref{it5} of section \ref{mutsect}), where any two fixed sides joined by an exterior arc are mutually wired together (i.e., constrained to exhibit the same state).  So with the top and bottom sides of $\mathcal{R}$ fixed, these two sides are mutually wired together in the first $\RCB$ event, $\RCB_1$, and independently wired (i.e., not constrained to exhibit the same state) in the second, $\RCB_2$.  For the Potts model (section \ref{FLBCsect}), we condition on the $\PMB$ event with the top and bottom sides of $\mathcal{R}$ mutually wired together and the left and right sides having the fluctuating BC.

Conditioned as usual on either of the two $\RCB$ events described above, there are $C_2=2$ distinct FK-cluster crossing events of the rectangle, the event $\RCX_1$ that a vertical crossing cluster joins the two fixed sides of $\mathcal{R}$, and the event $\RCX_2$ that there is no such cluster (items \ref{it1} and \ref{it4} of section \ref{mutsect}).  The top row of figure \ref{RecXings} illustrates these crossing events with interior arc rectangle diagrams (item \ref{it4} of section \ref{mutsect}).  There are also $C_2=2$ distinct spin-cluster crossing events of the rectangle, $\PMX_1$ and $\PMX_2$.  These events are similarly defined, but for spin clusters rather than FK clusters.  Because the  probabilities of the first and second crossing events sum to one, our simulations measure the probability of only the first, the vertical crossing event $\RCX_1$ or $\PMX_1$.

We used the Swendsen-Wang algorithm \cite{sw}, with bond activation probability $p=p_c^{\text{sqr.}}$ (\ref{criticalpt}), to generate $2^{22}=4,194,304$ samples of the critical random cluster model and the critical Potts model in each rectangle.  The time required to generate all samples for one rectangle, using a single 2GHz processor, was about five days.  For each model, we seeded the FK clusters in a way that respects the corresponding side-alternating BC:
\begin{enumerate}
\item\label{SimRec1} \textbf{The $Q$-color random cluster model:}  $Q\in\{2,3,4\}$.  In each sample, we seeded the FK boundary clusters that touch a fixed side of $\mathcal{R}$ before we seeded any other FK cluster.  The $\RCB$ event requires that we activate all FK bonds within the top and bottom sides of $\mathcal{R}$.  If these sides were mutually wired together, then we assigned them the same color, and if they were wired independently of each other, then we did not impose this constraint. 
\item\label{SimRec2} \textbf{The $Q$-state Potts model:}  $Q\in\{2,3,4\}$.  In each sample, we seeded the spin boundary clusters that touch a fixed side of $\mathcal{R}$ first, giving them the first spin state.  Then we seeded the spin clusters that touched a fluctuating side (item \ref{BCitem2} of the introduction \ref{intro}) of $\mathcal{R}$ next, giving each of them any but the first spin state with uniform probability.  Finally, we seeded all other spin clusters, giving each any spin state with uniform probability.
\end{enumerate}
The wiring within the fixed sides of $\mathcal{R}$ ensured that one FK (resp.\ spin) boundary cluster touched the bottom side of $\mathcal{R}$ in case \ref{SimRec1} (resp.\ case \ref{SimRec2}).  If this cluster touched the top side of $\mathcal{R}$ too, then the sample exemplified a vertical crossing event $(\varsigma=1)$, and if otherwise, then it did not.  We tallied the number of samples with a vertical crossing and divided it by the total number of samples to measure the vertical crossing probability.

Formula (\ref{xing}) with $x_1=0$, $x_2=m$, $x_3=1$, and $x_4=\infty$ gives the probability $P_{\varsigma|\vartheta}$ of the $\varsigma$th FK (resp.\ spin) cluster crossing event $\RCX_\varsigma$ (resp.\ $\PMX_\varsigma$) conditioned on the $\vartheta$th $\RCB$ event $\RCB_\vartheta$ (resp.\ the one $\PMB$ event, with $\vartheta=1$) for the rectangle $\mathcal{R}$ with aspect ratio $R$ (bottom to left side) (\ref{RecExplicitR}).  (There, we have $\lambda=m$.  Also, section \ref{rectxingsummary} casts this formula in the simpler form (\ref{N2explicit1}, \ref{N2explicit2}).)  Then according to figure \ref{RecXings}, $P_{1|1}$ (resp. $P_{1|2}$) is the probability of the vertical crossing event in $\mathcal{R}$, conditioned on the mutual wiring (resp.\ independent wiring) BC event.

Figure \ref{RecXingPlots} compares our simulation measurements of these rectangle crossing probabilities to our predictions (\ref{N2explicit1}, \ref{N2explicit2}, \ref{confxing}).  We plot the probabilities as functions of $x=(R-1)/(R+1)\in(0,1)$, where $R$ is the rectangle's aspect ratio (bottom to left side).  Our plots show excellent agreement for $Q\in\{2,3\}$ and more modest agreement for $Q=4$.  (Actually, we did not need to test the $Q=2$ cases because  their crossing probability formulas (\ref{chelkak}, \ref{ourform}) \cite{kozd,chelsmir} have been proven rigorously.  But for completeness, we include them here.)  Table \ref{ErrorSummaryTable} gives the average relative error of our measurements and the standard  deviation of our relative errors from the average relative error.  (See the last two paragraphs of section \ref{HexSimSect} for an explanation of these results.)

\begin{figure}[b]
\centering
\includegraphics[scale=0.27]{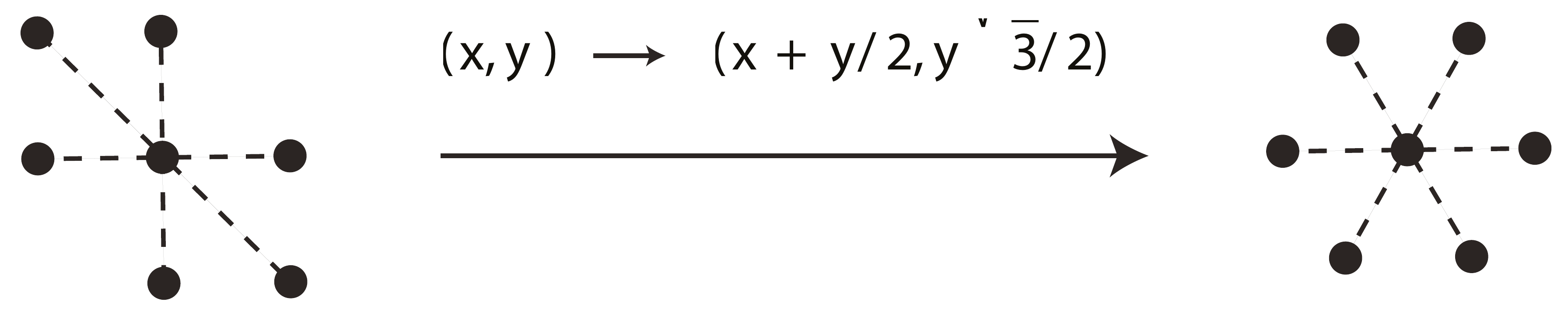}
\caption{The transformation from a square lattice to a triangular lattice.  Dotted lines connect the center site with its nearest neighbors.}
\label{TriLat}
\end{figure}

\subsection{Simulation results for hexagon crossing probability}\label{HexSimSect}

In this section, we present computer simulation measurements of FK cluster (resp.\ spin cluster) crossing probabilities for the critical $Q\in\{2,3,4\}$ random cluster (resp.\ Potts) model on a triangular lattice inside an equiangular hexagon $\mathcal{H}$.  (We presented  simulation measurements of critical percolation ($Q=1$) crossing probabilities for the hexagon in \cite{fzs}.)  Our simulations sampled thirty-three hexagons, each with two successive side lengths: $\ell$ for the length of (i.e., number of lattice sites in) the bottom and upper left/right sides, and $\ell'$ for the length of the remaining sides.  For the $i$th hexagon, we chose
\be\label{ell}\ell_i=\lfloor 60.2353 + 28.2353(i-1)\rfloor,\quad \ell'_i\approx1024-\ell_i\approx\ell_{34-i},\quad \text{for $i\in\{1,2,\ldots,33\}.$}\ee
These side-lengths (\ref{ell}) are almost uniformly distributed between $\ell_1$ and $\ell_{33}$, and with $R_i:=\ell_i/\ell_i'$ the side-length ratio of the $i$th hexagon, we have $R_1\approx1/16,$ $R_{17}=1$, and $R_{33}\approx16$.  

In our simulations, the triangular lattice is represented by a square lattice with six nearest neighbors to each lattice site, as figure \ref{TriLat} shows.  Also, the hexagon is not equiangular, but its  interior angles measure $3 \pi/4$, $\pi/2$, $ 3 \pi/4$, $ 3 \pi/4$, $\pi/2$, and $ 3 \pi/4$ radians respectively, in counterclockwise order starting from the bottom-left vertex.  These  angles  ensure that the hexagon's sides follow along nearest-neighbor bonds.  The transformation of figure \ref{TriLat} takes us from this setup to the usual triangular lattice inside the equiangular hexagon $\mathcal{H}$.

\begin{figure}[b]
\centering
\includegraphics[scale=0.27]{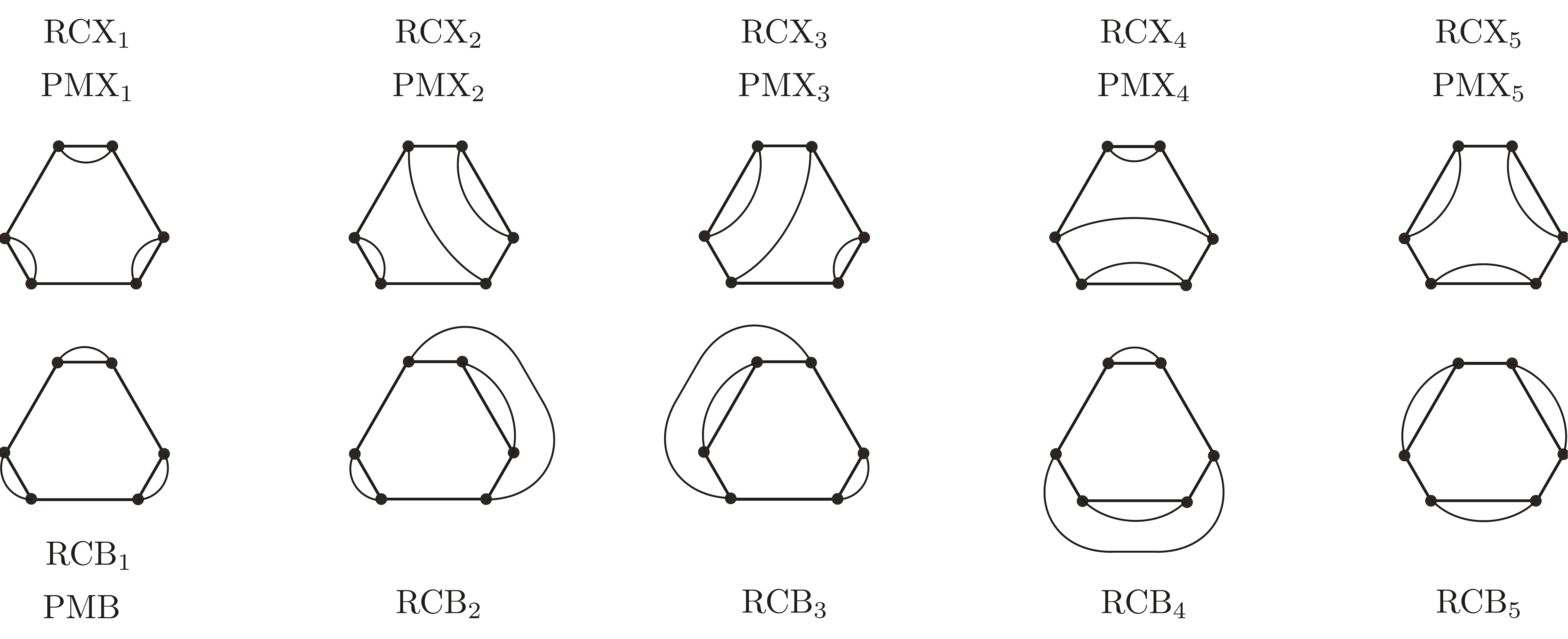}
\caption{ Hexagon crossing events (top row) and  side-alternating BC events (bottom row).  With its  bottom and upper-left/right sides fixed, the bottom-left (resp.\ bottom-right) hexagon has mutual (resp.\ independent) wiring, and bottom-middle hexagons have mixed wiring.}
\label{HexXings}
\end{figure}

The crossing probability formula (\ref{xing}) presumes that we condition the system inside the hexagon on a side-alternating BC event, as described in items \ref{BCitem1} or \ref{BCitem2} of the introduction \ref{intro}.  In our present situation, these BCs are as follows:
\begin{enumerate}[wide, labelwidth=!, labelindent=0pt]
\item\label{SimHex1} \textbf{The $Q$-color random cluster model:} Item \ref{BCitem1} of the introduction \ref{intro} and items \ref{it1}--\ref{it8} of section \ref{mutsect} describe the available side-alternating BC events.  They are $\RCB$ events, and there are $C_3=5$ of them, $\RCB_1$, $\RCB_2,\ldots, \RCB_5 $, enumerated as follows: in $\RCB_1$, the three fixed sides of $\mathcal{H}$ are mutually wired together (i.e., constrained to exhibit the same state).  In $\RCB_5$, the three sides are independently wired (i.e., not constrained to exhibit the same state). Finally, in $\RCB_2,$ $\RCB_3$, and $\RCB_4$, two fixed sides are mutually wired together while the remaining fixed side is wired independently of the others.  Because the hexagon is invariant under rotations of $2\pi/3$  (thanks  to its alternating side-lengths), we may identify these last three  events as  one.  All in all, our simulations sampled the $\RCB$ events $\RCB_1$, $\RCB_3$, and $\RCB_5$, which we respectively call ``mutual wiring," ``independent wiring," and ``mixed wiring."
\item\label{SimHex2} \textbf{The $Q$-state Potts model:} Item \ref{BCitem2} of the introduction \ref{intro} and section \ref{FLBCsect} describe the one available side-alternating BC event. It is the event $\PMB$ with all fixed sides of $\mathcal{H}$ mutually wired together and the remaining three sides of $\mathcal{H}$ exhibiting the fluctuating BC.
\end{enumerate}
In our simulations, we imposed the fixed BC on the bottom and upper-left/right sides of $\mathcal{H}$.  As such, the remaining three sides of $\mathcal{H}$ exhibited either the free BC or the fluctuating BC.  The bottom row of figure \ref{HexXings} illustrates these side-alternating BCs with exterior arc hexagon diagrams (item \ref{it5} of section \ref{mutsect}). There, any two fixed sides that are joined by an exterior arc are mutually wired together.

There are $C_3=5$ distinct FK-cluster crossing events of the hexagon, $\RCX_1$, $\RCX_2,\ldots, \RCX_5 $ (items \ref{it1} and \ref{it4} of section \ref{mutsect}), and we enumerate them as follows: in $\RCX_1$, a single cluster connects all fixed sides of $\mathcal{H}$.  In $\RCX_5$, no cluster connects any two fixed sides of $\mathcal{H}$.  Finally, in $\RCX_2$, $\RCX_3$, and $\RCX_4$, a single cluster connects two fixed sides of $\mathcal{H}$ but does not touch the remaining fixed side.  The top row of figure \ref{HexXings} illustrates these crossing events with interior arc hexagon diagrams (item \ref{it4} of section \ref{mutsect}).  

%%%%%%%%%%%%%%%%%%%%%%%%%%%%%%%%%%%%%%%%%%%%%%%%%%%%%%%%%%%%%%%%%%%%%%%%%%%
%%%%%%%%%%%%%%%%%%%%%%%%%%%%%%%%%%%%%%%%%%%%%%%%%%%%%%%%%%%%%%%%%%%%%%%%%%%

Thanks to some symmetries of the hexagon, we only needed to measure probabilities of some of these crossing events if we conditioned on an $\RCB$ event (item \ref{SimHex1}).  For example, because the mutual (resp.\ independent) wiring $\RCB$ event $\RCB_1$ (resp.\ $\RCB_5$) is invariant under rotations of $2\pi/3$, the probabilities of the FK-cluster crossing events $\RCX_2$, $\RCX_3$, and $\RCX_4$ conditioned on this $\RCB$ event are equal.  As such, we only measured the conditioned probability of the event $\RCX_3$ in which a single FK cluster connects the bottom and upper-right sides of $\mathcal{H}$.  Although the mixed wiring $\RCB$ event $\RCB_3$ does not have this rotational symmetry, it is evident that the probabilities of $\RCX_2$ and $\RCX_4$ conditioned on $\RCB_3$ are equal, so we only measured the conditioned probability of the former event.  

There are also $C_3=5$ distinct spin-cluster crossing events of the hexagon, $\PMX_1$, $\PMX_2,\ldots, \PMX_5 $ (items \ref{it13}--\ref{it23} of section \ref{FLBCsect}).  In the event $\PMX_\varsigma$, spin clusters join the fixed sides of $\mathcal{H}$ as do FK clusters in the event $\RCX_\varsigma$.  Again, thanks  to symmetries of the hexagon, we only needed to measure probabilities of events $\PMX_1$, $\PMX_3$, and $\PMX_5$ when we condition on the one available $\PMB$ event (item \ref{SimHex2}).

We used the Swendsen-Wang algorithm \cite{sw}, with bond activation probability $p=p_c^{\text{tri.}}$ (\ref{pc2}), to generate $2^{22}=4,194,304$ samples of the critical random cluster model and the critical Potts model in every hexagon.  The time required to generate all samples for one hexagon, using a 2GHz processor, was about five days.  We seeded the FK (resp.\ spin) clusters of case \ref{SimHex1} (resp.\ case \ref{SimHex2}) for the critical random cluster (resp.\ Potts) model inside the hexagon, as we did in case \ref{SimRec1} (resp.\ case \ref{SimRec2}) for the same model inside the rectangle.  As we generated the FK (resp.\ spin) boundary clusters of a sample, we noted which fixed sides of $\mathcal{H}$ these clusters touched to determine the crossing pattern of the sample.  We tallied the number of samples with the $\varsigma$th crossing pattern and divided this total by the total number of samples to measure the $\varsigma$th FK (resp.\ spin) cluster crossing probability.

%%%%%%%%%%%%%%%%%%%%%%%%%%%%%%%%%%%%%%%%%%%%%%%%%%%%%%%%%%%%%%%%%%%%%%%%%%%%%%%%%%%%%%%%%%%%%%%%%%%%%%%%%%%%%%%%%%%%%%%%%%%%%%%%%%%%%%%%%%%%%%%%
\begin{figure}[t]
\centering
\includegraphics[scale=0.27]{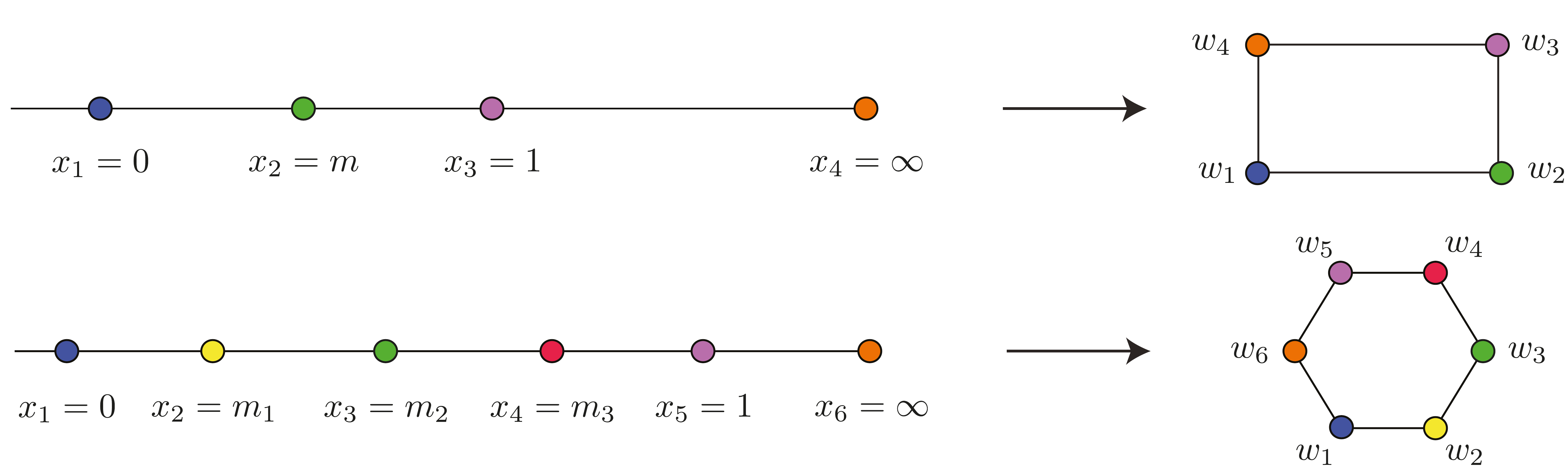}
\caption{The top (resp.\ bottom) line illustrates the Schwarz-Christoffel map sending the upper half-plane onto the interior of the rectangle (resp.\ hexagon).  The $j$th vertex $w_j\in\mathbb{C}$ of the rectangle (resp.\ hexagon) is the image of $x_j$ under this map.}
\label{Swchmap}
\end{figure}
%%%%%%%%%%%%%%%%%%%%%%%%%%%%%%%%%%%%%%%%%%%%%%%%%%%%%%%%%%%%%%%%%%%%%%%%%%%%%%%%%%%%%%%%%%%%%%%%%%%%%%%%%%%%%%%%%%%%%%%%%%%%%%%%%%%%%%%%%%%%%%%%

The formula (\ref{xing}) for the crossing probability is given in terms of six coordinates $x_j$, corresponding to the images of the six vertices $w_j$ of the hexagon $\mathcal{H}$ under a conformal bijection from the interior of $\mathcal{H}$ onto the upper half-plane $\mathbb{H}$.  We chose the inverse of this bijection to be the Schwarz-Christoffel transformation $f$ (\ref{sch}) with parameters (\ref{choices1}) having $C=1$ (appendix \ref{transformxing}).  With these choices,
\be\label{ftrans}f(z)=\frac{2}{3}\sideset{}{_0^z}\int\zeta^{-1/3}(m_1-\zeta)^{-1/3}(m_2-\zeta)^{-1/3}(m_3-\zeta)^{-1/3}(1-\zeta)^{-1/3}\,{\rm d}\zeta.\ee
This map (\ref{ftrans}) sends the boundary points (figure \ref{Swchmap})
\be\label{thepoints} x_1=0<x_2=m_1<x_3=m_2<x_4=m_3<x_5=1<x_6=\infty\ee
counterclockwise to the hexagon's vertices $w_1,$ $w_2,\ldots,w_6$ respectively.  With $w_1=f(0)=0$, $w_2=f(m_1)>0$, and $w_3=f(m_2)$ in $\mathbb{H}$, the hexagon resides in $\mathbb{H}$ with its base flush against the positive-real axis.  

According to (\ref{crossratio}), the crossing probability actually depends on just three parameters, $\lambda_j=m_j$ (\ref{crossratio}) for $j\in\{1,2,3\}$, that determine the shape of $\mathcal{H}$.  Restricting to
\be\label{m1m3}m_1=\frac{m_2^2}{1-m_2+m_2^2},\qquad m_3=\frac{m_2}{1-m_2+m_2^2}\ee
ensures that the side-lengths of $\mathcal{H}$ alternate.  Let us briefly explain why: it is easy to see that the following conformal bijection $\phi(m_1\,|\,\cdot):\mathbb{H}\longrightarrow\mathbb{H}$ continuously extended to $\mathbb{R}$ and thus sending $m_1\mapsto0$, $0\mapsto\infty$, and $\infty\mapsto1$,
\be \phi(m_1\,|\,z):=\frac{z-m_1}{z}\ee
rotates the hexagon $\mathcal{H}$ from (\ref{ftrans}) by $-\pi/3$ radians.  If the side-lengths of $\mathcal{H}$ alternate, then $\mathcal{H}$ must be invariant under two consecutive such rotations, corresponding to $z\mapsto\phi(\phi(m_1\,|\,m_2)\,|\,\phi(m_1\,|\,z))$.  This implies the system of equations
\begin{gather} 0=\phi(\phi(m_1\,|\,m_2)\,|\,\phi(m_1\,|\,m_2)),\qquad m_1=\phi(\phi(m_1\,|\,m_2)\,|\,\phi(m_1\,|\,m_3)),\qquad m_2=\phi(\phi(m_1\,|\,m_2)\,|\,\phi(m_1\,|\,1)),\\
m_3=\phi(\phi(m_1\,|\,m_2)\,|\,\phi(m_1\,|\,\infty)),\qquad 1=\phi(\phi(m_1\,|\,m_2)\,|\,\phi(m_1\,|\,0)),\qquad \infty=\phi(\phi(m_1\,|\,m_2)\,|\,\phi(m_1\,|\,m_1)).\end{gather}
Solving this system gives (\ref{m1m3}).  This reduces the number of independent parameters that control the shape of $\mathcal{H}$ from three to one.  Hence, if $\ell$ (resp. $\ell'$) is the length of the bottom (resp.\ top) side, then the side-length ratio $R=\ell/\ell'$ is
\be\label{ratioside}\ell=f(m_1),\quad \ell'=|f(m_2)-f(m_1)|\quad\Longrightarrow\quad R=\frac{f(m_1)}{|f(m_2)-f(m_1)|}.\ee
Relations (\ref{ftrans}--\ref{ratioside}) put $R\in(0,\infty)$ in one-to-one correspondence with $m_2\in(0,1)$.  We can manipulate these relations using the symmetry of the hexagon to find
\be\label{explicitR}
R=\dfrac{\displaystyle{\int_0^1} \left[\zeta(1-\zeta)\left(1-\frac{m_2^2\,\zeta}{1-m_2+m_2^2}\right)(1-m_2 \zeta)\left(1-\frac{m_2\,\zeta}{1-m_2+m_2^2}\right)\right]^{-1/3}\,{\rm d}\zeta}{\displaystyle{\int_0^1} \left[\zeta(1-\zeta)\left(1-\left(1-\frac{m_2}{1-m_2+m_2^2}\right)\zeta\right)(1-(1-m_2) \zeta)\left(1-\left(1-\frac{m_2^2}{1-m_2+m_2^2}\right)\zeta\right)\right]^{-1/3}\,{\rm d}\zeta}.
\ee

Formula (\ref{xing}) with (\ref{thepoints}, \ref{m1m3}) gives the probability $P_{\varsigma|\vartheta}$ of the $\varsigma$th FK (resp.\ spin) cluster crossing event $\RCX_\varsigma$ (resp.\ $\PMX_\varsigma$) conditioned on the $\vartheta$th $\RCB$ event $\RCB_\vartheta$ (resp.\ the one $\PMB$ event with $\vartheta=1$) for the hexagon $\mathcal{H}$ with alternating side lengths and side-length ratio $R$ (\ref{explicitR}).  Items \ref{step1}--\ref{step5} of section \ref{xingsummary} determine formulas for the functions $\mathcal{F}_\vartheta$ and $\Pi_\varsigma$ that appear in this formula.  However, simpler formulas are available for the hexagon.  Indeed, we may replace the formula (\ref{Fexplicit}) for $\mathcal{F}_\vartheta$ by its alternative (\ref{Fother}) with one less integration contour.  With connectivities labeled as in figure \ref{Connectivities}, we have
\begin{align}\label{Fc1}\mathcal{F}_1&=\text{(\ref{Fother}) with $\Gamma_1=\mathscr{P}(x_2,x_3)$, $\Gamma_2=\mathscr{P}(x_4,x_5)$, and $c=6$},\\
\label{Fc3}\mathcal{F}_3&=\text{(\ref{Fother}) with $\Gamma_1=\mathscr{P}(x_2,x_3)$, $\Gamma_2=\mathscr{P}(x_5,x_6)$, and $c=4$},\\
\label{Fc5}\mathcal{F}_5&=\text{(\ref{Fother}) with $\Gamma_1=\mathscr{P}(x_1,x_2)$, $\Gamma_2=\mathscr{P}(x_3,x_4)$, and $c=6$},\end{align}
where $\mathscr{P}(x_i,x_j)$ is the Pochhammer contour entwining points $x_i$ and $x_j$ (figure \ref{BreakDown}).  Our choices of $c$ values in (\ref{Fc1}--\ref{Fc3}) ensure that integration contours entwine adjacent branch points and that no integration contour arcs over or wraps around another contour or branch point.  This choice facilitates the numerical integration.  In addition, section \red{II C} of \cite{fsk} gives simpler formulas for the hexagon crossing weights $\Pi_1$ and $\Pi_2$.  These formulas use the functions
\begin{multline}\label{Iij}I_{ij}^{\scaleobj{0.85}{(c)}}(\kappa\,|\,m_1,m_2,m_3):=\Bigg(\prod_{\substack{j<k \\ j,k\neq c}}^6(x_k-x_j)^{2/\kappa}\Bigg)\Bigg(\prod_{\substack{k=1 \\ k\neq c}}^6|x_c-x_k|^{1-6/\kappa}\Bigg)\\
\int_{x_i}^{x_{i+1}}\int_{x_j}^{x_{j+1}}\mathcal{N}\bigg[(u_2-u_1)^{8/\kappa}\prod_{i=1}^2\prod_{j=1}^6(x_j-u_i)^{\beta_j}\bigg]\,{\rm d}u_1\,{\rm d}u_2,\qquad i\neq j,\quad i,j,c\in\{1,2,3,4,5,6\},\\
\text{with $x_1=0,$ $x_2=m_1,$ $x_3=m_2,$ $x_4=m_3,$ $x_5=1,$ $x_6=\infty$,}\,\,\beta_j=\begin{cases}-4/\kappa,& j\neq c, \\ 12/\kappa-2,&j=c,\end{cases},\,\,\text{and}\,\,\int_{x_6}^{x_7}:=\int_{-\infty}^0,\end{multline}
where $\mathcal{N}[\,\ldots\,]$ orders the differences inside its brackets so the integrand is real-valued over $[x_i,x_{i+1}]\times[x_j,x_{j+1}]$.  From the formulas for $\Pi_1$ and $\Pi_2$, we find formulas for the other crossing weights, $\Pi_\varsigma$ with $\varsigma\in\{3,4,5\}$, by rotating the hexagon to send either the first or second connectivity to the $\varsigma$th connectivity and then cyclically permuting the arguments of $\Pi_1$ or $\Pi_2$ and the value of $c$ consistently with this rotation.

Next, we adjust the formulas for the crossing weights given in the previous paragraph in order to minimize the number of contour integrals that must be computed numerically.  Assuming $\kappa>4$ (so we may replace all Pochhammer contours with simple curves via figure \ref{PochhammerContour}), these adjusted formulas are
\begin{align}\label{Pi1For}\Pi_1(\kappa\,|\,x_1,x_2,\ldots,x_6)&=\mathcal{P}(\kappa\,|\,x_1,x_2,\ldots,x_6)[I_{13}\super{2}-n I_{12}\super{2}](\kappa\,|\,x_1,x_2,\ldots,x_6),\\
\label{Pi2N3}\Pi_2(\kappa\,|\,x_1,x_2,\ldots,x_6)&=\mathcal{P}(\kappa\,|\,x_1,x_2,\ldots,x_6)I_{12}\super{2}(\kappa\,|\,x_1,x_2,\ldots,x_6),\\
\Pi_3(\kappa\,|\,x_1,x_2,\ldots,x_6)&=\mathcal{P}(\kappa\,|\,x_1,x_2,\ldots,x_6)I_{34}\super{4}(\kappa\,|\,x_1,x_2,\ldots,x_6),\\
\Pi_5\label{Pi5For}(\kappa\,|\,x_1,x_2,\ldots,x_6)&=\mathcal{P}(\kappa\,|\,x_1,x_2,\ldots,x_6)[I_{24}\super{4}-n I_{34}\super{4}](\kappa\,|\,x_1,x_2,\ldots,x_6),\end{align}
where
\be\mathcal{P}(\kappa\,|\,x_1,x_2,\ldots,x_6):=n(\kappa)^2 \, \frac{\Gamma(2-8/\kappa)^2}{\Gamma(1-4/\kappa)^4}\Bigg(\prod_{\substack{i<j \\ i,j\neq 2}}^6(x_j-x_i)^{2/\kappa}\Bigg)\Bigg(\prod_{k\neq2}^6|x_2-x_k|^{1-6/\kappa}\Bigg).\ee
Indeed, (\ref{Pi1For}) gives a formula for $\Pi_1$  that is apparently even simpler than the one presented in section \red{II C} of \cite{fsk}.  Also, we use (\ref{Pi2N3}) for our formula for $\Pi_2$, which is slightly different from the formula given in section \red{II C} of \cite{fsk}; this difference arises from a changed value of $c$ in (\ref{Iij}).  Using the techniques presented in \cite{fsk}, it is easy to verify that (\ref{Pi1For}--\ref{Pi5For}) are indeed formulas for the crossing weights $\Pi_1,$ $\Pi_2$, $\Pi_3$, and $\Pi_5$.

 Upon inserting (\ref{Fc1}--\ref{Fc5}, \ref{Pi1For}--\ref{Pi5For}) into (\ref{xing}) with $N=3$ and (\ref{thepoints}), we find explicit formulas for crossing probabilities in the hexagon.  Altogether, for the hexagon with mutual wiring, the probability of the $\varsigma$th FK (resp.\ spin) cluster crossing event with $\varsigma\in\{1,3,5\}$ is
\begin{align}
\label{firstprob}P_{1|1}^\mathcal{H}(\kappa\,|\,m_1,m_2,m_3)&=\bigg[\frac{n^2I_{13}^{\scaleobj{0.85}{(2)}}-n^3I_{12}^{\scaleobj{0.85}{(2)}}}{I_{24}^{\scaleobj{0.85}{(6)}}}\bigg](\kappa\,|\,m_1,m_2,m_3),\\
P_{3|1}^\mathcal{H}(\kappa\,|\,m_1,m_2,m_3)&=\bigg[\frac{nI_{34}^{\scaleobj{0.85}{(4)}}}{I_{24}^{\scaleobj{0.85}{(6)}}}\bigg](\kappa\,|\,m_1,m_2,m_3),\\
\label{lastmutprob}P_{5|1}^\mathcal{H}(\kappa\,|\,m_1,m_2,m_3)&=\bigg[\frac{I_{24}^{\scaleobj{0.85}{(4)}}-nI_{34}^{\scaleobj{0.85}{(4)}}}{I_{24}^{\scaleobj{0.85}{(6)}}}\bigg](\kappa\,|\,m_1,m_2,m_3),\end{align}
where $\kappa$ is related to $Q$ via (\ref{dense}) (resp.\ (\ref{dilute}), denoted there as $\hat{\kappa}$).  For the hexagon with mixed wiring, the probability of the $\varsigma$th FK-cluster crossing event with $\varsigma\in\{1,2,3,5\}$ is given by
\begin{align}
\label{firstmixprob} P_{1|3}^\mathcal{H}(\kappa\,|\,m_1,m_2,m_3)&=\bigg[\frac{nI_{13}\super{2}-n^2I_{12}\super{2}}{I_{25}^{\scaleobj{0.85}{(4)}}}\bigg](\kappa\,|\,m_1,m_2,m_3),\\
\label{secondmixprob}P_{2|3}^\mathcal{H}(\kappa\,|\,m_1,m_2,m_3)&=\bigg[\frac{I_{12}^{\scaleobj{0.85}{(2)}}}{I_{25}^{\scaleobj{0.85}{(4)}}}\bigg](\kappa\,|\,m_1,m_2,m_3),\\
\label{thirdmixprob}P_{3|3}^\mathcal{H}(\kappa\,|\,m_1,m_2,m_3)&=\bigg[\frac{n^2I_{34}^{\scaleobj{0.85}{(4)}}}{I_{25}^{\scaleobj{0.85}{(4)}}}\bigg](\kappa\,|\,m_1,m_2,m_3),\\
\label{lastmixprob} P_{5|3}^\mathcal{H}(\kappa\,|\,m_1,m_2,m_3)&=\bigg[\frac{nI_{24}^{\scaleobj{0.85}{(4)}}-n^2I_{34}^{\scaleobj{0.85}{(4)}}}{I_{25}^{\scaleobj{0.85}{(4)}}}\bigg](\kappa\,|\,m_1,m_2,m_3),\end{align}
where, again, $\kappa$ is related to $Q$ via (\ref{dense}).  Finally, for the hexagon with independent wiring, the probability of the $\varsigma$th FK-cluster crossing event with $\varsigma\in\{1,3,5\}$ is 
\begin{align}\label{firstindprob} 
P_{1|5}^\mathcal{H}(\kappa\,|\,m_1,m_2,m_3)&=\bigg[\frac{I_{13}^{\scaleobj{0.85}{(2)}}-nI_{12}^{\scaleobj{0.85}{(2)}}}{I_{13}^{\scaleobj{0.85}{(6)}}}\bigg](\kappa\,|\,m_1,m_2,m_3),\\
P_{3|5}^\mathcal{H}(\kappa\,|\,m_1,m_2,m_3)&=\bigg[\frac{nI_{34}^{\scaleobj{0.85}{(4)}}}{I_{13}^{\scaleobj{0.85}{(6)}}}\bigg](\kappa\,|\,m_1,m_2,m_3),\\
\label{lastprob}P_{5|5}^\mathcal{H}(\kappa\,|\,m_1,m_2,m_3)&=\bigg[\frac{n^2I_{24}^{\scaleobj{0.85}{(4)}}-n^3I_{34}^{\scaleobj{0.85}{(4)}}}{I_{13}^{\scaleobj{0.85}{(6)}}}\bigg](\kappa\,|\,m_1,m_2,m_3),\end{align}
where, again, $\kappa$ is related to $Q$ via (\ref{dense}).  
Restricted to (\ref{m1m3}), these formulas (\ref{firstprob}--\ref{lastprob}) are easy to integrate numerically for $\kappa>4$ and values of $m_2\in(0,1)$ sufficiently far from zero and one, using Mathematica.  However, numerical integration can give poor results for the first ($R\approx1/16$, $m_2\approx0$) and last ($R\approx16,$ $m_2\approx1$) two hexagons in (\ref{ell}).  

\begin{figure}[b]
\centering
\includegraphics[scale=0.27]{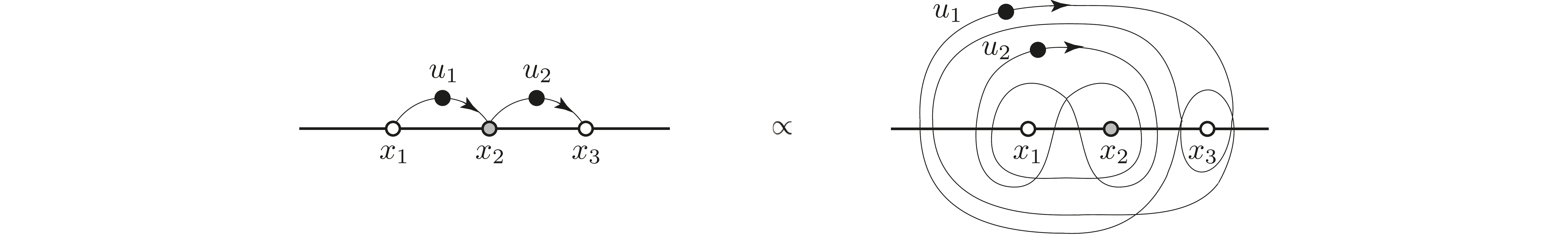}
\caption{Integrating (\ref{Iij}) with $c=2$ around the nested pair of Pochhammer contours (right) gives a result that is proportional to performing the first integration along the left interval and the second integration along the right interval (left).}
\label{ContourId1}
\end{figure}

The improper integrals $\smash{I_{ij}\super{c}(\kappa)}$ (\ref{Iij}) diverge for all $\kappa\in(0,4]$, which includes the $Q\in\{2,3,4\}$ critical Potts models (\ref{dilute}).  To resolve this issue, we replaced all simple integration contours with Pochhammer contours as in figure \ref{PochhammerContour} and divided each Pochhammer contour integral by the normalization factor on the right side of the equation in that figure.  In order to numerically integrate the result, we decomposed each Pochhammer contour as in figure \ref{BreakDown}.  The integrals $\smash{I_{34}\super{4}}$, $\smash{I_{12}\super{2}}$, having two simple integration contours that span adjacent intervals, need to be evaluated with nested Pochhammer contours as in figure \ref{ContourId1}.  These nested contours correspond to the ``rainbow" configurations discussed in \cite{fsk}.  We evaluated the outer Pochhammer contour of figure \ref{ContourId1} as we do in figure \ref{BreakDown}, but with the left point $x_i$ of the latter figure replaced by the inner Pochhammer contour of figure \ref{ContourId1}, so $\beta_i=-4/\kappa+12/\kappa-2+8/\kappa=16/\kappa-2$.  Thus, we have
\begin{multline}\label{AltI12} I_{12}\super{2}(\kappa\,|\,x_1,x_2,\ldots,x_6)=\left(\frac{1}{4e^{-16\pi i/\kappa}\sin(-4\pi/\kappa)\sin(12\pi/\kappa)}\right)\left(\frac{1}{4e^{20\pi i/\kappa}\sin(16\pi/\kappa)\sin(-4\pi/\kappa)}\right)\\
\oint_{\Gamma_1}\oint_{\Gamma_2}(u_1-x_1)^{-4/\kappa}(u_1-x_2)^{12/\kappa-2}(u_2-x_1)^{-4/\kappa}(x_2-u_2)^{12/\kappa-2}(u_1-u_2)^{8/\kappa}\prod_{i=1}^2\prod_{j=3}^6(x_j-u_i)^{-4/\kappa}\,{\rm d}u_1\,{\rm d}u_2,\end{multline}
where $\Gamma_2=\mathscr{P}(x_1,x_2)$ and $\Gamma_1=\mathscr{P}(\Gamma_2,x_3)$ is the Pochhammer contour with its left (resp.\ right) cycle exclusively surrounding $\Gamma_2$ (resp.\ $x_3$).  The first (resp.\ second) prefactor on the right side of (\ref{AltI12}) arises from dividing out by the prefactor on the right side of the equation in figure \ref{PochhammerContour} associated with $\Gamma_1$ (resp.\ $\Gamma_2$), as mentioned above.  Using the techniques of section \red{II E} in \cite{fsk}, we may verify that inserting the right side of (\ref{AltI12}) gives an alternative formula for $\Pi_2(\kappa)$ that holds for all $\kappa\in(0,8)$.  Thus, the right side of (\ref{AltI12}) is the analytic continuation of $\smash{I_{12}\super{2}(\kappa)}$ to $\kappa\in(0,4]$.

%%%%%%%%%%%%%%%%%%%%%%%%%%%%%%%%%%%%%%%%%%%%%%%%%%%%%%%%%%%%%%%%%%%%%%%%%%%%%%%%%%%%%%%%%%%%%%%%%%%%%%%%%%%%%%%%%%%%%%%%%%%%%%%%%%%%%%%%%%%%%%%

\begin{figure}[p]
\centering
\includegraphics[scale=0.5]{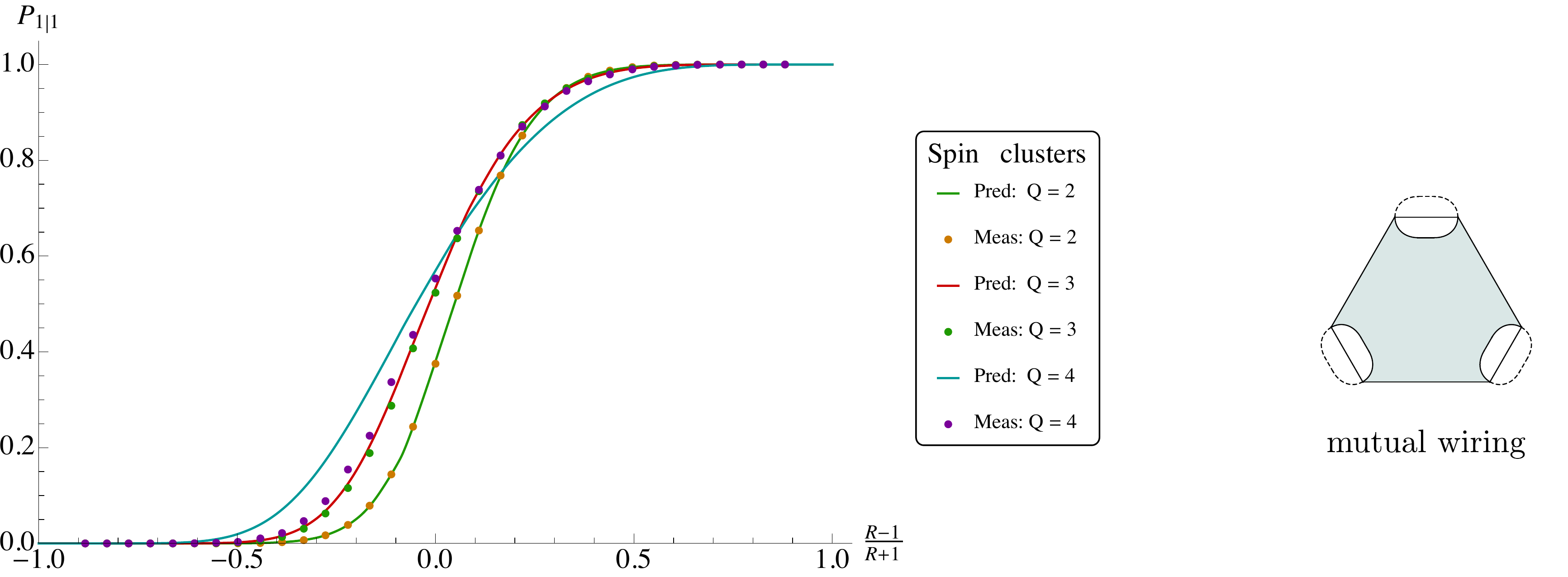}
\vspace{1cm}\\
\includegraphics[scale=0.5]{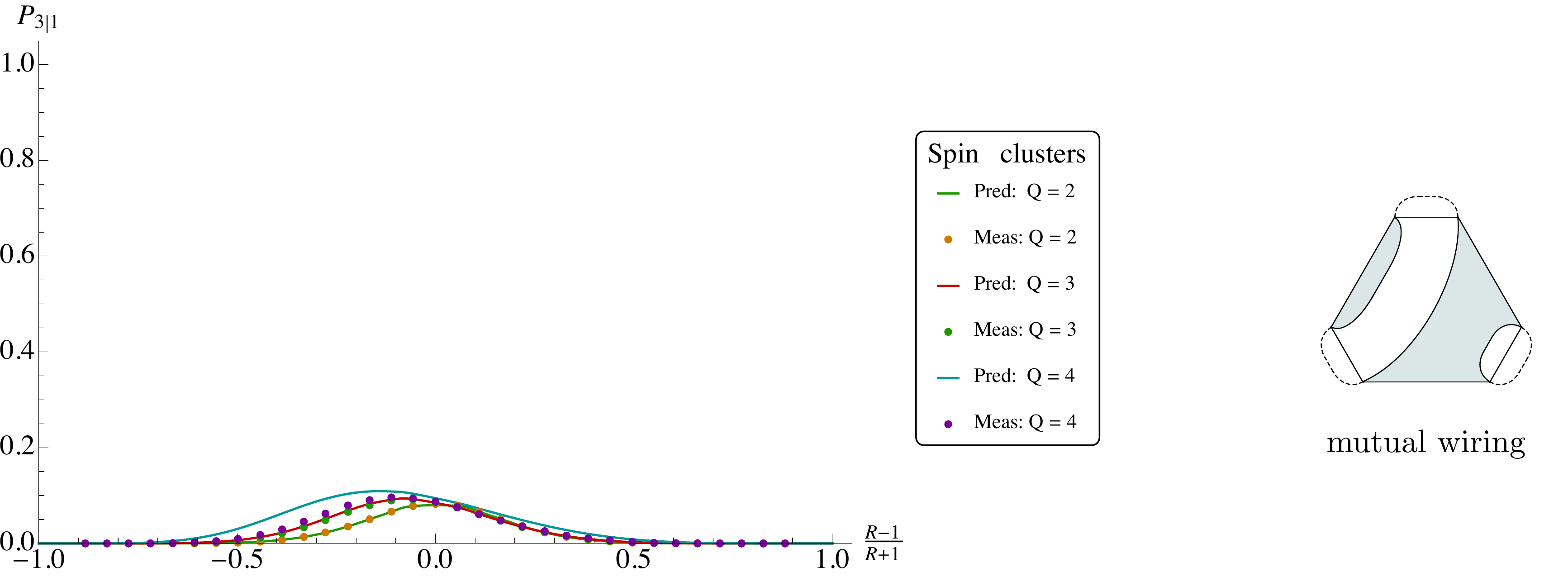}
\vspace{1cm}\\
\includegraphics[scale=0.5]{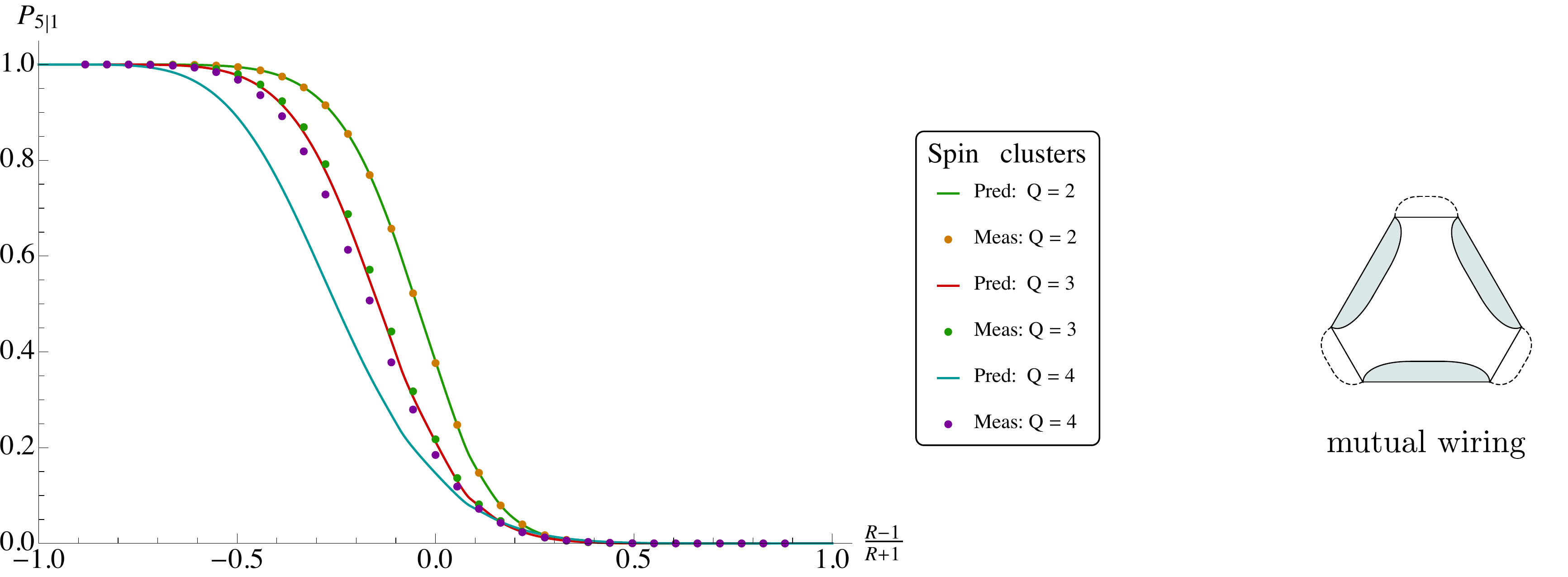}
\vspace{1cm}\\
\caption{Comparison of computer simulation measurements to our prediction (\ref{xing}, \ref{firstprob}--\ref{lastmutprob}) of spin-cluster crossing probability in a hexagon with side-length ratio (bottom side to bottom-left side) $R$.  Item \ref{BCitem2} of the introduction \ref{intro} defines the BCs and crossing events for Potts model spin clusters.  The hexagon to the right of each graph shows the $\varsigma$th crossing event whose probability $P_{\varsigma|1}$ appears in the graph.  In it, all boundary clusters are colored gray, sides touching a boundary cluster are fixed, and all fixed sides, with their endpoints joined together by dashed  exterior  arcs, are mutually wired together.}
\label{HexSpinXingMixed}
\end{figure}

\begin{figure}[p]
\centering
\includegraphics[scale=0.5]{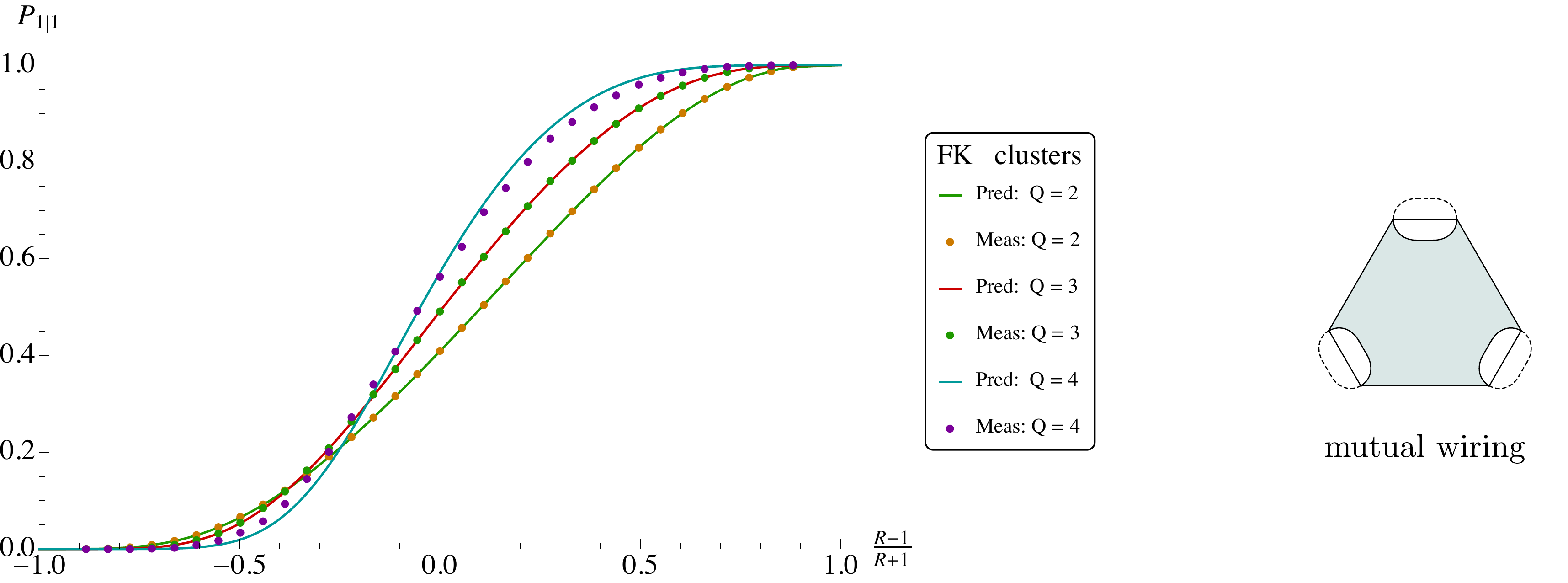}
\vspace{1cm}\\
\includegraphics[scale=0.5]{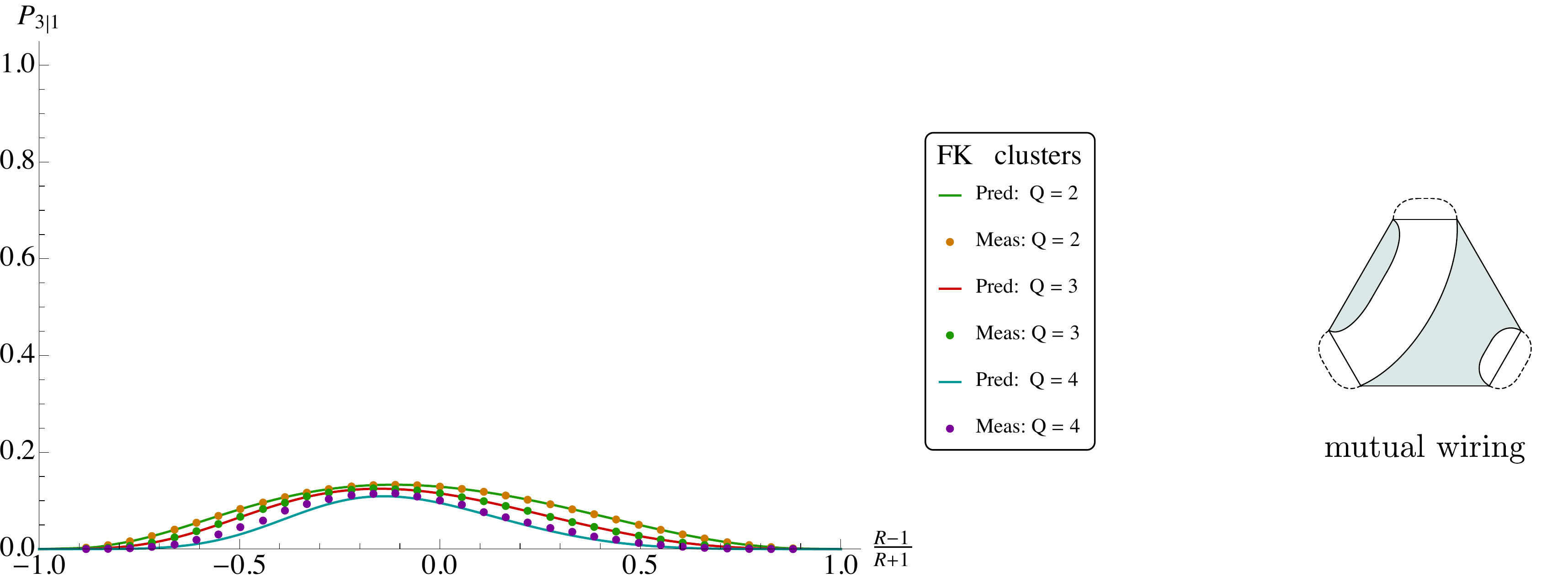}
\vspace{1cm}\\
\includegraphics[scale=0.5]{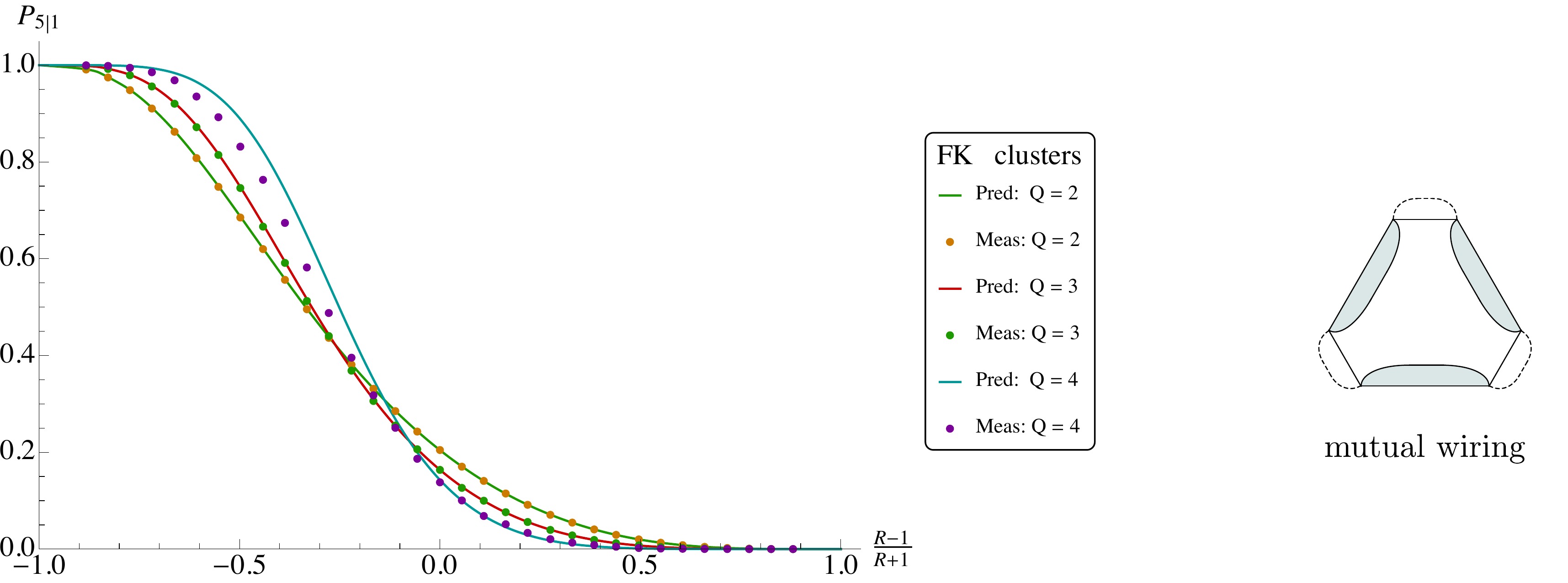}
\vspace{1cm}\\
\caption{Comparison of computer simulation measurements to our prediction (\ref{xing}, \ref{firstprob}--\ref{lastmutprob}) of FK-cluster crossing probability in a hexagon with side-length ratio (bottom side to bottom-left side) $R$.  Item \ref{BCitem1} of the introduction \ref{intro} defines the BCs and crossing events for random cluster model FK clusters.  The hexagon to the right of each graph shows the $\varsigma$th crossing event whose probability $P_{\varsigma|1}$ appears in the graph.  In it, all boundary clusters are colored gray, sides touching a boundary cluster are fixed, and all fixed sides, with their endpoints joined together by dashed  exterior  arcs, are mutually wired together.}
\label{HexFkXingMutual}
\end{figure}

\begin{figure}[p]
\centering
\includegraphics[scale=0.5]{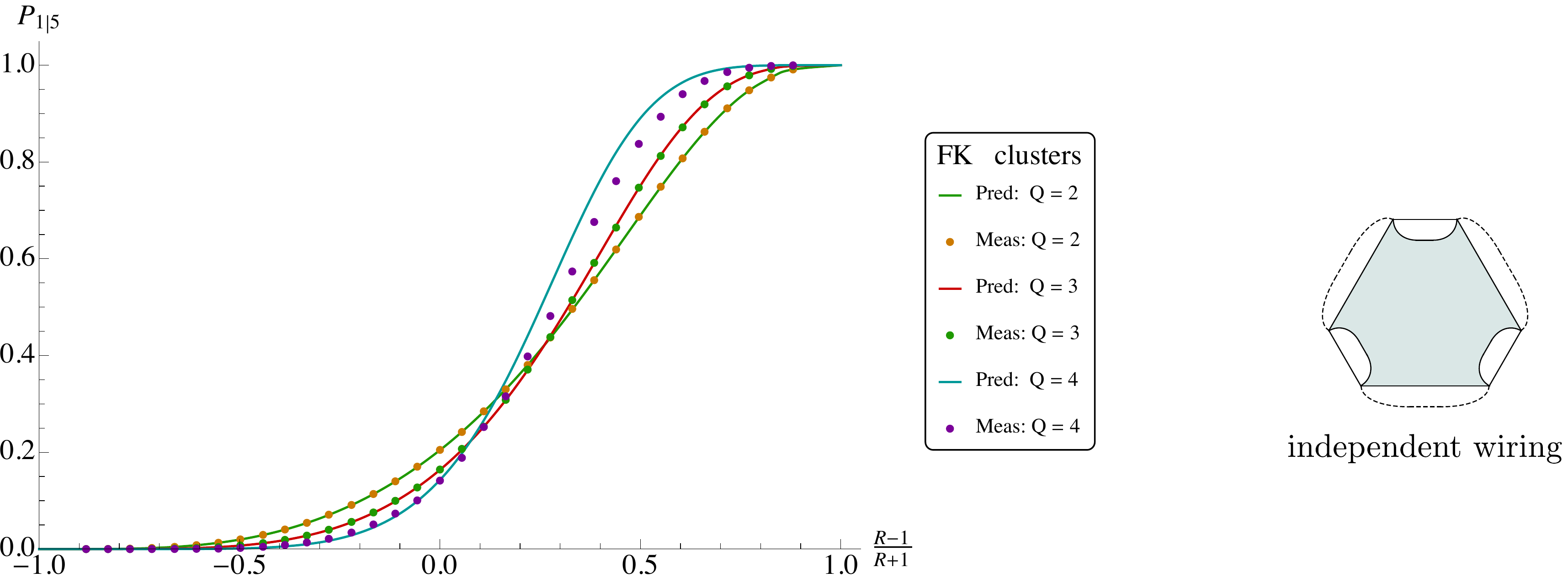}
\vspace{1cm}\\
\includegraphics[scale=0.5]{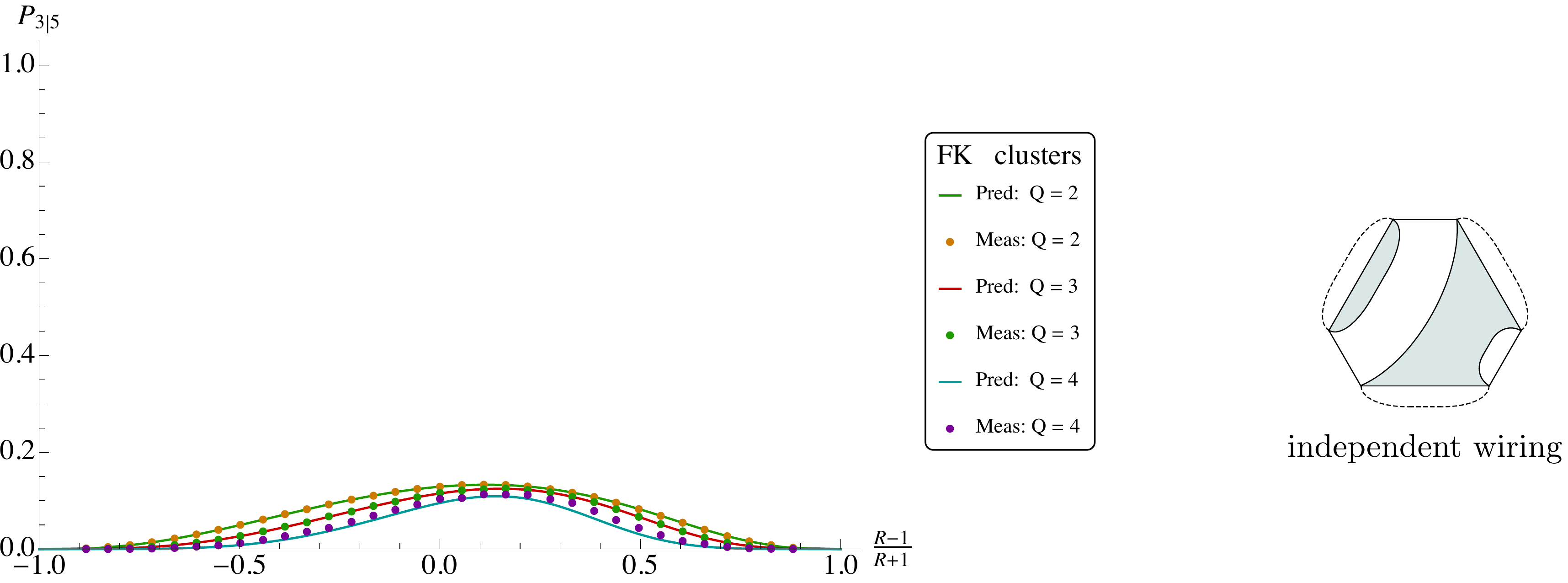}
\vspace{1cm}\\
\includegraphics[scale=0.5]{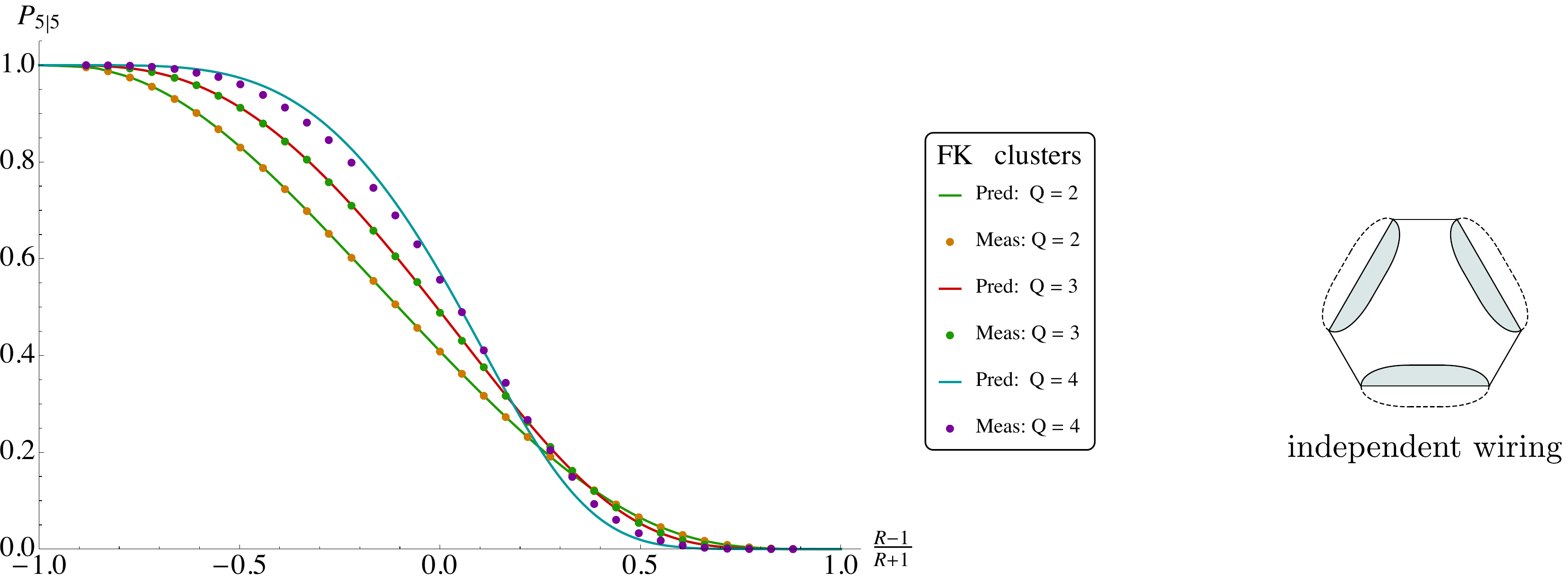}
\vspace{1cm}\\
\caption{Comparison of computer simulation measurements to our prediction (\ref{xing}, \ref{firstindprob}--\ref{lastprob}) of FK-cluster crossing probability in a hexagon with side-length ratio (bottom side to bottom-left side) $R$.  Item \ref{BCitem1} of the introduction \ref{intro} defines the BCs and crossing events for random cluster model FK clusters.  The hexagon to the right of each graph shows the $\varsigma$th crossing event whose probability $P_{\varsigma|5}$ appears in the graph.  In it, all boundary clusters are colored gray, sides touching a boundary cluster are fixed, and fixed sides, each with endpoints joined by a dashed  exterior  arc, are wired independently of each other.}
\label{HexFkXingIndep}
\end{figure}

\begin{figure}[p]
\centering
\includegraphics[scale=0.5]{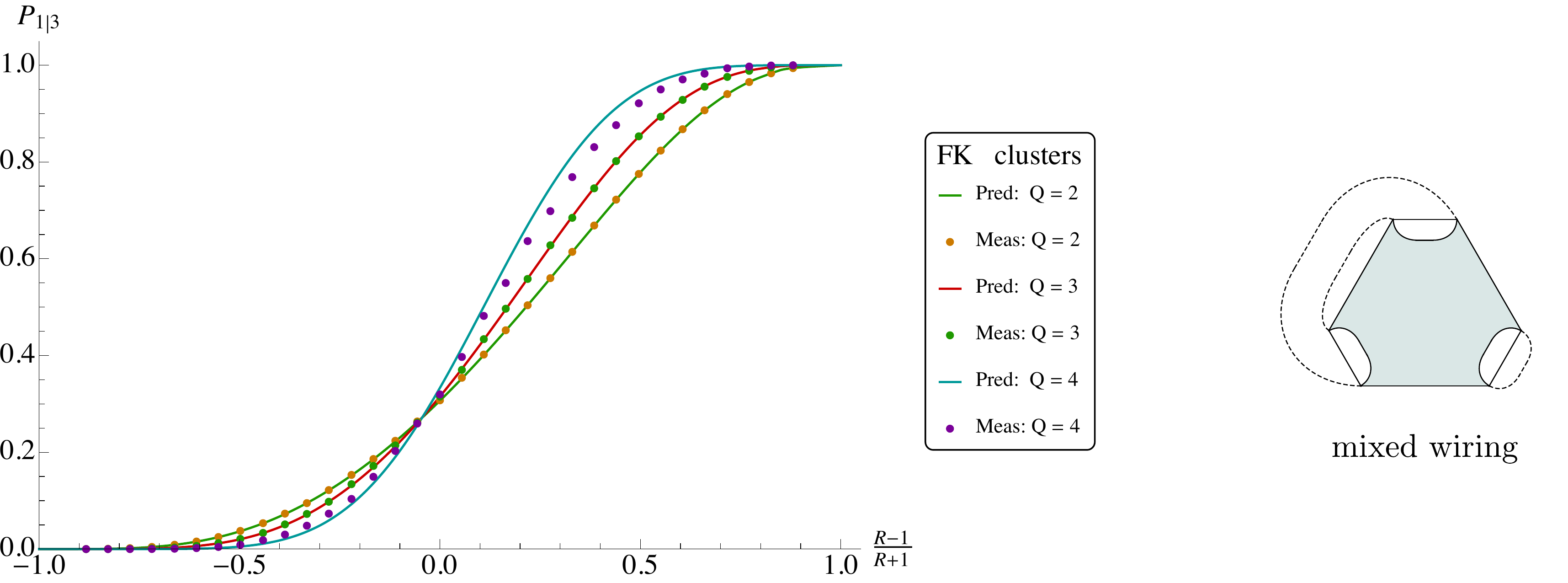}
\vspace{1cm}\\
\includegraphics[scale=0.5]{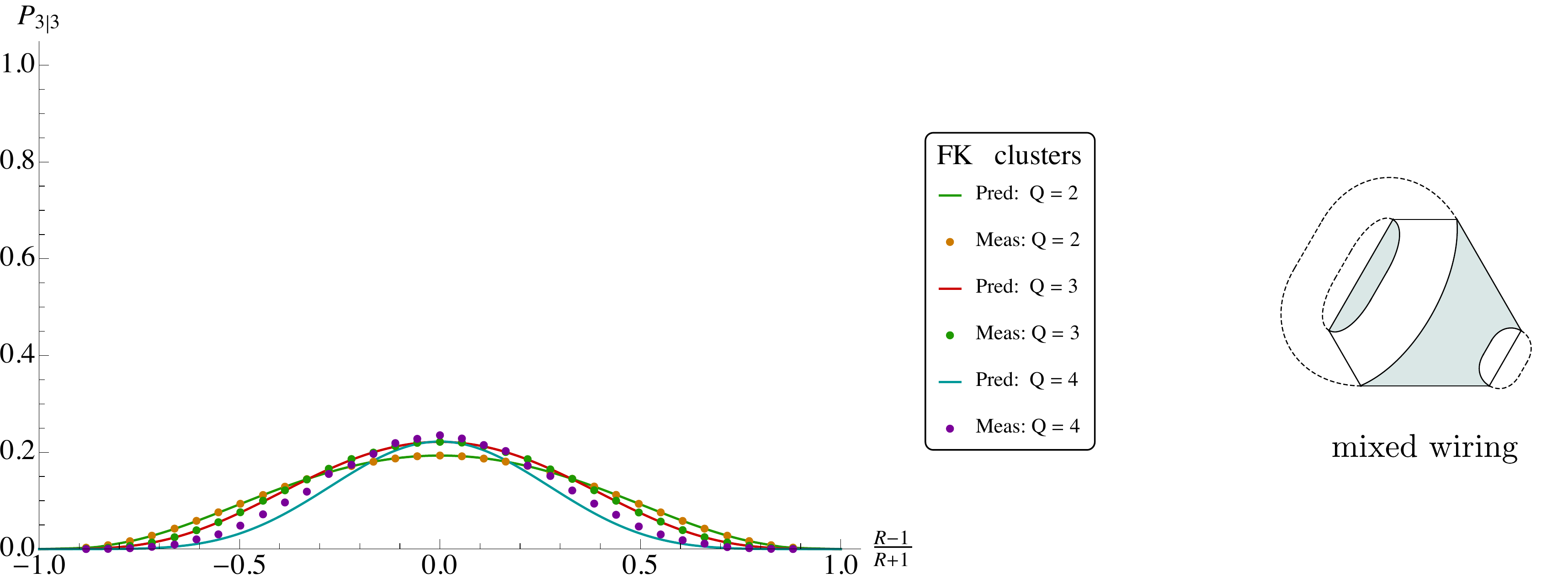}
\vspace{1cm}\\
\includegraphics[scale=0.5]{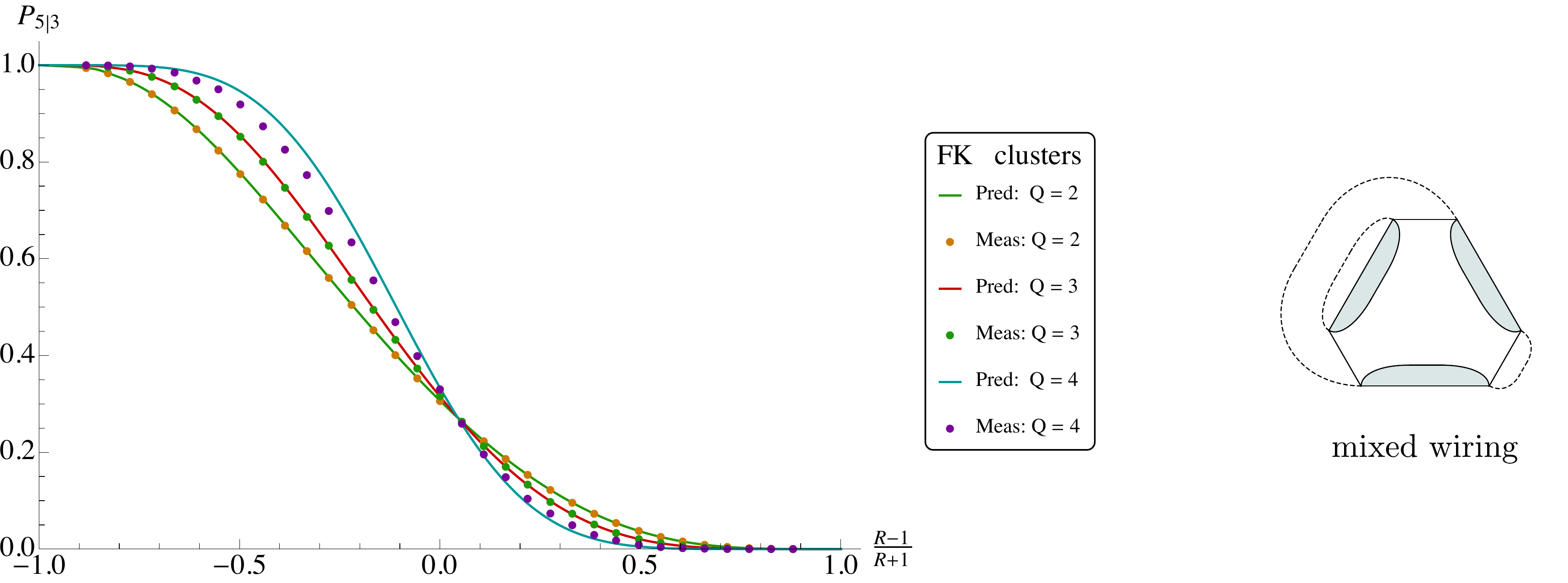}
\vspace{1cm}\\
\caption{Comparison of computer simulation measurements to our prediction (\ref{xing}, \ref{firstmixprob}--\ref{lastmixprob}) of FK-cluster crossing probability in a hexagon with side-length ratio (bottom side to bottom-left side) $R$.  Item \ref{BCitem1} of the introduction \ref{intro} defines the BCs and crossing events for random cluster model FK clusters.  The hexagon to the right of each graph shows the $\varsigma$th crossing event whose probability $P_{\varsigma|3}$ appears in the graph.  In it, all boundary clusters are colored gray, sides touching a boundary cluster are fixed, and any two fixed sides (one fixed side) with endpoints joined by a dashed  exterior  arc are mutually wired together (resp.\ wired independently of other fixed sides).  (Figure \ref{HexFkXingMixed2} shows the graph for the second crossing probability $P_{2|3}$.)}
\label{HexFkXingMixed}
\end{figure}

\begin{figure}[t]
\centering
\includegraphics[scale=0.5]{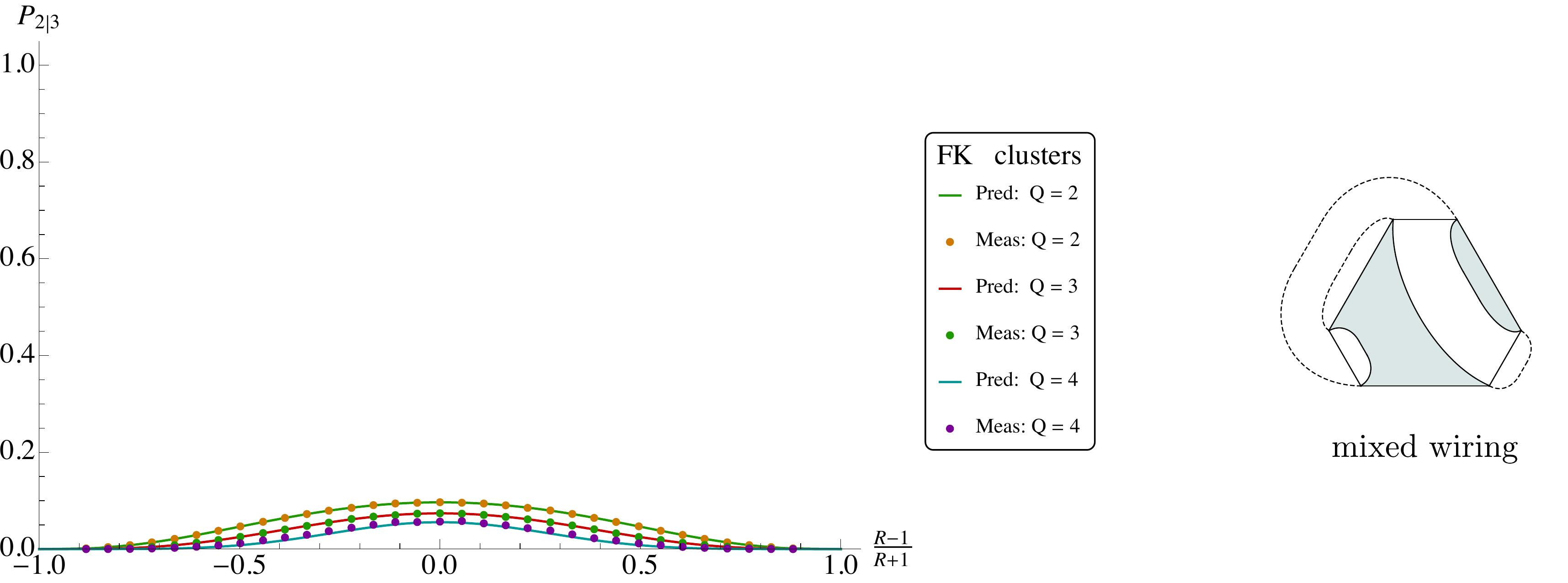}
\caption{Comparison of computer simulation measurements to our prediction (\ref{xing}, \ref{secondmixprob}) of FK-cluster crossing probability $P_{2|3}$ in a hexagon with side-length ratio (bottom side to bottom-left side) $R$.  See the caption of figure \ref{HexFkXingMixed} for further description.}
\label{HexFkXingMixed2}
\end{figure}

%%%%%%%%%%%%%%%%%%%%%%%%%%%%%%%%%%%%%%%%%%%%%%%%%%%%%%%%%%%%%%%%%%%%%%%%%%%

This formula (\ref{AltI12}) for $\smash{I_{12}\super{2}(\kappa)}$ has an inconvenience: among the values of $\kappa$ in (\ref{dense}, \ref{dilute}), it is singular at $\kappa=16/3$ (corresponding to $Q=2$ in the random cluster model).  Fortunately, this singularity is removable.  To see this, we first observe that because $\kappa=16/3>4$, we may replace $\Gamma_2$ with the simple contour $[x_1,x_2]$ via figure \ref{PochhammerContour}, so (\ref{AltI12}) simplifies to 
\begin{multline}\label{AltI121} I_{12}\super{2}(\kappa\,|\,x_1,x_2,\ldots,x_6)=\left(\frac{1}{4e^{20\pi i/\kappa}\sin(16\pi/\kappa)\sin(-4\pi/\kappa)}\right)\oint_{\Gamma_1}\int_{x_1}^{x_2}(u_1-x_1)^{-4/\kappa}(u_1-x_2)^{12/\kappa-2}\\
(u_2-x_1)^{-4/\kappa}(x_2-u_2)^{12/\kappa-2}(u_1-u_2)^{8/\kappa}\prod_{i=1}^2\prod_{j=3}^6(x_j-u_i)^{-4/\kappa}\,{\rm d}u_1\,{\rm d}u_2,\end{multline}
where $\Gamma_1=\mathscr{P}([x_1,x_2],x_3)$ is the Pochhammer contour with its left (resp.\ right) cycle exclusively surrounding $[x_1,x_2]$ (resp.\ $x_3$).  After decomposing $\Gamma_1$ into two loops via figure \ref{BreakDown} with $x_i\mapsto[x_1,x_2]$, $\beta_i=16/\kappa-2$, $x_j=x_3$, and $\beta_j=-4/\kappa$ (here, we give the straight contour in figure \ref{BreakDown} length zero, so its corresponding term does not contribute anything and the two loops mentioned are tangent to each other at a point on the real axis), we find
\begin{multline}\label{AltI122} I_{12}\super{2}(\kappa\,|\,x_1,x_2,\ldots,x_6)=\left[\frac{e^{-2\pi i(-4/\kappa)}-1}{4e^{20\pi i/\kappa}\sin(16\pi/\kappa)\sin(-4\pi/\kappa)}\oint_{\Gamma}\,\,-\,\,\frac{e^{2\pi i(20/\kappa)}(e^{-2\pi i(16/\kappa)}-1)}{4e^{20\pi i/\kappa}\sin(16\pi/\kappa)\sin(-4\pi/\kappa)}\oint_{\Gamma'}\right]\\
\int_{x_1}^{x_2}(u_1-x_1)^{-4/\kappa}(u_1-x_2)^{12/\kappa-2}(u_2-x_1)^{-4/\kappa}(x_2-u_2)^{12/\kappa-2}(u_1-u_2)^{8/\kappa}\prod_{i=1}^2\prod_{j=3}^6(x_j-u_i)^{-4/\kappa}\,{\rm d}u_1\,{\rm d}u_2,\end{multline}
where $\Gamma$ (resp.\ $\Gamma'$) is a counterclockwise-oriented simple loop exclusively surrounding $[x_1,x_2]$ (resp.\ $x_3$).  The limit as $\kappa\rightarrow16/3$ of the term with integration around $\Gamma'$ obviously exists.  To see that the limit as $\kappa\rightarrow16/3$ of the term with integration around $\Gamma$ exists too, we note that at $\kappa=16/3$, the integrand does not acquire a phase factor as $u_1$ traces $\Gamma$.  Thus, $\Gamma$ closes (i.e., its start-point and endpoint lie on the same Riemann sheet of the integrand), so the function
\begin{multline} F(\kappa\,|\,x_1,x_2,\ldots,x_6)= \Bigg(\prod_{\substack{i<j \\ i,j\neq 2}}^6(x_j-x_i)^{2/\kappa}\Bigg)\Bigg(\prod_{k\neq2}^6|x_2-x_k|^{1-6/\kappa}\Bigg)\oint_{\Gamma}\int_{x_1}^{x_2}(u_1-x_1)^{-4/\kappa}(u_1-x_2)^{12/\kappa-2}\\
(u_2-x_1)^{-4/\kappa}(x_2-u_2)^{12/\kappa-2}(u_1-u_2)^{8/\kappa}\prod_{i=1}^2\prod_{j=3}^6(x_j-u_i)^{-4/\kappa}\,{\rm d}u_1\,{\rm d}u_2,\end{multline}
is an element of $\mathcal{S}_N$ if $\kappa=16/3$, according to theorem \red{2} of \cite{florkleb3}.  However, $[x_1,x_2]$, $[x_3,x_4]$, $[x_4,x_5]$, and $[x_5,x_6]$ are two-leg intervals (definition \red{13} of \cite{florkleb}) of $F(\kappa=16/3)$, so we have $[\mathscr{L}_\varsigma]F(\kappa=16/3)=0$ for all equivalence classes $\varsigma\in\{1,2,3,4,C_3=5\}$.  (We describe $[\mathscr{L}_\varsigma]$ beneath (\ref{allowablelim}).  See definitions \red{9} and \red{11} of \cite{florkleb} for a formal definition.)  As a consequence of this fact, lemma \red{15} of \cite{florkleb} implies that $F(\kappa=16/3)=0$, so because $F$ is an analytic function of $\kappa$, we must have $F(\kappa)=O(\kappa-16/3)$ as $\kappa\rightarrow16/3$.  Altogether, this implies that the limit as $\kappa\rightarrow16/3$ of the term with integration around $\Gamma$ in (\ref{AltI122}), and therefore of $\smash{I_{12}\super{2}(\kappa)}$, exists, which is what we wanted to show.

The case with $\kappa=4$, corresponding to the $Q=4$ critical random cluster model (\ref{dense}) and the $Q=4$ critical Potts model (\ref{dilute}), also requires special treatment.  In this case, the improper integrals of (\ref{Iij}) diverge, so we must regularize them by replacing all simple integration contours with Pochhammer contours, as in figure \ref{PochhammerContour}.  But if $\kappa=4$, then the branch points $x_j$ of these contour integrals are, in fact, simple poles.  Hence, we may use the Cauchy integral formula to explicitly evaluate all of these contour integrals and find algebraic expressions for all crossing probabilities.  (See appendix \ref{equivappendix} for an example of such an evaluation.)

%%%%%%%%%%%%%%%%%%%%%%%%%%%%%%%%%%%%%%%%%%%%%%%%%%%%%%%%%%%%%%%%%%%%%%%%%%%%%%%%%%%%%%%%%%%%%%%%%%%%%%%%%%%%%%%%%%%%%%%%%%%%%%%%%%%%%%%%%%%%%%%%
\begin{sidewaystable}[p]
\centering
\renewcommand{\tabcolsep}{.1cm}
\begin{tabular}{llc|cccccc|cccccc}
\multicolumn{3}{c}{} & \multicolumn{6}{c}{Random cluster model (FK clusters)} & \multicolumn{6}{c}{Potts model (spin clusters)} \\ [0.3cm]
Polygon & BC & $\varsigma|\vartheta$ & \parbox{1.3cm}{$Q=2$ rel.\ err.} & \parbox{1.3cm}{$Q=2$ std.\ dev.} & \parbox{1.3cm}{$Q=3$ rel.\ err.} & \parbox{1.3cm}{$Q=3$ std.\ dev.} & \parbox{1.3cm}{$Q=4$ rel.\ err.} & \parbox{1.3cm}{$Q=4$ std.\ dev.} & \parbox{1.3cm}{$Q=2$ rel.\ err.} & \parbox{1.3cm}{$Q=2$ std.\ dev.} & \parbox{1.3cm}{$Q=3$ rel.\ err.} & \parbox{1.3cm}{$Q=3$ std.\ dev.} & \parbox{1.3cm}{$Q=4$ rel.\ err.} & \parbox{1.3cm}{$Q=4$ std.\ dev.} 

%%%%%%%%%%%%%%%%%%%%%%%%%%%%%%%%%%%%%%%%%%%%%%%%%%%%%%%%%%%%%%%%%%%%%%%%%%%%%Insert RecErrorSummary.txt from just below here to next commented part:

\\
[0.3cm]
\hline
&
&
&
&
&
&
&
&
&
&
&
&
\\
\multirow{2}{*}{Rectangle\hspace{.2cm}}
&
mut.
&
$2|1$
&
$
-0.002065
$
&
$
\hphantom{-}0.003407
$
&
$
-0.002349
$
&
$
\hphantom{-}0.006693
$
&
$
-0.071972
$
&
$
\hphantom{-}0.161850
$
&
$
\hphantom{-}0.001901
$
&
$
\hphantom{-}0.003128
$
&
$
\hphantom{-}0.022325
$
&
$
\hphantom{-}0.039424
$
&
$
\hphantom{-}0.133829
$
&
$
\hphantom{-}0.241286
$
\\
[0.3cm]
&
indep.
&
$2|2$
&
$
-0.001054
$
&
$
\hphantom{-}0.001852
$
&
$
-0.004215
$
&
$
\hphantom{-}0.006788
$
&
$
-0.049402
$
&
$
\hphantom{-}0.128972
$
&
&
&
&
&
&
\\
[0.3cm]

%%%%%%%%%%%%%%%%%%%%%%%%%%%%%%%%%%%%%%%%%%%%%%%%%%%%%%%%%%%%%%%%%%%%%%%%%%%%%Insert HexErrorSummary.txt from just below here to \end{tabular}:

\hline
&
&
&
&
&
&
&
&
&
&
&
&
\\
\multirow{10}{*}{Hexagon\hspace{.2cm}}
&
\multirow{3}{*}{mut.}
&
$1|1$
&
$
-0.003044
$
&
$
\hphantom{-}0.004332
$
&
$
-0.004179
$
&
$
\hphantom{-}0.008948
$
&
$
-0.069314
$
&
$
\hphantom{-}0.175866
$
&
$
-0.000357
$
&
$
\hphantom{-}0.002310
$
&
$
\hphantom{-}0.023562
$
&
$
\hphantom{-}0.038542
$
&
$
\hphantom{-}0.154072
$
&
$
\hphantom{-}0.038542
$
\\
&
&
$3|1$
&
$
-0.002468
$
&
$
\hphantom{-}0.004573
$
&
$
-0.007783
$
&
$
\hphantom{-}0.013127
$
&
$
-0.227979
$
&
$
\hphantom{-}0.208866
$
&
$
-0.000516
$
&
$
\hphantom{-}0.005806
$
&
$
\hphantom{-}0.029516
$
&
$
\hphantom{-}0.027194
$
&
$
\hphantom{-}0.365309
$
&
$
\hphantom{-}0.027194
$
\\
&
&
$5|1$
&
$
\hphantom{-}0.001773
$
&
$
\hphantom{-}0.002153
$
&
$
\hphantom{-}0.003414
$
&
$
\hphantom{-}0.005716
$
&
$
\hphantom{-}0.007292
$
&
$
\hphantom{-}0.099452
$
&
$
\hphantom{-}0.001041
$
&
$
\hphantom{-}0.004195
$
&
$
-0.018427
$
&
$
\hphantom{-}0.019032
$
&
$
-0.102192
$
&
$
\hphantom{-}0.019032
$
\\
[0.3cm]
&
\multirow{4}{*}{mixed}
&
$1|3$
&
$
-0.003425
$
&
$
\hphantom{-}0.004505
$
&
$
-0.005796
$
&
$
\hphantom{-}0.010392
$
&
$
-0.058427
$
&
$
\hphantom{-}0.174879
$
&
&
&
&
&
&
\\
&
&
$2|3$
&
$
-0.002258
$
&
$
\hphantom{-}0.004794
$
&
$
-0.003753
$
&
$
\hphantom{-}0.009949
$
&
$
-0.158441
$
&
$
\hphantom{-}0.111775
$
&
&
&
&
&
&
\\
&
&
$3|3$
&
$
-0.001923
$
&
$
\hphantom{-}0.004898
$
&
$
-0.007019
$
&
$
\hphantom{-}0.008183
$
&
$
-0.278074
$
&
$
\hphantom{-}0.226929
$
&
&
&
&
&
&
\\
&
&
$5|3$
&
$
\hphantom{-}0.001861
$
&
$
\hphantom{-}0.001454
$
&
$
\hphantom{-}0.001402
$
&
$
\hphantom{-}0.007845
$
&
$
-0.046882
$
&
$
\hphantom{-}0.156690
$
&
&
&
&
&
&
\\
[0.3cm]
&
\multirow{3}{*}{indep.}
&
$1|5$
&
$
-0.002535
$
&
$
\hphantom{-}0.003067
$
&
$
-0.003441
$
&
$
\hphantom{-}0.006427
$
&
$
-0.008541
$
&
$
\hphantom{-}0.116411
$
&
&
&
&
&
&
\\
&
&
$3|5$
&
$
-0.001637
$
&
$
\hphantom{-}0.006212
$
&
$
-0.004639
$
&
$
\hphantom{-}0.007255
$
&
$
-0.218566
$
&
$
\hphantom{-}0.156977
$
&
&
&
&
&
&
\\
&
&
$5|5$
&
$
\hphantom{-}0.001780
$
&
$
\hphantom{-}0.001654
$
&
$
-0.000397
$
&
$
\hphantom{-}0.005720
$
&
$
-0.064035
$
&
$
\hphantom{-}0.166325
$
&
&
&
&
&
&
\end{tabular}
\caption{Summary of errors in crossing probability measurements for the rectangle with mutual wiring ($\vartheta=1$) and independent wiring ($\vartheta=2$), and for the hexagon with mutual wiring ($\vartheta=1$), mixed wiring ($\vartheta=3$), and independent wiring ($\vartheta=5$).  We averaged each relative error over all thirty-three rectangles and hexagons, except that we dropped the points for rectangles and hexagons with crossing probability less than 0.01 because our simulations may have generated too few samples of such rare crossing events, resulting in artificially high relative errors.  We averaged the deviation of each relative error from the average relative error over all thirty-three rectangles and hexagons, but again, we dropped the points for the same rectangles and hexagons as before for the same reason as before.  Figure \ref{RecXings} (resp.\ figure \ref{HexXings}) illustrates the $\varsigma$th crossing event for the rectangle (resp.\ hexagon) respectively, with the top and bottom sides (resp.\ bottom and upper left/right sides) fixed.}
\label{ErrorSummaryTable}
\end{sidewaystable}
%%%%%%%%%%%%%%%%%%%%%%%%%%%%%%%%%%%%%%%%%%%%%%%%%%%%%%%%%%%%%%%%%%%%%%%%%%%%%%%%%%%%%%%%%%%%%%%%%%%%%%%%%%%%%%%%%%%%%%%%%%%%%%%%%%%%%%%%%%%%%%%%

The formulas given in (\ref{firstprob}--\ref{lastprob}) hold whenever $0<m_1<m_2<m_3<1$.  However, as a result of the $2\pi/3$ rotation invariance of the hexagon that arises from (\ref{m1m3}), some of the ostensibly different contour integrals in (\ref{Iij}) are in fact equal.  For example, we have $\smash{I_{12}\super{2}(\kappa\,|\,m_1,m_2,m_3)=I_{34}\super{4}(\kappa\,|\,m_1,m_2,m_3)}$ whenever this rotational symmetry (\ref{m1m3}) holds.  Using this and similar equalities, we minimized the number of different contour integrals needed to numerically evaluate  (\ref{firstprob}--\ref{lastprob}).

A final alteration exploited the M\"obius symmetry of (\ref{halfplaneconformalinvar}) to use the points $x_1 = 1-1/(1-m_2)$, $x_2 =0,$ $x_3 =1-m_2,$ $x_4 =1,$ $x_5 =1/m_2,$ and $x_6 =\infty$ rather than (\ref{thepoints}, \ref{m1m3}). This is equivalent to the previous choice (\ref{thepoints}, \ref{m1m3}) up to a scaling and translation. This change was useful because the nested Pochhammer contours of figure \ref{ContourId1} were difficult to evaluate in the coordinates of (\ref{thepoints}, \ref{m1m3}) whenever $m_2$ approached 0 or 1 because groups of four points approached each other, with integration contours between them.

Figure \ref{HexSpinXingMixed} compares our simulation measurements of the critical Potts model spin-cluster crossing probabilities for the hexagon with the mutual wiring $\PMB$ to our predictions (\ref{firstprob}--\ref{lastmutprob}).  Figure \ref{HexFkXingMutual} (resp.\ figures \ref{HexFkXingMixed} and \ref{HexFkXingMixed2}, resp.\ figure \ref{HexFkXingIndep}) compares our simulation measurements of the critical random cluster model FK-cluster crossing probabilities for the hexagon with the mutual wiring (resp.\ mixed wiring, resp.\ independent wiring) $\RCB$ to our predictions (\ref{firstprob}--\ref{lastmutprob}) (resp. (\ref{firstmixprob}--\ref{lastmixprob}), resp.\ (\ref{firstindprob}--\ref{lastprob})).  We plot the probabilities as functions of $x=(R-1)/(R+1)\in(0,1)$, where $R$ is the hexagon's ratio of alternating side-lengths (bottom to top side).  

Table \ref{ErrorSummaryTable} gives the relative error of our measurements, averaged over all thirty-three rectangles and hexagons, except that we drop the points for hexagons and rectangles with crossing probability less than 0.01 because our simulations may have generated too few samples of such rare crossing events, resulting in artificially high relative errors.  For $Q\in\{2,3\}$, the relative errors are all small, on the order of $10^{-3}$, except the $Q=3$ critical Potts model with relative error on the order of $10^{-2}$.  These low relative errors support our prediction of the formula (\ref{xing}) for crossing probability in rectangles and hexagons.  For $Q=4$, the relative errors are significantly greater, on the order of $10^{-1}$.  These greater relative errors exemplify the well-known fact \cite{car1,cziff} that the convergence of the discrete $Q=4$ critical random cluster model and critical Potts model to their continuum limits is much slower than it is for $Q\in\{2,3\}$.  To compensate this effect and improve our numerical results  would require sampling  the model many more times on a much larger lattice.  

Table \ref{ErrorSummaryTable} also gives the  standard  deviation of our relative errors from the average relative error.   (We omit the same rectangles and hexagons as in the previous paragraph.)  Again, for $Q\in\{2,3\}$, these deviations are all small, typically on the order of $10^{-3}$, and in a few cases, $10^{-2}$.  For $Q=4$, the deviations are significantly higher, on the order of $10^{-1}$.  This reflects the fact that  for $Q=4$,  the deviations between theory and simulation are  often much larger for  ``middle" data points than than they are for ``extreme end" data points (figures \ref{HexFkXingMutual}--\ref{HexSpinXingMixed}). Indeed, the extreme end data points correspond to very irregular hexagons whose shapes approximate an equilateral triangle, with sides (resp.\ corners) corresponding to the very long (resp.\ very short) sides of the hexagons.  For these hexagons, the crossing probabilities are always near zero or one, so they are easier to measure.  (Indeed, if the long sides of the  very irregular hexagons are fixed, then the probability that a single boundary cluster joins all of these fixed sides nearly equals one, and the other crossing probabilities nearly equal zero.  This is because inside an equilateral triangle with the fixed BC on all three of its sides, the system strongly favors configurations in which those sides are in the same state, if they are not already required to be so via mutual wiring.  Because the three sides of the triangle touch each other, their sites are trivially always part of a boundary cluster that joins all three sides.  (One should not confuse this with what happens inside a triangle with free BCs on all sides.  For example, for critical site percolation on a triangular lattice, a simple duality argument shows that the probability of a cluster connecting all three free sides of a triangle equals $1/2$.)  On the other hand, if the short sides of the very irregular hexagons are fixed, then the probability that any cluster joins two or more of these fixed sides nearly equals zero, and the probability that no such cluster joins any pair of these sides nearly equals one.  This is because inside a triangle with free BCs on all three of its sides, the probability that a single cluster touches all three corners equals zero in the continuum limit.)

\section{Summary}

The purpose of this article is to generalize previous results on crossing probabilities in planar geometries. Specifically,  we predict, via boundary conformal field theory (CFT) the formula (\ref{xing}) for the probability $P_{\varsigma|\vartheta}$ of the $\varsigma$th crossing event in the critical random cluster model (where crossing paths pass through boundary FK clusters), the critical Potts model (where crossing paths pass through boundary spin clusters), and the loop-gas model (where the crossing paths follow ``boundary loops") on a lattice inside a $2N$-sided polygon $\mathcal{P}$, conditioned on a side-alternating boundary condition (BC).  Our prediction is exact in the continuum limit.  In the formula (\ref{xing}) for $P_{\varsigma|\vartheta}$, $\varsigma\in\{1,2,\ldots,C_N\}$ indexes the $C_N$ possible crossing patterns (figure \ref{OctXingConfigs}), where $C_N$ is the $N$th Catalan number (\ref{catalan}), and $\vartheta\in\{1,2,\ldots,C_N\}$ indexes the side-alternating BCs that we exclusively consider in this article.  Items \ref{BCitem1}--\ref{BCitem3} of the introduction \ref{intro} explain the crossing events and the side-alternating BCs for each model, and section \ref{partfuncpoly} gives a full exposition.

Formula (\ref{xing}) is expressed in terms of these four quantities: the crossing weight $\Pi_\varsigma$ (which, in the context of multiple SLE$\kappa$, is also called the ``pure SLE$_\kappa$ partition function"), the Coulomb gas (contour integral) function $\mathcal{F}_\vartheta$, which arises from the solution of the partial differential equations (PDE) describing the affiliated CFT correlation functions (see the discussion of section \ref{CFTsect} beginning around (\ref{Upsilondefn})), the number of closed loops $l_{\varsigma,\vartheta}$ in the $(\varsigma,\vartheta)$th product diagram (see the discussion around  (\ref{pformula}) for the definition and meaning of $l_{\varsigma,\vartheta}$), and the loop fugacity $n(\kappa)$ of the related loop-gas model (item \ref{step1} of section \ref{xingsummary}), parameterized by the SLE$_\kappa$ parameter $\kappa\in(8/3,8)$.

This article begins with an introduction \ref{intro}, which surveys the large body of research and recent advances on crossing probabilities for the models of interest.  Section \ref{xingsummary}, which is self-contained,  gives a detailed explanation  of the formula (\ref{xing}) (but does not derive it), and  section \ref{rectxingsummary} contains its application to   rectangles, reproducing the rectangle crossing probability for critical percolation (\ref{RectCross}), \cite{c3,smirnov}, the critical $Q=2$ random cluster (i.e., FK-Ising) model (\ref{chelk}) \cite{chelsmir}, and the critical $Q=2$ Potts (i.e., Ising) model (\ref{AubinFormula}) \cite{argaub,kozd,bbk,dub,argaub2}.  Section \ref{partfuncpoly} gives formal expressions for the discrete random cluster model and Potts model partition functions.  We consider the case of the free BC first, and then subsections \ref{mutsect} and \ref{RCxingSect}  treat, respectively, the random cluster model with side-alternating BCs and crossing events in it; then subsection \ref{FLBCsect} considers the Potts model.   Also in section \ref{partfuncpoly}, we give formal expressions  (\ref{chixing}, \ref{chixing2}) for the crossing probabilities $\smash{P_{\varsigma|\vartheta}^\mathcal{P}}$ and $\smash{P_{\varsigma|1}^\mathcal{P}}$ of the random cluster model and the Potts model respectively, inside $\mathcal{P}$.  The former involves the number $\mathcal{C}_{\varsigma,\vartheta}$ of independent boundary clusters (i.e., clusters that touch a fixed side of $\mathcal{P}$) inside $\mathcal{P}$, and using basic graph theory, we  find an equation (\ref{pformula}) that relates $\mathcal{C}_{\varsigma,\vartheta}$ to the number $l_{\varsigma,\vartheta}$ of closed loops formed by the intersection of external and internal arcs in the $(\varsigma,\vartheta)$th product diagram (figure \ref{PolyGraph}).  We use this last result in section \ref{Onmodel} to express the random cluster model partition function as a certain partition function of a loop-gas model.
  
Section \ref{Onmodel} begins with a definition of the loop-gas model on a lattice inside $\mathcal{P}$ and its partition functions  $\smash{Z_{\vartheta}^\mathcal{P}}$ and $\smash{Z_{\varsigma,\vartheta}^\mathcal{P}}$, summing exclusively over events that we interpret as analogous to crossing and side-alternating BC events of the random cluster model.  More precisely, for the loop-gas model, the $\varsigma$th crossing event is the event that certain loops, called ``boundary loops," join the vertices of $\mathcal{P}$ through its interior in the $\varsigma$th connectivity (corresponding to the $\varsigma$th crossing pattern, see figure \ref{OctXingConfigsBdyArcs}), and the $\vartheta$th side-alternating BC event is the event that those same loops join the vertices of $\mathcal{P}$ through its exterior in the $\vartheta$th connectivity.  Next, we show that for $Q\in\{1,2,3,4\}$, these loop-gas model partition functions are asymptotic, in the continuum limit, to partition functions of random cluster models and Potts models (\ref{asympXY}) summing exclusively over certain side-alternating BC events.  Finally, we give a formal expression (\ref{masterxing}) for the probability $\smash{P_{\varsigma|\vartheta}^\mathcal{P}}$ of the $\varsigma$th loop-gas model crossing event conditioned on the $\vartheta$th loop-gas model side-alternating BC event in $\mathcal{P}$.  In the continuum limit, this probability equals the corresponding crossing probabilities of the random cluster model and the Potts model, discussed above, when evaluated at appropriate values of $\kappa$.

The main purpose of section \ref{CFTsect} is to combine results from previous sections with CFT arguments to derive the crossing probability formula (\ref{xing}), which is the main result of this article.  First, we relate the loop-gas model partition functions $Z_{\vartheta}^\mathcal{P}$, in the continuum limit, to the boundary CFT correlation function $\Upsilon_{\vartheta}^\mathcal{P}$ (\ref{corrfunc}), known to satisfy a particular system of  $2N+3$ PDEs.  The series of articles \cite{florkleb,florkleb2,florkleb3,florkleb4} determines the dimension  of, as well as explicit formulas for all elements of, the system's solution space $\mathcal{S}_N$.   (As part of the definition for $\mathcal{S}_N$, we also  require that the solutions are dominated by a certain product (\ref{powerlaw}) of power functions.)  Then just above (\ref{contscriptZsig}), we identify the continuum limit of $\smash{Z_{\varsigma,\vartheta}^\mathcal{P}}$ with the ``crossing weight" $\Pi_\vartheta$ (or, equivalently, the ``connectivity weight," \cite{florkleb4,fsk} or the ``pure SLE$_\kappa$ partition function" \cite{bbk,kype}, see the remarks just below (\ref{BN})).  This identification is the main step that leads to our prediction of the explicit formula (\ref{xing}) for polygon crossing probability.  The crossing weights are elements of $\mathcal{S}_N$, linearly related to the correlation function $\smash{\Upsilon_{\vartheta}^\mathcal{P}}$ through the meander matrix $M_N$ \cite{florkleb3,florkleb4,fgg,fgut,difranc,franc}.  This fact and the realization that $\Upsilon_\vartheta = \mathcal{F}_\vartheta$ \eqref{UpsF} combine to determine explicit formulas for all crossing weights and thus, for all crossing probabilities (\ref{xing}) of every models considered in this article, in any $2N$-sided polygon $\mathcal{P}$.  These probabilities are conditioned on the BC events of items \ref{BCitem1}--\ref{BCitem3} in the introduction \ref{intro}, and they are conformally invariant.  Finally, the discussion surrounding (\ref{xingSLE2}) interprets this formula as an instance of the conjectured formula for the probability of curve connectivity events in multiple SLE$_\kappa$ \cite{dub2,graham,kl,sakai}.

To support our predictions, we present high-resolution computer simulation results in section \ref{simxing}.  The simulations measure crossing probabilities of the $Q\in\{2,3,4\}$ random cluster (resp.\ Potts) model on a large square lattice inside a rectangle and on a large triangular lattice inside a hexagon, using the side-alternating BCs of item \ref{BCitem1} (resp.\ item \ref{BCitem2}) in the introduction \ref{intro}.  We compare these measurements with our prediction (\ref{xing}), finding very good agreement for $Q\in\{2,3\}$ and reasonable agreement for $Q=4$.  (Similar results for the $Q=1$ critical random cluster model, i.e., critical percolation, have already appeared in \cite{fzs,c3}.)

Three appendices accompany this article.  In appendix \ref{transformxing}, we use CFT ``corner operators" to regularize the conformal transformation of the boundary CFT correlation functions in section \ref{CFTsect} from the upper half-plane onto the polygon $\mathcal{P}$.  The need for regularization arises because this transformation is  not conformal  at the corners of $\mathcal{P}$.  In appendix \ref{Pottsappendix},  we study random cluster model partition functions that sum exclusively over ``color schemes," or side-alternating BC events where every other side of $\mathcal{P}$ is fixed to a particular color while the other sides are free.   Such ``color scheme partition functions" are natural because any partition function that sums over a free/fixed side-alternating BC event is necessarily a sum of the former.  However, instead of using color scheme partition functions here, we mainly use ``basic partition functions," which correspond to simpler loop-gas partition functions in the continuum limit.  In appendix \ref{Pottsappendix}, we show that in the continuum limit and for $Q\in\{1,2,3\}$, the span of all color scheme partition functions equals the span of all basic partition functions, as functions of the shape of $\mathcal{P}$.  In appendix \ref{equivappendix}, we give some simpler formulas  for the Coulomb gas function $\mathcal{F}_\vartheta(\kappa)$ (\ref{Fexplicit}) appearing in the denominator of the crossing probability formula (\ref{xing}).  They include an alternative formula (\ref{Fother}) used in \cite{florkleb3,florkleb4} for all $\kappa>0$ and explicit algebraic (i.e., without contour integrals) formulas for $\mathcal{F}_\vartheta(\kappa)$ that hold only at certain rational values of $\kappa$.  This latter result leads to a rigorous if indirect evaluation of some rather complicated nested Coulomb gas (contour) integrals.

\section{Acknowledgements}

The authors thank K.\ Kyt\"ol\"a, E.\ Peltola, K.\ Izyurov, Y.\ Saint-Aubin, and D.\ Ridout for insightful conversations and C.\ Townley Flores for carefully proofreading the manuscript.

This work was supported by National Science Foundation Grants Nos.\ PHY-0855335 (SMF), DMR-0536927 (PK and SMF), and DMS--0553487 (RMZ) and Academy of Finland grant ``Algebraic structures and random geometry of stochastic lattice models'' (SMF).

\appendix{}

\section{Transformation to the polygon via CFT corner operators}\label{transformxing}

The smeared loop-gas partition functions $Z_\vartheta(\{\epsilon_j\})$ and $Z_{\varsigma,\vartheta}(\{\epsilon_j\})$ (and the related crossing probabilities $P_{\varsigma|\vartheta}$ (\ref{2ndchi})) are expected to be invariant with respect to conformal transformations of the system domains.  At the same time, the universal part, $\Upsilon_\vartheta$ and $\Upsilon_{\varsigma,\vartheta}$ respectively, of their asymptotic behavior (\ref{simeqdefn}, \ref{contZsigthet}) as $\epsilon_j\rightarrow0$ is expected to be covariant with respect to these transformations.  In this appendix, we explain how the two different transformation laws for these related functions are consistent with each other, and we use the CFT corner operator approach to ``regularize" the conformal covariance law for $\Upsilon_\vartheta$ and $\Upsilon_{\varsigma,\vartheta}$ at corner points on the boundary of the transformed system domain.

We consider a loop-gas model on  a lattice  inside a Jordan domain $\mathcal{D}\subset\mathbb{C}$, and we let $f$ be a conformal bijection from the upper half-plane $\mathbb{H}$ onto $\mathcal{D}$ whose continuous extension to the real axis is conformal at $2N$ marked boundary points $x_1<x_2\ldots x_{2N}$.  As usual, we require that $N$ boundary loops exit and then re-enter $\mathcal{D}$ through its boundary points $w_j:=f(x_j)$, with exactly one boundary loop passing once within distance $\delta_j$ from $w_j$.  Identical to the case (\ref{Upsilondefn})  where  $\mathcal{D}=\mathbb{H}$, we expect the asymptotic behavior
\be\label{Upsilondefn3}\lim_{a\rightarrow0}Z_\vartheta^{\mathcal{D}}(\{w_j\};\{\delta_j\})/Z_f^\mathcal{D}\underset{\delta_j\rightarrow0}{\sim}
\delta_1^{\theta_1}\delta_2^{\theta_1}\dotsm\delta_{2N}^{\theta_1}\Upsilon^{\mathcal{D}}_\vartheta(\{w_j\}),\ee
where $\smash{\Upsilon^{\mathcal{D}}_\vartheta}$ is some universal function.  In (\ref{Upsilondefn3}), the superscript $\mathcal{D}$ indicates the system domain. (Also, if $\mathcal{D}=\mathbb{H}$, then we do not include this superscript, and (\ref{Upsilondefn3}) with $w_j\mapsto x_j$ and $\delta_j\mapsto\epsilon_j$ reduces to (\ref{Upsilondefn}).)  

Now we examine how to transform these partition functions and universal functions to their half-plane versions via $f$.  The supposition that the partition functions of (\ref{Upsilondefn3}) are conformally invariant amounts to 
\be\label{ZTransIt}Z_\vartheta^{\mathcal{D}}(\{w_j\};\{\delta_j\})=Z_\vartheta(\{x_j\};\{\epsilon_j\}),\qquad Z_f^\mathcal{D}=Z_f,\ee
where $\delta_j=|\partial f(x_j)|\epsilon_j$.  (Indeed, under a conformal map, the radius $\epsilon_j$ of the ball centered on $x_j$ dilates by a factor of $|\partial f(x_j)|$.)  Meanwhile, after we identify the universal function $\Upsilon_\vartheta^\mathcal{D}$ with the boundary CFT correlation function 
\be\label{UpsTransIt} \Upsilon_\vartheta^\mathcal{D}(w_1,w_2,\ldots,w_{2N})=\langle\psi_1(w_1)\psi_1(w_2)\dotsm\psi_1(w_{2N})\rangle_\vartheta^\mathcal{D},\ee
this universal function inherits the conformal covariance transformation law of its associated correlation function.  By this, we mean that for any conformal bijection $f:\mathbb{H}\rightarrow\mathcal{D}$ (and with $\theta_1$ (\ref{theta1}) the conformal weight of $\psi_1$), we have
\be\label{Upstrans}\Upsilon_{\vartheta}^{\mathcal{D}}(\{w_j\})=|\partial f(x_1)|^{-\theta_1}|\partial f(x_2)|^{-\theta_1}\dotsm|\partial f(x_{2N})|^{-\theta_1}\Upsilon_{\vartheta}(\{x_j\}).\ee

These rules (\ref{ZTransIt}, \ref{Upstrans}) completely determine $Z^\mathcal{D}_\vartheta$ and $\Upsilon^{\mathcal{D}}_\vartheta$ in terms of their half-plane versions $Z_\vartheta$ and $\Upsilon_\vartheta$ respectively.  Now, it is easy to see that the transformation rules (\ref{ZTransIt}, \ref{Upstrans}) imply one another.  Indeed, we have
\be\label{derivelaw}\text{invariance rule (\ref{ZTransIt}) for $Z_\vartheta$}\qquad\stackrel{(\ref{Upsilondefn}, \ref{Upsilondefn3})}{\underset{\delta_j=|\partial f(x_j)|\epsilon_j}{\Longleftrightarrow}}\quad\text{covariance rule (\ref{Upstrans}) for $\Upsilon_\vartheta$.}\ee

In this article, the system domain $\mathcal{D}$ is often a $2N$-sided polygon $\mathcal{P}$, and here, we encounter a subtlety: the continuous extension to the real axis of any conformal bijection $f:\mathbb{H}\rightarrow\mathcal{P}$, necessarily a Schwarz-Christoffel transformation
\begin{gather}\label{sch}f(z)=A\sideset{}{_{x_1}^{z}}\int(x_1-\zeta)^{\alpha_1/\pi-1}(x_2-\zeta)^{\alpha_2/\pi-1}\dotsm(x_{2N}-\zeta)^{\alpha_{2N}/\pi-1}\,{\rm d}\zeta+B\\
 \text{for some $A,B\in\mathbb{C},$ $x_1<x_2<\ldots<x_{2N},$ and } \alpha_j:=\left\{\parbox{4.5cm}{interior angle measure at the $j$th corner $w_j:=f(x_j)$ of $\mathcal{P}$,}\right\}\in[0,2\pi),\end{gather}
is not conformal at $x_j$ if $\alpha_j\neq\pi$.  As such, $\partial f(x_j)$ either vanishes or is infinite, so (\ref{Upstrans}) is not applicable.  Also, the new radius $\delta_j=|\partial f(x_j)|\epsilon_j$ either vanishes or is infinite, which is physically unacceptable.

To further investigate this singular behavior, we consider the point $\omega_j$ on the $j$th side $[w_j,w_{j+1}]$ of the polygon, offset from the $j$th vertex by  a  small distance $\varepsilon_j>0$, and its preimage $\xi_j\in\mathbb{R}$:
\be\omega_j:=w_j+\varepsilon_j\exp i\bigg(\arg(A)+\sum_{k=1}^j(\pi-\alpha_k)\bigg)=f(\xi_j).\ee
After replacing $w_j\mapsto\omega_j$ and $x_j\mapsto\xi_j$ into (\ref{Upstrans}), we send $\varepsilon_j\rightarrow0$.  Then as the product of derivatives on the right side of (\ref{Upstrans}) blows up or vanishes in this limit, so too does $\Upsilon^\mathcal{D}_\vartheta(\{\omega_j\})$ on the right side.   This is physically unacceptable.

In order to eliminate these singularities from $\Upsilon_\vartheta^\mathcal{D}$, which is a physical quantity, we invoke CFT corner operators.  To do this, we first note that the asymptotic behavior of the dilation factor $|\partial f(\xi_j)|$ as $\varepsilon_j\rightarrow0$ is
\be\label{derivasym}|\partial f(\xi_j)|\underset{\varepsilon_j\rightarrow 0}{\sim}\left(\frac{\alpha_j}{\pi}\varepsilon_j^{1-\pi/\alpha_j}\right)\left(\frac{|A|\pi}{\alpha_j}\right)^{\pi/\alpha_j}\prod_{k\neq j}^{2N}|x_k-x_j|^{(\alpha_k-\pi)/\alpha_j}.\ee
Evidently, (\ref{derivasym}) either blows up or vanishes as $\varepsilon_j\rightarrow0$ if $\alpha_j\neq\pi$.  To eliminate this singularity, we introduce the following replacements for each $j\in\{1,2,\ldots,2N\}$:
\begin{align}\label{SubIt1}\delta_j:=|\partial f(x_j)|\epsilon_j&\quad\longmapsto\quad\tilde{\delta}_j:=\epsilon_j\lim_{\varepsilon_j\rightarrow0}|\partial f(\xi_j)|\left(\frac{\pi}{\alpha_j}\varepsilon_j^{\pi/\alpha_j-1}\right)\\
\label{tilUps} \Upsilon_\vartheta^\mathcal{P}(w_1,w_2,\ldots,w_{2N})&\quad\longmapsto\quad\smash{\tilde{\Upsilon}_{\vartheta}^{\mathcal{P}}}(w_1,w_2,\ldots,w_{2N})=\lim_{\varepsilon_1\rightarrow0}\lim_{\varepsilon_2\rightarrow0}\dotsm\lim_{\varepsilon_{2N}\rightarrow0}\left(\frac{\alpha_1}{\pi}\varepsilon_1^{1-\pi/\alpha_1}\right)^{\theta_1}\\
&\nonumber\qquad\qquad\times\left(\frac{\alpha_2}{\pi}\varepsilon_2^{1-\pi/\alpha_2}\right)^{\theta_1}\dotsm\left(\frac{\alpha_{2N}}{\pi}\varepsilon_{2N}^{1-\pi/\alpha_{2N}}\right)^{\theta_1}\Upsilon_{\vartheta}^\mathcal{P}(\omega_1,\omega_2,\ldots,\omega_{2N}).\end{align}
Without the limit $\varepsilon_j\rightarrow0$, the inserted regularization factors of (\ref{SubIt1}, \ref{tilUps}) cancel each other on the right side of (\ref{Upsilondefn3}) with $\mathcal{D}=\mathcal{P}$.  The  advantage is  that the new quantities (\ref{SubIt1}, \ref{tilUps}) are finite in the limit $\varepsilon_j\rightarrow0$.

We formalize these substitutions (\ref{SubIt1}, \ref{tilUps}) with CFT ``corner operators" \cite{jcar,simkleb}.  If two boundary segments join at an angle with vertex $w\in\mathbb{C}$ and measure $\alpha$ and $\phi_h(w)$ is a primary boundary operator of conformal weight $h$, then
\be\label{corner}\phi_h^c(w):=\lim_{\varepsilon\rightarrow0}\left(\frac{\alpha}{\pi}\varepsilon^{1-\pi/\alpha}\right)^h\phi_h(w+\varepsilon)\ee
is the \emph{primary corner operator} derived from $\phi_h$.  (Replacing $w\mapsto\lambda w$ and $\varepsilon\mapsto\lambda\varepsilon$ for any complex $\lambda$ reveals that its conformal weight equals $\pi h/\alpha$.)   Using the CFT expression (\ref{UpsTransIt}) for $\Upsilon_\vartheta^\mathcal{P}$ and (\ref{tilUps}), we may write this substitution as
\begin{multline}\Upsilon_\vartheta^\mathcal{P}(w_1,w_2,\ldots,w_{2N})=\langle\psi_1(w_1)\psi_1(w_2)\dotsm\psi_1(w_{2N})\rangle_\vartheta^\mathcal{P}\\\longmapsto\qquad \smash{\tilde{\Upsilon}_\vartheta^\mathcal{P}}(w_1,w_2,\ldots,w_{2N})=\langle\psi_1^c(w_1)\psi_1^c(w_2)\dotsm\psi_1^c(w_{2N})\rangle_\vartheta^\mathcal{P}.\end{multline}

To finish, we find an explicit expression for $\smash{\tilde{\Upsilon}_\vartheta^\mathcal{P}}$  if $\mathcal{P}$ is an equiangular $2N$-sided polygon (so all $2N$ interior angles of $\mathcal{P}$ measure $\alpha=(N-1)\pi/N$).  For any $C>0$, we choose the following parameters for the map (\ref{sch}):
\begin{gather}\label{choices1} x_1 =0,\qquad \text{$x_j=m_{j-1}$ for $j\in\{2,3,\ldots,2N-2\}$},\qquad x_{2N-1}=1,\qquad x_{2N}\rightarrow\infty,\\ 
\label{choices2}A=\frac{\alpha}{\pi C}x_{2N}^{1-\alpha/\pi}e^{-i(\pi-\alpha)},\qquad B=0,\qquad C>0,\qquad \text{$\alpha_j=\dfrac{(N-1)\pi}{N}$ for  all  $j\in\{1,2,\ldots,2N\}$}.\end{gather}
Then the vertices $w_j:=f(x_j)$ of $\mathcal{P}$ are such that $w_1=0$, $w_2>0$, and $w_j\in\mathbb{H}$ for all $j\in\{3,4,\ldots,2N\}$, so $\mathcal{P}$ resides in the upper half-plane with its base $[w_1,w_2]$ flush against the positive-real axis.  From (\ref{sch}) with (\ref{choices1}, \ref{choices2}), we have
\begin{gather}\label{x1'}\xi_1\sim(C\varepsilon_1)^{N/(N-1)}\prod_{k=1}^{2N-3} m_k^{1/(N-1)},\qquad\xi_{2N-1}\sim1+(C\varepsilon_{2N-1})^{N/(N-1)}\prod_{k=1}^{2N-3} (1-m_k)^{1/(N-1)},\\
\label{x2N'}\xi_{2N}\sim(C\varepsilon_{2N})^{-N/(N-1)},\qquad\xi_j\sim m_{j-1}+(C\varepsilon_j)^{N/(N-1)}\prod_{k\neq j}^{2N-3} |m_k-m_j|^{1/(N-1)}\quad \text{for all $j\in\{2,3,\ldots,2N-2\}$},\end{gather}
as we send $\varepsilon_j\rightarrow0$.  In particular, we find the asymptotic behavior of $\xi_{2N}$ (\ref{x2N'}) after substituting $\zeta(t)=\xi_{2N}/t$ into the map (\ref{sch}) with the parameters (\ref{choices1}, \ref{choices2}).  This substitution also gives
\be\label{finding2}|\partial f(\xi_{2N})|\underset{\varepsilon_{2N}\rightarrow0}{\sim} C^{N/(N-1)}\varepsilon_{2N}^{2N/(N-1)}\left(\frac{\alpha}{\pi}\varepsilon_{2N}^{1-\pi/\alpha}\right),\quad\text{where $\alpha=(N-1)\pi/N$}.\ee
After inserting these results (\ref{x1'}--\ref{finding2}) into (\ref{tilUps}), we find  that  the regularized universal function $\smash{\tilde{\Upsilon}_\vartheta^\mathcal{P}}$ of the equiangular $2N$-sided polygon $\mathcal{P}$ generated by the conformal bijection $f:\mathbb{H}\rightarrow\mathcal{P}$ (\ref{sch}) with (\ref{choices1}, \ref{choices2})  is given by
\begin{multline}\label{tildeform}\tilde{\Upsilon}^{\mathcal{P}}_{\vartheta}(w_1,w_2,\ldots,w_{2N})=\Bigg[\Bigg(\prod_{k=1}^{2N-3}m_k(1-m_k)\Bigg)\Bigg(\prod_{j<k}^{2N-3}(m_j-m_k)\Bigg)\Bigg]^{(1-6/\kappa)/(N-1)}\\
\times C^{N(1-6/\kappa)/(N-1)}\lim_{x\rightarrow\infty}x^{6/\kappa-1}\Upsilon_\vartheta(0,m_1,m_2,\ldots,m_{2N-3},1,x).\end{multline}
(Thanks to its M\"obius covariance, we know that this limit (\ref{tildeform}) of $\Upsilon_\vartheta$ exists.)   Similarly, we may  define a regularized version of the universal function $\Upsilon_{\varsigma,\vartheta}$ (\ref{contZsigthet}) by replacing the index $\vartheta$ in (\ref{tildeform}) with the double index $\varsigma,\vartheta$.

\section{Color scheme partition functions for the random cluster model}\label{Pottsappendix}

In this Appendix, we consider the ``color scheme partition function" $\smash{W_c^\mathcal{P}( p,Q )}$ defined in the first paragraph of section \ref{mutsect}.  These are natural quantities with which to construct an arbitrary $\RCB$ partition function in the critical random cluster model.  However, the purpose of this appendix is to show that for most $N\in\mathbb{Z}^+$ and $Q\in\{1,2,3,4\}$, the color scheme partition functions are linear combinations of the basic partition functions, defined in (\ref{BasicX}),
\be\label{Xpar}\smash{X_1^\mathcal{P}},\smash{X_2^\mathcal{P}},\ldots,\smash{X_{C_N}^\mathcal{P}}.\ee
Thus, we may work with the latter rather than the former without loss of generality.  The latter are more convenient because they converge to natural loop-gas model partition functions in the continuum limit (section \ref{Onmodel}).

Now obviously, any random cluster model $\RCB$ event is a union of (disjoint) color schemes, so any $\RCB$ partition function is a sum of color scheme partition functions.  Therefore, if we view $\RCB$ partition functions as functions of the shape of the polygon $\mathcal{P}$, then
\be\label{SetOfAll} \{\text{the set of all $\RCB$ partition functions}\}\subset\text{span}\,\{\smash{W_c^\mathcal{P}}( p_c,Q )\,|\,\text{$c$ a color scheme}\}.\ee
The specific purpose of this appendix is to determine for which $(Q,N)\in\{1,2,3,4\}\times\mathbb{Z}^+$ we have
\be\label{PreEq}\text{span}\,\{X_\vartheta^\mathcal{P}(\text{$p$}_c,Q )\,|1\leq\vartheta\leq C_N\}=\text{span}\,\{\smash{W_c^\mathcal{P}}( p_c,Q )\,|\,\text{$c$ a color scheme}\}\ee
in the continuum limit.  If (\ref{PreEq}) is true, then in light of (\ref{SetOfAll}), restricting our attention to the basic partition functions (\ref{Xpar}) does not result in loss of generality because any $\RCB$ partition function is a linear combination of them.

%%%%%%%%%%%%%%%%%%%%%%%%%%%%%%%%%%%%%%%%%%%%%%%%%%%%%%%%%%%%%%%%%%%%%%%%%%%%%%%%%%%%%%%%%%%%%%%%%%%%%%%%%%%%%%%%%%%%%%%%%%%%%%%%%%%%%%%
\begin{figure}[t]
\centering
\includegraphics[scale=0.27]{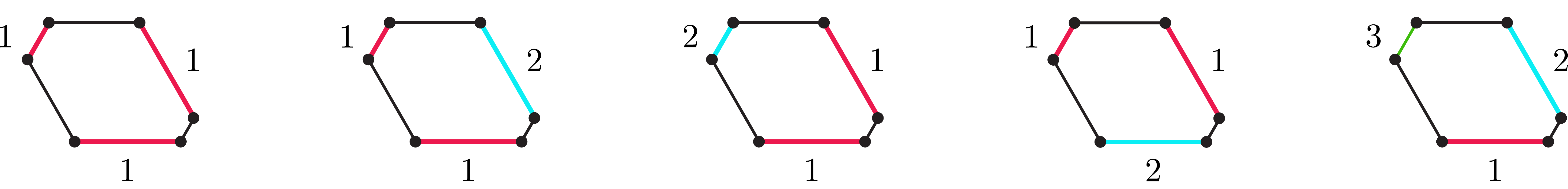}
\caption{If $Q=2$ (resp.\ $Q=3$), then the four  left (resp.\ five) hexagons  show the $s_3(2)=4$ (resp.\ $ s_3(3)=5$)  (\ref{EquivClassW}) ways that we may assign one of $Q$ colors to  the odd sides of the hexagon, regarding only differences in color (\ref{duplication}) between these sides. }
\label{PolyColors}
\end{figure}
%%%%%%%%%%%%%%%%%%%%%%%%%%%%%%%%%%%%%%%%%%%%%%%%%%%%%%%%%%%%%%%%%%%%%%%%%%%%%%%%%%%%%%%%%%%%%%%%%%%%%%%%%%%%%%%%%%%%%%%%%%%%%%%%%%%%%%%

Evidently, (\ref{SetOfAll}) implies that (\ref{PreEq}) holds if we replace its equality $=$ with an inclusion $\subset$.  To show that equality holds for certain $(Q,N)\in\{1,2,3,4\}\times\mathbb{Z}^+$, we prove that the dimensions of the spaces in (\ref{PreEq}) agree for these values of $(Q,N)$ by showing that for these values, we have
\be\label{PreEq2}\text{rank}\,\{X_\vartheta^\mathcal{P}(\text{$p$}_c,Q )\,|1\leq\vartheta\leq C_N\}=|\{\smash{W_c^\mathcal{P}}( p_c,Q )\,|\,\text{$c$ a color scheme}\}|.\ee
To find the rank of $\{\smash{X_\vartheta^\mathcal{P}}( p_c,Q )\,|1\leq\vartheta\leq C_N\}$ in the continuum limit, we restore the prefactor $\smash{Q^{(-\ell_{1,\vartheta}+1)/2}}$ in (\ref{Xgas}) (omitted from (\ref{Xgas1}, \ref{Xgas2}, \ref{asympXY}) for convenience) to the right side of both left equations in the top line of (\ref{preasympXY}), obtaining
\be\label{Xins} X_\vartheta^\mathcal{P}(\text{$p$}_c,Q )\underset{a\rightarrow0}{\sim} Q^{(-\ell_{1,\vartheta}+1)/2}Z^\mathcal{P}_\vartheta(\kappa),\quad \text{where $Q=4\cos^2(4\pi/\kappa)$ and $\kappa\in(4,8)$.}\ee
In order to match previous results in the literature \cite{jcar} (beneath (\ref{EqFinal})), we insert an extra but physically irrelevant factor of $\smash{Q^{N/2}}$ on the right side of (\ref{Xins}) too.  After doing this and recalling (\ref{simeqdefn}, \ref{UpsF}), we have
\be\label{Xins2} X_\vartheta^\mathcal{P}(\text{$p$}_c,Q )\underset{a\rightarrow0}{\sim} Q^{(N-\ell_{1,\vartheta}+1)/2}Z^\mathcal{P}_\vartheta(\kappa),\quad \text{where $Z_\vartheta(\kappa,\epsilon)\simeq\mathcal{F}_\vartheta(\kappa)$ as $\epsilon\downarrow0$ for the smeared version of $Z^\mathcal{P}_\vartheta(\kappa)$}.\ee
From (\ref{Xins2}), we infer that in the continuum limit the rank of $\{\smash{X_\vartheta^\mathcal{P}}( p_c,Q )\,|1\leq\vartheta\leq C_N\}$ equals the rank of $\mathcal{B}_N$ (\ref{TLBN}).  Recalling the meander matrix $M_N$ (\ref{F=MPi}) from section \ref{CFTsect}, by corollary \red{7} of \cite{florkleb3}, we have 
\be\label{Rank2}\text{rank}\,\{X_\vartheta^\mathcal{P}(\text{$p$}_c,Q )\,|1\leq\vartheta\leq C_N\}=\text{rank}\,\mathcal{B}_N=\text{rank}\,M_N.\ee
The meander matrix $M_N$ has full rank $C_N$ if $Q=4$ ($n=2$, $\kappa=4$) \cite{fgg}, so $\text{rank}\,\mathcal{B}_N=C_N$ in this case.  If $Q\in\{1,2,3\}$, then $M_N$ does not have full rank, and we use (\red{53}, \red{54}) of \cite{florkleb3} and (\red{5.23}, \red{5.24}) of \cite{fgg} to compute its rank.   This gives 
\be\label{dense2}\left\{\begin{array}{lllll} Q=1&\quad\stackrel{(\ref{dense})}{\Longrightarrow}\quad&\quad \kappa=6, & \quad\text{rank}\,\mathcal{B}_N=1&\text{$=C_N$ for $N=1$, $<C_N$ otherwise,} \\
Q=2&\quad\Longrightarrow\quad & \quad \kappa=16/3, & \quad\text{rank}\,\mathcal{B}_N=2^{N-1}&\text{$=C_N$ for $N\leq2$, $<C_N$ otherwise,}\\ 
Q=3&\quad\Longrightarrow\quad & \quad \kappa=24/5, & \quad\text{rank}\,\mathcal{B}_N=(3^{N-1}+1)/2&\text{$=C_N$ for $N\leq4$, $<C_N$ otherwise,}\\  
Q=4&\quad\Longrightarrow\quad & \quad \kappa=4, &\quad\text{rank}\,\mathcal{B}_N=C_N&\text{$=C_N$ for all $N\in\mathbb{Z}^+$}.\end{array}\right.\ee

Next, we determine the right side of (\ref{PreEq2}).  Although there are $Q^N$ color schemes, there are  fewer distinct color scheme partition functions, as the latter depend only on differences in color.  Indeed, we have (figure \ref{PolyColors}),
\be\label{duplication}W_b^\mathcal{P}( p_c,Q )=W_c^\mathcal{P}( p_c,Q )\quad\text{if}\quad b_i=b_j\quad\Longleftrightarrow\quad c_i=c_j\quad\text{for all $i,j\in\{1,2,\ldots,N\}.$}\ee
We say that color schemes $b$ and $c$ are equivalent and write $b\sim c$ if they satisfy the right relation in (\ref{duplication}), let $[b]=[c]$ denote their common equivalence class, and write $W_{[c]}^\mathcal{P}( p_c,Q )$ for $W_c^\mathcal{P}( p_c,Q )$.  Then the right side of (\ref{PreEq2}) equals
\be\label{EquivClassW} |\{\smash{W_c^\mathcal{P}}( p_c,Q )\,|\,\text{$c$ a color scheme}\}|=s_N(Q):=\left\{\parbox{2.8in}{number of equivalence classes for color schemes $c=(c_1,c_2,\ldots,c_N)$ with $c_i\in\{1,2,\ldots,Q\}$}\right\}.\ee
Thus, in light of (\ref{Rank2}), in order to determine for what $(Q,N)$ (\ref{PreEq2}) holds, we must determine for what $(Q,N)$ we have $s_N(Q)=\text{rank}\,\mathcal{B}_N$.  For this purpose, we let
\begin{align} \#c&=\text{the number of distinct colors in color scheme $c$}, \\
|[c]|&=\text{the cardinality of the equivalence class for color scheme $c$}.\end{align}
By assigning any one of the $Q$ colors to the first collection of sides with the same color in $c$, then assigning any one of the $Q-1$ remaining colors to the second collection of sides with the same color, etc., we find that
\be\label{relat} |[c]|=Q(Q-1)\dotsm(Q-\#c+1).\ee
The total number $Q^N$ of color schemes obviously equals the sum of the cardinalities (\ref{relat}) of all equivalence classes.  Because the number of equivalence classes $[c]$ having $\#c=P$ colors equals $s_N(P)-s_N(P-1)$, this total is
\be\label{Qid} Q^N=\sum_{P=1}^Q[s_N(P)-s_N(P-1)]Q(Q-1)\dotsm(Q-P+1),\quad\text{with $s_N(0)=0$.}\ee
We may use (\ref{Qid}) to compute $s_N(Q)$ for all $Q\in\mathbb{Z}^+$.  In this article, we restrict our attention to $Q\in\{1,2,3,4\}$.  For these values, we find
\begin{align}\label{ses} s_N(1)&=1,\\
s_N(2)&=2^{N-1},\\
s_N(3)&=(3^{N-1}+1)/2,\\
\label{seslast}s_N(4)&=[(2)(4^{N-2})+(3)(2^{N-2})+1]/3.\end{align}
Upon comparing these values (\ref{ses}--\ref{seslast}) with (\ref{dense2}), we discover that $s_N(Q)=\text{rank}\,\mathcal{B}_N$ for all $N\in\mathbb{Z}^+$ if $Q\in\{1,2,3\}$ and for all $N\in\{1,2,3\}$ if $Q=4$.  Thus, (\ref{PreEq}) is true for these values of $(Q,N)\in\{1,2,3,4\}\times\mathbb{Z}^+     $:
\be\label{EqFinal}\text{span}\,\{X_\vartheta^\mathcal{P}(\text{$p$}_c,Q )\,|1\leq\vartheta\leq C_N\}=\text{span}\,\{\smash{W_{[c]}^\mathcal{P}}( p_c,Q )\,|\,\text{$c$ a color scheme}\},\qquad\begin{cases}\text{if $Q\in\{1,2,3\}$, or} \\ \text{if $Q=4$ and $N\in\{1,2,3\}.$}\end{cases}\ee
Hence, we infer from (\ref{SetOfAll}, \ref{EqFinal}) that for any $Q\in\{1,2,3\}$ or for $Q=4$ and $N\in\{1,2,3\},$ any $\RCB$ partition function is a linear combination of the basic partition functions (\ref{Xpar}), so restricting our attention to the latter, as we do in this article, does not result in loss of generality.  On the other hand, if $Q=4$ and $N\geq4$, then $\text{rank}\,\mathcal{B}_N=C_N<s_N(Q)$ (\ref{dense2}, \ref{ses}).  Hence, not every $\RCB$ partition function is necessarily a linear combination of the basic partition functions (\ref{Xpar}), so restricting our attention to the latter, as we do in this article, does result in loss of generality.

\begin{figure}[t]
\centering
\includegraphics[scale=0.27]{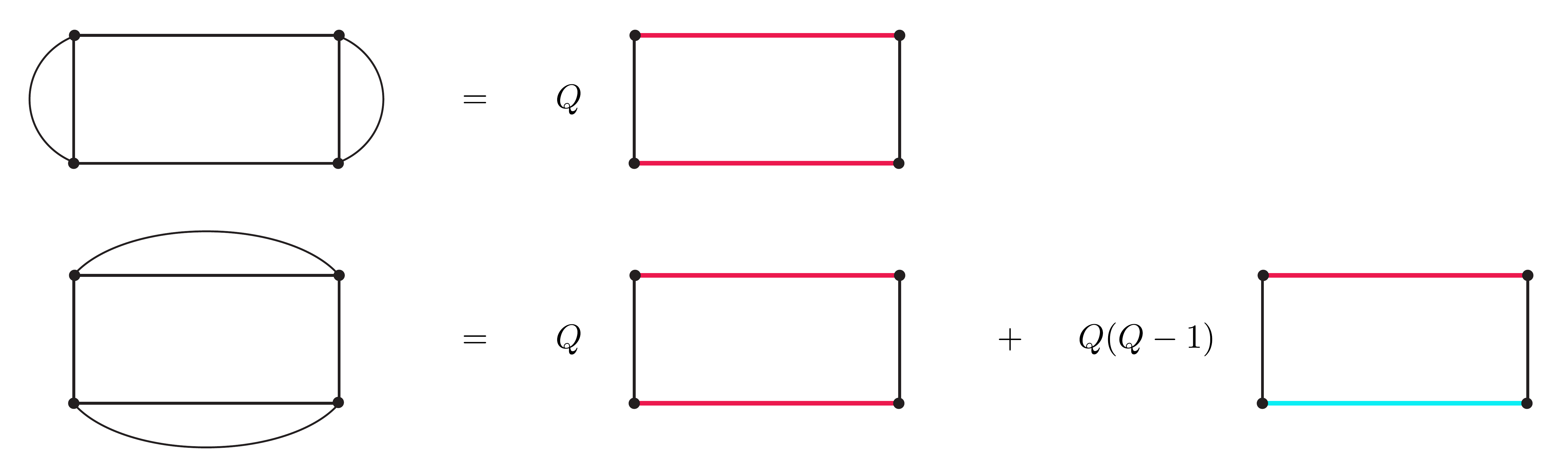}
\caption{The upper-left (resp.\ lower-left) rectangle, with top and bottom sides mutually wired together (resp.\ wired independently of each other) samples all configurations with these sides having the same color (resp.\ same color or different colors).}
\label{PolyColors2}
\end{figure}

We end this appendix by comparing our work for the case $N=2$ to similar results in \cite{jcar}.  To begin, we introduce some convenient notation: we write $c\leftrightsquigarrow\vartheta$ if $c\leftrightsquigarrow\RCB_\vartheta$ (\ref{squiqequiv}).  That is,
\be\label{equiv2} c=(c_1,c_2,\ldots,c_N)\leftrightsquigarrow\vartheta  \qquad\Longrightarrow\qquad\parbox{2.6in}{if a cluster joins the $i$th and $j$th fixed sides of $\mathcal{P}$ in the $\vartheta$th connectivity, then $c_i=c_j$.}\ee
Furthermore, if $b,c\in[c]$, then $b\leftrightsquigarrow\vartheta$ clearly implies $c\leftrightsquigarrow\vartheta$, so we simply write $[c]\leftrightsquigarrow\vartheta$ to mean all of these relations.  Then using (\ref{relat}), we may write (\ref{XEsum}) as
\begin{align}\label{preXsum}X_\vartheta^\mathcal{P}(\text{$p$}_c,Q )&=\sum_{c\leftrightsquigarrow\vartheta} W_c^\mathcal{P}( p_c,Q )\\
\label{preXsum2}&=\sum_{[c]\leftrightsquigarrow\vartheta} Q(Q-1)\dotsm(Q-\#c+1)W_{[c]}^\mathcal{P}( p_c,Q ).\end{align} 
If $N=2$, then there are two inequivalent color schemes: the top and bottom sides of a rectangle have the same color ($c=(1,1)$), or they have different colors ($c=(1,2)$).  The corresponding basic partition functions are (figure \ref{PolyColors2})
\begin{alignat}{2}\label{Xmutind}&\text{mutual wiring $(\vartheta=1,$ section \ref{rectxingsummary})}:\qquad &&X_1^\mathcal{P}=QW_{1,1}^\mathcal{P},\\
\label{Xmutind2}&\text{independent wiring $(\vartheta=2,$ section \ref{rectxingsummary})}:\qquad &&X_2^\mathcal{P}=QW_{1,1}^\mathcal{P}+Q(Q-1)W_{1,2}^\mathcal{P}.\end{alignat}
Now, we let $W_c(\epsilon)$ denote the smeared half-plane version of the color scheme partition function $W_c^\mathcal{P}$ (section \ref{CFTsect}) for $c=(1,1)$ and $c=(1,2)$.  With the meaning of $\simeq$ given by (\ref{Upsilondefn}, \ref{simeqdefn}), we identify $W_c(\epsilon)$ with a universal function $\Phi_c$:
\be\label{Phidefn}W_c(\epsilon)\simeq\Phi_c\quad\text{as $\epsilon\downarrow0$}.\ee
With $\smash{\psi_1^{fc}(x)}$ (resp.\ $\smash{\psi_1^{cf}(x)}$) denoting a BCC operator that changes a critical random cluster model BC from free (resp.\ fixed with color $c\in\{1,2,\ldots,Q\}$) for $y<x$ to fixed with color $c\in\{1,2,\ldots,Q\}$ (resp.\ free) for $y>x$, one writes 
\be\Phi_c(x_1,x_2,\ldots,x_{2N})=\langle\psi_1^{fc_1}(x_1)\psi_1^{c_1f}(x_2)\psi_1^{fc_2}(x_3)\psi_1^{c_2f}(x_4)\,\,\dotsm\,\,\psi_1^{fc_N}(x_{2N-1})\psi_1^{c_Nf}(x_{2N})\rangle.\ee 
Like $\psi_1(x)$, these BBC operators are primary CFT operators with conformal weights equaling the $(1,2)$ Kac weight (\ref{theta1}) from a CFT with central charge $c$ given by combining (\ref{dense}, \ref{central}).  Now, (\ref{Xins2}, \ref{preXsum2}, \ref{Phidefn}) combine to give
\be\label{Frelate} Q^{(N-\ell_{1,\vartheta}+1)/2}\mathcal{F}_\vartheta(\kappa)=\sum_{[c]\leftrightsquigarrow\vartheta} Q(Q-1)\dotsm(Q-\#c+1)\Phi_c(\kappa).\ee
If $N=2$, then the right side of (\ref{Frelate}) equals the right side of (\ref{Xmutind}, \ref{Xmutind2}) (with $\smash{W_c\super{P}}\mapsto\Phi_c$) for $\vartheta=1$ and $\vartheta=2$ respectively, and the left side of (\ref{Frelate}) equals $Q^{1/2}\mathcal{F}_1(\kappa)$ and $Q\mathcal{F}_2(\kappa)$ for $\vartheta=1$ and $\vartheta=2$ respectively.  Hence,
\be\begin{aligned}\label{FQ} Q^{1/2}\mathcal{F}_1(\kappa)&=Q\Phi_{1,1}(\kappa) \\
Q\mathcal{F}_2(\kappa)&=Q\Phi_{1,1}(\kappa)+Q(Q-1)\Phi_{1,2}(\kappa),\end{aligned}\qquad\Longrightarrow\qquad \begin{aligned}\Phi_{1,1}(\kappa)&=Q^{-1/2}\mathcal{F}_1(\kappa)\\ \Phi_{1,2}(\kappa)&=(Q-1)^{-1}[\mathcal{F}_2-Q^{-1/2}\mathcal{F}_1](\kappa)=\Pi_2(\kappa).\end{aligned}\ee
After inserting the formula (\ref{FN2}) for $\mathcal{F}_1$, now expressed in terms of hypergeometric functions with (\ref{Iinvar}, \ref{Ivar}), and the formula (\ref{Pivarsigma}) for $\Pi_2$ into (\ref{FQ}), we find these explicit formulas (with $x_{ij}:=x_i-x_j$ and $\lambda=x_{21}x_{43}/x_{31}x_{42}$):
\begin{align}\label{Phi11}\Phi_{1,1}(\kappa\,|\,x_1,x_2,x_3,x_4)&=n(\kappa)[x_{31}x_{42}]^{1-6/\kappa}\lambda^{2/\kappa}(1-\lambda)^{1-6/\kappa}\,_2F_1\bigg(\frac{4}{\kappa},1-\frac{4}{\kappa};2-\frac{8}{\kappa}\,\bigg|\,1-\lambda\bigg),\hspace{1cm}\\
\label{Phi12}\Phi_{1,2}(\kappa\,|\,x_1,x_2,x_3,x_4)&=\frac{\Gamma(12/\kappa-1)\Gamma(4/\kappa)}{\Gamma(8/\kappa)\Gamma(8/\kappa-1)}[x_{31}x_{42}]^{1-6/\kappa}\lambda^{1-6/\kappa}(1-\lambda)^{2/\kappa}\,_2F_1\bigg(\frac{4}{\kappa},1-\frac{4}{\kappa};\frac{8}{\kappa}\,\bigg|\,1-\lambda\bigg).
\end{align}
In \cite{jcar}, J.\ Cardy gives formulas for $\Phi_{1,1}$ and $\Phi_{1,2}$.  They are respectively (\red{50}, \red{51}) of that article.  In order to cast them in our notation, we must set, in (\red{50}, \red{51}) of \cite{jcar},
\be Z_{aa}=\Phi_{1,1},\qquad Z_{ab}=\Phi_{1,2},\qquad x=\theta_1\overset{(\ref{theta1})}{:=}(6-\kappa)/2\kappa,\qquad \eta=\lambda,\qquad \zeta=(\eta x_{32}x_{41})^{-1}.\ee
After applying the identity $_2F_1(a,b;c\,|\,z):=(1-z)^{c-b-a}\,_2F_1(c-a,c-b;c\,|\,z)$ and the Euler reflection formula for the Gamma function, Cardy's formula for $Z_{aa}$ matches our formula (\ref{Phi11}) for $\Phi_{1,1}$, and  (after correcting what seems to be a typo in the factor $\Gamma(1+x)$ of $Z_{ab}$ to $\Gamma(1+4x)$), Cardy's formula for $Z_{ab}$ matches our formula (\ref{Phi12}) for $\Phi_{1,2}$.

%%%%%%%%%%%%%%%%%%%%%%%%%%%%%%%%%%%%%%%%%%%%%%%%%%%%%%%%%%%%%%%%%%%%%%%%%%%%%%%%%%%%%%%%%%%%%%%%%%%%%%%%%%%%%%%%%%%%%%%%%%%%%%%%%%%%%%%
%%%%%%%%%%%%%%%%%%%%%%%%%%%%%%%%%%%%%%%%%%%%%%%%%%%%%%%%%%%%%%%%%%%%%%%%%%%%%%%%%%%%%%%%%%%%%%%%%%%%%%%%%%%%%%%%%%%%%%%%%%%%%%%%%%%%%%%

\section{Alternative formulas for Coulomb gas functions}\label{equivappendix}

In this appendix, we present some simpler formulas for the Coulomb gas function $\mathcal{F}_\vartheta$ (\ref{Fexplicit}), and we obtain algebraic formulas for $\mathcal{F}_\vartheta$ in certain cases.   

Now, there are two types of simpler formulas for the Coulomb gas function: the  alternative formula presented in section \ref{FirstSubsection} agrees with \eqref{Fexplicit} for all $\kappa>0$  (we thank K.\ Kyt\"ol\"a and E.\ Peltola for sharing this fact with us), and the alternative formulas presented in section \ref{SecondSubsection} hold only at special values of $\kappa$.

In principle, a direct proof of equality between these different formulas entails performing  some  of the integrations in (\ref{Fexplicit}).  However, because this integration appears to be impossible to do in closed form for most values of $\kappa$, we prove equality indirectly via these two steps:
\begin{enumerate}
\item\label{app1}  We  show that the functions of either formula are elements of $\mathcal{S}_N$, that is, they satisfy (\ref{nullstate}, \ref{wardid}, \ref{powerlaw}).
\item\label{app2}  We  show that the images of these functions under the following vector space isomorphism \cite{florkleb3} are equal:
\be\label{LinearMap} v:\mathcal{S}_N\rightarrow\mathbb{R}^{C_N},\qquad v(F)_\varsigma=[\mathscr{L}_\varsigma]F,\quad\text{for all $\varsigma\in\{1,2,\ldots,C_N\}$}.\ee
\end{enumerate}
(Here, $[\mathscr{L}_\varsigma]$ is an equivalence class of a sequence of limits, defined somewhat vaguely in (\ref{allowablelim}).  See definitions \red{9} and \red{11} of \cite{florkleb} to make this precise.)  Thanks to item \ref{app2}, the desired equality follows 
from item \red{4} of theorem \red{8} in \cite{florkleb3}.  These results lead to algebraic expressions (\ref{Intkappa=6}, \ref{Intkappa=3}, \ref{Intkappa=16/3}) for the contour integral in (\ref{Fexplicit}) at special values of $\kappa$.

Parts of this appendix assume familiarity with the results of \cite{florkleb,florkleb2,florkleb3,florkleb4}, rigorously proven only for $\kappa\in(0,8)$, and the  required main results  are the following: the collection $\mathcal{S}_N$ of all functions satisfying (\ref{nullstate}, \ref{wardid}, \ref{powerlaw}) has dimension $C_N$ (\ref{catalan}), and the collection $\mathcal{B}_N\subset\mathcal{S}_N$ (\ref{TLBN}) of Coulomb gas functions (\ref{Fexplicit}) is a basis for $\mathcal{S}_N$  if  $\kappa$ is not among (\ref{exceptional}) with $q\leq N+1$.  Also, the set $\mathscr{B}_N^*:=\{[\mathscr{L}_1],[\mathscr{L}_2],\ldots,[\mathscr{L}_{C_N}]\}$ (\ref{BNstar}) (see definitions \red{9} and \red{11} of \cite{florkleb} for a definition of $[\mathscr{L}_\varsigma]$) is a basis for the dual space $\mathcal{S}_N^*$, which allows us to define a dual basis $\mathscr{B}_N=\{\Pi_1,\Pi_2,\ldots,\Pi_{C_N}\}$ (\ref{BN}) of crossing weights for $\mathcal{S}_N$.  Finally, the linear map $v$ (\ref{LinearMap}) is a vector space isomorphism.

\subsection{An alternative formula for $\mathcal{F}_\vartheta$ (\ref{Fexplicit})}\label{FirstSubsection}
With $x_1<x_2<\ldots<x_{2N}$ an ordered collection of  points on  the real axis and $\boldsymbol{x}:=(x_1,x_2,\ldots,x_{2N})$, \cite{florkleb3} defines a Coulomb gas function $\mathcal{F}_{c,\vartheta}(\kappa\,|\,\boldsymbol{x})$ for each $c\in\{1,2,\ldots,2N\}$, $\vartheta\in\{1,2,\ldots,C_N\}$, and $\kappa>0$ by the explicit formula
\begin{multline}\label{Fother}\mathcal{F}_{c,\vartheta}(\kappa\,|\,\boldsymbol{x})=n(\kappa)\left[\frac{n(\kappa)\Gamma(2-8/\kappa)}{4\sin^2(4\pi/\kappa)\Gamma(1-4/\kappa)^2}\right]^{N-1}\Bigg(\prod_{\substack{j<k \\ j,k\neq c}}^{2N}(x_k-x_j)^{2/\kappa}\Bigg)\Bigg(\prod_{\substack{k=1 \\ k\neq c}}^{2N}|x_c-x_k|^{1-6/\kappa}\Bigg)\oint_{\Gamma_{N-1}}{\rm d}u_{N-1}\dotsm\\ 
\dotsm\oint_{\Gamma_2}{\rm d}u_2\,\,\oint_{\Gamma_1}{\rm d}u_1\,\,\mathcal{N}\Bigg[\Bigg(\prod_{\substack{l=1 \\ l\neq c}}^{2N}\prod_{m=1}^{N-1}(x_l-u_m)^{-4/\kappa}\Bigg)\Bigg(\prod_{m=1}^{N-1}(x_c-u_m)^{12/\kappa-2}\Bigg)\Bigg(\prod_{p<q}^{N-1}(u_p-u_q)^{8/\kappa}\Bigg)\Bigg],\end{multline}
where $n(\kappa)$ is given by (\ref{fugacity}), $\Gamma_m$ is the Pochhammer contour $\mathscr{P}(x_i,x_j)$ (figure \ref{PochhammerContour}) that shares its endpoints $x_i$ and $x_j$ with the $m$th arc of the $\vartheta$th connectivity on $x_1<x_2<\ldots<x_{2N}$ (item \ref{step2} of section \ref{xingsummary}), and no contour shares its endpoints with the $N$th arc of this connectivity, which has an endpoint at $x_c$.  Also, the symbol $\mathcal{N}$ selects the branch of the logarithm for each power function in its integrand so $\mathcal{F}_{c,\vartheta}$ is real-valued for $\kappa>0$ \cite{florkleb3}.  Last, if $x_i<x_j<x_k<x_l$, then the contour $\mathscr{P}(x_i,x_l)$ arcs over $x_j$ and $x_k$ (and $\mathscr{P}(x_j,x_k)$ if present) in the upper half-plane, and we reverse the orientation of $\mathscr{P}(x_i,x_j)$ if this contour arcs over $x_c$.  (See definition \red{4} and figure \red{5} of \cite{florkleb3} for further elaboration.)

The Coulomb gas function $\mathcal{F}_{c,\vartheta}$ is an example of a solution to the system (\ref{nullstate}, \ref{wardid}) that we produce via the Coulomb gas formalism of CFT \cite{fms,henkel,df1,df2}.  This method invokes a ``charge neutrality condition," which produces the distinguished point $x_c$ in the formula (\ref{Fother}).  However, corollary \red{9} of \cite{florkleb3} implies that the distinguished point is, in fact, not so special.  Indeed, the corollary states that $\mathcal{F}_{c,\vartheta}=\mathcal{F}_{c',\vartheta}$ for all $c,c'\in\{1,2,\ldots,2N\}$, $\vartheta\in\{1,2,\ldots,C_N\}$, and $\kappa\in(0,8)$.  In light of this fact, we drop the subscript $c$ and denote this Coulomb gas function by $\mathcal{F}_\vartheta$.

This independence of (\ref{Fother}) on our choice of $c$ suggests that an alternative formula for $\mathcal{F}_\vartheta$, one in which all points $x_j$ appear with equal footing, may exist.  Indeed, it does, and it is given by (\ref{Fexplicit}).  While this formula has greater aesthetic appeal and appears more compact, it is actually more complicated, with a longer integrand and one more integration contour than (\ref{Fother}).  For this reason, we use (\ref{Fother}) over (\ref{Fexplicit}) as our formula for $\mathcal{F}_\vartheta$ in most practical situations.

The CFT Coulomb gas formalism also produces the more complicated, yet more symmetric-appearing, formula (\ref{Fexplicit}) if we replace the charge neutrality condition with the condition that the total charge of the associated CFT Coulomb gas correlation function equals zero instead.  Forsaking the charge neutrality condition puts us at risk of generating functions that, while satisfying the null-state PDEs (\ref{nullstate}) and  translation invariance, violate the last two conformal Ward identities  of (\ref{wardid}).  Somewhat surprisingly,  this violation does not occur if we require a zero total charge; the functions thus generated (\ref{Fexplicit}) still satisfy the entire system (\ref{nullstate}, \ref{wardid}).  K.\ Kyt\"ol\"a and E.\ Peltola brought this fact to our attention, and they proved it in their recent article \cite{kype2}.  (See their lemma \red{4.14} and proposition \red{4.15}.)

%%%%%%%%%%%%%%%%%%%%%%%%%%%%%%%%%%%%%%%%%%%%%%%%%%%%%%%%%%%%%%%%%%%%%%%%%%%%%%%%%%%%%%%%%%%%%%%%%%%%%%%%%%%%%%%%%%%%%%%%%%%%%%%%%%%%%%%
\begin{figure}[t]
\centering
\includegraphics[scale=0.27]{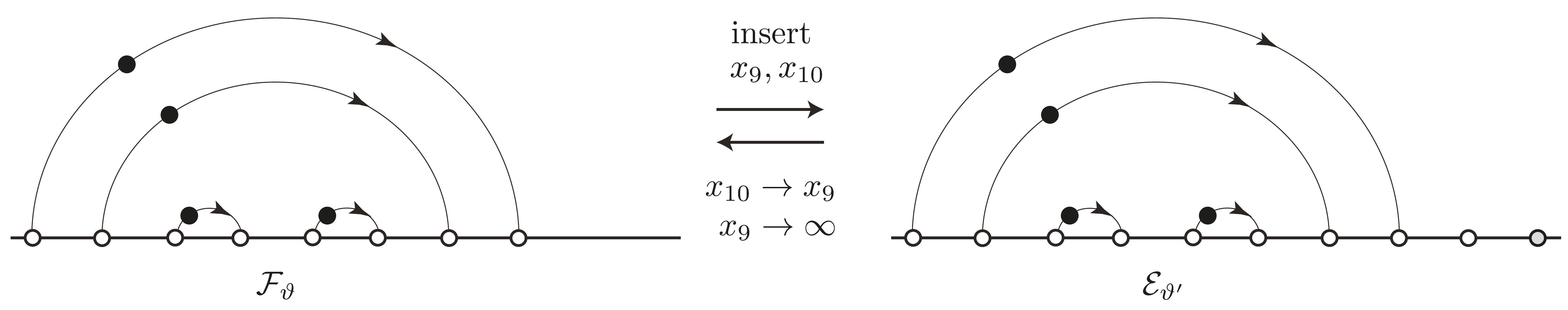}
\caption{A Coulomb gas function $\mathcal{F}_\vartheta\in\mathcal{S}_4$ (\ref{Fexplicit}), with the points $x_1,$ $x_2,\ldots,x_8$ (white circles) and integration contours (arcs) shown, and the corresponding function $\mathcal{E}_{\vartheta'}\in\mathcal{S}_5$ (\ref{Fother}) with the two inserted points $x_9$ and $x_c=x_{10}$ (gray circle) shown.}
\label{InsertionLimit}
\end{figure}
%%%%%%%%%%%%%%%%%%%%%%%%%%%%%%%%%%%%%%%%%%%%%%%%%%%%%%%%%%%%%%%%%%%%%%%%%%%%%%%%%%%%%%%%%%%%%%%%%%%%%%%%%%%%%%%%%%%%%%%%%%%%%%%%%%%%%%%

Now we move to the main purpose of this section: to rigorously prove that (\ref{Fother}, \ref{Fexplicit}) are different formulas for the same function $\mathcal{F}_\vartheta$ for all $\kappa>0$.  The proof invokes the two steps \ref{app1}, \ref{app2} described in the introduction of this appendix:
\begin{enumerate}[wide, labelwidth=!, labelindent=0pt]
\item First, we show that the functions given by (\ref{Fother}, \ref{Fexplicit}) are elements of $\mathcal{S}_N$.  In section \red{V} of \cite{dub}, J.\ Dub\'edat proves this claim for the first formula (\ref{Fother}), and as mentioned above, \cite{kype2} proves this claim for the second formula (\ref{Fexplicit}).

Before moving to the next step, we offer what is perhaps a simpler proof, using the results of \cite{florkleb3}, that the function $\mathcal{F}_\vartheta$ defined by the second formula (\ref{Fexplicit}) is an element of $\mathcal{S}_N$ if $\kappa\in(0,8)$.  The idea is to promote $\mathcal{F}_\vartheta$ to another Coulomb gas function $\mathcal{E}_{\vartheta'}\in\mathcal{S}_{N+1}$ by inserting two new points ($x_{2N+1}$,  $x_{2N+2}$) with $x_{2N} < x_{2N+1}<x_{2N+2}$.  Here, the subscript $\vartheta'$ indicates the $\vartheta'$th connectivity, obtained from the $\vartheta$th by joining the two inserted endpoints $x_{2N+1}<x_{2N+2}$ with an arc.  Then, we define $\mathcal{E}_{\vartheta'}$ by the first formula (\ref{Fother}) with $N\mapsto N+1$ and with $c=2N+2$ (so no integration contour of $\mathcal{E}_{\vartheta'}$ surrounds $x_{2N+2}$ or its partner $x_{2N+1}$) (figure \ref{InsertionLimit}).  Next, we consider  this limit:
\be\label{Elim}\lim_{x_{2N+2}\rightarrow x_{2N+1}}(x_{2N+2}-x_{2N+1})^{6/\kappa-1}\mathcal{E}_{\vartheta'}(\kappa\,|\,x_1,x_2,\ldots,x_{2N+2}),\qquad \mathcal{E}_{\vartheta'}\in\mathcal{S}_{N+1}.\ee
We may obtain a closed expression for this limit (\ref{Elim}) from the explicit formula of $\mathcal{E}_{\vartheta'}$, and this expression appears to depend on $x_{2N+1}$.  But according to lemma \red{4} of \cite{florkleb}, because $\mathcal{E}_{\vartheta'}\in\mathcal{S}_{N+1}$, this limit (\ref{Elim}) is actually independent of $x_{2N+1}$.  Therefore, sending $x_{2N+1}\rightarrow\infty$ after taking the limit in (\ref{Elim}) does not alter that function.  After taking this second limit, we find that
\be\label{EFlim} \text{(\ref{Elim})}\quad=\lim_{x_{2N+1}\rightarrow \infty}\lim_{x_{2N+2}\rightarrow x_{2N+1}}(x_{2N+2}-x_{2N+1})^{6/\kappa-1}\mathcal{E}_{\vartheta'}(\kappa\,|\,x_1,x_2,\ldots,x_{2N+2})=\mathcal{F}_\vartheta(\kappa\,|\,x_1,x_2,\ldots,x_{2N})\,\,(\ref{Fexplicit}).\ee
That is, we recover $\mathcal{F}_\vartheta$ (\ref{Fexplicit}) from the limit (\ref{Elim}) of $\mathcal{E}_{\vartheta'}\in\mathcal{S}_{N+1}$ after sending $x_{2N+1}\rightarrow\infty$ in the latter.  Furthermore, lemma \red{5} of \cite{florkleb} says that the limit (\ref{Elim}) is an element of $\mathcal{S}_N$.  In light of this fact, it follows from (\ref{EFlim}) that the function $\mathcal{F}_\vartheta$ defined by the second formula (\ref{Fexplicit}) is an element of $\mathcal{S}_N$ if $\kappa\in(0,8)$.

\item Next, we prove that (\ref{Fother}) equals (\ref{Fexplicit}) if $\kappa\in(0,8)$ by showing that their images are identical under the isomorphism (\ref{LinearMap}).  We compute $v(\mathcal{F}_\vartheta)$, with $\mathcal{F}_\vartheta$ given by the first formula (\ref{Fother}), in the proof of lemma \red{6} in \cite{florkleb3}, finding (figure \ref{innerproduct})
\be\label{vF}v(\mathcal{F}_\vartheta)_\varsigma=n(\kappa)^{l_{\varsigma,\vartheta}},\quad\text{for all $\varsigma,\vartheta\in\{1,2,\ldots,C_N\}$}.\ee
By exactly repeating the analysis of that proof, we prove that (\ref{vF}) is also true if $\mathcal{F}_\vartheta$ is given by the second formula (\ref{Fother}).  (There are a few small changes that do not significantly affect the analysis.  Case \red{3} of the proof does not arise.  Also, the power $12/\kappa-2$, is present in the integrand of the first formula (\ref{Fother}) but absent from the integrand of the second formula (\ref{Fexplicit}).  And finally, infinity is not a branch point of the integrand of the first formula (\ref{Fother}) but is of the integrand of the second formula (\ref{Fexplicit}).)  Hence,  because $v$ is an isomorphism,  it follows that (\ref{Fother}) and (\ref{Fexplicit}) are different formulas for the same function $\mathcal{F}_\vartheta(\kappa)$ if $\kappa\in(0,8)$.  Finally, analytic continuation extends this equality to all $\kappa>0$.
\end{enumerate}

\subsection{Algebraic formulas for $\mathcal{F}_\vartheta(\kappa)$ at some special values of $\kappa$}\label{SecondSubsection}

If the SLE$_\kappa$ parameter $\kappa$ is one of the values related to the $Q$-color critical random cluster model with $Q\in\{1,2,4\}$ (\ref{dense}) or the $Q$-state Potts model with $Q\in\{2,4\}$ (\ref{dilute}), then $\mathcal{F}_\vartheta(\kappa)$ (\ref{Fexplicit}) has an algebraic (i.e., without contour integrals) formula for all $\vartheta\in\{1,2,\ldots,C_N\}$.  In the $Q=4$ case, we may obtain such a formula by explicit integration, but in the $Q\in\{1,2\}$ cases, it is not evident that we can explicitly perform the integration.  Instead, to prove that $\mathcal{F}_\vartheta(\kappa)$ has an algebraic formula, we observe that both $\mathcal{F}_\vartheta(\kappa)$ and the formula satisfy the same system of  PDEs (\ref{nullstate}, \ref{wardid}) and the same boundary conditions.  From these facts and theorem \red{8} of \cite{florkleb3}, we infer that $\mathcal{F}_\vartheta(\kappa)$ equals the algebraic formula.   Some of these algebraic formulas already appear in the literature ($Q=2$ \cite{guim,kype}), although not in direct connection with $\mathcal{F}_\vartheta(\kappa)$, and others are, to our knowledge, completely new ($Q\in\{1,4\}$).     The $Q=3$ case is ostensibly absent, and we do not know if algebraic formulas for $\mathcal{F}_\vartheta(\kappa)$ (\ref{Fexplicit}) exist in this case.  At least for $N=2$, the hypergeometric expressions (\ref{Iinvar}, \ref{Ivar}) for $\mathcal{F}_1(\kappa)$ and $\mathcal{F}_2(\kappa)$ seem to not have such formulas if $Q=3$ (so $\kappa=24/5$ (\ref{dense}) or $10/3$ (\ref{dilute})).  For $Q \in \{1,2\}$, our results in this subsection, while rigorous, are obtained in an indirect way, using previous results in \cite{florkleb,florkleb2,florkleb3,florkleb4} and other results in the literature \cite{guim,kype}. Thus we obtain algebraic formulas for some rather complicated contour integrals  based on a full analysis of the solutions of the system of PDEs (\ref{nullstate}, \ref{wardid}) that  $\mathcal{F}_\vartheta(\kappa)$ satisfies.
\begin{enumerate}[wide, labelwidth=!, labelindent=0pt]
\item\textbf{$Q=1$ critical random cluster model} (i.e., critical percolation) ($\kappa=6$):  Here, we have $n(\kappa)=1$ (\ref{fugacity}).  As such, (\ref{Fsum}) gives 
\begin{multline}\label{Fsum2} \text{$[\mathscr{L}_\varsigma]\mathcal{F}_\vartheta(\kappa=6)=1$ for all $\varsigma,\vartheta\in\{1,2,\ldots,C_N\}$}\\
\stackrel{(\ref{dual})}{\Longrightarrow}\qquad \text{$\mathcal{F}_\vartheta(\kappa=6)=[\Pi_1+\Pi_2+\dotsm+\Pi_{C_N}](\kappa=6)$ for all $\vartheta\in\{1,2,\ldots,C_N\}$}.\end{multline}
Moreover, it is trivial to see that $1\in\mathcal{S}_N$ (i.e., satisfies (\ref{nullstate}, \ref{wardid}, \ref{powerlaw})) if $\kappa=6$ (so $\theta_1=0$ (\ref{theta1})) and $[\mathscr{L}_\varsigma]1=1$ for all $\varsigma\in\{1,2,\ldots,C_N\},$ so $1=[\Pi_1+\Pi_2+\ldots+\Pi_{C_N}](\kappa=6)$.  Thus, (\ref{Fsum2}) implies that 
\begin{align}\label{Fkappa6} \mathcal{F}_\vartheta(\kappa=6)&=1\\
&=[\Pi_1+\Pi_2+\dotsm+\Pi_{C_N}](\kappa=6),\quad\text{for all $\vartheta\in\{1,2,\ldots,C_N\}$},\end{align}
which is our algebraic result for the $Q=1$ critical random cluster model.

We may use (\ref{Fkappa6}) to indirectly evaluate the Coulomb gas integral in (\ref{Fexplicit}) with $\kappa=6$.  Indeed, after inserting the formula (\ref{Fexplicit}) for $\mathcal{F}_\vartheta(\kappa=6)$ into (\ref{Fkappa6}) (and replacing Pochhammer contours with simple contours as in figure \ref{PochhammerContour}), we encounter  this  algebraic formula for the definite integral in (\ref{Fexplicit}) (item \ref{step3d} of section \ref{xingsummary} defines $\mathcal{N}[\,\,\ldots\,\,]$):
\begin{multline}\label{Intkappa=6}\int_{\Gamma_N}{\rm d}u_N\,\,\dotsm\int_{\Gamma_2}{\rm d}u_2\,\,\int_{\Gamma_1}{\rm d}u_1\,\,\mathcal{N}\Bigg[\Bigg(\prod_{l=1}^{2N}\prod_{m=1}^N(x_l-u_m)^{-2/3}\Bigg)\Bigg(\prod_{p<q}^N(u_p-u_q)^{4/3}\Bigg)\Bigg]\\
=\frac{\Gamma(1/3)^{2N}}{\Gamma(2/3)^N}\Bigg(\prod_{j<k}^{2N}(x_k-x_j)^{-1/3}\Bigg).\end{multline}
In (\ref{Intkappa=6}), the integration contours are nonintersecting simple contours in the upper half-plane that join their endpoints, all of them among $x_1,$ $x_2,\ldots,x_{2N}$, pairwise in the $\vartheta$th connectivity for $\vartheta\in\{1,2,\ldots,C_N\}$ (item \ref{step2} of section \ref{xingsummary}).  Thus, using results from \cite{florkleb,florkleb2,florkleb3,florkleb4}, we have indirectly evaluated the Coulomb gas integral in (\ref{Intkappa=6}).

\item\textbf{$Q=2$ critical Potts model} (i.e., critical Ising model) ($\kappa=3$): Here, we again have $n(\kappa)=1$ (\ref{fugacity}).  As such, (\ref{Fsum}) gives (\ref{Fsum2}) once again, but with $\kappa=3$ instead.  Now, we have
\begin{align}\label{Fkappa3}\mathcal{F}_\vartheta(\kappa=3\,|\,x_1,x_2,\ldots,x_{2N})&=\sum_\lambda\,\,\text{sgn}\,\lambda\,\bigg(\prod_{\{i,j\}\in\lambda}\frac{1}{x_i-x_j}\bigg)\\
\label{Fkappa32}&=[\Pi_1+\Pi_2+\dotsm+\Pi_{C_N}](\kappa=3\,|\,,x_1,x_2,\ldots,x_{2N}),\quad\text{for all $\vartheta\in\{1,2,\ldots,C_N\},$}\end{align}
where the sum in (\ref{Fkappa3}) is over all mutually exclusive pairings $\lambda$ of indices in the set $\{1,2,\ldots,2N\}$, and where $\text{sgn}\,\lambda$ is the sign of the product $\prod(i-k)(i-l)(j-k)(j-l)$ over pairs of distinct $\{i,j\},\{k,l\}\in\lambda$.  Indeed, proposition \red{4.6} of \cite{kype} states that the right side of (\ref{Fkappa3}) satisfies (\ref{nullstate}, \ref{wardid}) if $\kappa=3$, so because it obviously satisfies (\ref{powerlaw}) too, it is an element of $\mathcal{S}_N$.  Furthermore, the cascade property in proposition \red{4.6} of \cite{kype} implies that the action of $[\mathscr{L}_\varsigma]$ on the right side of (\ref{Fkappa3}) equals one for all $\varsigma\in\{1,2,\ldots,C_N\}$.  Thus, the right sides of (\ref{Fkappa3}, \ref{Fkappa32}) are equal, so because (\ref{Fsum2}) is true for $\kappa=3$, both sides of (\ref{Fkappa3}) are equal.  This is our algebraic result for the $Q=2$ critical Potts model.

We may use (\ref{Fkappa3}) to indirectly evaluate the Coulomb gas integral in (\ref{Fexplicit}) with $\kappa=3$.  Indeed, after inserting the formula (\ref{Fexplicit}) for $\mathcal{F}_\vartheta(\kappa=3)$ into (\ref{Fkappa3}), we find (item \ref{step3d} of section \ref{xingsummary} defines $\mathcal{N}[\,\,\ldots\,\,]$):
\begin{multline}\label{Intkappa=3}\oint_{\Gamma_N}{\rm d}u_N\,\,\dotsm\oint_{\Gamma_2}{\rm d}u_2\,\,\oint_{\Gamma_1}{\rm d}u_1\,\,\mathcal{N}\Bigg[\Bigg(\prod_{l=1}^{2N}\prod_{m=1}^N(x_l-u_m)^{-4/3}\Bigg)\Bigg(\prod_{p<q}^N(u_p-u_q)^{8/3}\Bigg)\Bigg]\\
=\left[\frac{3\Gamma(-1/3)^2}{\Gamma(-2/3)}\right]^N\Bigg(\prod_{j<k}^{2N}(x_k-x_j)^{-2/3}\Bigg)\sum_\lambda\,\,\text{sgn}\,\lambda\,\bigg(\prod_{\{i,j\}\in\lambda}\frac{1}{x_i-x_j}\bigg).\end{multline}
In (\ref{Intkappa=3}), the integration contours are nonintersecting Pochhammer contours in the upper half-plane, with endpoints among $x_1,$ $x_2,\ldots,x_{2N}$, and joining these points  pairwise in the $\vartheta$th connectivity for  any  $\vartheta\in\{1,2,\ldots,C_N\}$.  Thus, using results from \cite{florkleb,florkleb2,florkleb3,florkleb4}, we have indirectly found an algebraic formula for the Coulomb gas integral in (\ref{Intkappa=3}).
\item\textbf{$Q=2$ critical random cluster model} ($\kappa=16/3$): Here, we have $n(\kappa)=\sqrt{2}$ (\ref{fugacity}).  Moreover, we have that for all $\vartheta\in\{1,2,\ldots,C_N\}$ and all $i\in\{1,2,\ldots,2N-1\}$ (see the proof of lemma \red{6} in \cite{florkleb3}),
\begin{multline}\label{AsyF}\mathcal{F}_\vartheta(\kappa=16/3\,|\,x_1,x_2,\ldots,x_{2N})\underset{x_{i+1}\rightarrow x_i}{\sim} (x_{i+1}-x_i)^{-1/8}\\
\times\begin{cases}\sqrt{2}\mathcal{G}_{\varrho\,\,}(\kappa=16/3\,|\,x_1,x_2,\ldots,x_{i-1},x_{i+2},\ldots,x_{2N}),& \text{$x_i$ and $x_{i+1}$ are joined in the $\vartheta$th connectivity}, \\ \hphantom{\sqrt{2}}\mathcal{G}_\varpi(\kappa=16/3\,|\,x_1,x_2,\ldots,x_{i-1},x_{i+2},\ldots,x_{2N}), & \text{$x_i$ and $x_{i+1}$ are not joined in the $\vartheta$th connectivity},\end{cases}\end{multline}
where $\mathcal{G}_\varrho,\mathcal{G}_\varpi\in\mathcal{B}_{N-1}$, and where we form the $\varrho$th (resp.\ $\varpi$th) connectivity from the $\vartheta$th connectivity by dropping from the latter the one arc with its endpoints at $x_i$ and $x_{i+1}$ (resp.\ detaching in the latter the two arcs with their endpoints at $x_i$ or $x_{i+1}$ from those points and joining their dangling ends together to form one arc).  (Thus, $\varrho,\varpi\in\{1,2,\ldots,C_{N-1}\}$.)  Now, we have
\begin{multline}\label{Fkappa=16/3}\mathcal{F}_\vartheta(\kappa=16/3\,|\,x_1,x_2,\ldots,x_{2N})\,\,=\prod_{j=1\,\text{odd}}^{2N-1}(y_{j+1}-y_j)^{-1/8}\\ 
\times\,\left\{\sum_{\mu_1=\pm1}\sum_{\mu_3=\pm1}\dotsm\sum_{\mu_{2N-1}=\pm1}\prod_{j<k\,\text{odd}}^{2N-1}\Bigg(\frac{(y_k-y_j)(y_{k+1}-y_{j+1})}{(y_{k+1}-y_j)(y_k-y_{j+1})}\Bigg)^{\mu_j\mu_k/4}\right\}^{1/2},\end{multline}
where $y_j,y_{j+1}\in\{x_1,x_2,\ldots,x_{2N}\}$ with $y_j<y_{j+1}$ are the endpoints of the $j$th arc in the $\vartheta$th connectivity.  Indeed, \cite{guim} says that for $\vartheta=C_N$ (so $y_j=x_j$ for all $j\in\{1,2,\ldots,2N\}$), the right side of (\ref{Fkappa=16/3}) satisfies (\ref{nullstate}, \ref{wardid}) with $\kappa=16/3$.  By appropriately permuting the points $x_j$, which does not change the system (\ref{nullstate}, \ref{wardid}), we generate the right side of (\ref{Fkappa=16/3}) for all $\vartheta<C_N$ and see that it satisfies the system (\ref{nullstate}, \ref{wardid}) too.  Obeying (\ref{powerlaw}) as well, each solution (\ref{Fkappa=16/3}) is an element of $\mathcal{S}_N$.  So to prove the equality in (\ref{Fkappa=16/3}), it suffices to show that the left and right sides have identical images under the action of $[\mathscr{L}_\varsigma]$ for each $\varsigma\in\{1,2,\ldots,C_N\}$.  This immediately follows from the fact that the right side of (\ref{Fkappa=16/3}) has the asymptotic behavior (\ref{AsyF}), the same as of $\mathcal{F}_\vartheta(\kappa=16/3)$ (\ref{Fexplicit}), as $x_{i+1}\rightarrow x_i$ for all $i\in\{1,2,\ldots,2N-1\}$.

In addition to $\mathcal{F}_{C_N}$, that is (\ref{Fkappa=16/3}) with $y_j=x_j$ and $\vartheta=C_N$, (\red{6}) of \cite{guim} gives $2^{N-1}-1$ more solutions to the system (\ref{nullstate}, \ref{wardid}) with $\kappa=16/3$, and their formulas are identical to (\ref{Fkappa=16/3}) with $y_j=x_j$, except that a factor, equaling one or minus one and depending on $\mu_1,$ $\mu_3,\ldots,\mu_{2N-1}$, multiplies the product in (\ref{Fkappa=16/3}).  We let $\mathcal{D}_N$ denote the total collection of $2^{N-1}$ solutions in (\red{6}) of \cite{guim}.  Because the rank of $\mathcal{B}_N$ (\ref{TLBN}, \ref{Fkappa=16/3}) with $\kappa=16/3$ (\ref{dense2}) equals the  cardinality of $\mathcal{D}_N$, we anticipate that these two sets of solutions, both containing $\mathcal{F}_{C_N}$, span the same subspace of $\mathcal{S}_N$.  Indeed, this is true.  To see why, we note that the  solutions of $\mathcal{D}_N$ are  conformal blocks \cite{fms}.  Thus,  $\text{span}\,\mathcal{D}_N$  is closed under the braid group action that permutes the points $x_j$.  Now for any $\vartheta<C_N$, there exists a braiding that sends $\mathcal{F}_{C_N} \in\mathcal{D}_N$ to $\mathcal{F}_\vartheta \in\mathcal{B}_N$ \cite{fp}.  Thus,  $\mathcal{B}_N\subset\text{span}\,\mathcal{D}_N$.  But because $\text{rank}\,\mathcal{B}_N=|\mathcal{D}_N|$ (\ref{dense2}), we have $\text{span}\,\mathcal{B}_N=\text{span}\,\mathcal{D}_N$.  For $N\geq3$, this span is a proper subspace of $\mathcal{S}_N$, and we do not know if any solutions in $\mathcal{S}_N\setminus\text{span}\,\mathcal{B}_N$ have algebraic formulas too.  Eqn.\  (\ref{Fkappa=16/3}) is our algebraic result for the $Q=2$ critical random cluster model.

We may use (\ref{Fkappa=16/3}) to indirectly evaluate the Coulomb gas integral in (\ref{Fexplicit}) with $\kappa=16/3$.  Indeed, after inserting the formula (\ref{Fexplicit}) for $\mathcal{F}_\vartheta$ into (\ref{Fkappa=16/3}) (and replacing Pochhammer contours with simple contours as in figure \ref{PochhammerContour}), we obtain this algebraic formula for the definite integral in (\ref{Fexplicit}) (item \ref{step3d} of section \ref{xingsummary} defines $\mathcal{N}[\,\,\ldots\,\,]$):
\begin{multline}\label{Intkappa=16/3}\int_{\Gamma_N}{\rm d}u_N\,\,\dotsm\int_{\Gamma_2}{\rm d}u_2\,\,\int_{\Gamma_1}{\rm d}u_1\,\,\mathcal{N}\Bigg[\Bigg(\prod_{l=1}^{2N}\prod_{m=1}^N(x_l-u_m)^{-3/4}\Bigg)\Bigg(\prod_{p<q}^N(u_p-u_q)^{3/2}\Bigg)\Bigg]=\Bigg(\prod_{j<k}^{2N}(x_k-x_j)^{3/8}\Bigg)\\
\times\,\frac{\Gamma(1/4)^{2N}}{[2\Gamma(1/2)]^N}\prod_{j=1\,\text{odd}}^{2N-1}(y_{j+1}-y_j)^{-1/8}\left\{\sum_{\mu_1=\pm1}\sum_{\mu_3=\pm1}\dotsm\sum_{\mu_{2N-1}=\pm1}\prod_{j<k\,\text{odd}}^{2N-1}\Bigg(\frac{(y_k-y_j)(y_{k+1}-y_{j+1})}{(y_{k+1}-y_j)(y_k-y_{j+1})}\Bigg)^{\mu_j\mu_k/4}\right\}^{1/2}.\end{multline}
In (\ref{Intkappa=16/3}), the integration contours are nonintersecting simple contours in the upper half-plane, with endpoints among $x_1,$ $x_2,\ldots,x_{2N}$ and joining  these points  pairwise in the $\vartheta$th  connectivity, and we define $y_j$ under (\ref{Fkappa=16/3}).  Thus, using results from \cite{florkleb,florkleb2,florkleb3,florkleb4}, we have indirectly found an algebraic expression for the Coulomb gas integral in (\ref{Intkappa=16/3}).

\item\textbf{$Q=4$ critical random cluster model and critical Potts model} ($\kappa=4$, but more generally, we consider all $\kappa$ such that $4/\kappa\in\mathbb{Z}^+$):  If $\kappa=4$ then  $n(\kappa)=2$ (\ref{fugacity}), and  each $x_j$ is a simple pole of the integrand of $\mathcal{F}_\vartheta(\kappa=4)$ (\ref{Fexplicit}).  More generally, each $x_j$ is an order-$r$ pole of the integrand of $\mathcal{F}_\vartheta(\kappa)$ (\ref{Fexplicit}) if and only if $\kappa=4/r$ for $r\in\mathbb{Z}^+$.  In this case, we decompose each Pochhammer contour into a collection of loops that surrounds the contour's endpoints (figure \ref{BreakDown}, (\red{34}) of \cite{florkleb3}), and we perform the integration around these loops via the Cauchy integral formula.  We find
\begin{align}\begin{aligned}\label{Fkappa4} \mathcal{F}_\vartheta(\kappa=4/r\,|\,&x_1,x_2,\ldots,x_{2N})=(-1)^{N(r-1)}\left(\frac{(r-1)!}{(2r-2)!}\right)^N\,\prod_{i<j}^{2N}(x_j-x_i)^{r/2}\sum_{\mu_1=1}^2\sum_{\mu_2=3}^4\dotsm\\ 
&\dotsm\sum_{\mu_N=2N-1}^{2N}\omega(\mu_1,\mu_2,\ldots,\mu_N)\,\partial_{u_1}^{r-1}\partial_{u_2}^{r-1}\dotsm\partial_{u_N}^{r-1}\Bigg[\prod_{m=1}^N\prod_{l\neq\mu_m}^{2N}|x_l-u_m|^{-r}\prod_{p<q}^N(u_p-u_q)^{2r}\Bigg]_{u_m=x_{\mu_m}},\end{aligned}\\
\text{where}\quad\omega(\mu_1,\mu_2,\ldots,\mu_N):=\begin{cases}(-1)^{\mu_1+\mu_2+\dotsm+\mu_N+N}, & \text{$r$ odd}, \\ 1, & \text{$r$ even}.\end{cases}\end{align}
(In \cite{florkleb4}, we found an equivalent formula (\red{35}) by integrating the alternative formula (\ref{Fother}) for $\mathcal{F}_\vartheta$.)  Specializing to $\kappa=4$ ($r=1$), this becomes 
\begin{multline} \mathcal{F}_\vartheta(\kappa=4\,|\,x_1,x_2,\ldots,x_{2N})=\prod_{i<j}^{2N}(x_j-x_i)^{1/2}\sum_{\mu_1=1}^2\sum_{\mu_2=3}^4\dotsm\\ 
\dotsm\sum_{\mu_N=2N-1}^{2N}(-1)^{\mu_1+\mu_2+\dotsm+\mu_N+N}\Bigg[\prod_{m=1}^N\prod_{l\neq\mu_m}^{2N}|x_l-x_{\mu_m}|^{-1}\prod_{p<q}^N(x_{\mu_p}-x_{\mu_q})^2\Bigg].\end{multline}
This is our algebraic result for the $Q=4$ critical Potts model and the $Q=4$ critical random cluster model.
\end{enumerate}

\end{document}